\documentclass[12pt,a4paper]{article}
\usepackage{jheppub}
\pdfoutput=1
\usepackage{slashed}
\usepackage{amsmath}
\usepackage{graphicx}
\usepackage{caption}
\usepackage{subcaption} 
\usepackage{epstopdf}
\usepackage{IEEEtrantools}

\newcommand{\beq}{\begin{equation}}
\newcommand{\eeq}{\end{equation}}
\newcommand{\nn}{\nonumber}
\def\[{\left[}
\def\]{\right]}
\def\({\left(}
\def\){\right)}
\def\cC{{\cal C}}

\def\cS{{\cal S}}

\newcommand{\be}{\beta}

\def\sgn {\text{sgn}}

\def\Tr{\mathrm{Tr}}

\def\nn{\nonumber\\}

\def\sgn{\text{sgn}}

\def\Tr{\mathrm{Tr}}

\def\nn{\nonumber\\}

\def \be {\begin{equation}}
\def \ee {\end{equation}}
\def \bea {\begin{eqnarray}}
\def \eea {\end{eqnarray}}
\def \beal#1 {\begin{align}#1\end{align}}
\def\kappas{\kappa^{{\cal N}=2}}

\everymath{\displaystyle}

\preprint{}

\title{Flows, Fixed Points and Duality in Chern-Simons-matter theories}
\author[a]{Ofer Aharony,}
\author[b]{Sachin Jain,}
\author[c]{and Shiraz Minwalla}
\affiliation[a]{Department of Particle Physics and Astrophysics, Weizmann Institute of Science, Rehovot 7610001, Israel}
\affiliation[b]{Indian Institute of Science Education and Research, Homi Bhabha Rd, Pashan, Pune 411008, India}
\affiliation[c]{Department of Theoretical Physics, Tata Institute of Fundamental Research, Homi Bhabha road, Mumbai 400005, India}

\abstract{It has been conjectured that 3d fermions minimally coupled 
to Chern-Simons gauge fields are dual to  3d critical scalars, also 
minimally coupled  to Chern-Simons gauge fields. The large $N$ arguments for this duality can formally be used to  
show that Chern-Simons-gauged {\it critical} (Gross-Neveu) fermions are also dual to gauged `{\it regular}' scalars 
at every order in a $1/N$ expansion, provided both theories are well-defined (when one fine-tunes the two relevant parameters of each of these theories to zero). In the strict 
large $N$ limit these `quasi-bosonic' theories appear as fixed lines parameterized by 
$x_6$, the coefficient of a sextic term in the potential. While $x_6$ is an exactly marginal 
deformation at leading order in large $N$, it develops a non-trivial $\beta$ function at first subleading 
order in $1/N$. We demonstrate that the beta function is a cubic polynomial in $x_6$ at this order in $1/N$, 
and compute the coefficients of the cubic and quadratic terms as a function of the 't Hooft coupling. 
We conjecture that flows governed by this leading large $N$ beta function have three fixed points for 
$x_6$ at every non-zero value of the 't Hooft coupling, implying the existence of three distinct regular bosonic 
and three distinct dual critical fermionic conformal fixed points, at every value of the 't Hooft coupling. We analyze the phase structure of these fixed point theories at zero temperature.
We also construct dual pairs of large $N$ fine-tuned renormalization group
flows from supersymmetric ${\cal N}=2$ Chern-Simons-matter theories, such that one of the flows 
ends up in the IR at a regular boson theory while its dual partner flows to a critical fermion theory.
This construction suggests that the duality between these  theories persists at finite $N$, at least when $N$ is large.}

\begin{document}

\maketitle

\section{Introduction and summary}

This paper is devoted to the study of a web of five closely related classes of 
Chern-Simons-matter theories, with gauge groups $SO(N)$, $SU(N)$ or $U(N)$ and 
matter in the fundamental representation\footnote{One can also consider $USp(2N)$ theories, but in the large $N$ limit in which we work, they are identical to $SO(2N)$ theories.}.
The five classes of three dimensional quantum field theories that we study include several 
conformal field theories. More general theories are obtained by deforming 
these fixed points with relevant operators. Distinct theories in this web 
are related by renormalization group flows and by quantum-corrected 
Legendre transformations at large $N$. Less trivially, several 
distinct theories that we study are also conjectured to be related to  
each other by strong-weak coupling dualities that exchange bosons and fermions. 

There are many reasons to be interested in Chern-Simons-matter theories of the type
studied in this paper. Despite the fact that they are effectively solvable in the large $N$ limit,
these theories are dynamically very rich and display properties that are unusual
for quantum field theories. As already mentioned, these theories enjoy invariance under 
strong-weak coupling bosonization dualities even in the absence of supersymmetry. 
Moreover, both S matrices and thermal partition functions of these theories have very unusual 
properties, including modified transformations under crossing symmetry of the S matrix
(see below for references).\footnote{Several of these properties appear to have their origin in the fact 
that the coupling of fundamental excitations to Chern-Simons gauge fields  makes them 
effectively non-Abelian anyons. Conversely, the lessons from the study of the theories 
described above may well apply more generally to all systems with effectively anyonic 
excitations.} The effective solvability of these models at large $N$ permits the detailed 
study of these interesting phenomena. It is possible that the lessons learned from this 
study will apply more generally to all Chern-Simons-matter theories, even 
away from the large $N$ limit. 

The next set of motivations for the study of these theories comes from the 
AdS/CFT correspondence. The theories we study 
have been suggested \cite{Klebanov:2002ja, Sezgin:2002rt, Giombi:2009wh,
Giombi:2011kc,  Chang:2012kt} to have another dual description at large 
$N$, in terms of theories of classical high-spin gravity on $AdS_4$ (see \cite{Giombi:2016ejx} and references therein). 
While this duality is precise only in the strict large $N$ limit, the field theories 
are well-defined even at finite $N$, and provide the only known quantization of 
the bulk dual higher spin theories. 

Next, there are many known results for highly 
supersymmetric cousins of these theories, including conjectured field theory dualities and also conjectured dualities of some of 
these theories at strong coupling to bulk supergravity, string theory or M theory \cite{Aharony:2008ug}.
The combination of the large $N$ techniques used in this paper with the exact results 
from supersymmetry could lead to unanticipated synergies. 

Yet another set of motivations for the study of the theories considered in this paper comes 
from condensed matter physics. Finite $N$ versions of the theories studied herein 
have already found applications in condensed matter physics in the 
study of the quantum Hall effect (see below for some references). It does not seem implausible that more such 
applications will be found over the coming years. 

This self-contained introductory section is divided into two subsections. In 
subsection \ref{tp} we introduce the  Chern-Simons-matter theories studied in this paper\footnote{For simplicity of presentation, in the introduction we restrict 
attention to $U(N)$ theories with equal levels for the $SU(N)$ and $U(1)$ 
factors, and with a single scalar or fermion field in the fundamental 
representation. More general theories are described in the main text. }
and present a brief summary of known results about these theories. In 
subsection \ref{prp} we summarize the main results of this paper.

\subsection{The theories we study} \label{tp}
\subsubsection{Listing of theories}

The five classes of theories of interest to this paper are the following $U(N)$ gauge theories :

\begin{itemize}

\item The ${\cal N}=2$ supersymmetric (S) Chern-Simons-matter theory with a single chiral multiplet in the fundamental 
representation:
\beal{
S_S(\phi, \psi)  &= \int d^3 x  \biggl[i \varepsilon^{\mu\nu\rho} \frac{\kappa}{4 \pi}
\Tr( A_\mu\partial_\nu A_\rho -\frac{2 i}{3}  A_\mu A_\nu A_\rho)
+ D_\mu \bar \phi D^\mu\phi + \bar\psi \gamma^\mu D_\mu \psi \nn
& \qquad\qquad\qquad + \frac{4 \pi^2}{\kappa^2} (\bar\phi\phi)^3
+ \frac{4 \pi}{\kappa} (\bar\psi \psi) (\bar\phi\phi)
+ \frac{2 \pi}{\kappa} (\bar\psi\phi)( \bar\phi \psi)
 \biggl].
\label{saction}
}

\item The critical bosonic (CB) theory (the critical $U(N)$ model coupled to a Chern-Simons (CS) gauge field):
\be
S_{CB}(\phi, \sigma_B)  = \int d^3 x  \biggl[i \varepsilon^{\mu\nu\rho}\frac{\kappa_B}{4\pi}
\Tr( A_\mu\partial_\nu A_\rho -{2 i\over3}  A_\mu A_\nu A_\rho)
 + D_\mu \bar \phi D^\mu\phi+ \sigma_B\bar\phi \phi \biggl].
\label{cst1}
\ee

\item The regular fermion (RF) theory (fermions coupled to a CS gauge field):
\be
 S_{RF}(\psi) =\int d^3 x \bigg[ i \varepsilon^{\mu\nu\rho} {\kappa_F \over 4 \pi}
\Tr( A_\mu\partial_\nu A_\rho -{2 i\over3}  A_\mu A_\nu A_\rho)
+  \bar{\psi} \gamma_\mu D^{\mu} \psi \bigg].
\label{rft1}
\ee

\item The regular boson (RB) theory (scalars coupled to a CS gauge field):
\be
S_{RB}(\phi)=  \int d^3 x  \biggl[i \varepsilon^{\mu\nu\rho}{\kappa_B\over 4\pi}
\Tr( A_\mu\partial_\nu A_\rho -{2 i\over3}  A_\mu A_\nu A_\rho)
 + D_\mu \bar \phi D^\mu\phi  + \frac{(2\pi)^2}{\kappa_B^2} \left(x^B_6+1\right)( \bar \phi \phi)^3 \biggl].
\label{rbt1}
\ee

\item The critical fermion (CF) theory (the Gross-Neveu model coupled to a CS gauge field):
\be
S_{CF}(\psi,\sigma_F) =\int d^3 x \bigg[ i \varepsilon^{\mu\nu\rho} {\kappa_F \over 4 \pi}
\Tr( A_\mu\partial_\nu A_\rho -{2 i\over3}  A_\mu A_\nu A_\rho)
+  \bar{\psi} \gamma_\mu D^{\mu} \psi  - \frac{4\pi}{\kappa_F}\zeta \bar{\psi} \psi + \frac{(2\pi)^2}{\kappa_F^2}x^F_6 \zeta^3 \bigg].
\label{cft1}
\ee

\end{itemize}
The RF and CB theories were together called `quasi-fermionic' theories in  \cite{Maldacena:2011jn, Maldacena:2012sf}
while the RB and CF theories were referred to as `quasi-bosonic' theories. We will employ this 
nomenclature in the rest of this paper.

In the rest of this subsection we briefly review the definition and 
key properties of the five classes of theories listed above.

\subsubsection{Supersymmetric theories S}

The supersymmetric (SUSY) theories S are quite well understood. The Lagrangian \eqref{saction} (in the dimensional reduction regulation scheme) is known to define a superconformal fixed point \cite{Gaiotto:2007qi} with four supercharges. At least at weak coupling, this fixed 
point has three relevant and no marginal deformations.\footnote{In addition to  the three relevant deformations, the Lagrangian \eqref{saction} has 4 classically 
marginal deformations. It has been shown by explicit computation 
(see section \ref{susysec} for more details) that the anomalous dimensions of all these operators are positive at weak coupling. 
It follows that the only relevant operators about this fixed point  at weak coupling are its three  classically relevant deformations. It is possible 
that this result changes at strong coupling (though the strong-weak coupling duality of this theory constrains possible modifications). In any case at large $N$ and in the 't Hooft limit all anomalous dimensions are of order $\frac{1}{N}$ so the theory must have at least  three strongly relevant deformations at all values of the 't Hooft coupling. This is the 
result we use in this paper.}

 These superconformal fixed 
points  are 
conjectured \cite{Benini:2011mf, Park:2013wta, Aharony:2013dha} to enjoy invariance under a strong-weak coupling self-duality 
similar to the Giveon-Kutasov duality \cite{Giveon:2008zn} (see appendix \ref{mapping} for some details). 
This duality reshuffles bosons and fermions within supermultiplets 
\cite{Jain:2013gza, Inbasekar:2015tsa, Gur-Ari:2015pca}, and so can be thought of as
a bosonization duality. There is considerable evidence that this duality holds 
for all values of $N$ and $\kappa$ for which the theory has a supersymmetric vacuum. 
The evidence for this duality includes several calculational checks using the 
method of supersymmetric localization \cite{Kapustin:2010mh, Willett:2011gp, Kapustin:2011gh}, as well as the relation of this duality 
by flows to many other supersymmetric dualities (see, for instance, \cite{Intriligator:2013lca, Aharony:2013dha})\footnote{The superconformal theory described in this section may be rigorously defined by adding a 
supersymmetric Yang-Mills term to the action. The resulting theory is free at high energies, but 
reduces to \eqref{saction} at low energies (at which point the Yang-Mills coupling 
effectively diverges so the Yang-Mills term in the action is negligible). In other words,
theory S is the end point of a ${\cal N}=2$ SUSY renormalization group (RG) flow that starts in the asymptotically free ${\cal N}=2$ Yang-Mills Chern-Simons 
matter theory. In order to reach the theory S in the IR, we need to perform a one parameter tuning on the 
(one parameter) space of ${\cal N}=2$ RG flows. This tuning sets the coefficient of the ${\cal N}=2$ mass deformation 
about the theory S to zero. }.

\subsubsection{Quasi-Fermionic theories}

The critical boson and regular fermion theories are also relatively 
well understood.  The CB theory 
may be thought of as the $U(N)$ Wilson-Fisher theory gauged by a 
Chern-Simons gauge field. The RF theory is even simpler; 
it may be thought of as a collection of $N$ free complex fermions minimally 
coupled to a $U(N)$ Chern-Simons gauge field. Neither of the Lagrangians 
\eqref{cft1} or \eqref{rft1} has a continuous dimensionless parameter. It follows 
that the path integrals with these Lagrangians define isolated conformal field theories 
(in dimensional reduction regulation schemes) \cite{Aharony:2011jz, Giombi:2011kc} 
whenever they are well-defined.  The resultant conformal theories both have a single relevant deformation - 
a mass term for the bosons or fermions respectively\footnote{As in the case of the SUSY theories, these theories may properly be defined by adding a Yang-Mills 
term to the Lagrangians \eqref{cst1} or \eqref{rft1}, namely as the end-points
of RG flows that originate in the Yang-Mills-Chern-Simons-matter theories (in the case of \eqref{cst1} we need to also add a kinetic term for $\sigma_B$, or a $|\phi|^4$ term). We discuss this possible definition, for these theories and for the quasi-bosonic theories \eqref{rbt1} and \eqref{cft1}, in section \ref{ymcs} below. In order to reach the quasi-fermionic fixed points we should perform a one-parameter tuning of these RG flows in order to reach a theory without a mass gap. The quasi-fermionic theories exist whenever such a fixed point exists, and are unambiguous if this fixed point is unique. 
At finite $N$ is not clear for which values of $N$ and $\kappa$ these assumptions are correct 
(see \cite{Gaiotto:2017tne} and \cite{Gomis:2017ixy} for fascinating conjectures about these issues in closely related 
contexts). In the large $N$ 't~Hooft limit and in the weakly coupled large $\kappa$ limit, 
however, this can always be done.}.

The theories CB and RF -- the so called `quasi-fermionic theories' --
have been conjectured to be related to each other 
via a strong weak coupling duality that exchanges \eqref{cst1} and \eqref{rft1}.\footnote{The duality may also exchange $U(N)$ and $SU(N)$ gauge theories \cite{Radicevic:2015yla,Aharony:2015mjs}, as we review below.} 
The existence of such a duality was first suggested in \cite{Giombi:2011kc}; the first 
concrete conjecture for this duality was made in \cite{Aharony:2012nh} based partly on the 
results of \cite{Maldacena:2011jn, Maldacena:2012sf}. See \cite{Aharony:2015mjs} for a  recent and  relatively precise
statement of the conjectured dualities, and \cite{Seiberg:2016gmd} for an even more precise version.

Below we review a proposed `derivation' of this duality. Here we merely note that there exists 
substantial independent  calculational evidence for the conjectured duality between
theories CB and RF  in the 't~Hooft large $N$ limit $N \to \infty$, $\kappa \to \infty$  with the  
't~Hooft coupling $\lambda = N / \kappa$ held fixed. In this limit these theories all have high-spin symmetries which severely
 constrain their dynamics \cite{Maldacena:2011jn, Maldacena:2012sf}. The evidence for duality includes the matching 
of correlators \cite{Aharony:2012nh, GurAri:2012is, Bedhotiya:2015uga}, S matrices \cite{Jain:2014nza,  Dandekar:2014era, 
Yokoyama:2016sbx} and thermal partition 
functions \cite{Giombi:2011kc, Jain:2012qi, Aharony:2012ns, Jain:2013py,  Takimi:2013zca, Yokoyama:2013pxa, Yokoyama:2012fa}
on the two sides of the duality.  Independent evidence for these dualities at large but finite $N$ includes 
the matching of part of the baryon and monopole spectra between these theories 
\cite{Radicevic:2015yla, Aharony:2015mjs}. There is also some evidence that 
these dualities continue to hold at small values of $N$; in particular 
at specific small values of $N$ and $\kappa$ they may be related to independently conjectured  dualities that
show up in condensed matter systems \cite{Seiberg:2016gmd,Karch:2016sxi}. 
 
The CB and RF theories, and the conjectured duality between them, are reviewed in more detail in section 
\ref{csrf} below.

\subsubsection{A `derivation' of quasi-fermionic dualities from SUSY dualities} \label{cbrf}

In this subsubsection we review the `derivation' \cite{Jain:2013gza,Gur-Ari:2015pca} 
of the duality between the quasi-fermionic theories CB and RF starting from
the assumed self-duality of the supersymmetric theories S.

As mentioned above, the supersymmmetric theory S admits at least 3 relevant deformations at large $N$. 
It follows that there exists an (at least) 2 parameter set of renormalization group (RG) flows originating at this theory. 
The self-duality of theory S identifies pairs of naively different RG flows.

The authors of \cite{Jain:2013gza} identified a one parameter tuning of the large $N$ flows that originate at S, with 
the property that all these flows end up in the IR at the critical boson theory CB. They then demonstrated 
that the duals to these flows all end up at the regular fermion theory RF. At large $N$ the duality 
of the CB and RF theories thus follows as a consequence of the duality of theory S. 

Now consider two flows: the infinite $N$ flow $F$ of \cite{Jain:2013gza, Gur-Ari:2015pca} and 
a large but finite $N$ flow $F'$ that coincides with $F$ in the deep UV. As the $\beta$ functions that govern
$F'$ differ only slightly from the $\beta$ functions that govern $F$, the two flows will deviate only slightly 
from each other over RG flow `times' that are independent of $N$. It follows that the flow $F'$ will approach very near
to the quasi-fermion fixed point, before eventually being repelled away from it along the direction of its relevant operator. 
However any flow that approaches a neighborhood of the IR fixed point can generically be retuned to ensure that it 
actually ends up precisely at the fixed point, provided the number of parameters characterizing the
UV RG flows (in this case 2) is greater than or equal to the number of relevant operators about the IR fixed point 
(in this case 1).\footnote{This argument can fail only if the leading order flows are highly non generic, and one
can check by explicit calculation that this is not the case for the flows studied in this paper.} 
These considerations suggest that the duality between the CB and RF theories continues to hold for 
finite large values of $N$. It may hold also for smaller values.

\subsubsection{Quasi-Bosonic theories}

We now turn to the theories that are the main focus of our paper, namely the 
regular boson and critical fermion theories. 
These theories are harder to define than their quasi-fermionic counterparts. In order to explain why this is the case, 
we first review the situation with the simplest of these theories, the 
regular boson theory at $\kappa_B=\infty$, i.e. $\lambda_B=0$. To keep the action \eqref{rbt1} finite we need to define a new coupling $\lambda_6 = x_6^B \lambda_B^2$ which remains finite as $\lambda_B \to 0$.
In this special case the Lagrangian \eqref{rbt1} is free when the classically marginal parameter $\lambda_6=0$, and the two relevant operators $|\phi|^2$ and $|\phi|^4$ are tuned to zero. It 
is no longer free at $\lambda_6 \neq 0$. A one-loop computation (see e.g. \cite{Pisarski:1982vz}) 
establishes that the deformation about the free theory parameterized by $\lambda_6$, while classically 
marginal, is actually marginally irrelevant at positive values of $\lambda_6$ \footnote{While the theory can formally 
be defined at negative values of $\lambda_6$, it is presumably unstable and so uninteresting at these values.}. 
It follows that the theories at positive $\lambda_6$ cannot be defined 
by renormalization group flows away from the free theory, and have no obvious definition. 

The situation is, however, better in the large $N$ limit. The $\beta$ function of this theory 
(still at $\kappa_B=\infty$) was computed by Pisarski at first non-trivial order at large $N$ but at all values of $\lambda_6$, and turns out 
\cite{Pisarski:1982vz} to take the form 
\begin{equation}\label{betaf}
\beta_{\lambda_6} = \frac{1}{N} \left(a (\lambda_6)^2  - b (\lambda_6)^3 \right) +{\cal O}(1/N^2),
\end{equation}
 where $a$ and $b$ are known  positive constants of order unity. In addition to $\lambda_6=0$, the $\beta$ function 
 \eqref{betaf} vanishes at  $\lambda_6=\frac{a}{b}$. Moreover the operator $|\phi|^6$ is relevant about this new fixed point.
 RG flows that originate in this fixed point define the $\kappa_B=\infty$ large $N$ RB theory at every positive value of 
 $\lambda_6$. As the $\beta$ function of the theory at finite but large values of $N$ deviates only slightly from 
 \eqref{betaf} at values of $\lambda_6$ that are of order unity, it follows that this fixed point continues to exist, 
 continues to be repulsive and continues to define the $\kappa_B=\infty$ RB theory at finite but large $N$ \footnote{On the other hand the $\beta$ function of the theory at large but finite $N$ may deviate 
significantly from \eqref{betaf} at values of $\lambda_6$ that scale like a positive power of $N$. There may 
even exist new fixed points at such values, and for finite values of $N$ it is also possible that the corresponding Yang-Mills-Chern-Simons theories do not flow to any conformal field theories (for any value of their parameters), see section \ref{ymcs}. We will not analyze these issues in this paper. }.

In this paper (see the next subsection for more details) we compute the generalization of the leading large $N$ 
beta function \eqref{betaf} to non-zero $\lambda_B$,
and thereby provide a definition of the RB theories at finite but large values of $N$ and $\kappa_B$ \footnote{The $\beta$ functions we compute  may also be viewed 
as a generalization of the perturbative computation of \cite{Aharony:2011jz},
which was performed up to quadratic order in the couplings $\lambda_6$ and 
$\lambda_B$. }.

We now turn to the second quasi-bosonic theory, namely the critical fermion theory. In this case the theory \eqref{cft1} has no obvious free point\footnote{Nonetheless, considerations of parity can be used to demonstrate that $x_6^F=0$ 
is a fixed point when $\lambda_F=0$.}.
 To the best of our knowledge, the RG flows of $x_6^F$ in this model have not 
previously been studied at any value of $N$ or $\kappa_F$, not even in the effectively ungauged limit $\kappa_F\rightarrow\infty$. 
In this paper 
we compute the $\beta$ function for $x_6^F$ in the large 
$N$ limit, and use the results of our computation to propose 
a definition for the CF theories at all values of $\lambda_F$, 
even at finite (but large) values of $N$. Note that the results of our paper also give a precise definition of the large $N$ ungauged $3d$ Gross-Neveu  model.

\subsection{The principal results of this paper} \label{prp}

The principal new results of this paper are a computation of the 
leading large $N$ beta functions of the two quasi-bosonic theories, an analysis of their large $N$ zero temperature phase structure,
and the construction of dual pairs of RG flows from the ${\cal N}=2$ superconformal 
fixed points to the two different quasi-bosonic theories, supporting the conjectured existence and duality of these theories. In this subsection we review each of these results in turn. 
We discuss in this paper only the special case of a single matter field. 

\subsubsection{Beta functions for quasi-bosonic theories}

In this paper we analyze the leading large 
$N$ $\beta$ functions of $x_6$ in the two quasi-bosonic theories. We show that for all values of the 't Hooft coupling these beta functions are third-order polynomials in $x_6$, and we explicitly compute the coefficient of $x_6^3$. This is enough to understand the qualitative form of the renormalization group flows. The lower coefficients are known perturbatively, but not for all values of the 't Hooft coupling. One way to compute these coefficients more generally is by computing and summing the infinite number of leading non-planar Feynman diagrams in the quasi-bosonic theories that contribute to this beta function. Instead of doing this explicitly, we relate these coefficients to computations in the quasi-fermionic theories.

As we explain in detail below (see around \eqref{pirb}), each quasi-bosonic theory may be 
viewed as a (quantum) Legendre transform of its quasi-fermionic counterpart. The Legendre 
transform is taken with respect to the lowest dimension scalar operator $J_0$ of the corresponding
quasi-fermionic theory. Schematically
\begin{equation}\label{fermo} \begin{split}
Z_{\rm{reg ~boson}} &= \int D \zeta <e^{ \int \zeta J_0 +x_6 \zeta^3}>_{\rm{crit~boson}}\\
Z_{\rm{crit ~fermion}} & = \int D \zeta <e^{ \int \zeta J_0 +x_6 \zeta^3}>_{\rm{reg~fermion}}\\
\end{split}
\end{equation}
(see \eqref{rbt} and \eqref{cft} for non-cartoon versions of these equations).

In order to evaluate the quasi-bosonic partition functions on the left-hand side of \eqref{fermo} 
we first evaluate the expectation values in the integrand of the right-hand side. 
The result is an effective action for $\zeta$ whose $n^{th}$ order vertices are the $n$-point 
Green's functions of $J_0$ in the corresponding quasi-fermionic theory. This effective 
action is less formal than it might first seem, as its leading terms, related to the two, three and four point functions of 
$J_0$, are explicitly known at all values of $\lambda$ in the large $N$ limit, as we review in some 
detail below.

Note that the leading large $N$ 
effective action for $\zeta$ comes from integrating out $N$ fundamental fields, and so 
it is (in a natural normalization) proportional to $N$.
Note also that the correlators of $J_0$ 
 of the two different quasi-fermionic theories are already known (or conjectured) 
to map to each other under duality, so the effective
actions \eqref{fermo} for $\zeta$ are automatically duality-covariant.

At the next step in the computation of our $\beta$ function, we 
perform the path integral over the Lagrange multiplier fields $\zeta$. 
As the action for the Lagrange multiplier fields has an overall 
factor of $N$, the leading large $N$ contribution to the $\beta$ 
function is given only by one loop graphs from this effective 
action (see \cite{Pomoni:2008de} for a similar observation 
in a different context). This is true at all values of $\lambda$ and $x_6^B$ (or $x_6^F$). Evaluating 
the divergent pieces of these one loop graphs we find that the 
$\beta$ function for $x_6=x_6^B=x_6^F$, for flows towards the UV, is given by 
\begin{equation}\label{bfintro} \begin{split}
& \frac{d x_6}{d \ln (\Lambda)}
=  -a  + b g_3   
-  c g_3^3,    \\
&g_3\equiv 24 \pi^2 
\left( x_6 -\frac{4}{3} \cot^2 \left( \frac{ \pi \lambda_B}{2} \right) \right)= 24 \pi^2 \left( x_6 -\frac{4}{3} \tan^2 \left( \frac{ \pi \lambda_F}{2} \right)
\right).\\
\end{split}
\end{equation}
Note in particular that $g_3$ is linearly related  to $x_6$, and so the $\beta$ function in \eqref{bfintro} is a cubic 
function of $x_6$ at every value of $\lambda_B$ and $\lambda_F$, the 't~Hooft couplings of the RB and CF theories, respectively. Here
\begin{equation}\begin{split} \label{othdef}
c&= \frac{1}{24 \pi^2} 
\left( \frac{1}{2 \pi^2 \kappa_B g_2^3} \right),\\
g_2&\equiv \left( \frac{4 \pi}{\tan(\frac{\pi\lambda_B}{2})} \right) = -\left( \frac{4 \pi}{\cot(\frac{\pi\lambda_F}{2})} \right),\\
\kappa_B&=-\kappa_F, ~~~~\lambda_B= \lambda_F - {\rm sgn} (\lambda_F). \\
\end{split}
\end{equation}
Note that $c$ is positive for all values of $\lambda_B$ and $\lambda_F$ (ranging from $(-1)$ to $1$). 

Like $c$, the coefficients $a$ and $b$ are functions of $\lambda_B$ and $\lambda_F$, and are of order $1/N$, but they are 
independent of $x_6$. $a$ and $b$ are determined (see \eqref{finalbetafn}) in terms of 
coefficients that characterize particular kinematical limits of the leading large $N$
3-, 4-, and 5-point functions of the operator $J_0$ in the quasi-fermionic theories, together with two numbers that require a subleading-in-$1/N$ computation : 
the first correction to the anomalous dimension of $J_0$ in these theories, and the coefficient governing the splitting of the 3-point function (once this anomalous dimension is non-zero) into a contact term and a non-contact term. The leading large $N$ three and four-point functions are known exactly. However the  leading large $N$ five-point function\footnote{See however \cite{Yacoby:2018yvy}.} and the two sub-leading $1/N$ corrections referred to above are currently known only at small $\lambda_B$ or small $\lambda_F$. Practically speaking, therefore, 
$a$ and $b$ are currently known only at small $\lambda_B$ or $\lambda_F$. 

As $c$ is positive, the beta function \eqref{bfintro} is negative at large positive values of 
$x_6$ but is positive at large negative values of $x_6$ (this is true at every 
value of $\lambda$). In other 
words, flows towards the IR drive large positive values of $x_6$ 
to $\infty$ and large negative values of $x_6$ to $-\infty$.
\footnote{As we discuss in section \ref{phasestruct}, the resulting theories may not have a stable vacuum.}
It follows immediately that our cubic beta function generically
has either one unstable fixed point or two unstable and one stable fixed points. \footnote{We refer to a fixed point as stable if it is attractive for flows towards the IR, and unstable if 
it is repulsive for flows towards the IR.}
The explicit values of $a$ and $b$ at weak bosonic and fermionic coupling suggest 
-- and we conjecture  -- that the $\beta$ function in fact has 
three zeroes at every non-zero $\lambda_B$. Ordering the fixed points along 
the $x_6$ axis, the first and third of these fixed points are repulsive 
(for flows towards the IR) while the second fixed point is
attractive. 

\begin{figure}[h]
\begin{center}
\includegraphics[width=14.5cm,height=7.5cm]{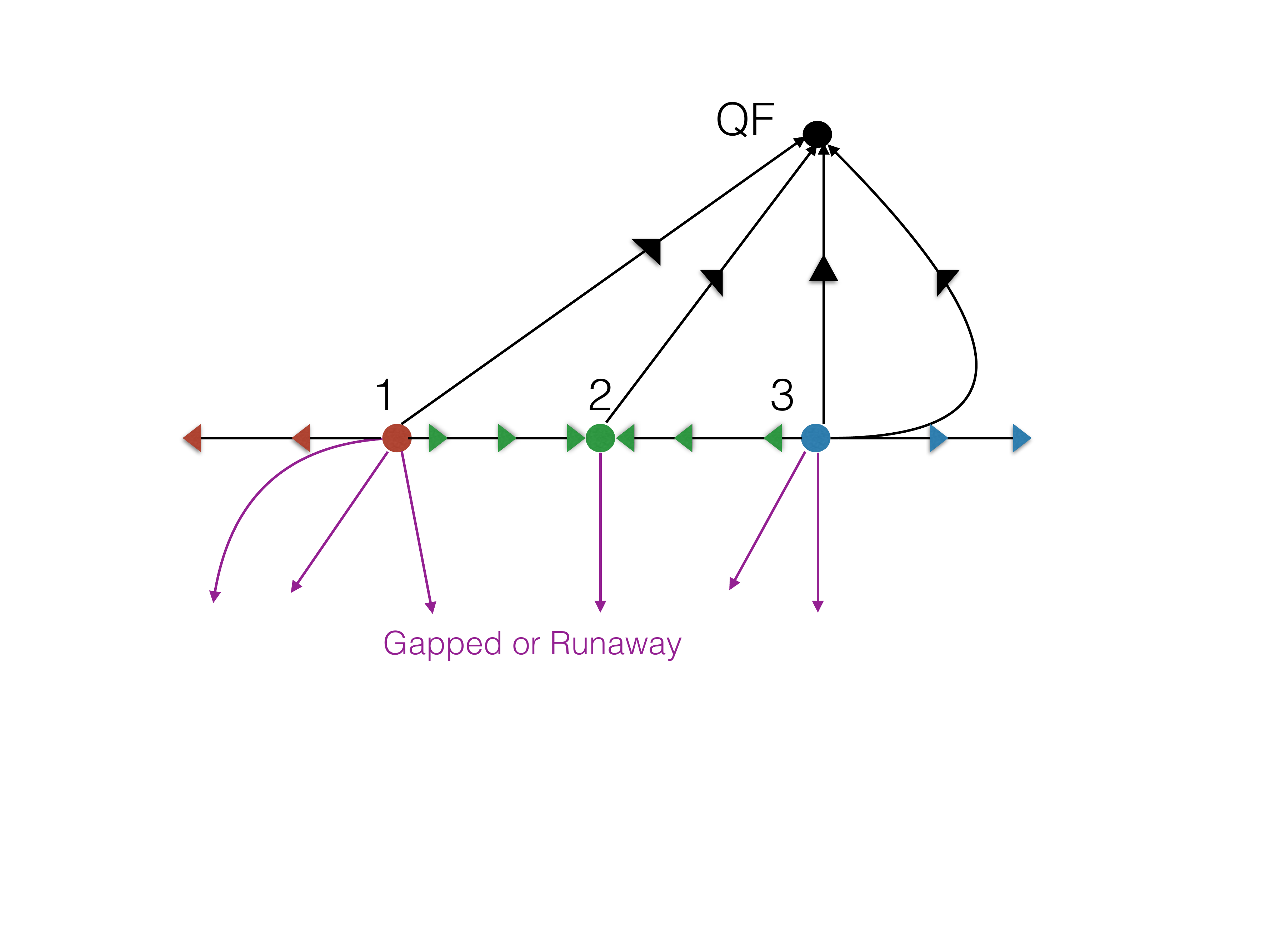}
  \caption{\label{Phsedg} Renormalization group flows of the quasi-bosonic theories as a function of $x_6$ (the horizontal axis) and of their second-most-relevant deformation (the vertical axis), when the most relevant operator is fine-tuned to zero.}
 \end{center}
\end{figure}

The structure of these RG flows of $x_6$ is depicted in 
Figure \ref{Phsedg}, where $x_6$ labels the horizontal direction. We have added in this figure also the expected behaviour of the quasi-bosonic theories when we turn on their second-most-relevant deformation (a $({\bar \phi}\phi)^2$ term in \eqref{rbt1}), whose coefficient labels the vertical axis, but still tune the most relevant operator to zero. We will discuss this further in section \ref{phasestruct}.
 
 Note that  the second fixed point has a total of  $2$ relevant operators (the mass term and $({\bar\phi}\phi)^2$) while the other two fixed points  have $3$ relevant operators; the additional operator is parametrized by $x_6$.

The $\beta$ function \eqref{bfintro} manifestly 
respects duality invariance: the beta function of the regular boson 
theory agrees with the beta function of the critical 
fermion theory under the standard bosonization duality map 
(see below for more details). As noted above, this 
feature is built in to our method of computation; it is 
a direct consequence of the duality covariance of the 
effective action for the Lagrange multiplier $\zeta$.

Our results allow us to give a clear definition  
of the space of quasi-bosonic theories, at finite (but large enough) values of $N$. 
These theories are defined by 
the space of RG flows away from the two repulsive fixed points 
\footnote{It is possible that our conjecture that there are exactly 
three fixed points at all values of $\lambda$ is not correct, and there
is a range of values of $\lambda$ about which the beta function \eqref{bfintro}
has one rather than three fixed points. In that case this single 
fixed point is necessarily repulsive, and the space of quasi-bosonic
theories is defined by RG flows away from this fixed point.}.

\subsubsection{Phase structure and stability of quasi-bosonic 
	theories}

In the 't Hooft large $N$ limit the $\beta$ function \eqref{bfintro} is of order  $\frac{1}{N_B}$. If we restrict our attention to the large $N_B$ limit, the 
quasi-bosonic theories are thus conformal at every value of $x_6$. 

At leading order in the large $N_B$ limit, the free energy of the quasi-bosonic theories at a finite temperature $T$ has been studied at every value of $x_6$ \cite{Aharony:2012ns,Minwalla:2015sca}. At the `conformal' point (i.e. at 
arbitrary $x_6$ but with all relevant deformations turned off)
the free energy is proportional to $T^3$ as expected from
conformal invariance. When relevant deformations (masses and $\phi^4$ couplings) are turned on, candidate phases are given 
by solutions to a known set of gap equations (see e.g. \cite{Minwalla:2015sca}). These gap equations typically have 
multiple solutions at any particular value  of the microscopic parameters and the temperature. The dominant phase is the solution with the lowest 
free energy. As microscopic parameters are varied, this dominant 
phase changes, giving rise to an intricate phase diagram. 

The gap equations of the quasi-bosonic theories continue to admit multiple 
solutions even at zero temperature, though they simplify greatly in this limit. In section 
\ref{phasestruct} below we compute the phase diagram of the 
RB theory at zero temperature by comparing the free energies 
of the various solutions to the gap equation.  Our final results are graphically summarized in the phase diagrams in Figures \ref{figctb}, \ref{figA1B1ph}, and \ref{figA3B3} below. 

The phase diagram of the RB theory changes as we change $x_6$. Interestingly, our final results are  qualitatively different 
depending on whether $x_6<\phi_1$, $\phi_1 <x_6<\phi_2$ or 
$x_6>\phi_2$, where $\phi_1$ and $\phi_2$ are critical values 
of $x_6$ given by 
\begin{equation}\label{phdefintro}
\phi_1\equiv \frac{4}{3}\left(\frac{1}{(2-|\lambda_B|)^2}-1\right),\qquad\qquad
\phi_2\equiv  \frac{4}{3}\left(\frac{1}{{\lambda}_B^2}-1\right).
\end{equation}
The upcoming paper \cite{new} explains this fact by computing an exact Landau Ginzburg potential for the variable ${\bar \phi} \phi$ 
in the RB theory. The phase diagram of the RB theory is obtained by finding all the extrema of this Landau Ginzburg potential and 
choosing the extremum with the minimum free energy. The 
differences in the phase structure of the RB theory in 
different intervals of $x_6$ is a consequence of the fact that 
this Landau Ginzburg potential is qualitatively different in the 
three regions mentioned above. In particular the Landau Ginzburg potential of \cite{new} has 
the property that it is unbounded from below when either $x_6 <\phi_1$ or $x_6>\phi_2$. When $x_6$ lies between $\phi_1$ and $\phi_2$, 
on the other hand,  the exact Landau Ginzburg potential 
is bounded below. 

The fact that the Landau Ginzburg action is unbounded from below 
when $x_6<\phi_1$ or $x_6> \phi_2$ suggests that the RB theory 
is unstable in this range of parameters. On the other hand 
the theory appears to be perfectly stable in the range 
$$\phi_1< x_6 < \phi_2.$$

Once we take finite $N$ effects into account, we have already seen that the RB theory is really well-defined only at the 
three fixed points of the $\beta$ function. The relationship
of these three fixed point values to $\phi_1$ and $\phi_2$ 
is graphically demonstrated in Figure \ref{figphse12a}, both 
at weak bosonic coupling and at weak fermionic coupling. 
Interestingly, the `middle' fixed point of the $\beta$ function 
(the stable fixed point about which $x_6$ is an irrelevant 
deformation) lies between $\phi_1$ and $\phi_2$ at both 
weak bosonic and weak fermionic coupling. We conjecture 
that this continues to be the case  at every value of $\lambda_B$, so that this fixed point 
is always well-defined with a stable vacuum.

\subsubsection{Flows from theory S to quasi-bosonic theories}

Above we discussed the RG flows of the parameter $x_6$ within the manifold of quasi-bosonic theories. 
In this subsubsection we turn our attention to a different class of flows: flows from the supersymmetric theory
S to the manifold of quasi-bosonic theories. The flows that we study in this section are non-trivial even at leading order in the large $N$ limit (unlike the $\beta$ functions of the previous subsubsection 
that were of order $\frac{1}{N}$), so they are `fast' flows in contrast to the `slow' flows of $x_6$ described above. 
At the end of the current section we will discuss the relationship between these two different classes of flows. 

\begin{figure}[h]
\begin{center}
\includegraphics[width=16.5cm,height=9.5cm]{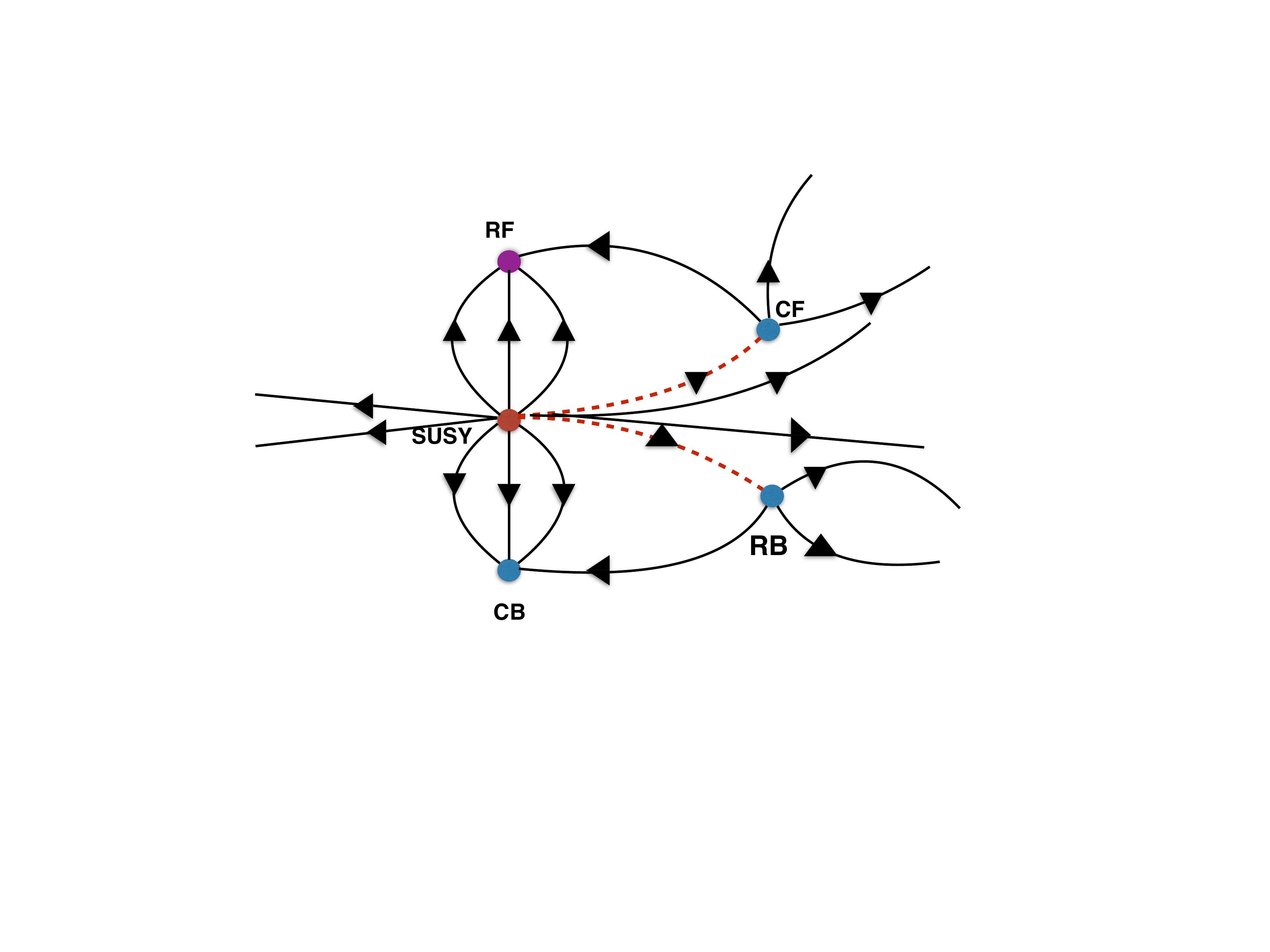}
  \caption{\label{surgflows} In this figure we present a qualitative sketch of the structure of
large $N$ critical flows from theory S, constructed in \cite{Jain:2013gza}, when the most relevant deformation is tuned to zero. We have added to their discussion the extra flows denoted by dashed red lines, which are further fine-tuned, from the supersymmetric theory to the quasi-bosonic theories.}
 \end{center}
\end{figure}

In subsection \ref{cbrf} we reviewed the one parameter fine-tuning \cite{Jain:2013gza,Gur-Ari:2015pca} of the two parameter class of dual pairs of RG flows starting at theory S, that end up in the deep IR in the CB and RF fixed points, respectively. While the discussion 
of \cite{Jain:2013gza,Gur-Ari:2015pca} is correct at generic parameters of the flows, it turns out to be possible to further fine tune 
the remaining parameter in the flows of \cite{Jain:2013gza,Gur-Ari:2015pca}. 
The special feature of the resulting RG flows is that they terminate 
on the manifold of RB and CF theories (i.e. on the manifold of dual pairs of 
quasi-bosonic theories), rather than at quasi-fermionic theories as was the case for the generic flows of
 \cite{Jain:2013gza,Gur-Ari:2015pca}. In Figure \ref{surgflows} below we present a qualitative sketch of the structure of
 critical flows constructed in \cite{Jain:2013gza}, including in it also the further fine 
 tuned flows (denoted by red lines) from the supersymmetric theory to the quasi-bosonic theories.

As we have emphasized above, the flows described in this subsubsection and in Figure \ref{surgflows} are large $N$ flows, governed by the
leading order beta functions that are of order unity in the large $N$ limit.   
The fact that we had to perform a two-parameter tuning to end up on the manifold of quasi-bosonic
theories is a consequence of the fact that there are exactly two relevant directions away from the manifold of quasi-bosonic theories
in the strict large $N$ limit \footnote{In the RB theory these two strongly relevant directions are the scalar mass and the $|\phi|^4$ coupling, and in the CF theory \eqref{cft1}  they may be written as $\zeta$ and $\zeta^2$.}.

At finite large $N$, the fast flows from theory S to the space of quasi-bosonic theories described in this subsection, 
generically do not end at true fixed points. Instead they terminate at some generic value of $x_6$ on the space of 
quasi-bosonic theories. After the fast flow is completed, the flow of $x_6$ continues at a much slower pace. The slow part of the 
flow is governed by the $\beta$ functions described in the previous subsubsection, and takes place over RG flow time scales of order $N$. 

It turns out that the fast part of the RG flows hits the quasi-bosonic manifold between the second and third fixed point when 
$|\lambda_B|$ is smaller than some critical value of order unity, which lies between $0$ and $1$. The subsequent slow flow then ends up in the 
second (attractive) quasi-bosonic fixed point. For $|\lambda_B|$ less than this critical value, the flows constructed  in this paper
may thus be thought of as a `derivation' of the duality between the second RB fixed point and the second CF fixed point, for finite large values of $N$, assuming 
the well-established duality of the supersymmetric theories S.

\begin{figure}[h]
\begin{center}
\includegraphics[width=14.5cm,height=7.5cm]{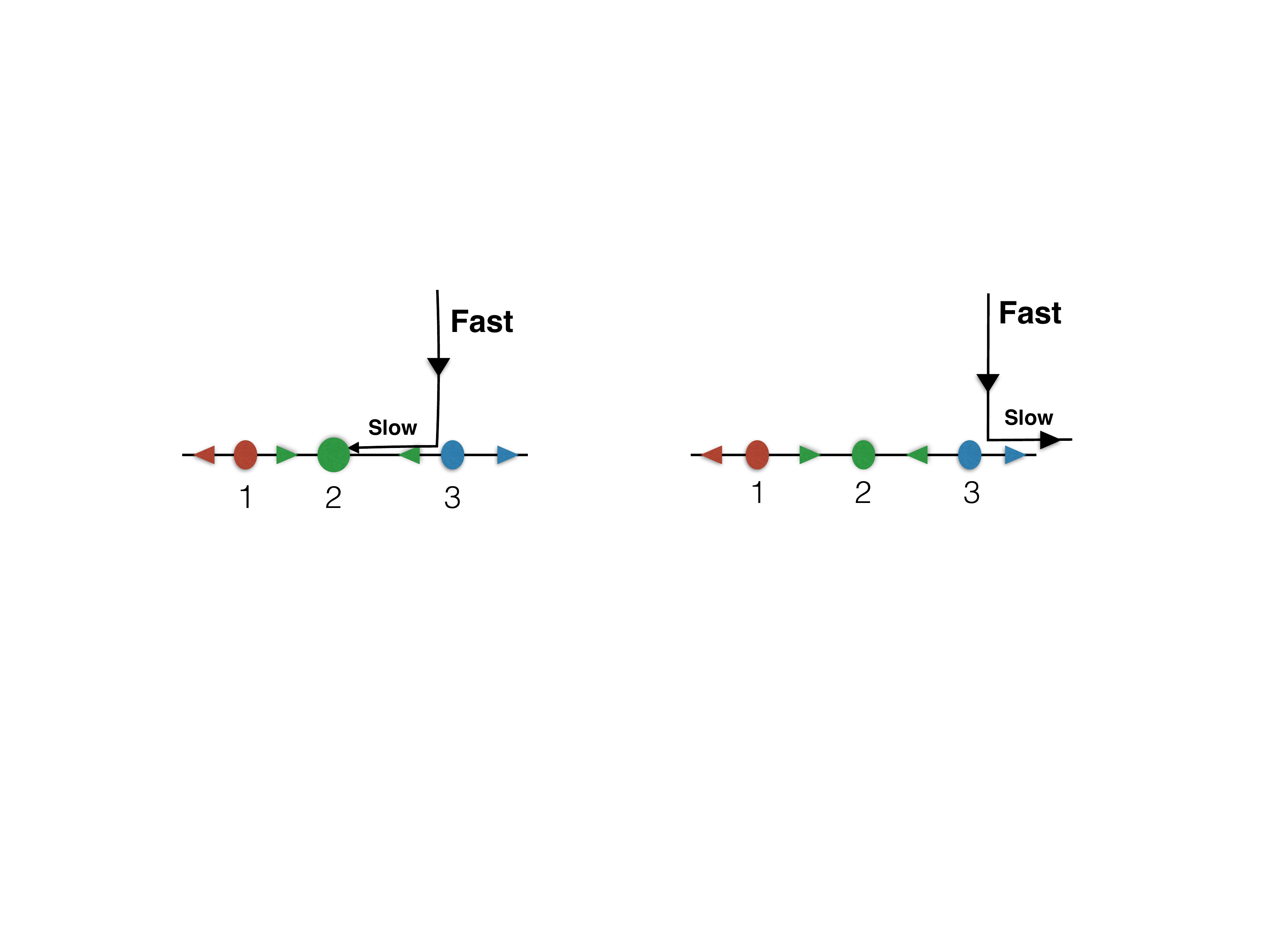}
  \caption{\label{fsfl} In this figure we depict two different classes of flows. The horizontal axis is $x_6$, and the vertical axis depicts the other parameters in the SUSY flow. The figure on the left, relevant for small $|\lambda_B|$, shows a fast flow arriving between the second and third fixed points, and a subsequent slow flow ending up in the 
second (attractive) quasi-bosonic fixed point. The figure on the right, relevant for large $|\lambda_B|$, shows a fast flow arriving  at a value of $x_6$ larger than the third fixed point, and a subsequent slow flow driving
 $x_6$ to infinity.}
 \end{center}
\end{figure}

For $|\lambda_B|$ greater than the critical value described in the last paragraph, the fast flows constructed in this paper 
hit the manifold of quasi-bosonic theories at a value of $x_6$ larger than the third fixed point. The subsequent slow flow 
drives $x_6$ to infinity.\footnote{Formally speaking, precisely 
at the critical value of $\lambda_B$ the fast flow ends up precisely at the third fixed point in the space of quasi-bosonic 
theories. However as the set of allowed values of $\lambda$ is discrete at any finite $N$ no matter how large, it is presumably not possible to end up at this fixed point at any finite value of $N$.} See Figure \ref{fsfl} for 
a sketch of these two different classes of flows. 

In the discussion above we have assumed the correctness of our conjecture that the beta function on the line of 
quasi-bosonic theories  has three zeroes for 
every value of $\lambda$. It is possible that this conjecture is incorrect and that there is a range of values of $\lambda$ for which 
the beta function has only a single zero. In this case the single fixed point is always repulsive. RG flows from the SUSY theory 
that are tuned to land on the line of quasi-bosonic theories will always flow away from this fixed point -- either to 
$x_6=\infty$ or to $x_6=-\infty$. We cannot generically tune a flow from the SUSY theory to hit such a fixed point, as it has $3$ relevant deformations while flows originating at the SUSY theory have just two dimensionless parameters.

We believe that the results of this subsection together with those of the previous subsection strongly suggest that the 
regular boson and critical fermion theories both exist, and are dual to each other, at all values of $\lambda$ and $x_6$ for infinite $N$, and also at least 
for large but finite $N$ and $\kappa$, when $x_6$ sits at its fixed points\footnote{There is substantial independent direct 
calculational evidence for this duality between the quasi-bosonic theories in the strict $N \to \infty$ limit. In this strict limit 
the beta function \eqref{bfintro} vanishes, and the RB and CF theories are both conformal at every 
value of $x_6$ . The high-spin symmetries continue to constrain the dynamics \cite{Maldacena:2011jn, Maldacena:2012sf}. 
Direct computational evidence for the duality in the large $N$ limit 
includes computations of correlators \cite{Aharony:2012nh, GurAri:2012is, Bedhotiya:2015uga}, S matrices \cite{Jain:2014nza,  Dandekar:2014era, 
Yokoyama:2016sbx} and thermal partition 
functions \cite{Giombi:2011kc, Jain:2012qi, Aharony:2012ns, Jain:2013py,  Takimi:2013zca, Yokoyama:2013pxa},
and they all yield results in perfect agreement with the conjectured duality between the full set of large $N$ RB and CF theories 
at arbitrary values of $x_6$.}.

In section \ref{ymcs} we analyze an alternative way to flow to the quasi-fermionic and quasi-bosonic theories, by starting from a high-energy Yang-Mills-Chern-Simons theory. This flow makes sense for arbitrary values of $N$ and $\kappa$; note that for finite values of these numbers it is hard to tell if the formal definitions of these theories, \eqref{cst1}-\eqref{cft1}, are well-defined or not. For finite values of $N$ and $\kappa$ it is hard to say when this flow ends at fixed points (after appropriate fine-tuning) and when it does not.  In any case, this flow suggests that whenever we can end up at a quasi-fermionic theory, we can also end up at the corresponding quasi-bosonic theory, with one additional fine-tuning. This is subject to the same caveats as above, that there is a stable fixed point for $x_6$, and that the `fast flow' governed by the Yang-Mills coupling brings us to the domain of attraction of this fixed point.

\section{Critical scalars and regular fermions} \label{csrf}

As we have explained in the introduction, the simplest and best established 
non-supersymmetric Bose-Fermi duality in $2+1$ dimensions is that between the 
CB and RF theories. In this section 
we review relevant aspects of these theories and their duality. 

\subsection{Critical scalars}

\subsubsection{Definitions} 

The critical scalar (CB) theory is a conformal field theory defined by the 
Lagrangian 
\beal{
S_{CB}(\phi, \sigma_B)  &= \int d^3 x  \biggl[i \varepsilon^{\mu\nu\rho}{\kappa_B\over 4\pi}
\Tr( A_\mu\partial_\nu A_\rho -{2 i\over3}  A_\mu A_\nu A_\rho)
+i \varepsilon^{\mu\nu\rho} {N_B \kappa'_B \over 4 \pi}
 B_\mu\partial_\nu B_\rho  \nn
&~~~~~~~~~~~~~
 + D_\mu \bar \phi D^\mu\phi+ \sigma_B\bar\phi \phi \biggl].
\label{cst}
}
Here $\kappa_B$ is the level of an $SU(N_B)$ gauge field $A_\mu$ while 
$N_B \kappa'_B$ is the level of a $U(1)$ gauge field $B_\mu$. The two join together into a $U(N_B) = (SU(N_B) \times U(1)) / Z_{N_B}$ gauge field. Both $\kappa_B$ and $\kappa'_B$ are non-zero integers\footnote{More precisely, we must have $ \kappa'_B =k_B+ m N_B$ where $m$ is an integer, for details see equation $(2.5)$ of \cite{Hsin:2016blu}.}. The fields 
$A_\mu$ are $N \times N$ traceless Hermitian matrices. The covariant derivative 
$D_\mu$ is given by 
\begin{equation}\label{covder}
D_\mu \phi=  \partial_\mu \phi + i A_\mu \phi + i B_\mu \phi.
\end{equation}

The theory based on the action \eqref{cst} is defined using the dimensional reduction
scheme \cite{Siegel:1979wq}. The same physical theory is obtained if one uses a Yang-Mills regulator 
and simultaneously replaces the levels $\kappa_B$ and $\kappa'_B$ by the levels 
$k_B$ and $k'_B$ given by (see Appendix \ref{levels})
\begin{equation}\label{kkmtb}
\kappa_B={\sgn}(k_B) \left( |k_B|+N_B \right), ~~~\kappa'_B=k'_B,
\end{equation}
so that we must have $|\kappa_B| > N_B$.
If we give a mass to the critical bosons (with a positive mass squared) and integrate them out, the 
resultant low energy theory is a pure $U(N_B)$ Chern-Simons theory, whose levels 
$k_B,k'_B$  can be identified with the levels of the related $U(N_B)$ WZW theories.

The critical boson theories defined so far are labeled by three integers: $N_B$, $\kappa_B$ and 
$\kappa'_B$. In the rest of this paper we will be principally interested in the following three two-integer subfamilies of critical scalar theories: 

\begin{itemize}
\item {\bf $SU(N_B)$ critical scalars}. These theories are defined by the Lagrangian 
\eqref{cst} with $B_\mu$ set to zero. These theories are labeled either by the 
pair of integers $(N_B, \kappa_B)$ or $(N_B, k_B)$ according to taste.
\item {\bf $U(N_B)$ critical scalars of type 1}. These are critical scalar theories with 
$\kappa'_B=\kappa_B$, $k'_B= {\sgn (k_B)}(|k_B|+N_B)$. 
These theories are also labeled by a pair of integers -- again
 either $(N_B, \kappa_B)$ or $(N_B, k_B)$ according to taste.  
\item {\bf $U(N_B)$ critical scalars of type 2}. These are critical scalar theories with 
$k'_B=k_B$, $\kappa_B= {\sgn (\kappa'_B)}(|\kappa'_B|+N_B)$.
 Once again these theories are labeled either by the integer pairs 
$(N_B, \kappa_B)$ or $(N_B, k_B)$ according to taste. 
\end{itemize}

For each of the three families of theories above, we define the 
't Hooft coupling
\begin{equation}\label{tcdef}
\lambda_B= \frac{N_B}{\kappa_B}.
\end{equation}
In the large $N_B$ limit, the theories above are most conveniently parameterized by 
the integer $N_B$ and the effectively continuous real parameter  $\lambda_B$, obeying $|\lambda_B| < 1$.

\subsubsection{Review of useful results at large $N_B$}

At leading order in the large $N_B$ limit the $SU(N_B)$,  type 1 and type 2 $U(N_B)$  theories 
are all identical. Several 
results for these theories have been established for all values of $\lambda_B$ in this limit. 
We will now present a brief review of those results that will be of interest to us later in this paper. 

Denote the lowest dimensional scalar operator in the critical boson theory by 
${\tilde J}_0$. In the dimensional regularization scheme (in which operators of distinct bare dimension 
cannot mix) ${\tilde J}_0 \propto \sigma_B$. We choose the normalization of ${\tilde J}_0$ as 
\begin{equation} \label{defjt}
{\tilde J}_0 = \sigma_B.
\end{equation}

The scaling dimension $\Delta_B(\lambda_B)$ of ${\tilde J}_0$ has the large $N_B$ expansion
\begin{equation}\label{sdopb}
\Delta_B(\lambda_B) = 2 +\frac{\delta_B(\lambda_B)}{\kappa_B} + {\cal O}(1/\kappa_B^2) .
\end{equation}
The function $\delta_B(\lambda_B)$ is as yet unknown. 

In the limit $\lambda_B \to 0$, which gives the critical $U(N_B)$ vector model, the anomalous dimension is \cite{Vasiliev:1982dc}
\begin{equation}\label{deltab}
\frac{\delta_B(\lambda_B)}{\kappa_B} =- \frac{16}{3 \pi^2 N_B}
\implies \delta_B(\lambda_B)= -\frac{16}{3 \pi^2 \lambda_B}.
\end{equation}

In the large $N_B$ limit, the two-point function of 
${\tilde J}_0$ is given by  \cite{Aharony:2012nh}
\begin{equation}\label{tpln} \begin{split}
&\langle {\tilde J}_0 (q)  {\tilde J}_0 (-q') \rangle 
=  \frac{1}{\kappa_B} (2 \pi)^3 \delta^3(q-q')  G_\sigma(p), \\
& G_\sigma(q)= n_{B} q^{ \left( 2 \Delta_B(\lambda_B) -3 \right) },  \\
&n_{B}= -4\frac{\pi }{\tan(\frac{\pi\lambda_B}{2})}  +  {\cal O}(1/\kappa_B). \\
\end{split}
\end{equation}

At leading order in large $N_B$ \eqref{tpln} reduces to 
\begin{equation}\label{tplnl}
\langle {\tilde J}_0 (q)  {\tilde J}_0 (-q') \rangle 
=  (2 \pi)^3 \delta^3(q-q') ~~ \frac{(-4 \pi) |q| }{\kappa_B \tan(\frac{\pi\lambda_B}{2})}.
\end{equation}
At small $\lambda_B$ this result further simplifies to 
\begin{equation}\label{tplnlw}
\langle {\tilde J}_0 (q)  {\tilde J}_0 (-q') \rangle 
=  (2 \pi)^3 \delta^3(q-q') ~~  \frac{(-8)|q|}{N_B} 
\left( 1 + {\cal O}(\lambda_B^2) \right).
\end{equation}

Note that
\begin{equation}\label{fte}
\int \frac{d^3 p}{(2 \pi)^3} |p| e^{-\epsilon |p|}  e^{i p\cdot r}
= \frac{1}{2 \pi^2 r} \int_0^\infty dp p^2 \sin (pr) e^{- \epsilon p} 
= - \frac{1}{\pi^2} \frac{r^2 - 3 \epsilon^2}{(r^2+\epsilon^2)^3}  .
\end{equation}
By taking the limit  $ \epsilon \to 0$ it follows that the Fourier 
transform of $-|p|$ is $\frac{1}{ \pi^2 r^4}$. It follows, in particular, 
that the two point functions \eqref{tplnl} and \eqref{tplnlw}
are all positive when reexpressed in position space, consistent with 
the expectations from unitarity.\footnote{Restated, the two point function of an operator of dimension 2 is positive in position space if and only if it is negative in momentum space.
On the other hand it is easily verified that 
\begin{equation}\label{ftte} 
\int \frac{d^3 p}{(2 \pi)^3} \frac{1}{|p|} e^{i p\cdot r}= \frac{1}{2 \pi^2 r^2},
\end{equation}
establishing that a two point function of an operator of dimension one -- of the sort we will encounter in the next section -- is unitary in position space if and only if its Fourier transform is positive in momentum space. }

The three point function of ${\tilde J}_0$ is given by  
\begin{equation}\label{thpln} \begin{split}
&\langle {\tilde J}_0(q_1) {\tilde J}_0(q_2) {\tilde J}_0 (q_3) 
\rangle = (2 \pi)^3 \delta^3(q_1+q_2 + q_3) {\tilde G}_3(q_1, q_2, q_3),\\
&{\tilde G}_3(q_1, q_2, q_3)= \frac{32\pi^2}{\kappa_B^2}  \left( c_B^\delta(\lambda_B)+ c_B(\lambda_B)
 F(|q_1|, |q_2| , |q_3|, \frac{\delta_B}{\kappa_B}) \right).\\
\end{split}
\end{equation} 
\eqref{thpln} simply asserts that the three point function of three ${\tilde J}_0(q_1)$ operators is the sum of a 
contact term and a piece proportional to $F(q_1, q_2, q_3, \frac{\delta_B}{\kappa_B} )$, 
the Fourier transform of the usual power law position space expression for a three point function of three operators of 
dimension  $2 + \frac{\delta_B}{\kappa_B}$. An explicit integral expression for $F(q_1, q_2, q_3, \frac{\delta_B}{\kappa_B})$ was 
presented in \cite{Bzowski:2013sza}; using that expression we demonstrate in Appendix \ref{confform} that in the normalization we use
\footnote{The fact that contact terms are absent in two point functions but are present 
in three point functions of ${\tilde J}_0$ is a simple consequence of 
dimensional analysis. The engineering dimension of ${\tilde J}_0(x)$ is 2. It follows that the 
engineering dimension of  ${\tilde J}_0(q)$ is $(-1)$, and so a constant times a 
momentum conserving delta function has the right dimension to appear 
in a three point function of ${\tilde J}_0$. Such a contribution to 
higher or lower point functions would have to be accompanied by a power of 
the cutoff, and explicit powers of the cutoff never appear in dimensional 
regularization. } 
\begin{equation} \label{mainf}
\frac{\delta_B}{|\kappa_B|}~ F(q_1,q_2,q_3,\frac{\delta_B}{\kappa_B}) = 1- 3 \frac{\delta_B}{\kappa_B}~ \ln\left( \frac{\Lambda}{q_1+q_2+q_3}\right)+{\cal O}(\frac{1}{\kappa_B^2}),
\end{equation}
note that $F(q_1,q_2,q_3,\frac{\delta_B}{\kappa_B})$ reduces to a constant at leading order at $1/N$ and deviates from the constant only at first sub-leading order. It follows that the contact piece in \eqref{thpln} and $F(q_1,q_2,q_3,\frac{\delta_B}{\kappa_B})$ are indistinguishable at leading order at large $N$.

The three point functions of ${\tilde J}_0$ have been computed at leading order in the large $N$ limit. Comparing \eqref{thpln} to 
the explicit results of this computation we find \cite{Aharony:2012nh}
\begin{equation}\begin{split} \label{thptdet} 
&c_B^\delta(\lambda_B)=  r_B(\lambda_B) \left( \frac{\cot^2(\frac{\pi\lambda_B}{2})}{\cos^2(\frac{\pi\lambda_B}{2})} -\frac{1}{4} \right)
+ {\cal O}(1/\kappa_B), \\
& c_B(\lambda_B)= (1-r_B(\lambda_B)) \frac{\delta_B}{\kappa_B} \left( \frac{\cot^2(\frac{\pi\lambda
_B}{2})}{\cos^2(\frac{\pi\lambda_B}{2})} -\frac{1}{4} \right)  
+ {\cal O}(1/|\kappa_B|^2),\\
\end{split}
\end{equation}
where $r_B(\lambda_B)$ is a  parameter which has not yet been determined, because the three point function has been explicitly evaluated only at leading order at large $N$ where $F(q_1,q_2,q_3,\frac{\delta_B}{\kappa_B})$ and the contact term  in \eqref{thpln} can not be distinguished as we have explained  above.  Expanding \eqref{thpln}  to first order in ${1\over \kappa_B}$  using \eqref{thptdet}, we find 
\begin{equation}\label{nonanag} \begin{split}
{\tilde G}_{3}(q_1,q_2,q_3)= & \frac{32 \pi^2}{\kappa_B^2}\left( \frac{\cot^2(\frac{\pi\lambda_B}{2})}{\cos^2(\frac{\pi\lambda_B}{2})}  -\frac{1}{4} \right) \cdot  \\
& \cdot \left( \left(1+{\cal O}(\frac{1}{\kappa_B})\right)+3 r_B(\lambda_B) \frac{\delta_{B}}{\kappa_B} \ln\left(\frac{\Lambda}{|q_1|+|q_2|+|q_3|}\right)+{\cal O}(\frac{1}{\kappa_B^2})\right).
\end{split}
\end{equation}
   Note that the non-analytic term in momenta in \eqref{nonanag} starts out at order ${\cal O}(\frac{1}{\kappa_B})$. 
   At leading order in the $\frac{1}{N_B}$ expansion
   \begin{equation}\label{thplnl} \begin{split}
&\langle {\tilde J}_0(q_1) {\tilde J}_0(q_2) {\tilde J}_0 (q_3) 
\rangle = (2 \pi)^3 \delta^3(q_1+q_2 + q_3)~~
\frac{32\pi^2}{\kappa_B^2}  
\left( \frac{\cot^2(\frac{\pi\lambda_B}{2})}{\cos^2(\frac{\pi\lambda_B}{2})} -\frac{1}{4} \right),
\end{split}
\end{equation}
  and so  the three point function is a pure contact term in accordance 
with general expectations \cite{Maldacena:2012sf}.  

Notice that the parameter $r_B(\lambda_B)$ that appears in \eqref{thptdet} disappears in the leading order result 
\eqref{thplnl}. This is the reason that $r_B(\lambda_B)$ cannot be determined by a comparison with explicit leading 
large $N_B$ results, and so is unknown. As the term proportional to $r_B(\lambda_B)$ in \eqref{nonanag} 
multiplies a factor of $\ln (\Lambda)$, this term will contribute to the $\beta$ function computed later in this 
paper and so will be of importance to us. In Appendix \ref{fbt} we present a formal expression for 
$r_B$ at $\lambda_B=0$.

At small $\lambda_B$ \eqref{thplnl} simplifies further to 
\begin{equation}\label{thplnlw} \begin{split}
&\langle {\tilde J}_0(q_1) {\tilde J}_0(q_2) {\tilde J}_0 (q_3) 
\rangle = (2 \pi)^3 \delta^3(q_1+q_2 + q_3)~~
\frac{128 }{N_B^2} \left( 1 + {\cal O}(\lambda_B^2) \right)  .
\end{split}
\end{equation}

The four point function of four scalar operators has also been 
recently computed at leading order in the large $N_B$ limit
\cite{Turiaci:2018nua} (see \cite{Aharony:2018npf, Yacoby:2018yvy} for further developments). 
\footnote{In what follows we restrict our attention to the leading order 
in the expansion in $\frac{1}{N_B}$.} The four point function takes the form 
\begin{equation}\label{fpln} 
\langle {\tilde J}_0(q_1) {\tilde J}_0(q_2) {\tilde J}_0 (q_3) {\tilde J}_0 (q_4) 
\rangle = (2 \pi)^3 \delta^3(q_1+q_2+q_3+q_4) {\tilde G}_4^0(q_1, q_2, q_3, q_4),
\end{equation}
where ${\tilde G}_4^0$ has mass dimension $(-1)$ (it is a homogeneous function 
of its arguments of homogeneity  
$(-1)$). It was demonstrated in \cite{Turiaci:2018nua} that the 
momentum dependence of the four point function is the same 
as the momentum dependence of the operator ${\bar \psi} \psi$ 
in a theory of free fermions. The prefactor of this momentum 
dependence was also determined in \cite{Turiaci:2018nua}. In
this paper we will be principally interested in the four point 
function in a particular kinematical limit, namely the limit 
 $\delta \to 0$ of ${\tilde G}_4^0(p, -p-\delta, k+\delta, -k)$. 
The $\delta \to 0$ limit of this correlator is smooth 
\footnote{More precisely we are interested in the momenta configurations 
	${\tilde G}_4^0(p, -p-\delta, k+\delta, -k)$ in the limit $\delta \to 0$. 
	It is an interesting fact that this limit is smooth whenever $\Delta_{min}$ -- the 
	dimension of the most relevant operator in the theory under study (in the current 
	context $\Delta_{min}$ is approximately 2) -- is greater 
	than $\frac{d}{2}$ (where $d$ is the spacetime dimension of the theory, in 
	the current context $d=3$). This  follows immediately from the study  
	of the OPE in the channel in which the two operators with momentum 
	$p$ and $-p-\delta$ are brought together. The contribution of an operator 
	of dimension $\Delta$ to this OPE scales like $\frac{1}{R^{2 \Delta}}$ 
	in position space (here $R$ is a measure of the distance between the composite
	operators with momenta $p$ and $-p-\delta$ and the composite operators with 
	momenta $k+\delta$ and $k$)  and so like like $\int d^d R \frac{e^{i \delta. R}}{R^{2 \Delta}}$
	in momentum space. The integrand in this expression is correct only at large $R$ 
	(at small $R$ it is cut off by the fuzz inherent in the the scale of the composite operators - set by $\frac{1}{k}$ and $\frac{1}{p}$ 
	respectively). The contribution to the integral from the IR (large $R$) is finite
	when $\delta \to 0$ provided $2 \Delta> 3$. If $2 \Delta <3$, on the other hand,  the 
	integral scales like $\frac{1}{\delta^{d-2 \Delta}}$. In particular 
	if $d=3$ and $\Delta=1$ then we find a divergence like $\frac{1}{\delta}$, as 
	we see explicitly in \eqref{fpvert} in the free boson theory.}
so the four point function in the kinematical limit of interest to us is simply ${\tilde G}_4^0(p, -p, k, -k)$. It turns out that 
\begin{equation}\label{fpfa}
\begin{split}
&{\tilde G}_4^0(p,-p,k,-k)= \\
& - 2^7 \left(\frac{2\pi}{\kappa_B}\right)^3 \frac{1}{(\tan(\frac{\pi\lambda_B}{2}))^2 \sin(\pi\lambda_B)}\left( \frac{1}{ 2 |k+p|}+ \frac{1}{ 2 |k-p|} -\frac{1}{2}\frac{|k|}{ |p|^2} -\frac{1}{2} \frac{(p\cdot k)^2}{|p|^4 |k|}  
+ {\cal O}(k/p^2) \right).
\end{split}
\end{equation}

In particular at $\lambda_B=0$ (see the explicit computation in  Appendix \ref{fbt} for a check)  
\begin{equation}\label{fpf}
{\tilde G}_4^0(p,-p,k,-k)= -\frac{8^4}{N_B^3} \left( \frac{1}{ 2 |k+p|}+ \frac{1}{ 2 |k-p|} -\frac{1}{2}\frac{|k|}{ |p|^2} -\frac{1}{2} \frac{(p\cdot k)^2}{|p|^4 |k|}  
+ {\cal O}(k/p^2) \right).
\end{equation}
In a Taylor series expansion in $\frac{|k|}{|p|}$ we have
\begin{equation}\label{formg41}
{\tilde G}^0_4= \frac{1}{|p| \kappa_B^3} \left({\tilde g}_{(4,1)}  + {\tilde g}_{(4,2)} \frac{|k|}{|p|} +  
{\tilde g}_{(4,3)} \frac{(p\cdot k)^2}{|p|^3|k|} + {\cal O}(k^2/p^2)\right),
\end{equation} 
where 
\begin{equation}\label{ginoth}
{\tilde g}_{(4,1)}= -2{\tilde g}_4, \qquad
{\tilde g}_{(4,2)}= {\tilde g}_4,\qquad
{\tilde g}_{(4,3)}= {\tilde g}_4,
\end{equation}
with 
\begin{equation}\label{tgf}
{\tilde g}_4 =  \frac{2^9\pi^3}{(\tan(\frac{\pi\lambda_B}{2}))^2 \sin(\pi\lambda_B)}.
\end{equation}

Specifically in the limit $\lambda_B \to 0$
\begin{equation}
{\tilde g}_4=\frac{1}{2} \frac{8^4}{\lambda_B^3}
\end{equation} 
so that
\begin{equation}\label{gfree}
{\tilde g}_{(4,1)}=-\frac{8^4}{\lambda_B^3},\qquad
{\tilde g}_{(4,2)}= \frac{1}{2}\frac{8^4}{\lambda_B^3},\qquad
{\tilde g}_{(4,3)}=\frac{1}{2} \frac{8^4}{\lambda_B^3}.~~~
\end{equation}

For our purpose we also need the five point function 
\begin{equation}\label{fivpln} 
\langle {\tilde J}_0(q_1) {\tilde J}_0(q_2) {\tilde J}_0 (q_3) {\tilde J}_0 (q_4) {\tilde J}_0 (q_5) 
\rangle = (2 \pi)^3 \delta^3(q_1+q_2+q_3+q_4+q_5) {\tilde G}_5^0(q_1, q_2, q_3, q_4,q_5) ,
\end{equation}
where ${\tilde G}_5^0$ has mass dimension $(-2)$ (it is a homogeneous function 
of its arguments of homogeneity  
$(-2)$). For our purpose, we are just interested in the function 
${\tilde G}_5^0(p, -p, 0, 0,0)$.\footnote{More precisely we are interested in $G(p, -p-\delta_1-\delta_2 -\delta_3, \delta_1, \delta_2, \delta_3)$
in the limit that all $\delta$'s are small. We will assume that this limit
exists and is unambiguous in what follows. } It follows from dimensional analysis that 
 this Green's function admits a  form  
\begin{equation}\label{formg5}
{\tilde G}_5^0= \frac{1}{|p|^2 \kappa_B^4} {\tilde g}_{(5,0)} ,
\end{equation} 
where ${\tilde g}_{(5,0)}$ is a number (which is a  function of $\lambda_B$), whose value at $\lambda_B=0$ is discussed in appendix \ref{quintterm}.

\subsubsection{$SO(N)$ theories}

Throughout this paper we primarily discuss theories with 
unitary gauge groups \eqref{cst}. It is, however, useful to note 
that the results presented above also apply (with minor modifications) 
at leading order in large $N_B$ to the $SO(N_B)$ gauged theory with action
\begin{equation}
S_{CB}(\phi, \sigma)  = \int d^3 x  \biggl[i \varepsilon^{\mu\nu\rho}{\kappa_B\over 8\pi}
\Tr( A_\mu\partial_\nu A_\rho -{2 i\over3}  A_\mu A_\nu A_\rho)
 + \frac{D_\mu  \phi D^\mu\phi}{2}+ \sigma_B \phi \phi \biggl],
\label{csto}
\end{equation}
where $A_\mu$ are now imaginary antisymmetric matrices -- i.e. generators of 
$SO(N_B)$. 
The 't Hooft coupling  is once again defined by 
$\lambda= \frac{N_B}{\kappa_B}$. 
The leading large $N_B$ $n$-point Green's functions of the theory \eqref{csto} at any 
particular value of the 't Hooft coupling are obtained
from the leading large $N_B$ Green's functions of the theory \eqref{cst}  
at the same value of the 't Hooft coupling using the translation formulae
\begin{equation}\label{corggt}
{\tilde G}_n^{SO(N)} = \frac{1}{2} {\tilde G}_n^{SU(N)}.
\end{equation}

\subsection{The regular fermion theory} 

The regular fermion (RF) theory is defined by the Lagrangian 
\begin{equation}
\label{csfnonlinear1}
 S_{CF} =\int d^3 x \bigg[ i \varepsilon^{\mu\nu\rho} {\kappa_F \over 4 \pi}
\Tr( A_\mu\partial_\nu A_\rho -{2 i\over3}  A_\mu A_\nu A_\rho)+i \varepsilon^{\mu\nu\rho} {\kappa'_F \over 4 \pi}
 B_\mu\partial_\nu B_\rho 
+  \bar{\psi} \gamma_\mu D^{\mu} \psi \bigg].
\end{equation}
Our conventions for levels are the same as in the previous subsection. In particular the 
path integral with the action \eqref{csfnonlinear1} is defined with the dimensional regulation scheme. 
We get the same physics with a Yang-Mills regulator if we modify the action above replacing 
$\kappa_F$ with $k_F$ and $\kappa'_F$ with $k'_F$ where
\begin{equation}\label{kkmtf}
\kappa_F={\sgn}(k_F) \left( |k_F|+N_F \right), ~~~\kappa'_F=k'_F.
\end{equation}
One difference with the theories of the previous subsection is that $\kappa_F$ and $k_F$ are 
half integers (numbers of the form $\frac{2 n+1}{2}$ with integer $n$) in our notation rather than integers. This can be understood 
in the following terms. If we add a mass term with mass parameter $m_F$ to the Lagrangian 
\eqref{csfnonlinear1} and integrate the fermion out then the resulting gauge theory, at long distances, 
is a pure Chern-Simons theory with Chern-Simons level (equal to the level of the dual WZW theory) equal 
to $(k_F +\frac{\sgn (m_F)}{2})$. This level is an integer, as it has to be, if and only if $k_F$ 
is a half integer, and this defines what we mean by half-integer levels.

It is convenient to define
\begin{equation}\label{tkf}
{\tilde \kappa}_F= \kappa_F+  \frac{\sgn(\kappa_F)}{2}.
\end{equation}
The quantity ${\tilde \kappa}_F$ agrees with the effective value of $\kappa$ for the low energy pure 
Chern-Simons theory obtained after integrating out the  fermion, when it is given a  mass of the same sign as
$\kappa_F$. 
We define the fermionic 't Hooft coupling by the equation 
\begin{equation}\label{ftc}
\lambda_F= \frac{N_F}{{\tilde \kappa}_F}.
\end{equation}
It follows that $\lambda_F$ is simply the standard definition of the 't Hooft coupling for the effective 
low energy Chern-Simons  theory obtained after integrating out the fermions with a mass of the same sign as
${\tilde \kappa_F}$. 

As in the previous subsection we have $SU(N_F)$, type 1 $U(N_F)$ and type 2 $U(N_F)$ regular fermion 
theories. The definition of these theories is the obvious fermionic analogue of the definitions of the 
previous subsection.

\subsubsection{Review of results at large $N_F$}

As in the previous subsection, the lowest dimension scalar field $J_0$ in 
this theory will play an important role in what follows. 
In the dimensional reduction scheme (and with an appropriate choice of normalization that we adopt)
\begin{equation}\label{jdf}
J^F_0=\frac{4 \pi {\bar \psi} \psi}{{\tilde \kappa_F}}. 
\end{equation}
The dimension of this operator is given by 
\begin{equation}\label{sdopf}
\Delta_F(\lambda_F) = 2 - \frac{\delta_F(\lambda_F)}{{\tilde \kappa_F}} + {\cal O}(1/{\tilde \kappa_F}^2) 
\end{equation}
where $\delta_F(\lambda_F)=-\frac{4}{3}\lambda_F+O(\lambda_F^2)$ (see equation (5.32) of \cite{Giombi:2016zwa}).

In the large $N_F$ limit the two point function of 
$J^F_0$ is given by \cite{Aharony:2012nh}
\begin{equation}\label{tplnf} \begin{split}
&\langle J^F_0 (q)  J^F_0 (-q') \rangle 
=  \frac{1}{{\tilde \kappa_F}} (2 \pi)^3 \delta^3(q-q') n_{F} q^{ \left( 2 \Delta_F(\lambda_F) -3 \right) },  \\
&n_{F}= -4\pi \tan\left( \frac{\pi \lambda_F}{2}  \right)
 +  {\cal O}(1/{\tilde \kappa_F}). \\
\end{split}
\end{equation}
At leading order in large $\kappa_F$, the small $\lambda_F$ expansion of \eqref{tplnf} gives
\begin{equation} \label{tplnfw}
\langle J^F_0 (q)  J^F_0 (-q') \rangle = -(2 \pi)^3 \delta^3(q-q') 2\pi^2 |q| N_F\frac{1}{{\tilde \kappa}_F^2}+{\cal O}(\lambda_F^2).
\end{equation}

Three point functions of $J_0^F$ take the form \cite{Aharony:2012nh}
\begin{equation}\label{thplnf} \begin{split}
&\langle { J}^F_0(q_1) { J}^F_0(q_2) { J}^F_0 (q_3) 
\rangle = \frac{(2\pi)^3}{{\tilde \kappa_F}^2}  \delta^3(q_1+q_2 + q_3) \left( c_F^\delta(\lambda_F)+ c_F(\lambda_F)
 F(|q_1|,|q_2|,|q_3|,\frac{\delta_F}{{\tilde \kappa_F}}) \right),\\
&c_F^\delta(\lambda_F)= r_F(\lambda_F) 32\pi^2\tan^2\left(\frac{\pi\lambda_F}{2}\right)+ {\cal O}(1/{\tilde \kappa_F}), \\
& c_F(\lambda_F)=\frac{\delta_F}{{\tilde \kappa_F}}\left(1-r_F(\lambda_F)\right) 32\pi^2\tan^2\left(\frac{\pi\lambda_F}{2}\right) + {\cal O}(1/|{\tilde \kappa_F}|^2).
\end{split}
\end{equation}
The small $\lambda_F$ expansion of \eqref{thplnf} gives
\begin{equation}\label{thplnfw}
\langle { J}^F_0(q_1) { J}^F_0(q_2) { J}^F_0 (q_3) 
\rangle = (2 \pi)^3 \delta^3(q_1+q_2 + q_3)8\pi^2 N_F\frac{\pi^2 }{{\tilde \kappa_F}^3} \lambda_F+{\cal O}(\lambda_F^4).
\end{equation}
Note, in particular, that the three point function vanishes in the limit 
$\lambda_F \to 0$. This is a consequence of the fact that the operator 
${\bar \psi} \psi = \frac{\kappa_F}{4 \pi} J_0^F$ is odd under parity 
transformations in the limit $\lambda_F \to 0$. As the left-hand side of \eqref{thplnfw} is odd under parity, while the $\delta$ function 
on the right-hand side is even, parity invariance forces the coefficient to vanish.

As in the previous subsection we restrict our discussion to the four point function leading order in the 
large $N_F$ limit. At this order the four point function is 
completely known. The momentum dependence of this function 
is precisely that of the free fermi theory 
\cite{Turiaci:2018nua} (see \cite{Bedhotiya:2015uga} for related 
earlier work). As in the previous subsection we define 
\begin{equation}\label{fplnfer} 
\langle { J}_0(p_1) { J}_0(p_2) { J}_0(p_3) { J}_0 (p_4)
\rangle = (2 \pi)^3 \delta^3(p_1+p_2 +p_3 +p_4) { G}_4^0(p_1, p_2, p_3, p_4) 
\end{equation}
where $G_4^0$ has mass dimension $(-1)$. Again 
we are interested in $ G_4^0(p, -p, k, -k)$  in the limit 
$|k| \ll |p|$. Expanding the Greens function in a Taylor series 
expansion in $\frac{|k|}{|p|}$ we have 
\begin{equation}\label{formg42}
{ G}^0_4= -\frac{1}{|p| {\kappa}_F^3} \left( {g}_{(4,1)} + {g}_{(4,2)} \frac{|k|}{|p|} +  
{g}_{(4,3)} \frac{(p\cdot k)^2}{|p|^3|k|} + {\cal O}(k^2/p^2)\right),
\end{equation} 
\begin{equation}\label{ginothf}
{ g}_{(4,1)}= -2{g}_4, \qquad
{g}_{(4,2)}= {g}_4,\qquad
{g}_{(4,3)}= {g}_4,
\end{equation}
with 
\begin{equation}\label{tgft}
{g}_4 = -\frac{2^9\pi^3}{(\cot(\frac{\pi\lambda_F}{2}))^2 \sin(\pi\lambda_F)}
\end{equation}

In particular in the limit $\lambda_F=0$ 
(see Appendix \ref{fft} for an explicit check) we find 
\begin{equation}\label{fpfer}
G_4^0(p,-p,k,-k)=-N_F\left(\frac{4\pi}{\kappa_F}\right)^4 \left( \frac{1}{ |p|} -\frac{1}{2}\frac{|k|}{ |p|^2} -\frac{1}{2} \frac{(p\cdot k)^2}{|p|^4 |k|}  
+ {\cal O}(k^2/p^2) \right),
\end{equation}
from which it follows that at leading order in small $\lambda_F$
$$g_4=-\frac{1}{2} (4\pi)^4 \lambda_F$$
so that
\begin{equation}\label{gfreef}
g_{(4,1)}=(4\pi)^4 \lambda_F,\qquad
g_{(4,2)}=-\frac{1}{2} (4\pi)^4 \lambda_F,\qquad
g_{(4,3)}=-\frac{1}{2} (4\pi)^4 \lambda_F.
\end{equation}

For our purpose we also need the five point function 
\begin{equation}\label{fivplnf} 
\langle J^F_0(q_1) J^F_0(q_2) J^F_0 (q_3) J^F_0 (q_4) J^F_0 (q_5) 
\rangle = (2 \pi)^3 \delta^3(q_1+q_2+q_3+q_4+q_5)  G_5^0(q_1, q_2, q_3, q_4,q_5) 
\end{equation}
where $G_5^0$ has mass dimension $(-2)$ (it is a homogeneous function 
of its arguments of homogeneity  
$(-2)$). In the specific case
$G_5^0(p, -p, 0, 0,0)$ this Green's function takes the  form  
\begin{equation}\label{formg43}
G_5^0= \frac{1}{|p|^2 \kappa_F^4} {g}_{(5,0)}  
\end{equation} 
where $g_{(5,0)}$ is a number (which is a  function of $\lambda_F$). 
In the limit $\lambda_F \to 0$ 
parity forces
\begin{equation}\label{gff}
{g}_{(5,0)}=0.
\end{equation}

\subsection{Duality} \label{du}

The critical boson and regular fermion theories have been conjectured to be related to each other 
via three different dualities (written here for a single matter field):
\begin{itemize}
\item $SU(N)$ regular fermion theories at (Yang-Mills regulated) level 
$k$ are dual to type 2 $U(|k|+\frac{1}{2})$ critical boson theories at level 
$-{\sgn (k)} N$. 
\item Type 2 $U(N)$ regular fermion theories at (Yang-Mills regulated) level 
$k$ are dual to $SU(|k|+ \frac{1}{2})$ critical boson theories at level 
$ - {\sgn (k)} N$. 
\item Type 1 $U(N)$ regular fermion theories at (Yang-Mills regulated) 
level $k$ are dual to $U(|k| + \frac{1}{2})$ Type 1 critical boson theories at 
(Yang-Mills regulated) level $-{\sgn (k)} N$. 
\end{itemize} 
At leading order in the 't Hooft large $N$ limit, the duality maps described above become identical, and can all be restated in the following form:
\begin{equation}
 \label{lndm}
\lambda_F =-\sgn(\lambda_B)(1- |\lambda_B|), \qquad\qquad
\frac{N_F}{|\lambda_F|} = \frac{N_B}{|\lambda_B|}, \qquad\qquad
{\tilde \kappa}_F= - \kappa_B.
\end{equation}
Under this duality map  the correlators of $J^F_0$ are identical at leading order 
to the correlators of ${\tilde J}_0$, up to a contact term in the three point function. Specifically
\begin{equation}
\begin{split}
&\langle {J}^F_0(q_1) { J}^F_0(q_2)   \rangle = \langle {\tilde J}_0(q_1) {\tilde J}_0(q_2)  \rangle, \\
&\langle {J}^F_0(q_1) { J}^F_0(q_2)   { J}^F_0(q_3)\rangle = \langle {\tilde J}_0(q_1) {\tilde J}_0(q_2) {\tilde J}_0(q_3)  \rangle - \frac{6 (2\pi)^2}{\kappa_B^2}  ~  (2 \pi)^3 \delta(q_1+q_2 + q_3),\\
&\langle {J}^F_0(q_1) { J}^F_0(q_2)   { J}^F_0(q_3){ J}^F_0(q_4)\rangle = \langle {\tilde J}_0(q_1) {\tilde J}_0(q_2) {\tilde J}_0(q_3){\tilde J}_0(q_4)\rangle. \\
\label{jcorrs}
\end{split}
\end{equation}
The shift in contact terms between three point functions listed in \eqref{jcorrs} follows by comparing \eqref{thptdet} and \eqref{thplnf} and using \eqref{rel1} below. \footnote{Recall that in the strict large $N$ limit we could not distinguish the power and contact parts of 
the three point function. As we have explained above, we do not yet know how much of the computed three-point function in either the 
critical boson or the regular fermion theory is to be attributed to the contact term and how much to the power law part of 
the correlators (see \eqref{thptdet}; the ambiguity is parameterized by $r_B(\lambda_B)$ in that equation). Even though we cannot 
disentangle the contact and power law contributions in the bosonic and fermionic theories individually, if we assume that the 
duality between these theories is valid, it follows that the difference between the three point functions in these theories 
is purely in the contact term, as the duality asserts that the power law part of the correlators between these theories 
must match. This leads to \eqref{jcorrs} and \eqref{rel1}.}

The conjectured duality between these two theories suggests that the match of correlation functions 
reported above persists -- up to contact terms and the shift of $k$ above -- to all orders in the $1/N$ expansion. 
The contact term appearing in \eqref{jcorrs} was computed only in the large $N$ limit. More generally the fermionic and bosonic three point functions are conjectured to be related via 
\begin{equation}\label{ctdiff}
\langle {J}^F_0(q_1) { J}^F_0(q_2)   { J}^F_0(q_3)\rangle-
\langle {\tilde J}_0(q_1) {\tilde J}_0(q_2) {\tilde J}_0(q_3)  \rangle
=-\frac{3! (2\pi)^2}{\kappa_B^2} (1+{\cal O} (\frac{1}{\kappa_B}) ) ~  (2 \pi)^3 \delta(q_1+q_2 + q_3).
\end{equation}

At leading order at large $N$ bosonic and fermionic answer map to each other under duality. In particular, under the duality map it is easily verified that 
\begin{equation}\label{rel}
{\tilde g}_{(4,1)}= {g}_{(4,1)},\qquad
{\tilde g}_{(4,2)}= {g}_{(4,2)},\qquad
{\tilde g}_{(4,3)}= {g}_{(4,3)}.
\end{equation}

Although five point functions have not explicitly been computed 
on either the bosonic or fermionic sides, the conjectured duality 
between the two theories leads us to expect that 
\begin{equation}\label{rel1}
{\tilde g}_{(5,0)}= {g}_{(5,0)}.
\end{equation}

The fact that three point functions must agree at separated points implies that
\begin{equation}\label{rel1}
(1-r_B(\lambda_B)) \left( \frac{\cot^2(\frac{\pi\lambda
_B}{2})}{\cos^2(\frac{\pi\lambda_B}{2})} -\frac{1}{4} \right) =  \left(1-r_F(\lambda_F)\right) \tan^2\left(\frac{\pi\lambda_F}{2}\right).
\end{equation}
\eqref{rel} is to be understood as follows. All terms of the left-hand side are evaluated 
at an arbitrary value of  $\lambda_B$. All terms on the right-hand side are evaluated 
at an arbitrary value of $\lambda_F$. The equality in \eqref{rel} holds 
provided $\lambda_B$ and $\lambda_F$ are related by \eqref{lndm}. 

The assumption of duality also implies that the 
anomalous dimensions of ${\tilde J}_0$ and $J_0^F$ are related; specializing 
to first subleading order in the $1/N$ expansion this implies that 
\begin{equation}\label{delrel}
\delta_B(\lambda_B)=  \delta_F(\lambda_F),
\end{equation}
provided $\lambda_B$ and $\lambda_F$ are related by \eqref{lndm}.

More generally, using \eqref{ctdiff}, the duality between the CB and RF 
theories implies the following relationship between the sourced partition functions of the bosonic and fermionic theories
\begin{equation}\label{dualitymap}
\int D\phi D \sigma e^{-S_{cb}(\phi, \sigma) +\int  {\tilde J}_0(x) \zeta(x)
- \frac{(2 \pi)^2}{\kappa_B^2}( 1 +{\cal O}(\frac{1}{\kappa_B} )) \int \zeta^3(x) }
= \int D\psi e^{-S_{rf}(\psi) +\int  J_0^F(x) \zeta(x) }.
\end{equation}

In this subsection we have, so far, discussed the $U(N)$ theory. There is a similar duality for $SO(N)$ theories (see \cite{Aharony:2016jvv} for details), and using 
\eqref{corggt} it follows that for $SO(N)$ 
\begin{equation}\label{dualitymapso}
\int D\phi D \sigma e^{-S_{cb}(\phi, \sigma) +\int  {\tilde J}_0(x) \zeta(x)
- \frac{(2 \pi)^2}{2 \kappa_B^2}( 1 + {\cal O}(\frac{1}{\kappa_B} )) \int \zeta^3(x) }
= \int D\psi e^{-S_{rf}(\psi) +\int  J_0^F(x) \zeta(x) }.
\end{equation}

\subsection{Duality from thermal partition functions} \label{pf}

The spectrum of operators of the CB $\sim$ RF 
theories includes a single relevant scalar operator ${\tilde J}_0 \sim J_0^F$
(here $\sim$ denotes equality under duality). It follows that the RG flow
that originates at the CB $\sim $ RF theory 
is unique, and corresponds to deforming the conformal 
theory by  $m {\tilde J}_0$.\footnote{More precisely, 
there are two such flows corresponding to $m$ positive or $m$ negative.}

It is possible to study these RG flows as a function of scale
by computing the free energy of the mass-deformed theories described above
as a function of the temperature. This calculation may 
be performed in the large $N$ limit, as we now review (see \cite{Aharony:2012ns,Jain:2013py,Takimi:2013zca} for details). 

Consider the mass-deformed critical boson theory defined by the action 
\begin{equation} \label{amd}
S=S_{CB}(\phi, \sigma) +\frac{N_B}{4\pi} \int m_B^{cri} \sigma= S_{CB} + \frac{N_B}{4\pi}\int m_B^{cri} {\tilde J}_0.
\end{equation}
Similarly consider the mass-deformed regular fermion theory 
\begin{equation} \label{amf} 
S=S_{RF}(\psi) +  \int m_F^{reg} \bar \psi \psi
= S_{RF}(\psi) +\frac{{\tilde \kappa}_F}{4\pi} \int m_F^{reg} {J}^F_0.
\end{equation}
The conjectured duality between the critical boson and regular fermion theories leads us to expect that equations \eqref{amd} and \eqref{amf} define the same theory provided that\footnote{Our theories are, throughout, defined using dimensional 
regularization. With this scheme it turns out that ${m}_B^{cri}$ is 
the pole mass of the critical boson at zero temperature. On the other hand 
the pole mass of the fermionic theory at zero temperature, $c_{F,0}$, is 
given by 
$c_{F,0}=\frac{{m}_F^{reg}}{\sgn(\lambda_F) -\lambda_F}.$ It follows that 
\eqref{relm} ensures that the dual bosonic and fermionic theories have equal 
pole masses.}
\begin{equation}\label{relm}
\frac{N_B}{4\pi} m_B^{cri}=\frac{{\tilde \kappa}_F}{4\pi} m_F^{reg}
\qquad \implies \qquad m_{F}^{reg}=-\lambda_B m_{B}^{cri}.
\end{equation}
We will now review evidence  that this is indeed the case.

The finite temperature  partition function of these theories on $S^2$ was 
computed, as a function of holonomies around the thermal circle,
in \cite{Jain:2013py}. As we are in the large $N$ limit, the result 
depends on eigenvalues only through an eigenvalue density function, 
$\rho_B(\alpha)$ (in the case of the bosonic theory) and $\rho_F(\alpha)$
(in the case of the fermionic theory). This computation proceeds  
as follows. One first sums Feynman diagrams to determine `offshell' 
partition functions
\begin{equation}\label{offshellpf} 
Z_B= e^{-T^2 V_2 F_B[ \rho_B(\alpha), c_B]}, ~~~  Z_F= e^{-T^2 V_2 F_F[ \rho_F(\alpha), c_F]}.
\end{equation} 
These partition functions are offshell because they 
depend on the additional variables $c_B$ and $c_F$, which have physical 
interpretations as the thermal pole masses of the bosonic and fermionic 
theories, respectively, in units of the temperature. 
The actual partition function is given by 
extremizing $F_B$ and $F_F$ with respect to $c_B$ and $c_F$, respectively, and plugging 
these extremized values into \eqref{offshellpf}. 
\footnote{This procedure gives the partition function as a functional of the 
eigenvalue density function $\rho(\alpha)$. The final partition function  
is obtained by integrating $Z[\rho(\alpha)]$ over holonomy eigenvalues with
the appropriate measure, as explained in detail in \cite{Jain:2013py}. 
 The requisite integrals 
can be evaluated using saddle point methods in the large $N$ limit
 \cite{Jain:2013py}.} 

The explicit results 
for $F_B$ and $F_F$, obtained by summing the appropriate infinite class of 
Feynman diagrams, are (see, for example, equations (3.7) and (3.12) of \cite{Jain:2013gza} and also the recent paper \cite{Choudhury:2018iwf} for the bosonic computation 
in the Higgsed phase)\footnote{In the equation below we have dropped 
the $c_B$ and $c_F$ independent  `zero temperature counter terms' included in  \cite{Jain:2013gza}. These counter terms were included 
by hand in  \cite{Jain:2013gza} to set the vacuum energy of both field theories to zero. The counter terms are field and temperature independent, 
and so do not impact thermodynamics. However they are mass dependent, and impact the computation of 
the quantum effective action of the theory as a function of ${\tilde J}_0$ (which is naively given by the 
Legendre transform of \eqref{offshellfe}). This ambiguity may lie at the heart of our confusions below
 concerning the stability of these theories with respect to condensation of ${\tilde J}_0$.}
\begin{equation} \label{offshellfe}
\begin{split}
F_B[ \rho_B(\alpha),c_B] &=\frac{N_B}{6\pi} {\Bigg[}  -
\frac{ \left(\lambda_B-{\rm sgn}(\lambda_B) - {\rm sgn} (X_B) \right) }{\lambda_B}|c_B|^3 +\frac{3}{2} {\hat m}_B^{cri} c_B^2\\
&+3 \int_{-\pi}^{\pi} \rho_B(\alpha) d\alpha\int_{|c_B|}^{\infty}dy y\left(\ln\left(1-e^{-y-i\alpha}\right)+\ln\left(1-e^{-y+i\alpha}\right)  \right)
 {\Bigg]},\\
 F_F[ \rho_B(\alpha),c_F] &=\frac{N_F}{6\pi} {\Bigg[}  |c_F|^3 \frac{\left(\lambda_F-\sgn(X_F)\right)}{\lambda_F} +\frac{3}{2 \lambda_F} {\hat m}_F^{reg} c_F^2\\
&-3 \int_{-\pi}^{\pi} \rho_F(\alpha) d\alpha\int_{|c_F|}^{\infty}dy y\left(\ln\left(1+e^{-y-i\alpha}\right)+\ln\left(1+e^{-y+i\alpha}\right)  \right)
 {\Bigg]},\\
 X_F&=2\lambda_F \cC+{\hat m}_F^{reg},\\
 X_B&= 2 \lambda_B \cS -{\rm sgn}(\lambda_B)|c_B| - m_B^{{\rm cri}} \lambda_B,
\end{split}
\end{equation}
where ${\hat m}_B^{cri}$ and ${\hat m}_F^{reg}$ are the masses divided by the temperature.
\footnote{Note that the terms in \eqref{offshellfe} that are independent of $|{\hat c}_B|$ and 
$|{\hat c}_F|$ are both proportional to $\frac{1}{T^3}$. A shift in these terms thus shifts 
the partition functions in \eqref{offshellpf} by terms proportional to $e^{-a \beta}$ 
and so represents a shift of the zero of energy of the theory in question by $a$. It follows 
that these constant terms are convention dependent and have no absolute physical significance. 
In the absence of a physical principle that determines their value, these terms can be retained or dropped at will.
}

As explained above, in order to evaluate the actual partition function of our 
theory we are instructed to extremize $F_B$ ($F_F$)
with respect to $c_B$ and $c_F$. The condition that $F$ be extremized gives 
us an equation -- called a gap equation -- that can be used to determine 
$c_B$ and $c_F$, respectively. The gap equation for the bosonic theory takes 
the form
\footnote{When $X_B$ and $\lambda_B$ have opposite signs, 
the second term on the left-hand side of \eqref{mdcb} vanishes. In this 
so called `unHiggsed' phase the bosonic gap equation simplifies
to
\begin{equation}
\label{uhge}
2 \cS	= {\hat m}_B^{{\rm cri} }.
\end{equation}
But when \eqref{uhge} is obeyed $X_B$ reported in 
\eqref{offshellfe} always has the opposite sign from $\lambda_B$. 
In other words every solution of \eqref{uhge} is a solution 
of the bosonic gap equations. On the other hand when 
$X_B$ and $\lambda_B$ have the same sign the bosonic theory 
is in the so called `Higgsed' phase and the bosonic gap equation 
becomes 
\begin{equation}\label{mdcbf}
2 \cS - \frac{2 }{|\lambda_B|} |{\hat c}_B|= {\hat m}_B^{{\rm cri} },
\end{equation}
in which case $X_B$ reported in \eqref{offshellfe} automatically 
has the same sign as $\lambda_B$. In other words, every solution
of \eqref{mdcbf} is also a solution of the bosonic gap equations. 
In other words the space of solutions of the bosonic gap 
equations is the union of the legal solutions to \eqref{uhge} 
and \eqref{mdcbf}.}
\begin{equation}\label{mdcb}
2 \cS - \frac{ \left( {\rm sgn}(\lambda_B) + {\rm sgn}(X_B) \right) }{\lambda_B} |{\hat c}_B|= {\hat m}_B^{{\rm cri} }.
\end{equation}

The gap equation for the fermionic theory is 
\begin{equation}\label{mdrf}
|c_F|=\sgn(X_F) \left( 2 \lambda_F \cC +{\hat m}_F^{{\rm reg}}\right),
\end{equation}
where
\begin{equation}\begin{split}
\cC =&\frac{1}{2} \int d\alpha\rho_F(\alpha)   \( \log(2 \cosh (\frac{|c_F| +i\alpha }{2}))+ \log(2 \cosh (\frac{|c_F| - i\alpha }{2})) \),\\
\cS =&\frac{1}{2}\int d\alpha\rho_B(\alpha)   \( \log(2 \sinh (\frac{|c_B| +i\alpha }{2}))+ \log(2 \sinh (\frac{|c_B| - i\alpha }{2})) \).
\label{ss}
\end{split}\end{equation}
The bosonic and fermionic holonomy eigenvalue distribution functions 
are related to each other by the formula (see \cite{Jain:2013py})
\begin{equation}\label{lrtr}
|\lambda_B| \rho_B (\alpha)+ |\lambda_F| \rho_F(\pi - \alpha) = \frac{1}{2 \pi}.
\end{equation}
When \eqref{lrtr} holds it is easily verified that 
\begin{equation}\label{dualofq1}
\lambda_B \cS=-\frac{\sgn(\lambda_F)}{2}|c_F|+\lambda_F \cC,\qquad\qquad
\lambda_F \cC=-\frac{\sgn(\lambda_B)}{2}|c_B|+\lambda_B \cS,
\end{equation}
where $\lambda_B$ and $\lambda_F$ are related as in \eqref{lndm}.

Using \eqref{dualofq1} and the first line of \eqref{lndm} it is easily verified
that the bosonic and fermionic offshell free energies \eqref{offshellfe} 
-- and so the gap equations \eqref{mdcb} and \eqref{mdrf} that follow 
from their extremization -- turn into each other (up to the addition of the physically insignificant cosmological constant counter-terms mentioned above) when the couplings of the 
two theories are identified by \eqref{lndm} and the masses of the two 
theories are related by \eqref{relm}. 
\footnote{Solutions of the fermionic theory that obey 
\begin{equation}\label{dualcondi}
\sgn(\lambda_F)\sgn(X_F)\ge 0
\end{equation}
map to solutions of the bosonic theory in the `unHiggsed' 
phase, while solutions of the fermionic theory that obey 
the converse of \eqref{dualcondi} map to solutions of the 
bosonic theory in the `Higgsed' phase. See \cite{Choudhury:2018iwf} for further discussion of this
point.}
This agreement gives powerful 
independent evidence for the duality between the CB and RF theories. 

\subsection{Phase Structure at 
	zero temperature}

The analysis of the previous 
section allows us immediately -- and very simply -- 
to determine the phase 
structure of the RF and 
CB theories at zero temperature.
This analysis is facilitated 
by the fact that the zero  temperature limit of the gap equations and offshell free energies reported in the previous subsection are particularly simple.  In this limit 
$|c_B|$ and $|c_F|$ both tend to infinity and 
\begin{equation}\label{zt}
\cC=\frac{|c_F|}{2}, ~~~~\cS=\frac{|c_B|}{2}, ~~~X_F=\lambda_F |c_F|+{\hat m}_F^{reg}.
\end{equation} 
After dropping constant terms, the  expressions \eqref{offshellfe} for the free energy reduce to 
\begin{equation} \label{offshellfezt}
\begin{split}
F_B[ \rho_B(\alpha),c_B] &=\frac{N_B}{6\pi} \left( - 
\frac{ \left(\lambda_B-{\rm sgn}(\lambda_B) - {\rm sgn} (X_B) \right) }{\lambda_B}|c_B|^3 +\frac{3}{2} {\hat m}_B^{cri} c_B^2 \right),\\
F_F[ \rho_B(\alpha),c_F] &=\frac{N_F}{6\pi} \left( |c_F|^3 \frac{\left(\lambda_F-\sgn(X_F)\right)}{\lambda_F} +\frac{3}{2 \lambda_F} {\hat m}_F^{reg} c_F^2 \right).
\end{split}
\end{equation}
The bosonic gap equation (which can be obtained either as the zero temperature limit of \eqref{mdcb} or from the 
variation of the first of \eqref{offshellfezt}) simplifies to 
\begin{equation}\label{mdcbzt}
\frac{ \left(\lambda_B-{\rm sgn}(\lambda_B) - {\rm sgn} (X_B) \right) }{\lambda_B}|c_B|= {\hat m}_B^{cri}.
\end{equation}
When $X_B \lambda_B < 0$ (i.e. in the unHiggsed phase) this 
equation simplifies to 
\begin{equation}\label{stuh}
|c_B|= {\hat m}_B^{cri};
\end{equation}
in other words the unHiggsed gap equation has 
exactly one solution when 
$m_B^{\rm cri}$ is positive, but no solutions
when $m_B^{\rm cri}$ is negative. Let us now 
turn to Higgsed solutions. In terms of the variable
\begin{equation}
{\hat \lambda}_B=-{\rm sgn}(\lambda_B)
(2 -|\lambda_B|) 
\end{equation}
the Higgsed zero temperature gap equation 
reduces to 
\begin{equation}\label{sthe}
\frac{{\hat \lambda}_B}{\lambda_B}|c_B|= m_B^{{\rm cri}}.
\end{equation}
As $\frac{{\hat \lambda}_B}{\lambda_B}$ is 
always negative, it follows that there exists 
exactly one Higgsed vacuum whenever 
$m_B^{{\rm cri}}$ is negative, and no 
Higgsed vacua when $m_B^{{\rm cri}}$ is 
positive. 

In summary the bosonic theory has a unique unHiggsed vacuum when $m_B^{{\rm cri}}$ is positive and a unique Higgsed vacuum whenever $m_B^{{\rm cri}}$ is negative. It follows that the CB theory undergoes
a (second order) phase transition, from a
unHiggsed to the Higgsed phase, as $m_B^{{\rm cri}}$ passes from positive to negative.

The fermionic gap equation (which can be obtained either as the zero temperature limit of \eqref{mdrf} or from the 
variation of the second of \eqref{offshellfezt}) 
simplifies to 
\begin{equation}\label{recast}
|c_F|= \frac{|m_F|}{1-|\lambda_F|} 
\end{equation}
when
\begin{equation}\label{sfc}
{m_F^{\rm reg}}
\geq 0,
\end{equation}
and to 
\begin{equation}\label{ztgen}
|c_F|=\frac{|{\hat m}_F^{reg}|}{1+|\lambda_F|}
\end{equation}
when
\begin{equation}\label{oppsgn}
{\hat m}_F^{reg}\lambda_F <0.
\end{equation}
Under the duality map, the 
condition \eqref{sfc} maps
to the $m_B^{{\rm cri}}\geq 0$ condition
and \eqref{recast} maps to 
\eqref{stuh}, while the converse
condition \eqref{oppsgn} maps
to the condition  $m_B^{{\rm cri}}< 0$ and \eqref{ztgen} 
maps to \eqref{sthe}. In other 
words the Fermionic theory 
in the parametric regime 
\eqref{sfc} maps to the 
critical boson theory 
in its unHiggsed phase, while 
the fermionic theory in the 
regime \eqref{oppsgn} maps 
to the CB theory in the 
Higgsed phase.

We end this section with a brief discussion of a confusing point. Focusing on the bosonic theory, we  have 
explained above that the large $N$ free energies in our theories are obtained 
by extremizing an offshell free energy with respect to $|c_B|$. This fact may 
tempt the reader to view the offshell free energy as a Landau Ginzburg 
free energy for the `order parameter' $|c_B|$ or $|c_F|$. In our view it is 
unclear that this is a correct viewpoint, or even what 
precisely this viewpoint 
might mean.

To see this, note that the zero temperature offshell free energy reported in \eqref{offshellfezt}
can be more explicitly rewritten as  
\begin{equation} \label{osbzt}
\begin{split}
&F_B[ \rho_B(\alpha),c_B] \\
&=\frac{N_B}{6\pi} \left( -|c_B|^3 +\frac{3}{2} {\hat m}_B^{cri} |c_B|^2 \right)  ~~~{\rm when} ~{\hat m}_B^{cri}>0, ~~~{\rm ~(unhiggsed)}\\
&=\frac{N_B}{6\pi} \left( |c_B|^3 \frac{\left(2-|{ \lambda}_B|\right)}{{ \lambda}_B} +\frac{3}{2} |{\hat m}_B^{cri}| |c_B|^2 \right) ~~~
{\rm when}~{\hat m}_B^{cri}<0. ~~~{\rm  ~(Higgsed) }
\end{split}
\end{equation}

It is clear from this equation that the zero mass 
limit of the off shell potential for $|c_B|$ is different depending 
on whether the mass approaches zero from above or below\footnote{In particular the coefficient of the cubic term in the second line of 
	\eqref{osbzt} is positive -- suggesting that the CB theory is stable -- while the same coefficient is negative in the first line, suggesting that the CB theory is unstable. }, while a genuine 
`Landau Ginzburg' potential should be well-defined at every value of the mass
including zero.  

Conservatively one should regard the offshell free energy as no more than an intermediate
device to be used for the computation of the onshell free energy. It would be 
interesting to find a genuine Landau Ginzburg potential for these theories and use it to analyze their stability etc. 
We leave this interesting task for future work.
 \footnote{The fact 
that the effective action for $c_B$ is unbounded from below would appear to 
imply that the quantum effective action for $\sigma_B$ is unbounded from 
below, naively suggesting that the bosonic theory is unstable at all 
values of $\lambda_B$ including $\lambda_B=0$. At $\lambda_B=0$, however, 
our theory is the much studied large $N$ Wilson Fisher theory, a theory 
which shows no evidence of instability. We think it is most likely that the
naive reasoning which suggests this instability is invalidated by a subtlety,
possibly related to contact terms as mentioned in the footnote under 
\eqref{offshellfe}. We leave a detailed clarification of this point 
to the future. See below for more comments.}

\section{Regular scalars and critical fermions and their RG flows} \label{betafunc}

\subsection{Definitions}

The regular bosonic theory (RB) is defined by the action 
\begin{equation}\label{rbt1n} 
S_{RB}(\phi, \sigma, \zeta) = S_{CB}(\phi, \sigma) -\int \sigma(x) \zeta(x)
+ \frac{(2 \pi)^2}{\kappa_B^2} \left(x_6^B+1 \right)\int \zeta^3(x) .
\end{equation}
Equivalently
\begin{equation}\label{rbt} 
S_{RB}(\phi, \sigma, \zeta) = S_{CB}(\phi, \sigma) -\int  {\tilde J}_0(x) \zeta(x)
+ \frac{(2 \pi)^2}{\kappa_B^2} \left(x_6^B+1 \right)\int \zeta^3(x) .
\end{equation}
Here $S_{CB}(\phi, \sigma)$ is the action \eqref{cst} for the critical bosonic theory,  $\zeta$ 
is a new dynamical field and $x_6^B$ is a parameter.\footnote{In \eqref{rbt} $x_6^B$ is shifted by $1$ in order to account for the difference in the contact term between the bosonic and fermionic theory, see \eqref{ctdiff}. We have accounted for this difference at leading order in large $N$ limit. In order to define the regular boson theory at finite $N$, this shift should be replaced by the corrected shift $(1+\delta_6)$ between the contact terms in the two theories.}

If we insert \eqref{cst} into \eqref{rbt} and integrate out $\sigma$ we
find that the action in \eqref{rbt} reduces to the regular boson action\footnote{Note, in particular, that $\zeta$ is an operator of order $N$, explaining the 
factor of $\frac{1}{\kappa_B^2}$ in the coefficient of $\zeta^3$ in 
\eqref{rbt}.} 
\begin{equation}\label{frebs}\begin{split}
& S_{RB}(\phi)=  \int d^3 x  \biggl[i \varepsilon^{\mu\nu\rho}{\kappa_B\over 4\pi}
\Tr( A_\mu\partial_\nu A_\rho -{2 i\over3}  A_\mu A_\nu A_\rho)
+i \varepsilon^{\mu\nu\rho} {N_B \kappa'_B \over 4 \pi}
 B_\mu\partial_\nu B_\rho  \\
&\qquad\qquad\qquad\qquad
 + D_\mu \bar \phi D^\mu\phi  + \frac{(2 \pi)^2}{\kappa_B^2} \left(x_6^B+1\right) ( \bar \phi \phi)^3 \biggl],
\end{split}
\end{equation} 
since we have
\begin{equation}\label{wiz}
\zeta= {\bar \phi}{\phi}.
\end{equation}

In a similar manner the critical fermionic (CF) theory is defined by 
the action\footnote{ The reader may find our definition of the regular boson and critical fermion theories 
suspiciously formal. The equivalence of
\eqref{rbt} and \eqref{frebs} makes clear, however, that the regular boson theory is just a usual quantum field theory
written in a complicated manner. At least in the $1/N$ expansion, bosonization duality then ensures that the 
same is true for the critical fermion theory. }
 \begin{equation}\label{cft}
S_{CF}(\psi, \zeta)=S_{RF}(\psi) -\int  J_0^F(x) \zeta(x) + \frac{(2 \pi)^2}{{\tilde \kappa}_F^2} 
x_6^F\int \zeta^3(x).
\end{equation}
Note that the relevant operators in this theory are $\zeta$ and $\zeta^2$; the latter operator is proportional to $J_0^F$ by the equation of motion of $\zeta$.

It follows from \eqref{dualitymap} that \eqref{rbt} and \eqref{cft} define 
identical theories when $x_6^B = x_6^F$. This can be seen explicitly as follows. Let us define two new actions,
$S^{eff}_{RB}(\zeta)$ and $S^{eff}_{CF}(\zeta)$, using the following definitions:
\begin{equation}\label{seffdef}
\begin{split}
&\int D \phi D \sigma e^{-S_{RB}(\phi, \sigma, \zeta)}=e^{-S^{eff}_{RB}(\zeta)}, \\
&\int D \psi e^{-S_{CF}(\psi, \zeta)}=e^{-S^{eff}_{CF}(\zeta)} .
\end{split}
\end{equation}
Then the conjectured duality between the critical boson and regular fermion theory implies that for $x_6^B = x_6^F$
\begin{equation}\label{effactrel}
S^{eff}_{RB}(\zeta; \kappa_B, \lambda_B) = S^{eff}_{CF}(\zeta; {\tilde \kappa}_F, \lambda_F),
\end{equation}
provided ${\tilde \kappa}_F$ and $\lambda_F$ are related to $\kappa_B$ and $\lambda_B$ via 
\eqref{lndm} (and the values of $x_6$ agree between the two theories). More explicitly
\begin{equation} \label{eare}
S^{eff}_{RB}(\zeta; \kappa_B,\lambda_B) = S^{eff}_{CF}(\zeta; -\kappa_B,\lambda_B-\sgn(\lambda_B)).
\end{equation}
Similar relations hold also for finite values of $N$ and $\kappa$ (related as in the previous section).

It is easy to obtain formal expressions for the two effective actions described above in terms of 
the generators of correlation functions of the critical boson and regular fermion theories. Note that these effective actions are highly non-local, since $\zeta$ is not dynamical. In 
particular 
\begin{equation}\label{rbtzeta}
\begin{split}
&S^{eff}_{RB}(\zeta)= \frac{(2 \pi)^2}{\kappa_B^2} \left(x_6^B+(1+ \delta_6)\right)\int \zeta^3(x) 
-\sum_{n=2}^{\infty} \frac{1}{n!} \int  dp_n 
\langle {\tilde J}_0(-p_1) \ldots {\tilde J}_0(-p_n) \rangle_{CB} \zeta(p_1) \ldots \zeta(p_n), ~
 \\
&S^{eff}_{CF}(\zeta)= \frac{(2 \pi)^2}{{\tilde \kappa}_F^2} x_6^F \int \zeta^3(x)-\sum_{n=2}^{\infty} \frac{1}{n!} \int dp_n
\langle J^F_0(-p_1) \ldots J^F_0(-p_n)  \rangle_{RF} \zeta(p_1) \ldots \zeta(p_n), \\
 &dp_n\equiv (2 \pi)^3 \delta(\sum_{i=1}^n p_i) \left( \prod_{i=1}^n \frac{ d^3 p_i}{(2 \pi)^3} \right).
\end{split}
\end{equation}
Equation \eqref{effactrel} follows from \eqref{rbtzeta} together with the CB-RF duality, which implies that 
$\langle {\tilde J}_0(p_1) \ldots {\tilde J}_0(p_n) \rangle_{CB}$ (the correlators of the CB
theory) agree with $\langle {J}^F_0(p_1) \ldots {J}^F_0(p_n) \rangle_{RF}$ 
(the correlators of the RF theory) up to the map \eqref{lndm} between the 
parameters of the two theories.

Modulo issues related to divergences that we will examine below, the RB theory is defined 
by the path integral 
\begin{equation}\label{pirb}
 \int d \zeta e^{-S^{eff}_{RB}(\zeta)}.
\end{equation}
We denote the quantum effective action generating the 1PI correlation functions associated with this path integral by 
$S^{1PI}_{RB}(\zeta)$. In a similar way the critical fermion theory is defined by the path integral 
\begin{equation}\label{picf}
\int d \zeta e^{-S^{eff}_{CF}(\zeta)},
\end{equation}
and we denote the 1PI effective action associated with this path integral by 
$S^{1PI}_{CF}(\zeta)$. It follows from \eqref{effactrel} that 
 \begin{equation}\label{effactop}
 S^{1PI}_{RB}(\zeta)=S^{1PI}_{CF}(\zeta),
 \end{equation}
where the two IPI effective actions are evaluated at equal values of $x_6=x_6^B=x_6^F$ but at values 
of ${\tilde \kappa}_F$ and $\lambda_F$ that are given in terms of $\kappa_B$ and $\lambda_B$ 
by \eqref{lndm}.

\subsection{$SO(N)$ theory}

In the case of the $SO(N)$ theory, we use the following modified actions 
to define the regular boson and critical fermionic theories:
\begin{equation}\label{rbtso} 
S_{RB}(\phi, \sigma, \zeta) = S_{CB}(\phi, \sigma) -\int  {\tilde J}_0(x) \zeta(x)
+ \frac{(2 \pi)^2}{2 \kappa_B^2} \left(x_6^B+1\right)\int \zeta^3(x) ,
\end{equation}
\begin{equation}\label{cftso}
S_{CF}(\psi, \zeta)=S_{RF}(\psi) -\int  J_0^F(x) \zeta(x) + \frac{(2 \pi)^2}{ 2{\tilde \kappa}_F^2} 
x_6^F\int \zeta^3(x).
\end{equation}
Note the additional factor of $\frac{1}{2}$ in the term proportional to 
$\zeta^3$ in comparison with \eqref{rbt} and \eqref{cft}. With these 
definitions \eqref{effactrel} and \eqref{effactop} still apply. Note that
at leading order in the large $N$ expansion 
\begin{equation} \label{sesu}
S^{eff}_{RB}(\zeta)_{SO(N)}= \frac{1}{2} S^{eff}_{RB}(\zeta)_{SU(N)}
\end{equation}

\subsection{The $\beta$ function for $x_6$}

The path integrals \eqref{seffdef}  are well-defined, and conjectured to be equal, only after renormalization. 
If these path integrals are computed with a momentum space UV cutoff 
$\Lambda$, then, in order to obtain sensible results for correlators at fixed 
distance,  $x_6$ has to be chosen to be a function of $\Lambda$ 
as $\Lambda$ is taken to infinity. 
More precisely, as $\Lambda$ is taken to infinity
we must choose $x_6$ to simultaneously flow according to the equation
\begin{equation}\label{bf}
\frac{ d x_6}{d \ln (\Lambda)}= \beta(x_6).
\end{equation}

In the next few subsections we compute the $\beta$ function 
$\beta(x_6)$. In order to do this we compute the 
1PI effective action for $\zeta$, rescale $\zeta$ appropriately to 
eliminate $\ln (\Lambda)$ from quadratic terms in the action and then 
read off the $\beta$ function in \eqref{bf} from the requirement that cubic terms in 
\eqref{effactop} are also independent of $\ln (\Lambda)$. 

Let us first note that the $\Lambda$ dependence of all terms in 
$S^{eff}$ is tightly controlled by conformal invariance. Recall that 
the coefficient of $\zeta^n$ in $S^{eff}(\zeta)$ is simply $G_n$, 
the correlator of $n$ ${\tilde J}_0$ operators. It follows from 
conformal invariance that the power law (i.e. non-contact) 
parts of these correlators scale like $\Lambda^{ \frac{\delta_B}{|\kappa_B|}}$.\footnote{The reader might think that this $\Lambda$ dependence is 
relatively trivial, and  can be removed in one fell swoop by the field redefinition 
$\zeta = \zeta' \Lambda^{-\frac{\delta_B(\lambda_B)}{|\kappa_B|}}$. 
This is not accurate. While this field redefinition does remove the 
$\Lambda$ dependence from power law contributions to $G_n$, 
it introduces spurious $\Lambda$ dependence into (previously 
$\Lambda$ independent) contact term contributions to $G_n$. }
On the other hand contact contributions are independent of $\Lambda$.
It follows that $S^{eff}(\zeta)$ is independent of $\ln (\Lambda)$ 
at leading order in the large $N$ limit. The dependence of 
$S^{eff}$ on $\Lambda$ at first subleading order in $\frac{1}{N}$ 
is non-trivial, but very simple. It arises entirely from the 
expansion of $\Lambda^{ \frac{\delta_B}{|\kappa_B|}}$ in a Taylor
series expansion in the anomalous dimension $\delta_B$. 

In addition to the explicit factors of $\ln (\Lambda)$ in $S^{eff}(\zeta)$, 
$S^{1PI}$ has additional $\Lambda$ dependence from divergences in loops. 
As we have explained in the introduction, the dynamics governed by $S^{eff}_{RB}(\zeta)$ is weakly 
coupled at large $N$; in a canonical normalization $N$ sits outside 
$S^{eff}_{RB}(\zeta)$ as an overall factor. It follows that $\frac{1}{N}$ 
is a loop counting parameter for loops generated by $S^{eff}_{RB}(\zeta)$. 
In this paper we restrict attention to first order in the $\frac{1}{N}$ 
expansion; at this order the 1PI effective action for $\zeta$ receives 
contributions only from one loop graphs generated by $S^{eff}_{RB}(\zeta)$.
These one loop graphs are sometimes divergent, and generate a second
source of $\Lambda$ dependence. 

Adding the explicit $\ln (\Lambda)$ dependence in $S^{eff}$ to the additional 
$\ln (\Lambda)$ dependence from loops, we have the full $\Lambda$
dependence of the 1PI effective action for $\zeta$, from which we extract the $\beta$ function. 

Note that $x_6$ appears as a coefficient in the leading order
part of $S^{eff}_{RB}$. On the other hand all divergences appear only at first subleading order in $\frac{1}{N}$. It follows immediately that $\beta(x_6)$ is of order $\frac{1}{N}$, and in 
particular it vanishes in the strict large $N$ limit. 

In addition to the `slow' flows described so far, the regular boson/critical 
fermion theories also have faster RG flows, seeded by operators that 
are relevant even at large $N$. These operators are $\zeta^2$ 
and $\zeta$. In the later part of this section we will turn to a discussion 
of these faster RG flows.

\subsection{Explicit form of $S^{eff}_{RB}(\zeta)$}

Equation \eqref{rbtzeta} gives a definition of the function 
$S^{eff}_{RB}(\zeta)$. In this brief subsection we use the 
explicit results for two, three and four point functions 
in the previous section to obtain an explicit expression for 
$S^{eff}_{RB}(\zeta)$, subject to the following limitations:
\begin{itemize} 
\item We only list those terms in the effective action that are 
of quintic or lower order in $\zeta$. Terms in the effective action 
of order $\zeta^6$ or higher are ignored in our listing.
\item We list  quintic and quartic terms in the effective action only at leading 
order in $\frac{1}{N}$, and that too only in the kinematical regime that we 
focused on in the previous section (see below for more details).
\item We list quadratic and cubic terms in this effective action 
only at leading and first subleading order in $\frac{1}{N}$. Moreover 
we only list those subleading terms that depend on the UV coupling 
$\Lambda$; we ignore first subleading corrections in $\frac{1}{N}$ 
that are finite as $\Lambda \to \infty$.
\end{itemize}

The rational for the rather strange set of restrictions listed above is that
the set of terms chosen by the rules described above are precisely the 
terms that contribute to the $\beta$ function of $x_6$ at order $\frac{1}{N}$.

Subject to all the restrictions described above we have 
\begin{equation}\label{termlisy} \begin{split}
&S^{eff}_{RB}=  \frac{g_2}{2 \kappa_B}
\int \frac{d^3q}{(2 \pi)^3} |q|\left( 
\frac{|q|}{\Lambda} \right)^{\frac{2 \delta_B(\lambda_B)}{\kappa_B}}
\zeta(q) \zeta (-q) \\
&+ \frac{g_3}{6 \kappa_B^2} 
\int \frac{d^3 q_1}{(2 \pi)^3}
\frac{d^3 q_2}{(2 \pi)^3}
\frac{d^3 q_3}{(2 \pi)^3} (2 \pi)^3 \delta(q_1+q_2 +q_3) 
 \zeta(q_1) \zeta(q_2) \zeta(q_3) \\
 &+ \frac{{\tilde g}_3}{6 \kappa_B^2} 
\int \frac{d^3 q_1}{(2 \pi)^3}
\frac{d^3 q_2}{(2 \pi)^3}
\frac{d^3 q_3}{(2 \pi)^3} (2 \pi)^3  \frac{\delta_{B}}{\kappa_B} \ln\left(\frac{\Lambda}{|q_1|+|q_2|+|q_3|}\right) \delta(q_1+q_2 +q_3) 
 \zeta(q_1) \zeta(q_2) \zeta(q_3) \\
&- \frac{1}{24 \kappa_B^3} \int  \prod_{i=1}^{4} \frac{d^3 q_i}{(2 \pi)^3}(2 \pi)^3 \delta(q_1+q_2 +q_3 + q_4 ) 
\kappa_B^3 {\tilde G}^0_4(q_1, q_2, q_3, q_4)
\zeta(q_1) \zeta(q_2) \zeta(q_3) \zeta(q_4) \\
&- \frac{1}{5! \kappa_B^4} \int \prod_{i=1}^{5} \frac{d^3 q_i}{(2 \pi)^3} 
(2 \pi)^3 \delta(q_1+q_2 +q_3 + q_4 +q_5) 
\kappa_B^4 {\tilde G}^0_5(q_1, q_2, q_3, q_4,q_5)
\zeta(q_1) \zeta(q_2) \zeta(q_3) \zeta(q_4)\zeta(q_5),
\end{split}
\end{equation}
with the definitions
\begin{equation}\begin{split} \label{termdef}
g_2&= \left( \frac{4 \pi}{\tan(\frac{\pi\lambda_B}{2})} \right) = -\left( \frac{4 \pi}{\cot(\frac{\pi\lambda_F}{2})} \right),\\
g_3&= (24 \pi^2) 
\left( 1+ x_6^B -\frac{4}{3} \left( \frac{\cot^2(\frac{\pi\lambda_B}{2})}{\cos^2(\frac{\pi\lambda_B}{2})} -\frac{1}{4} \right)  \right)\\
&= (24 \pi^2) 
\left( x_6^B -\frac{4}{3} \cot^2 \left( \frac{ \pi \lambda_B}{2} \right) \right)= (24 \pi^2) \left( x_6^F -\frac{4}{3} \tan^2 \left( \frac{ \pi \lambda_F}{2} \right)
\right),\\
{\tilde g}_3&=-3 r_B(\lambda_B) 32 \pi^2 \left( \frac{\cot^2(\frac{\pi\lambda_B}{2})}{\cos^2(\frac{\pi\lambda_B}{2})} -\frac{1}{4} \right) \\
&=-3\cdot 32 \pi^2 \left( \frac{3}{4}+ r_F(\lambda_F) \tan^2\left(\frac{\pi\lambda_F}{2}\right)\right),\\
\kappa_B^3 {\tilde G}^0_4(p, -p, k, -k)&= \frac{1}{|p|} \left({\tilde g}_{(4,1)}  + {\tilde g}_{(4,2)} \frac{|k|}{|p|} +  
{\tilde g}_{(4,3)} \frac{(p\cdot k)^2}{|p|^3|k|} + {\cal O}(k^2/p^2)\right),\\
{\tilde g}_{(4,1)}&= -2{\tilde g}_4, \qquad
{\tilde g}_{(4,2)}= {\tilde g}_4,\qquad
{\tilde g}_{(4,3)}= {\tilde g}_4, ~~~
{\tilde g}_4 =  \frac{2^9\pi^3}{(\tan(\frac{\pi\lambda_B}{2}))^2 \sin(\pi\lambda_B)} \\
\kappa_B^4 {\tilde G}^0_5(p, -p, 0,0,0)&= \frac{1}{p^2} {\tilde g}_{(5,0)}. \\
\end{split}
\end{equation}

\subsection{Computation of $S^{1PI}_{RB}(\zeta)$}

At leading order at large $N_B$ $S^{1PI}_{RB}(\zeta)$ agrees with 
$S^{eff}_{RB}(\zeta)$. At first subleading order in $\frac{1}{N_B}$, 
$S^{1PI}_{RB}(\zeta)$ has corrections over $S^{eff}_{RB}(\zeta)$
coming from one loop diagrams with the leading large $N_B$ part of 
$S^{eff}_{RB}(\zeta)$ thought of as the classical action. In this 
subsection we will compute the relevant one loop graphs to find 
the first correction to the quantum effective action -- i.e. to determine 
$(S^{1PI}_{RB}(\zeta) - S^{eff}_{RB}(\zeta))$ at order $\frac{1}{N_B}$. 
We will only be interested in those corrections to the effective action 
that depend on the UV cutoff scale $\Lambda$, and will ignore corrections 
that are finite as $\Lambda \to \infty$. 

\subsubsection{Computation of $S^{1PI}_{RB}(\zeta) - S^{1PI}_{RB}(\zeta)$ 
at quadratic order}

\begin{figure}[h]
\begin{center}
\includegraphics[width=9.5cm,height=3.5cm]{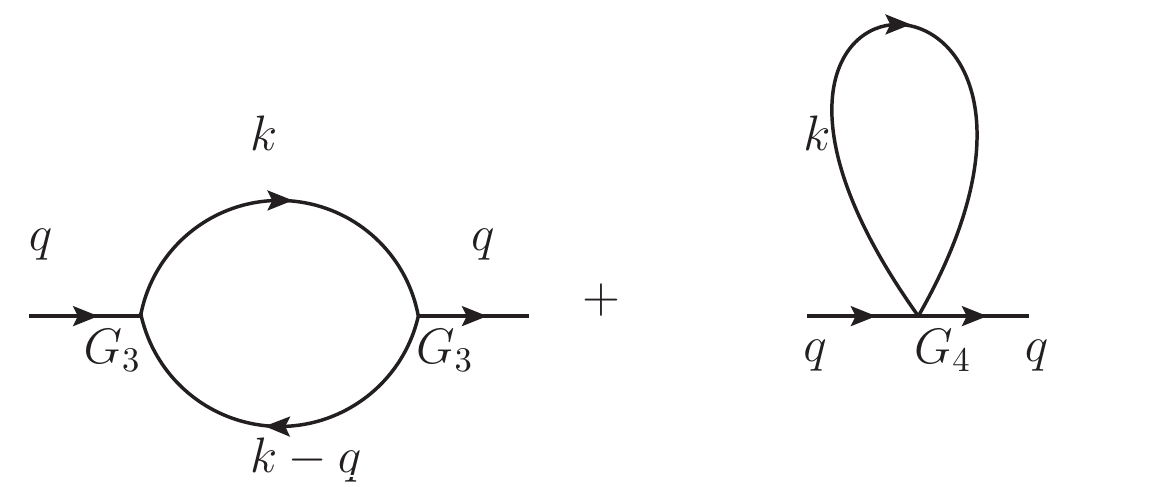}
  \caption{\label{2pt1loop} Diagrams that contribute at one loop order to the quadratic term \eqref{sopi} in the effective action. }
 \end{center}
\end{figure}
At one loop order and restricting attention to terms quadratic in 
$\zeta$ we have (see figure \ref{2pt1loop})
\begin{equation}\label{sopi} \begin{split} 
& S^{1PI}_{RB}(\zeta) - S^{eff}_{RB}(\zeta)= \frac{1}{2 \kappa_B^2} 
\int  \frac{d^3q}{(2 \pi)^3} \delta \Gamma_2(q) \zeta(q) \zeta(-q),\\
& \delta \Gamma_2 (q)= -\frac{g_3^2}{2 g_2^2} 
\int \frac{d^3k}{(2 \pi)^3} \frac{1}{|k||k-q|} - 
\frac{1}{2 g_2} \int \frac{d^3k}{(2 \pi)^3} 
\frac{ {\tilde G}^0_4( q, -q, k, -k)}{|k|}.
\end{split}
\end{equation}
The integral in the first term in the second line of \eqref{sopi} 
is easily evaluated in dimensional regularization: 
\begin{equation}\label{firsint}
\int \frac{d^3k}{(2 \pi)^3} \frac{1}{|k||k-q|} = -\frac{|q|}{4\pi^2}.
\end{equation}
Note in particular that the integral \eqref{firsint} has a linear divergence but not a logarithmic divergence, 
so it does not contribute to the $\beta$ functions we compute below. 
The second term on the second line of \eqref{sopi} cannot be evaluated
in general, as we do not know the full Green's function 
${\tilde G}^0_4( q, -q, k, -k)$. However logarithmic 
divergences in this integral at large $k$ can only arise from terms in ${\tilde G}^0_4$ 
that have the homogeneity of $\frac{|q|}{k^2}$. The relevant terms are 
those proportional to ${\tilde g}_{(4,2)}$ and
${\tilde g}_{(4,3)}$.  Using the fact that $\cos^2 (\theta)$ averages to 
$\frac{1}{3}$ on the two-sphere, the divergent
part of $\delta \Gamma_2$ is 
\begin{equation}\label{dgttwo}
\delta \Gamma_2 (q)= - 
\left( \frac{{\tilde g}_{(4,2)} + \frac{{\tilde g}_{(4,3)}}{3} }{2 g_2} \right) 
\int \frac{d^3k}{(2 \pi)^3} \frac{|q|}{|k|^3} = 
- \left(  \frac{3 {\tilde g}_{(4,2)} + {\tilde g}_{(4,3)} }{12 \pi^2  g_2} \right) 
|q| \ln \left(\frac{\Lambda}{|q|}\right).
\end{equation}
Adding this piece to the quadratic part of $S^{eff}_{RB}(\zeta)$ (see 
\eqref{termlisy}) we conclude that the divergent part of the quadratic terms in the 
quantum effective action, accurate to one loop, is given by 
\begin{equation}\label{divpee} \begin{split}
S^{1PI}_{RB}& =  \frac{g_2}{2 \kappa_B}
\int \frac{d^3q}{(2 \pi)^3} |q|\left( 
\frac{|q|}{\Lambda} \right)^{\frac{2 \delta'_B(\lambda_B) }{\kappa_B}}
\zeta(q) \zeta (-q), \\
2 \delta'_B(\lambda_B) &= 2 \delta_B + \frac{3 {\tilde g}_{(4,2)} + {\tilde g}_{(4,3)} }{12 \pi^2  g_2^2}.\\
\end{split}
\end{equation} 

Using 
\begin{equation}
\label{valoq}
 \frac{3 {\tilde g}_{(4,2)} + {\tilde g}_{(4,3)} }{12 \pi^2  g_2^2}=\frac{32}{3 \pi \sin \pi \lambda_B}
\end{equation}
we conclude that 
\begin{equation}\label{divpeen}
\delta_B'-\delta_B= \frac{16}{3 \pi \sin \pi \lambda_B}
\end{equation}

Note that \eqref{divpee} implies that 
\begin{equation}\label{zetp}
\langle \zeta(p) \zeta(-p') \rangle \propto 
(2 \pi)^3 \delta^3( p-p') \frac{1}{|q|^{1+2 \delta'_B / \kappa_B}}.
\end{equation}
It follows that the scaling dimension of $\zeta$, $\Delta_\zeta$, is 
given by 
\begin{equation}\label{zsd}
\Delta_\zeta= 1- \frac{\delta'_B}{\kappa_B}.
\end{equation}
Recall on the other hand that 
\begin{equation}\label{ssd}
\Delta_\sigma= 2+ \frac{\delta_B}{\kappa_B}.
\end{equation}
The formula \eqref{divpeen} thus asserts that at order $\frac{1}{N}$, the difference between minus of the anomalous dimension of the (approximately) dimension
one operator of the regular boson theory and the 
anomalous dimension of the  (approximately)
dimension 2 operator of the critical boson theory is given by 
$\frac{32}{ 3 \pi^2 {\tilde N}}$ (where the quantity 
	${\tilde N}$ is the abstract parameter in terms of which the authors of \cite{Maldacena:2012sf} determined three point functions
	of the theory). It is possible that this elegant formula 
	has a simple explanation.

\subsubsection{Check at small $\lambda_B$}

As a check of our formulae  we now use 
\eqref{divpee} to compute $\delta'_B$ in the limits $\lambda_B \to 0$ and 
$\lambda_F \to 0$. 

In the limit $\lambda_B=0$ explicit computations yield 
(see \eqref{gfree} and \eqref{deltab})
\begin{equation}\label{boval}
\delta_B= -\frac{16}{3 \pi^2 \lambda_B},\qquad
\end{equation}

It follows from \eqref{divpeen} that $\delta_B'=0$. 
The fact that $\delta_B'$ vanishes at $\lambda_B=0$ 
is a consistency check of our formalism, as 
the regular boson theory is free when $\lambda_B=0$ (and when $g_3$ also vanishes). 

\subsubsection{Check at small $\lambda_F$}

Recall that the regular fermion theory is free at $\lambda_F=0$; 
it thus follows that $\delta_B=0$ when $\lambda_F=0$.   \eqref{divpeen} then implies that 
\begin{equation}\label{deff}
\delta'_B= - \frac{16}{3 \pi^2 \lambda_F}.
\end{equation}
In other words (see \eqref{zsd}) the dimension of the approximately dimension one operator in the ungauged 3d 
Gross-Neveu model is predicted by \eqref{divpeen} to be 
\begin{equation}\label{ds}
\Delta_\zeta= 1+ \frac{16}{3 \pi^2 \lambda_F \kappa_B}
= 1-\frac{16}{3 \pi^2 N_F},
\end{equation}
where we have used $\kappa_F= - \kappa_B$. The prediction 
\eqref{ds} is correct; it matches the anomalous dimension independently computed in,  for example, \cite{Giombi:2016zwa} and references therein.

\subsubsection{$SO(N)$ theory} \label{son} 

As we have already noted, at leading order in the large $N$ limit $S^{eff}_{SO(N)}(\zeta)=\frac{1}{2}S^{eff}_{SU(N)}(\zeta)$, 
which implies that to get $S^{eff}_{SO(N)}$ we need to take
\begin{equation}
g_i\rightarrow \frac{1}{2}g_i ~~{\rm{for}}~i=2,3,4,\cdots.
\end{equation}
For the $SO(N)$ theory the anomalous dimension of $\zeta$ that follows from this is \eqref{divpee}
\begin{equation} \begin{split}
2 \delta^{',SO(N)}_B(\lambda_B) = 2 \delta_B^{SO(N)} + 2\left(\frac{3 {\tilde g}_{(4,2)} + {\tilde g}_{(4,3)} }{12 \pi^2  g_2^2}\right).\\
\end{split}
\end{equation} 
Recall that $\delta_B^{SO(N)}= 2 \delta_B^{SU(N)}$, and thus it follows that 
\begin{equation}
\delta^{',SO(N)}_B=2  \delta^{',SU(N)}_B, 
\end{equation}
hence the anomalous dimension of $\zeta$ in the $SO(N)$ theory is twice that of the $SU(N)$ theory.

It follows that for the $SO(N_F)$ theory 
\begin{equation}\label{dsso}
\Delta_\zeta
= 1-\frac{32}{3 \pi^2 N_F}.
\end{equation}
\eqref{dsso} agrees with the anomalous dimension of
$\zeta$ in the $SO(N_F)$ critical fermion theory at $\lambda_F=0$ 
(see e.g. equation (3.12) in \cite{Giombi:2017rhm}).

\subsubsection{Computation of $S^{1PI}_{RB}(\zeta) - S^{eff}_{RB}(\zeta)$ 
at cubic order}

\begin{figure}[h]
\begin{center}
\includegraphics[width=9.5cm,height=3.5cm]{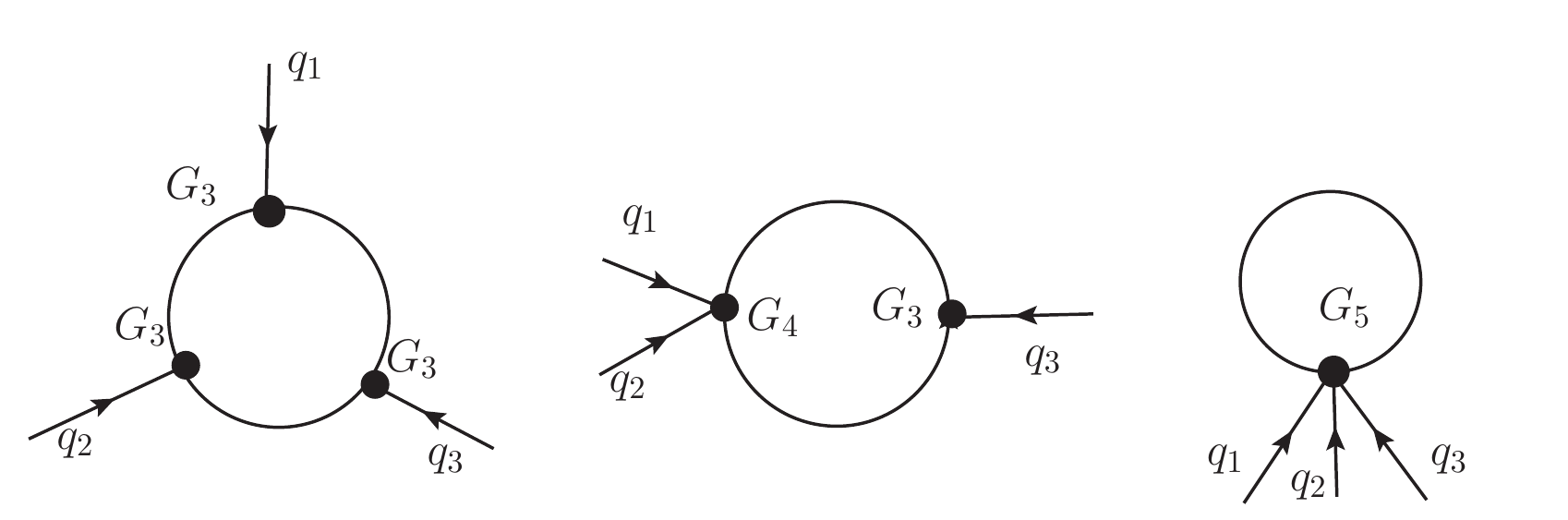}
  \caption{\label{3pt1loop} Diagrams that contribute to the one-loop correction \eqref{sopith} to the effective action. }
 \end{center}
\end{figure}

The diagrams that contribute to the one loop renormalization of the cubic part of the effective action are depicted
in Figure \ref{3pt1loop}. Evaluating these diagrams we find 
\begin{equation}\label{sopith} \begin{split} 
S^{1PI}_{RB}(\zeta) - S^{eff}_{RB}(\zeta)&= \frac{1}{6 \kappa_B^3} 
\int  \left( \prod_i \frac{d^3q_i}{(2 \pi)^3} \right) \delta \Gamma_3(q_1, q_2, q_3) 
(2 \pi)^3 \delta(q_1 +q_2 + q_3)
\zeta(q_1) \zeta(q_2) \zeta(q_3)\\
\delta \Gamma_3 (q_1, q_2, q_3)
&= \int \frac{d^3 q}{(2\pi)^3} \frac{G_3^S(q_1,q,-q-q_1) G_3^S(q_2,-q+q_3,q+q_1) 
G_3^S(q_3,-q,q-q_3)}{ G_2^0(q+q_1) G_2^0(q-q_3) G_2(q)}\\
&\qquad-\frac{3}{2}\int \frac{d^3 q}{(2\pi)^3} \frac{{\tilde G}_4^0(q_1,q_2,-q+q_3,q) G_3^S(q-q_3,-q,q_3) }{G_2^0(q) G_2^0(q-q_3)} \\
&\qquad+\frac{1}{2}\int \frac{d^3 q}{(2\pi)^3}   \frac{{\tilde G}_5^0(q_1,q_2,q_3,q,-q) }{G_2^0(q)}\\
&= \frac{g_3^3}{ g_2^3} 
\int \frac{d^3q}{(2 \pi)^3} \frac{1}{|q||q-q_3| |q+q_1|} +\\
&\qquad \frac{3g_3}{2 g_2^2} \int \frac{d^3q}{(2 \pi)^3} 
\frac{ {\tilde G}^0_4( q_1, q_2, q, q_3-q)}{|q_3-q||q|}-
\frac{1}{2 g_2} \int \frac{d^3q}{(2 \pi)^3} 
\frac{ {\tilde G}^0_5( q, -q, 0,0,0)}{|q|}\\
&\simeq \frac{1}{2 \pi^2 } \left( \frac{g_3^3}{ g_2^3} + 
\frac{3g_3 {\tilde g}_{(4,1)} }{2 g_2^2}-\frac{{\tilde g}_{(5,0)}}{2 g_2} \right) \ln (\Lambda),\\
G_3^S(q_1, q_2, q_3)&\equiv -\frac{g_3}{6 \kappa_B^2} (2\pi)^3
\delta(q_1+q_2+q_3),\\
\end{split}
\end{equation}
where $g_3$ was listed in \eqref{termdef} and where we have retained only divergent terms. It follows 
that the one loop corrected cubic term in the quantum effective action 
relevant for the computation of the $\beta$ function is given by 
\begin{equation}\label{divpeeth} \begin{split}
S^{1PI}_{RB}& =  
 \frac{g'_3}{6 \kappa_B^2} 
\int \frac{d^3 q_1}{(2 \pi)^3}
\frac{d^3 q_2}{(2 \pi)^3}
\frac{d^3 q_3}{(2 \pi)^3} (2 \pi)^3 \delta(q_1+q_2 +q_3) 
 \zeta(q_1) \zeta(q_2) \zeta(q_3), \\
g'_3&= g_3 +  \frac{1}{2 \pi^2 \kappa_B} \left( 2 \pi^2 \sgn(\kappa_B) \delta_B {\tilde g}_3 
 +\frac{g_3^3}{ g_2^3} + 
\frac{3g_3 {\tilde g}_{(4,1)} }{2 g_2^2} -\frac{{\tilde g}_{(5,0)}}{2 g_2} \right) \ln (\Lambda).
\end{split}
\end{equation}

\subsection{The $\beta$ function for $x_6$ at first subleading order 
in $\frac{1}{N_B}$}

Keeping track only of those one loop corrections that scale like $\ln (\Lambda)$, 
and truncating to quadratic and cubic order in $\zeta$, we have 
\begin{equation}\label{ccee} \begin{split}
S^{1PI}_{RB}& =   \frac{g_2}{2 \kappa_B}
\int \frac{d^3q}{(2 \pi)^3} \frac{ |q|^{1+ 2 \frac{\delta'_B}{\kappa_B}}}
{\Lambda^{2 \frac{\delta'_B}{\kappa_B}}}\zeta(q) \zeta (-q) \\
&\quad + \frac{g'_3}{6 \kappa_B^2} 
\int \frac{d^3 q_1}{(2 \pi)^3}
\frac{d^3 q_2}{(2 \pi)^3}
\frac{d^3 q_3}{(2 \pi)^3}  (2 \pi)^3 \delta(q_1+q_2 +q_3) 
 \zeta(q_1) \zeta(q_2) \zeta(q_3) . \\
\end{split}
\end{equation} 
The variable change
\begin{equation} \label{varchange} 
{\tilde \zeta}= \frac{\zeta}{\Lambda^\frac{\delta'_B}{\kappa_B}}
\end{equation} 
allows us to eliminate $\Lambda$ from the quadratic part of the action. 
In terms of the new variable we have 
\begin{equation}\label{cceenv} \begin{split}
S^{1PI}_{RB}& =   \frac{g_2}{2 \kappa_B}
\int \frac{d^3q}{(2 \pi)^3} { |q|^{1+ 2 \frac{\delta'_B}{\kappa_B}}} 
{\tilde \zeta}(q) {\tilde \zeta} (-q) \\
&+  \frac{g''_3}{6 \kappa_B^2} 
\int \frac{d^3 q_1}{(2 \pi)^3}
\frac{d^3 q_2}{(2 \pi)^3}
\frac{d^3 q_3}{(2 \pi)^3} (2 \pi)^3 \delta(q_1+q_2 +q_3) 
 {\tilde \zeta}(q_1) {\tilde \zeta}(q_2) {\tilde \zeta}(q_3), \\
2 \delta'_B&= 2 \delta_B + \frac{ 3{\tilde g}_{(4,2)} + {\tilde g}_{(4,3)} }{12 \pi^2  g_2^2},\\
g''_3&= g_3' \Lambda^{ \frac{3 \delta_B'}{\kappa_B}}= 
 g_3 +  \left(  \frac{{\tilde g}_3 
 \delta_B}{\kappa_B}+g_3   \left( \frac{3{\tilde g}_{(4,1)} }
{4 \pi^2 \kappa_B g_2^2} + \frac{3 \delta_B'}{\kappa_B} \right) 
+  g_3^3 \left( \frac{1}{2 \pi^2 \kappa_B g_2^3}  \right) -\frac{{\tilde g}_{(5,0)}}{4 \pi^2 \kappa_B g_2} 
 \right) \ln (\Lambda).
\end{split}
\end{equation} 
It follows that the dynamics defined by the action \eqref{cceenv} is 
independent of the cutoff scale $\Lambda$ if and only if
\begin{equation}\label{finalbetafn} \begin{split}
\frac{d g_3}{d \ln (\Lambda)}&= 
(24 \pi^2) \frac{d x_6^{B,F}}{d \ln (\Lambda)}=(24 \pi^2) \beta, \\
(24 \pi^2) \beta &=   \left(\frac{{\tilde g}_{(5,0)}}{4 \pi^2 \kappa_B g_2}
-\frac{{\tilde g}_3 
 \delta_B}{ \kappa_B}\right) - 
 g_3   \left( \frac{3 {\tilde g}_{(4,1)} }
{4 \pi^2 \kappa_B g_2^2} + \frac{3 \delta_B'}{\kappa_B} \right) 
-  g_3^3 \left( \frac{1}{2 \pi^2 \kappa_B g_2^3} \right),  \\
\end{split}
\end{equation}
where in the first line we used \eqref{termdef}.

\eqref{finalbetafn} is the beta function of the $SU(N)$ theory. Using the reasoning outlined in 
subsubsection \ref{son} it follows that the $\beta$ function of the $SO(N)$ theory is given by
\begin{equation}\label{finalbetafnson} \begin{split}
\frac{1}{2}\frac{d g_3}{d \ln (\Lambda)}&= 
\frac{(24 \pi^2)}{2} \frac{d x_6^{B,F}}{d \ln (\Lambda)}=   \left(\frac{{\tilde g}_{(5,0)}}{4 \pi^2 \kappa_B g_2}
-\frac{{\tilde g}_3 
 \delta_B}{ \kappa_B}\right) -
 g_3   \left( \frac{3 {\tilde g}_{(4,1)} }
{4 \pi^2 \kappa_B g_2^2} + \frac{3 \delta_B'}{\kappa_B} \right) 
-  g_3^3 \left( \frac{1}{2 \pi^2 \kappa_B g_2^3} \right) . \\
\end{split}
\end{equation}
In other words, the beta function of $x_6$ in $SO(N)$ is exactly twice the $\beta$ function for $x_6$ in the 
$SU(N)$ theory. 

In summary, the $SU(N)$ beta function listed in \eqref{finalbetafn} is a cubic polynomial of the form 
\begin{equation}\label{cpf}
\beta(g_3)= -a + b g_3 - c g_3^3,
\end{equation}
with coefficients that are functions of $\lambda$, given by
\begin{equation} \label{abcd}
\begin{split}
a& =  -\left(\frac{{\tilde g}_{(5,0)}}{4 \pi^2 \kappa_B g_2}
-\frac{{\tilde g}_3 
 \delta_B}{ \kappa_B}\right),\\
b&=-\left( \frac{3 {\tilde g}_{(4,1)} }
{4 \pi^2 \kappa_B g_2^2} + \frac{3 \delta_B'}{\kappa_B} \right),\\
c&=\left( \frac{1}{2 \pi^2 \kappa_B g_2^3} \right), \\
g_3&= (24 \pi^2) 
\left( x_6^B -\frac{4}{3} \cot^2 \left( \frac{ \pi \lambda_B}{2} \right) \right)= 
(24 \pi^2) \left( x_6^F -\frac{4}{3} \tan^2 \left( \frac{ \pi \lambda_F}{2} \right)
\right).\\
\end{split}
\end{equation}
\eqref{abcd} is one of the central results of this paper.
Recall that $g_2$ and ${\tilde g}_{(4,1)}$ are both known functions of $\lambda_B$ (both are listed in \eqref{termdef}). 
The quantity ${\tilde g}_{(5,0)}$ characterizes a particular kinematical limit of the leading large $N_B$ five point function of ${\tilde J}_0$ operators (see \eqref{termdef} for 
a precise definition). This quantity is currently not known 
as a function of $\lambda_B$ (but may well be possible to evaluate in the near future using the techniques of \cite{Yacoby:2018yvy}). The quantity $\delta_B$ parameterizes 
the anomalous dimension of ${\tilde J}_0$ at first subleading 
order in $\frac{1}{N_B}$. It is precisely defined in 
\eqref{sdopb}, and is currently not known as a function of 
$\lambda_B$. The quantity ${\tilde g}_3$ parameterizes the split 
of the three point function of three ${\tilde J}_0$ operators into contact and non-contact pieces and was precisely defined in 
\eqref{termdef} and \eqref{thptdet}. This quantity is also currently not known as a function of $\lambda_B$. Finally 
$\delta'_B$ is given in terms of $\delta_B$ by \eqref{divpeen}.

In the rest of this section we will first study the behaviour of the beta function \eqref{cpf} and \eqref{abcd} 
in the small $\lambda_B$ and $\lambda_F$ limits, and then study the global properties of the flows described by 
\eqref{cpf}. 

\subsection{The limit $\lambda_B \to 0$} \label{fsu}

To analyze the $\lambda_B \to 0$ limit, we introduce a rescaled coupling $\lambda_6$ defined by
\begin{equation} \label{scx6}
x_6^B = \frac{\lambda_6}{\lambda_B^2},
\end{equation}
and work at fixed $\lambda_6$ in the limit $\lambda_B \to 0$. 
In this limit the regular boson action \eqref{frebs} simplifies to\footnote{As $x_6^B \sim \frac{1}{\lambda_B^2}$, we can take $x_6+1 \approx x_6$ in \eqref{frebs} .}
\begin{equation}\label{acrb1}
S_{RB}(\phi)=  \int d^3 x  \left( D_\mu \bar \phi D^\mu\phi  + \frac{(2 \pi)^2}{N_B^2}\lambda_6 ( \bar \phi \phi)^3\right).
\end{equation}

Plugging \eqref{gfree} into \eqref{finalbetafn}, using the fact that $\delta_B=0$, and using
the results of Appendix \ref{vipb} we find 
\begin{equation}\label{finalbetafunction2}
 \frac{d \lambda_6}{d \ln (\Lambda)} = \frac{1}{N_B}9\lambda_6^2\left(1-\frac{\pi^2}{16} \lambda_6\right).
 \end{equation}


Similarly, if we plug \eqref{scx6} into the action for the $SO(N)$ regular boson theory we obtain the action  
\begin{equation} \label{actson1}
S= \int d^3x \left( \frac{1}{2} \partial_{\mu}\phi \partial^\mu \phi +\frac{(2 \pi)^2}{2N_B^2}\lambda_6 \phi^6   \right)
\end{equation}
(note that $\phi$ are now real fields). 
The beta function for this theory is simply twice \eqref{finalbetafunction2}, i.e. 
\begin{equation}\label{finalbetafunction1son}
 \frac{1}{2}\frac{d \lambda_6}{d \ln (\Lambda)} =  \frac{1}{N_B}9\lambda_6^2\left(1-\frac{\pi^2}{16} \lambda_6\right).
 \end{equation}

Let us now compare the beta function in \eqref{finalbetafunction1son} with the beta function reported in \cite{Pisarski:1982vz}.
The action in \cite{Pisarski:1982vz} is given by
\begin{equation} \label{actson11}
S= \int d^3x \left( \frac{1}{2} \partial_{\mu}\phi \partial^\mu \phi +\frac{\pi^2}{3}\lambda_6^{Pisarski} \phi^6   \right),
\end{equation}
which matches with \eqref{acrb1} if we make the identification
\begin{equation}\label{pskbeta}
\lambda_6^{our}=\frac{N_B^2}{6}\lambda_6^{Pisarski}.
\end{equation}

In \cite{Pisarski:1982vz} the leading order in large $N_B$ beta function is given by
\begin{equation}\label{psB}
 \frac{d \lambda_6^{Pisarski}}{d \ln (\Lambda)}= 3 N_B( \lambda_6^{Pisarski})^2-N_B^3\frac{\pi^2}{32} ( \lambda_6^{Pisarski})^3.
\end{equation}
It is easy to check that \eqref{finalbetafunction1son} reduces to  \eqref{psB} with the identification \eqref{pskbeta}, showing the consistency of our beta function with that of \cite{Pisarski:1982vz}.

\subsection{Flows at order $\lambda_B^2$}

In \cite{Aharony:2011jz} the $SO(N_B)$ regular boson theory was studied in the 
perturbative regime, i.e. with $\lambda_B$ and $\lambda_6$ both small, and 
with $\lambda_6 \sim \lambda_B^2$. The variables used in \cite{Aharony:2011jz}
differ from those we have used in their normalization; the translation dictionary 
is given by 
\begin{equation} \label{trans} \begin{split}
\lambda_6^{there}&= 96 \pi^2 \lambda_6,\\
\lambda^{there}&=4 \pi^2 \lambda_B.
\end{split}
\end{equation} 
The $\beta$ function reported in  \cite{Aharony:2011jz} is
\begin{equation} \label{smallbet}
\frac{d \lambda_6^{there}}{d \ln (\Lambda)} = \frac{1}{N_B} 
\left(  \frac{33 (N_B-1)}{32N_B \pi^2}(\lambda^{there})^4 
- \frac{5 (N_B-1)}{4N_B \pi^2}(\lambda^{there})^2 ~\lambda_6^{there} +   \frac{3N_B+22}{16N_B \pi^2}
(\lambda_6^{there})^2
\right).
\end{equation}
Using 
\eqref{trans}, \eqref{smallbet} turns into 
\begin{equation} \label{smallbetourfiniten}
\frac{d \lambda_6}{d \ln (\Lambda)} = \frac{1}{N_B} 
\left(  \frac{11 (1 -\frac{1}{N_B})}{4}\lambda_B^4 
- 20 (1-\frac{1}{N_B}) \lambda_B^2 ~\lambda_6 + 
\left(18 + \frac{198}{N_B} \right) \lambda_6^2
\right) .
\end{equation}

The results presented in this subsection so far apply at all values 
of $N_B$. Specializing to the large $N_B$ limit we obtain 
\begin{equation} \label{smallbetour}
\frac{d \lambda_6}{d \ln (\Lambda)} = \frac{1}{N_B} 
\left(  \frac{11}{4}\lambda_B^4 
- 20 \lambda_B^2 ~\lambda_6 + 18 \lambda_6^2
\right) 
\end{equation}
for the $SO(N_B)$ theory, and 
\begin{equation} \label{smallbetoursu}
2\frac{d \lambda_6}{d \ln (\Lambda)} = \frac{1}{N_B} 
\left(  \frac{11}{4}\lambda_B^4 
- 20 \lambda_B^2 ~\lambda_6 + 18 \lambda_6^2
\right)
\end{equation}
for the $SU(N_B)$ theory. Note that the first term on the right-hand side of \eqref{finalbetafunction1son} agrees with the 
last term on the right-hand side of \eqref{smallbetour}, establishing the consistency between
these two formulae. 

\eqref{smallbetoursu} carries qualitatively important information as we now explain. 
The beta function \eqref{finalbetafunction2} has  a zero $\lambda_6= \frac{16}{\pi^2}$ together 
with a double zero at $\lambda_6=0$. The zero at $\lambda_6= \frac{16}{\pi^2}$ is robust; 
small $\lambda_B$ corrections will change the precise value of the zero but cannot destabilize it. 
On the other hand a double zero is unstable to small corrections; corrections of one sign could 
cause the double zero to split into two single zeroes, while the opposite correction could remove 
the zeroes altogether. It is easy to check that the corrections \eqref{smallbetour} split the double 
zero into two single zeroes located (for large $N_B$) at 
\begin{equation} \label{betzer}
\lambda_6= \frac{20 \pm \sqrt{202}}{36} \lambda_B^2,
\end{equation}
or
\begin{equation}\label{betzero}
x_6+1=  \frac{20 \pm \sqrt{202}}{36}.
\end{equation}
The larger of the two roots (the attractive fixed point in the sense of flow
to the IR) is at 
\begin{equation}
x_6+1= .95035\cdots,
\end{equation}
while the smaller root (the repulsive fixed point in the sense of 
flow towards the IR)  lies at
\begin{equation}
x_6+1=.16076\cdots.
\end{equation}
It follows that at small $\lambda_B$ the approximate domain of attraction of the 
attractive fixed point is the large range 
\begin{equation}\label{doa}
x_6+1  \in (.16076, \frac{16}{\pi^2 \lambda_B^2} ).
\end{equation}

\subsection{The limit  $\lambda_F \to 0$}

As we have seen above, in the limit  $\lambda_F \to 0$
\begin{equation}\label{dbp}
\delta'_B(\lambda_F)=-\frac{16}{3\pi^2 \lambda_F}.
\end{equation}
We now turn to the $\beta$ function in this limit.
Plugging \eqref{dbp}, \eqref{fermfn} and \eqref{gfreef} into \eqref{finalbetafn} we obtain
\begin{equation} \label{bffs}
\begin{split}
 \frac{d x_6}{d \ln (\Lambda)} &= \frac{32 x_6}{\kappa_F \pi^2 \lambda_F}- \frac{36 x_6^3}{\kappa_F \pi^4 \lambda_F^3}.
\end{split}
\end{equation}

The $\beta$ function \eqref{bffs} can be rewritten in terms of the 
variable $\lambda_6^F$ 
\begin{equation}\label{lsx}
\lambda_6^F=\frac{x_6}{\lambda_F},
\end{equation}
in which it has a finite $\lambda_F \to 0$ limit
\begin{equation}\label{nbfn}
\frac{d \lambda^F_6}{d \ln (\Lambda)} =\frac{1}{N_F} 
\left( \frac{32 \lambda_6^F}{ \pi^2}-\frac{36 \left(\lambda_6^F\right)^3}{ \pi^4} \right) .
\end{equation}
The zeros of this beta function occur at $\lambda_6^F=0$ and $\lambda_6^F=\pm\frac{  2\sqrt{2} \pi}{3}$, 
  or at 
\begin{equation} \label{sls}
x_6=\left(0, \pm  \frac{2\sqrt{2} \pi \lambda_F}{3}\right).
\end{equation}
Notice that all three zeroes of this beta function are of order $\lambda_F$. 

The scalings in \eqref{nbfn} - and in particular the emergence of the 
natural variable $\lambda_6^F$ in the limit $\lambda_F \to 0$ is easy 
to understand. 
Recall that the critical fermion theory is defined by the action 
\begin{equation}\label{cfta}
S_{CF}(\psi, \zeta)=S_{RF}(\psi) -\int  
\frac{4 \pi {\bar \psi}{\psi}}{{\tilde \kappa}_F} \zeta(x) 
+ \frac{(2 \pi)^2}{{\tilde \kappa}_F^2} 
x_6 \int \zeta^3(x).
\end{equation}
The variable change 
\begin{equation}\label{varchferm}
\zeta= -\frac{{\tilde \kappa}_F}{4 \pi} \sigma_F
\end{equation} 
recasts this Lagrangian into the form 
\begin{equation}\label{cftar}
S_{CF}(\psi, \zeta)=S_{RF}(\psi) +\int  
{\bar \psi}{\psi} \sigma_F
- \frac{N_F \lambda_6^F}{8 \pi} 
\int \sigma_F^3(x),
\end{equation}
which has a finite limit as ${\tilde \kappa}_F \to \infty$, explaining the structure of \eqref{nbfn}.

The theory \eqref{cftar} with $\lambda_F = \lambda_6^F = 0$ has a parity symmetry, under which $\sigma_F$ is odd. This implies that both $\lambda_F$ and $\lambda_6^F$ change sign under a parity transformation. This is consistent with the form of \eqref{nbfn}, and implies that at weak coupling $\lambda_6^F=0$ must be a fixed point, and that the two other zeroes have equal magnitude and opposite sign, as we find.

\subsection{Global structure of the flows}

Let us now study the properties of the flows described by equation 
\eqref{cpf} at general values of $\lambda_B$ (or $\lambda_F$). 
We first note that the coefficient $c$ in \eqref{cpf} 
is positive at every value of $\lambda_B$ 
\footnote{This follows from the fact that 
$\kappa_B g_2$ is always positive.}. It follows that $\beta$ is negative at 
large positive values of $g_3$, but positive at large negative values of 
$g_3$. In other words all flows are repulsive
at large values of $|g_3|$, i.e. flows that start 
at large positive $g_3$ (or equivalently $x_6$) end up (in the deep IR) 
at $g_3 = \infty$ (equivalently $x_6 = \infty$). Similarly flows that start out at 
large negative values of $g_3$ (equivalently $x_6$) end up at $x_6= - \infty$. 

The polynomial \eqref{cpf} has exactly one real root whenever 
\begin{equation}\label{orr}
27 a^2 c > 4 b^3,
\end{equation}
and in this case the fixed point is repulsive (UV-stable).
On the other hand it has three real roots when 
\begin{equation}\label{orri}
27 a^2 c < 4 b^3.
\end{equation}
At the boundary between these two regimes there is one double zero at 
\begin{equation}
g_3= \sqrt{\frac{b}{2c}},
\end{equation}
and one additional root. 

When the condition \eqref{orri} is satisfied, the outer two roots are 
repulsive (in the sense of flow towards the IR) and represent fixed 
points about which $\zeta^3$ is a relevant operator. However the middle 
root is attractive, and represents a fixed point about which $\zeta^3$ is 
an irrelevant operator. 

Near $\lambda_B=0$ we have 
\begin{equation} \label{abcf}
a=-\frac{4096}{\kappa_B\pi^2 \lambda_B^3}, ~~~b=\frac{48 }{\kappa_B\pi^2 \lambda_B}, ~~~ c= \frac{ \lambda_B^3}{1024 \pi^2 \kappa_B},~~~ 
\end{equation} 
and $27a^2 c = 4b^3$, so we are at the boundary between the cases 
\eqref{orr} and \eqref{orri}. The beta function has a double root at 
\begin{equation}
g_3= -\frac{128}{\lambda_B^2},
\end{equation}
i.e. at $\lambda_6=0$, and a single root at
\begin{equation}
g_3= \frac{256}{\lambda_B^2},
\end{equation}
i.e. at 
\begin{equation}
\lambda_6=\frac{16}{\pi^2},
\end{equation}
in agreement with \eqref{finalbetafunction2}.
Turning on a small Chern-Simons coupling $\lambda_B$ takes us into the regime 
\eqref{orri}. As described above, the double root of the $\lambda_B=0$ theory 
splits into two roots, one repulsive and one attractive. The domain of 
attraction of the attractive root is listed in \eqref{doa} and becomes large as $\lambda_B \to 0$.

In the opposite limit $\lambda_F=0$ we have $a=0$ with 
$b$ and $c$ positive, so we lie squarely within the regime of \eqref{orri}. 
Once again we have two repulsive roots and one attractive root. The domain 
of attraction of the attractive root is the interval 
\begin{equation}\label{daar}
x_6 \in ( - \lambda_F \sqrt{ \frac{2}{3 \pi^2}},  
\lambda_F \sqrt{ \frac{2}{3 \pi^2}}),
\end{equation}
which becomes very small as $\lambda_F \to 0$.

At general values of $\lambda_B$ the coefficient $c(\lambda_B)$ is given by \eqref{abcd} and 
\eqref{termdef}. It follows from these exact results that $|\kappa_B| c$ is an even function 
of $\lambda_B$. This function starts out at zero when $\lambda_B=0$ and then increases 
monotonically with $\lambda_B$, reaching infinity as $\lambda_B$ approaches unity. As we have 
mentioned above, $c$ is positive at every value of $\lambda_B$. 

\begin{figure}[h]
\begin{center}
\includegraphics[width=12.5cm,height=9.5cm]{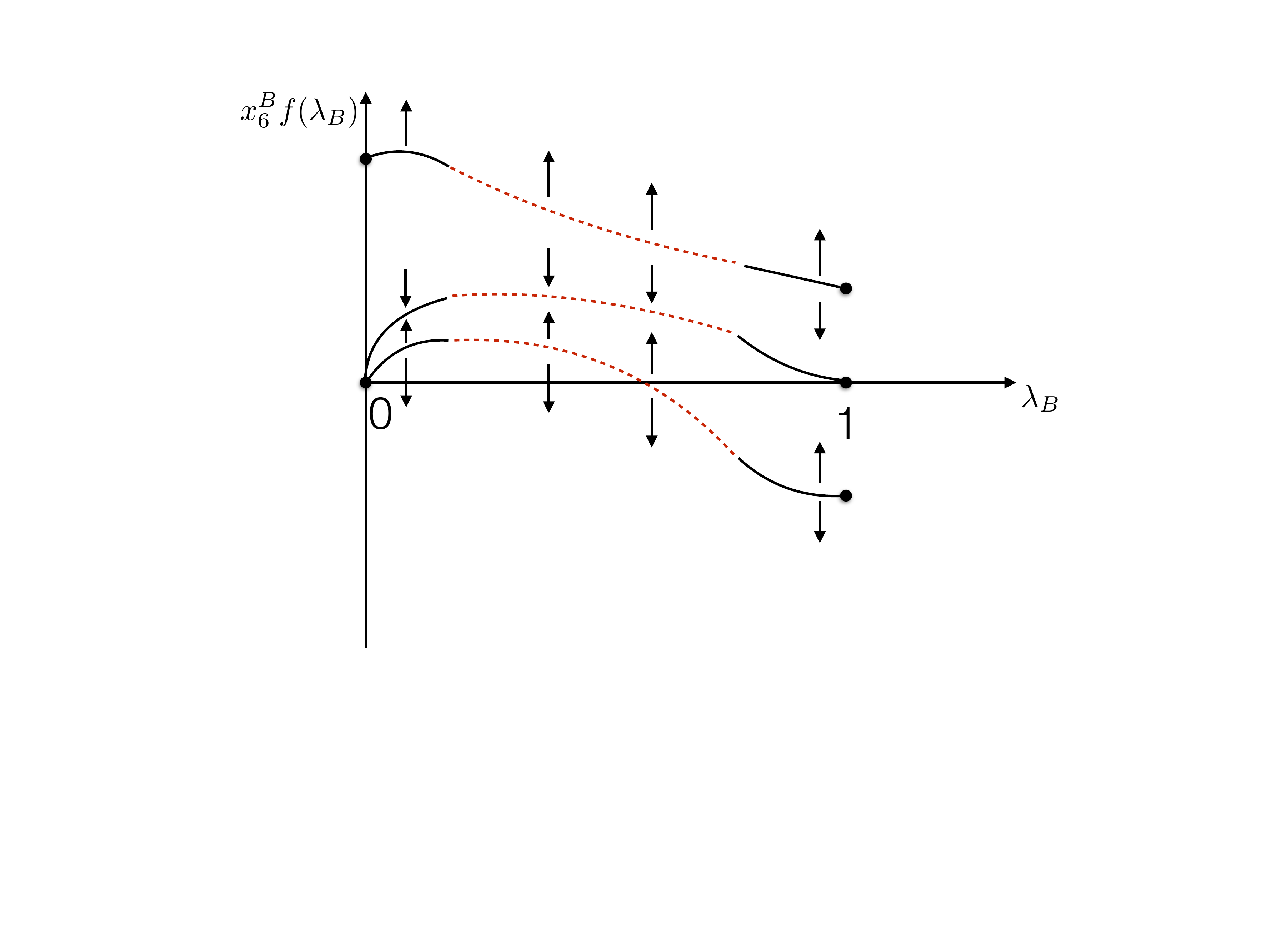}
  \caption{\label{r-g-flw} A schematic description of the large $N$ RG flows towards low energies described in section 3.10. The vertical axis is $x_6$ multiplied by a convenient function of $\lambda$, that keeps the fixed points at a finite separation both as $\lambda_B \to 0$ and as $\lambda_F \to 0$. The flows near the two limiting values of $\lambda_B$ (with the fixed points depicted as solid lines) are based on computations, while the flows in the middle (with the fixed points depicted as dashed lines) are conjectured, and it is possible that two of the lines actually meet and disappear for some value of $\lambda_B$ and then reappear at a larger value.}
 \end{center}
\end{figure}

$|\kappa_B|a$ starts out at $- \infty$ at $\lambda_B=0$ but vanishes at $\lambda_B=1$ (the vanishing is a direct consequence of the parity transformation properties of the free fermion theory). We do not 
know the value of $a$ at intermediate values of $\lambda_B$. The simplest possibility is that 
$a$, like $c$, increases monotonically as a function of $\lambda_B$, and in particular is always negative. 

$|\kappa_B| b$ starts out at $\infty$ at $\lambda_B=0$ and then decreases like $\frac{1}{\lambda_B}$
as $\lambda_B$ is increased. Near $\lambda_B=1$ this function once again blows up in a manner 
proportional to $\frac{1}{1-\lambda_B}$. We do not know the detailed functional form of $b$ at
intermediate values of $\lambda_B$. The simplest possibility consistent with these boundary 
conditions is that $|\kappa_B| b$ monotonically decreases from $\lambda_B=0$ to some $\lambda_B=\lambda_{min}$, 
after which it monotonically increases as $\lambda_B$ is increased from $\lambda_{min}$ to unity. 

As we have explained above, the global structure of the $\beta$ function is really determined by 
the ratio $\frac{27 a^2 c}{b^3}$. This ratio starts out at unity at $\lambda_B=0$, decreases 
as $\lambda_B$ is increased away from zero and reaches the value zero when $\lambda_B=1$. 
We do not know the detailed functional form of this ratio as a function of $\lambda_B$. The 
simplest possibility is that this function decreases monotonically as a function of $\lambda_B$ 
as $\lambda_B$ is varied from $0$ to unity. If this possibility is realized then the RG flows 
described in this paper have two repulsive and one attractive fixed point  at every value 
of $\lambda_B$ (see Figure \ref{r-g-flw} for a schematic description). Note that the size of the domain of attraction of this 
central stable fixed point (measured as the size of the interval of the variable $x_6$) 
is very large at small $\lambda_B$ but goes to zero as $\lambda_B \to 1$.

\subsection{Relevant deformations, thermodynamics and duality} \label{rd} 

So far in this section we have intensively studied RG flows of the 
parameter $x_6$. $x_6$ is special because the operator it multiplies,
$\zeta^3$, is marginal in the large $N$ limit. As a consequence,
flows of $x_6$ are controlled by a beta function of order $\frac{1}{N}$ 
and so are very slow, as we have seen earlier in this section. 

In contrast to $\zeta^3$, the operators $\zeta^2$ and $\zeta$ 
are relevant even in the strict large $N$ limit. 
The addition of one of these operators to the regular boson or 
critical fermion theory results in a `fast' RG flow. These fast flows 
take place over an `RG flow time' (i.e. logarithm of ratios of scales) of 
order unity rather than $N$. 

Flows induced by $\zeta^2$ and $\zeta$ in the large $N$ limit were studied in detail in 
\cite{Jain:2013gza}. The qualitative conclusion of this study was the following. 
A generic deformation in $\zeta^2, \zeta$ space ends up in a massive 
theory. A one parameter tuning (tuning the coefficient 
of $\zeta$ to zero in the dimensional regularization scheme) seeds a flow 
that ends up (for a specific sign of the $\zeta^2$ coefficient) in the CB theory (which by duality 
is the same as the RF 
theory) in the IR. The operator $\zeta^3$ is irrelevant about the 
new fixed point, so $x_6$ goes to zero and the dynamics of the fixed point is insensitive to 
the value of $x_6$ from which the flow starts. 

At a quantitative level, the results described in the previous paragraph
were obtained as follows. Consider the $\zeta$ and $\zeta^2$ deformed regular boson theory 
\beal{
S_B  &= \int d^3 x  \biggl[i \varepsilon^{\mu\nu\rho}{\kappa_B\over 4\pi}
\Tr( A_\mu\partial_\nu A_\rho -{2 i\over3}  A_\mu A_\nu A_\rho)
+i \varepsilon^{\mu\nu\rho} {\kappa_2 \over 4 \pi}
 B_\mu\partial_\nu B_\rho + D_\mu \bar \phi D^\mu\phi  \nn
&~~~~~~~~~~~~~
+m_b^2 \bar\phi \phi +  {4\pi b_4 \over \kappa_B}(\bar\phi \phi)^2
+ \frac{(2\pi)^2}{\kappa_B^2} \left( x_6^B+1\right)  (\bar\phi \phi)^3\biggl],
\label{rst}
}
and the $\zeta$ and $\zeta^2$ deformed critical fermion theory
\beal{~\label{csfnonlinear}
 S_F =&\int d^3 x \bigg[ i \varepsilon^{\mu\nu\rho} {\kappa_F \over 4 \pi}
\Tr( A_\mu\partial_\nu A_\rho -{2 i\over3}  A_\mu A_\nu A_\rho)+i \varepsilon^{\mu\nu\rho} {\kappa_2 \over 4 \pi}
 B_\mu\partial_\nu B_\rho 
+  \bar{\psi} \gamma_\mu D^{\mu} \psi \nn
&~~~~~~~~~~~~~
-\frac{4\pi}{\kappa_F}\zeta (\bar\psi \psi - {\kappa_F y_2^2\over4\pi} ) - { 4\pi y_4\over \kappa_F} \zeta^2 + \frac{(2\pi)^2}{\kappa_F^2} x_6^F \zeta^3\bigg]. 
}
The finite temperature partition functions of these theories were 
computed, in the large $N$ limit, in 
\cite{Aharony:2012ns,Jain:2013py,Minwalla:2015sca} by proceeding 
along the lines of the discussion of subsection \ref{du}. In the unHiggsed 
phase, the free energy  for the deformed regular boson 
turns out to take the form \cite{Minwalla:2015sca}
\begin{equation}\label{offb}
\begin{split}
F_B(\rho_B) &= F_{B,0}+\frac{N_B}{6\pi} \Bigg[ -{\hat c}_B^3+2\left({\hat c}_B^2-{\hat m}_b^2 \right){\tilde \cS}+2\lambda_B {\hat b_4} {\tilde \cS}^2 \\
&-3 \int_{-\pi}^{\pi} d\alpha \rho_{B}(\alpha)\int_{{\hat c}_B}^{\infty} dy ~y~\left( \ln\left(1-e^{-y-i\alpha}\right)  + \ln\left(1-e^{-y+i\alpha}\right)\right)\Bigg],
\end{split}
\end{equation}
where $F_{B,0}$ is a constant  independent of ${\hat c}_B$ and 
 the quantity ${\tilde \cS}$ is defined to be the solution to the equation 
\begin{align}
{{\hat c}_B^2} 
&= (4+3 x_6^B)\lambda_B^2 {\tilde \cS}^2 -4\lambda_B \hat b_4 {\tilde \cS}  +\hat m_b^2,
\label{scalargapequation}
\end{align}
and where $\hat b_4= { b_4 \over T}, \hat m_b= {m_b \over T}$, ${\hat c}_B = {c_B \over T}$, and ${\hat c}_{B,0} = \lim_{T\to 0} {\hat c}_B$
\cite{Minwalla:2015sca}.
The gap equation that follows from varying \eqref{offb} takes the form ${\tilde \cS}= \cS$  where $ \cS$ is defined in \eqref{ss}.
It follows in particular that on-shell
\begin{align}
{{\hat c}_B^2} 
&= (4+3 x_6^B)\lambda_B^2 { \cS}^2 -4\lambda_B \hat b_4 { \cS}  +\hat m_b^2.
\label{scalargapequation1}
\end{align}

In a soon to appear paper \cite{new} the regular boson free energy and gap equation have very recently been evaluated 
in the Higgsed phase as well (using the techniques developed in the recent paper \cite{Choudhury:2018iwf}). We refer the reader 
to \cite{new} for more details. Here we simply note that the final results of \cite{new} are in perfect agreement with the 
predictions of duality, and so can be deduced by applying 
the duality map to the fermionic results reviewed immediately 
below, as we will effectively do later in this paper.

The fermionic calculation 
can be done in two steps. We first integrate out $\psi$ at fixed $\zeta$. If we assume (as we do) that the saddle point value of $\zeta$ is a constant in space time, then the integral over fermions is precisely that performed in section \ref{csrf}.  We find the equation 
\begin{equation}\label{frecrf}
\begin{split}
F_F(\rho_F) &=\frac{N_F }{6\pi} \Bigg[ {\hat c}_F^3-2\lambda_F^2 {\tilde \cC}^3 -\frac{3}{2} \left({\hat c}_F^2 -\frac{16\pi^2}{\kappa_F^2}{\hat \zeta}_F^2 \right) {\tilde \cC} +\frac{6\pi {\hat y}_2^2}{\kappa_F\lambda_F}{\hat \zeta}_F-\frac{24 \pi^2 {\hat y}_4}{\kappa_F^2\lambda_F}{\hat \zeta}_F^2+\frac{24 \pi^3 x^F_6}{\kappa_F^3\lambda_F}{\hat \zeta}_F^3\\
&-3 \int_{-\pi}^{\pi}d\alpha \rho_{F}(\alpha)\int_{{\hat c}_F}^{\infty} dy ~y~\left( \ln\left(1-e^{-y-i\alpha}\right)  + \ln\left(1-e^{-y+i\alpha}\right)\right)\Bigg],
\end{split}
\end{equation}
with ${\tilde \cC}$ defined to be the solution to the equation
\be
{\hat c}_F^2 =({2\lambda_F} {\tilde \cC} -\frac{4\pi}{\kappa_F}\hat\zeta)^2,
\label{cf}
\ee
where $\hat \zeta \equiv {\zeta \over T}$, ${\hat c}_F \equiv {c_F \over T}$. \footnote{The relationship between the regular fermion theory 
of section \ref{csrf} and the equations just presented is the following. 
If we use \eqref{cf} to solve for ${\tilde \cC}$, make the replacement
$\frac {4\pi}{\kappa_F}{\hat \zeta} \rightarrow -m_F^{reg}$ in the solution,
and then plug the solution into \eqref{frecrf}, we obtain 
\eqref{offshellfe} up to terms independent of ${\hat c}_F$. }

$F_F$ in \eqref{frecrf} is a function is a function of ${\hat c}_F$ and $\hat \zeta$  and must must be extremized with respect to both of these variables. Extremizing with respect to ${\hat c}_F$ we get 
${\tilde \cC}={ \cC}$,
or equivalently
\be
{\hat c}_F^2 =({2\lambda_F} { \cC} -\frac{4\pi}{\kappa_F}\hat\zeta)^2,
\label{1cf}
\ee
where $\cC$ is defined in \eqref{ss}.
Extremizing with respect to $\hat \zeta$ we find 
\beal{
\label{sigsol}-\frac{3}{4} \left(\frac{4\pi}{\kappa_F}\right)^2 \hat\zeta^2 x_6^F-\frac{16\pi}{\kappa_F} \hat\zeta \lambda_F{\tilde \cC}+ \frac{8\pi}{\kappa_F} \hat\zeta \hat y_4+4 \lambda_F^2{\tilde \cC}^2-\hat y_2^2=0,
}
where $\hat y_4= {y_4 \over T}, \hat y_2 = {y_2 \over T}$.
\footnote{Of course \eqref{sigsol} is easily solved and yields
\beal{
\hat\zeta =\frac{\kappa_F}{\pi} { \hat y_4 - 2\lambda_F \cC \pm \sqrt{( \hat y_4 -  2\lambda_F\cC)^2 + {\frac{3}{4} x_6^F} ((2\lambda_F\cC)^2-{\hat y_2^2})} \over 3 x_6^F}.
\label{sigmasol}
}
}
\footnote{In performing the extremization of \eqref{frecrf} with respect to ${\hat c}_F$ and $\hat \zeta$ it is important to remember 
that ${\tilde \cC}$ is a function of both variables. One way to account for this fact is to use the chain rule, and to use 
\eqref{cf} to evaluate the 
partial derivatives of ${\tilde \cC}$ with respect to ${\hat c}_F$ and $\hat \zeta$.}

The gap equation \eqref{1cf} may be used to evaluate ${\hat \zeta}$ in terms of $|{\hat c}_F|$:
\begin{equation}\label{zetaitc}
\frac{4 \pi {\hat \zeta}}{\kappa_F} = 2 \lambda_F \cC-{\rm sgn}(X)|{\hat c}_F|,
\end{equation}
where 
\begin{equation} \label{defx}
X=2 \lambda_F \cC- \frac{4 \pi}{\kappa_F} {\hat \zeta} .
\end{equation}
When
\begin{equation}\label{condm}
{ \rm sgn}(X)= {\rm sgn} (\lambda_F)
\end{equation}
we may use \eqref{dualofq1} to reexpress \eqref{zetaitc} as 
\begin{equation}\label{zetaitca}
\frac{4 \pi {\hat \zeta}}{\kappa_F} = 2 \lambda_B \cS.
\end{equation} 
Inserting \eqref{zetaitca} into \eqref{sigsol} and once again using \eqref{dualofq1} to reexpress all remaining occurrences 
of $\cC$ in terms of $\cS$ we find that \eqref{sigsol} reduces to \eqref{scalargapequation1}
 provided we make the following identifications:
\be
\lambda_F=\lambda_B-{\rm sgn}(\lambda_B), \quad
x_6^F =x_6^B, \quad  
y_4 =  b_4, \quad y_2^2 = m_b^2.
\label{dualitytransform}
\ee
It follows that the bosonic and fermionic gap equations have the same solutions if we assume \eqref{condm}.
This agreement provides non-trivial support for the conjecture that 
the RB and CF theories are dual to each other, since they agree also 
after the relevant deformation. It was demonstrated in \cite{Minwalla:2015sca} that the agreement between gap equations 
extends to the off-shell free energies;  \eqref{frecrf} and \eqref{offb} are identical 
with the same identifications, under the assumption that \eqref{condm} holds. 

When
\begin{equation}\label{condm1}
{ \rm sgn}(X)= -{\rm sgn} (\lambda_F)
\end{equation}
we may use \eqref{dualofq1} to reexpress \eqref{zetaitc} as 
\begin{equation}\label{zetaitca1}
\frac{4 \pi {\hat \zeta}}{\kappa_F} = 2 \lambda_B \cS -2 |{\hat c}_B| {\rm sgn}(\lambda_B).
\end{equation} 
Inserting \eqref{zetaitca1} into \eqref{sigsol} and once again using \eqref{dualofq1} to reexpress all remaining occurrences 
of $\cC$ in terms of $\cS$  and using \eqref{dualitytransform},
we find that \eqref{sigsol} reduces to 
\begin{equation}\label{hggsedsclr}
3 |{\hat c}_B|^2 (1+x_6^B) + 4 b_4  {\rm sgn}(\lambda_B) |{\hat c}_B|+m_b^2+ (4+3 x_6^B)\lambda_B^2 \cS^2
-2 \lambda_B \cS \left(2 b_4 +|{\hat c}_B| {\rm sgn}(\lambda_B) (4+3 x_6^B) \right)=0.
\end{equation}
\eqref{hggsedsclr} is a prediction for the bosonic 
gap equation in the Higgsed phase. This prediction
has been verified in the soon to appear paper \cite{new}.

We expect the RB (CF) theory to flow to the CB (RF) theory when we turn on $b_4 > 0$ in \eqref{rst} (or $y_4 > 0$ in \eqref{csfnonlinear}), and indeed the free energy is consistent with this. Consider the limit
\begin{equation}\label{scl12}x_6^{B}=\rm{fixed},~ \hat b_4\rightarrow\infty,~m_b^2 \rightarrow\infty\end{equation} with $\frac{m_b^2}{2 b_4 \lambda_B}=m_B^{cri}$ fixed. In this limit, 
\eqref{scalargapequation} is easily solved and yields 
\begin{equation}\label{solS1}{\tilde S} =\frac{m_{B}^{cri} }{2}+ \frac{-4 c_B^2+ 4(m_B^{cri})^3 \lambda_B^3+ 3 (m_B^{cri})^3 x_6 \lambda_B } {8 m_b^2} +{\cal O}\left(\frac{1}{m_b^4}\right).
\end{equation} Plugging the solution back into \eqref{offb} we recover the first equation of \eqref{offshellfe} (up to $c_B$-independent constants), reflecting the fact
that the regular boson theory flows to the critical boson theory at low energies, provided we fine-tune so that the pole mass is held fixed. Of course similar results apply to the fermionic theory, and \eqref{scl12} represents a RG flow from the regular boson/critical fermion theories to the critical boson/regular fermion theory. This may be thought of as a derivation of the critical boson-regular fermion duality starting from the conjectured regular boson-critical fermion duality.

Surprisingly, there are also other limits in which the RB (CF) theories seem to flow to the CB (RF) theory, even though $b_4=0$ (but $x_6$ diverges). We will discuss these limits in section \ref{curious} below.

\section{Phase structure at zero temperature at large $N$}\label{phasestruct}

In this section we analyze the large $N$ phase structure of the `regular boson' (or equivalently `critical fermion') theories that are the subject of this paper, at zero temperature (a complete analysis of the phase structure at finite temperature, and especially 
on $S^2 \times S^1$, would be an interesting but intricate exercise). The `critical boson' (or equivalently `regular fermion') theories have only one relevant deformation (which one can call the fermion mass), and their phase structure is very simple. For either sign of this deformation they flow to a massive theory, which reduces to a topological theory at low energies. In the fermionic $U(N_F)_{k_F}$ theory, for one sign of the mass one obtains the $U(N_F)_{k_F+\frac{1}{2}}$ CS theory, and for the other sign the $U(N_F)_{k_F-\frac{1}{2}}$ CS theory. In the language of the bosonic $U(N_B)_{k_B}$ theory, for one sign one obtains the $U(N_B)_{k_B}$ CS theory, and for the other sign the $U(N_B-1)_{k_B}$ CS theory, and one interprets this as a Higgsing of the original gauge theory.

The phase structure of the `regular boson' theories is more elaborate, since they have two relevant operators, and another operator that is marginal in the large $N$ limit (and which becomes either relevant or irrelevant at finite $N$, depending on the precise fixed point one discusses). Generic flows from these theories also lead to massive theories, but some special flows lead to the `critical boson' theories.
At infinite $N$ the flows are labeled for each $x_6$ by the parameters $m_b$ and $b_4$ in \eqref{rst},  and on dimensional grounds it is clear that the phase structure only depends on the combination $m_b^2 / b_4^2$, and on the signs of $m_b^2$ and of $b_4$.
 In certain parameter ranges the gap equations \eqref{1cf} and \eqref{sigsol} admit multiple solutions, and then the true phase of the theory is the solution with the lowest free energy. As the parameters of the theory are varied our system can undergo a phase transition of one of two kinds. The first sort of transition 
occurs when the free energies of two competing branches of solutions can cross. The thermodynamically dominant 
solution changes discontinuously across such a line, giving rise to a first order phase transition. 
The second kind of transition happens when the dominant branch of solutions itself changes character across a codimension 
one wall in parameter space. In this situation we have a second order phase transition, characterized by a conformal field theory, which in our case will always be the `critical boson' theory.

In the zero temperature limit the gap equations and the free energy simplify considerably. Since these equations are invariant under the duality, we can describe them either in the bosonic or in the fermionic language. In this section we will mostly use the bosonic language since the interpretation of the phases is simpler in this language. As we have reviewed above, the necessary
bosonic computations were performed some time ago 
\cite{Minwalla:2015sca} in the unHiggsed phase, but only 
very recently in soon to be published work in the Higgsed phase 
\cite{new}. The final results of \cite{new} turn out to be 
perfectly consistent with the predictions of duality, and so 
can be deduced by applying the duality map to the published fermionic results which are known from direct computation in both 
phases (see the previous section). This is the strategy we will 
employ in this section: we take the fermionic results, map them 
to variables and then interpret them in bosonic language.

In the rest of this section we will carefully analyze all known solutions to the zero temperature gap equations as a function of the UV parameters; when multiple solutions exist we will carefully determine 
which solution has the lowest free energy. We then use this analysis to suggest a conjectured `phase diagram' of our theory. 
We would like to emphasize that the phase diagrams we propose below are tentative, since our analysis may miss some phases of the theory (e.g. phases in which translation invariance is broken, or phases with no stable vacuum). 

\subsection{The gap equations and their solutions}

In the zero temperature limit we use \eqref{zt} to simplify the gap equations. The gap equation \eqref{zetaitc} simplifies to 
\begin{equation}\label{zetaitczt}
\frac{4 \pi { \zeta}}{\kappa_F} = \left( \lambda_F -{\rm sgn}(X) \right)|{ c}_F|.
\end{equation}
It follows from \eqref{zetaitczt} that 
\begin{equation}\label{sgnz}
 -{\rm sgn} ( \frac{{ 4 \pi\zeta}}{\kappa_F})= {\rm sgn}(X).
\end{equation}
This equation states that ${\rm sgn}(X)$ may be identified with the sign of the bare mass of the fermionic excitations (see \eqref{csfnonlinear}). 

The fermionic gap equations have two classes of solutions; those for which \eqref{condm}
holds, and those for which the signs in \eqref{condm} are reversed. In the zero temperature limit, the first class 
of solutions are those for which the bare fermionic mass and $\lambda_F$ have the same sign, while the second class of solutions 
are those for which the bare fermionic mass and $\lambda_F$ have opposite signs. If the fermions are effectively massive they 
can be integrated out. In the first case the resultant low energy effective action is a pure Chern Simons theory with level 
$k^{eff}_F= {\rm sgn}(k_F) \left( |k_F| + \frac{1}{2} \right)$, and in the second case of level
$k^{eff}_F= {\rm sgn}(k_F) \left( |k_F| - \frac{1}{2} \right)$
(see Appendix \ref{levels}). 
Based on the translation of these low-energy theories to the bosonic language mentioned above, we
will refer to saddle points of the first type as unHiggsed solutions, and to saddle points of the second type 
as Higgsed solutions. In either case \eqref{zetaitczt} may be written as  
\begin{equation}\label{zetauh}
\frac{4 \pi { \zeta}}{\kappa_F}= {\tilde \lambda}_B |c_B|,
\end{equation}
where on the unHiggsed branch we 
have 
\begin{equation}\label{tlbo}
{\tilde \lambda}_B= \lambda_B ~~~({\rm unHiggsed}),
\end{equation} 
while on the Higgsed branch we have 
\begin{equation}\label{tlbt}
{\tilde \lambda}_B= {\hat \lambda}_B \equiv -{\rm sgn} (\lambda_B) (2- |\lambda_B|)={\rm sgn} (\lambda_F) (1+ |\lambda_F|) ~~~({\rm Higgsed}).
\end{equation} 
As in the previous section, it is easy to verify that both of these solutions are self-consistent; i.e. 
that \eqref{tlbo} implies that \eqref{condm} is obeyed, while \eqref{tlbt} implies that \eqref{condm1} is obeyed. 

In order to proceed we now plug \eqref{zetauh} into \eqref{sigsol} and use the zero temperature result ${\tilde \cC} \rightarrow \frac{|c_F|}{2}$. We present our results in the language of the bosonic theory, so we make the replacements $|c_F| \rightarrow 
|c_B|$ and \eqref{dualitytransform}. We also find it convenient to define the rescaled bosonic variables
\begin{equation}\label{defBf}
\begin{split}
 {\tilde B}_4&=b_4 {\tilde \lambda}_B, \\
 {\tilde A}&= \lambda_F^2 -2 \lambda_F {\tilde \lambda}_B - \frac{3}{4} x_6^F {\tilde \lambda}_B^2
 = 
 (\lambda_F-{\tilde \lambda}_B)^2- \left(1+\frac{3 x_6}{4} \right) {\tilde \lambda}_B^2.
 \end{split}
\end{equation}

In these variables \eqref{sigsol} takes the form 
\begin{equation}\label{nge}
{\tilde A}|c_B|^2 + 2 {\tilde B}_4|c_B|-m_b^2=0.
\end{equation}
The general solution to this gap equation is given by 
\begin{equation}\label{gse}
|c_B|= \frac{- {\tilde B}_4  \pm {\rm sgn}({\tilde B_4}) \sqrt{{\tilde B}_4^2+ {\tilde A} m_b^2}}{ {\tilde A}},
\end{equation}
where we chose the sign multiplying the square root for later convenience.
For each value of the parameters we can have up to four solutions of the gap equations, choosing the tilde variables to correspond either to the Higgsed or to the unHiggsed branches.
We will denote the two solutions in the unHiggsed branch by $\pm_u$, and similarly we have $\pm_h$ in the Higgsed branch.
Of course not all solutions to \eqref{gse} are legal; the quantity $|c_B|$ is intrinsically real and non-negative, while one or both of the roots \eqref{gse} may be complex or negative. How many of the roots of \eqref{gse} are legal depends on the
values of the parameters ${\tilde A}$ and ${\tilde B}_4$ in the two possible branches and on $m_b^2$.

When ${\tilde A} m_b^2 < -{\tilde B}_4^2$, there are no solutions to the gap equation on the corresponding branch, so we do not know the low-energy behaviour of the system. We will suggest below that in this range of parameters, we may have a runaway behaviour with no stable vacuum.
When ${\tilde B}_4$ and ${\tilde A}$ have the same sign, then only the solution with the $+$ sign  (the `$+$' solution) exists, for ${\tilde A} m_b^2 > 0$. As $m_b^2 \to 0$ (with this sign), this solution has $c_B \to 0$, indicating a second order phase transition there. When ${\tilde B}_4$ and ${\tilde A}$ have opposite signs, then for ${\tilde A} m_b^2 > 0$ only the `$-$' solution exists. This solution continues smoothly (with no phase transitions) all the way down to ${\tilde A} m_b^2 = -{\tilde B_4}^2$, where it merges with the `$+$' solution. This `$+$' solution exists in the range $-{\tilde B}_4^2 < {\tilde A} m_b^2 < 0$, and as $m_b^2 \to 0$ it has $c_B \to 0$, indicating a second order phase transition.

Note that the solutions that exhibit a second order phase transition are always the `$+$' solutions, with Higgsed and unHiggsed phases on the two sides of the transition. Thus it is natural to interpret those as sitting, in some sense, near the origin of the moduli space of the scalar fields, so that the scalars can condense and break the low-energy gauge group when going from one phase to the other. The `$-$' solutions can then be interpreted as solutions far from the origin, that only approach the origin when they merge with the `$+$' solutions. It would be interesting to support this interpretation by computing expectation values of scalar fields in these solutions.

Let us now see what this implies for specific values of the parameters $x_6$, $b_4$ and $m_b^2$. On the unHiggsed branch ${\tilde A}$ and ${\tilde B}_4$ take the values $A_u$ and $B_{4,u}$, where  
\begin{equation}\label{defBfu}
\begin{split}
 { B}_{4,u} &=b_4 { \lambda}_B, \\
 {A}_u &= 1 - \left( 1+\frac{3}{4} x_6^F \right)  {\lambda}_B^2.
 \end{split}
\end{equation}
On the other hand on the Higgsed branch they
take the values $A_h$ and $B_{4,h}$, where  
\begin{equation}\label{defBft}
\begin{split}
 B_{4,h}&= b_4 {\hat \lambda}_B= -{\rm sgn}(\lambda_B) b_4 (2-|\lambda_B|),  \\
 A_h &= 1-\left( 1+\frac{3}{4} x_6^F \right)   |{\hat \lambda}_B|^2.
  \end{split}
\end{equation}
Note that $B_{4,u}$ and $B_{4,h}$ always have opposite signs.

Let us define
\begin{equation}\label{phdef}
\phi_1\equiv \frac{4}{3}\left(\frac{1}{{\hat \lambda}_B^2}-1\right),\qquad\qquad
\phi_2\equiv  \frac{4}{3}\left(\frac{1}{{\lambda}_B^2}-1\right).
\end{equation}
Note that $\phi_2 > 0 > \phi_1$. It follows from \eqref{defBfu} and \eqref{defBft} 
that 
\begin{equation}\label{axc}
\begin{split}
&{\rm when} ~~x_6< \phi_1,~ ~~~~~~~~~A_u >0, ~~~A_h >0,\\
&{\rm when} ~~\phi_1<x_6< \phi_2, ~~~A_u >0, ~~~A_h<0,\\
&{\rm when} ~~x_6> \phi_2,~~~~~~~~~~A_u<0, ~~~A_h<0.\\
\end{split}
\end{equation}

\subsection{The phase diagram}
\label{phasediag}

In this section we present the main features of the phase diagram that follow from the equations above. More details and computations appear in appendix \ref{phasecomp}.

Let us begin with the middle range, $\phi_1 < x_6 < \phi_2$. For $B_{4,u} > 0$, the only solution to the gap equations for $m_b^2>0$ is $+_u$; this solution exists for all positive values of $m_b^2$. Similarly the only solutions to the gap equation for $m_b^2<0$ are the   $+_h$ solutions; these solutions exist for all negative values of $m_b^2.$ These two solutions join together along the positive $B_{4,u}$ axis. As we approach the axis from either side, $c_B\rightarrow0$, suggesting that there is a second order phase transition described by the `critical boson' fixed point there.  This is consistent with our analysis around \eqref{scl12}, which shows that the low-energy thermodynamic behaviour near this axis, namely when $m_B^{cri} = m_b^2 / 2 B_{4,u} \ll |m_b|, B_{4,u}$, is the same as that of the CB theory deformed by a mass $m_B^{cri}$.

On the other hand, for $B_{4,u} < 0$ four distinct solutions to the gap equation play a role. 
 The $-_u$ solution exists for $m_b^2 > -B_{4,u}^2/A_u$. The $+_u$ solution exists for $0>m_b^2 > -B_{4,u}^2/A_u$.
 The $-_h$ solution exists for $m_b^2 < -B_{4,h}^2 / A_h$; and the $+_h$ solution exists for $0<m_b^2 < -B_{4,h}^2 / A_h$.
 The value of $c_B$ goes to zero along the negative $B_{4,u}$ axis for both the $+_u$ and $+_h$ solutions, so these 
 two solutions can be thought of as a single branch of solutions that meet along a second order phase transition 
 line on the negative $B_{4,u}$ axis. 
 
 When $m_b^2 > -B_{4,u}^2/A_u$, $-_u$ is the only solution and so is the dominant phase. When 
 $m_b^2 < -B_{4,h}^2 / A_h$, $-_h$ is the only solution and so is the dominant phase. In the region 
 $ -B_{4,u}^2/A_u<m_b^2 < -B_{4,h}^2/A_h$ there are three distinct solutions to the gap equations. 
 All through this region, two of these solutions are $-_u$ and $-_h$. The third solution is either $+_u$ 
 or $+_h$, depending on whether $m_b^2$ is negative or positive. 
 We find (see Appendix \ref{phasecomp}) that one of the $-_h$ or the $-_u$ solution always has the lowest free energy and so 
 is the dominant phase. Recall that we have interpreted these solutions as  being far from the origin. Moreover we find that there is a first order phase transition between these two phases, in which $c_B$ jumps, somewhere in this range of values of $m_b^2$, see Figure \ref{figctb} below. The position of the phase transition line between these two phases 
 interpolates between $-B_{4,u}^2/A_u$ when $x_6=\phi_2$  and  $-B_{4,h}^2 / A_h$ when $x_6=\phi_1$. In 
 particular when $x_6=\phi_2$ only the top left quarter of the phase diagram in Figure \ref{figctb} is in the Higgsed phase, 
 while at $x_6=\phi_1$ only the top right quarter of the same phase diagram is in the unHiggsed phase. We reemphasize that the`$+$' solutions with their second order phase transition are never the lowest energy solutions in this regime of negative $B_{4,u}$.

So far we described the transitions as we change $m_b^2$; since the phase structure for specific signs of $m_b^2$ and $b_4$ depends only on $m_b^2
/ b_4^2$, we should examine also what happens near $b_4=0$ (for non-zero $m_b^2$). For ${\tilde A} m_b^2 > 0$ we have a single legal solution to \eqref{gse}, and otherwise none. In this region ${\tilde A}$ takes opposite signs in the Higgsed and unHiggsed theories, so  for either sign of $m_b^2$ we have a single solution near $b_4=0$ (a different solution for each of the two different signs). For $m_b^2 > 0$ the $+_u$ solution smoothly goes into the $-_u$ solution as $b_4$ crosses zero, with no phase transition, and for $m_b^2 < 0$ the $+_h$ solution smoothly goes over to the $-_h$ solution (the transitions are smooth because of the ${\rm sgn}({\tilde B_4})$ in \eqref{gse}).

The full phase structure we found for this range of values of $x_6$, including the low-energy topological theory in each region, is drawn in figure \ref{figctb}.

\begin{figure}[h]
\begin{center}
\includegraphics[width=12.5cm,height=9cm]{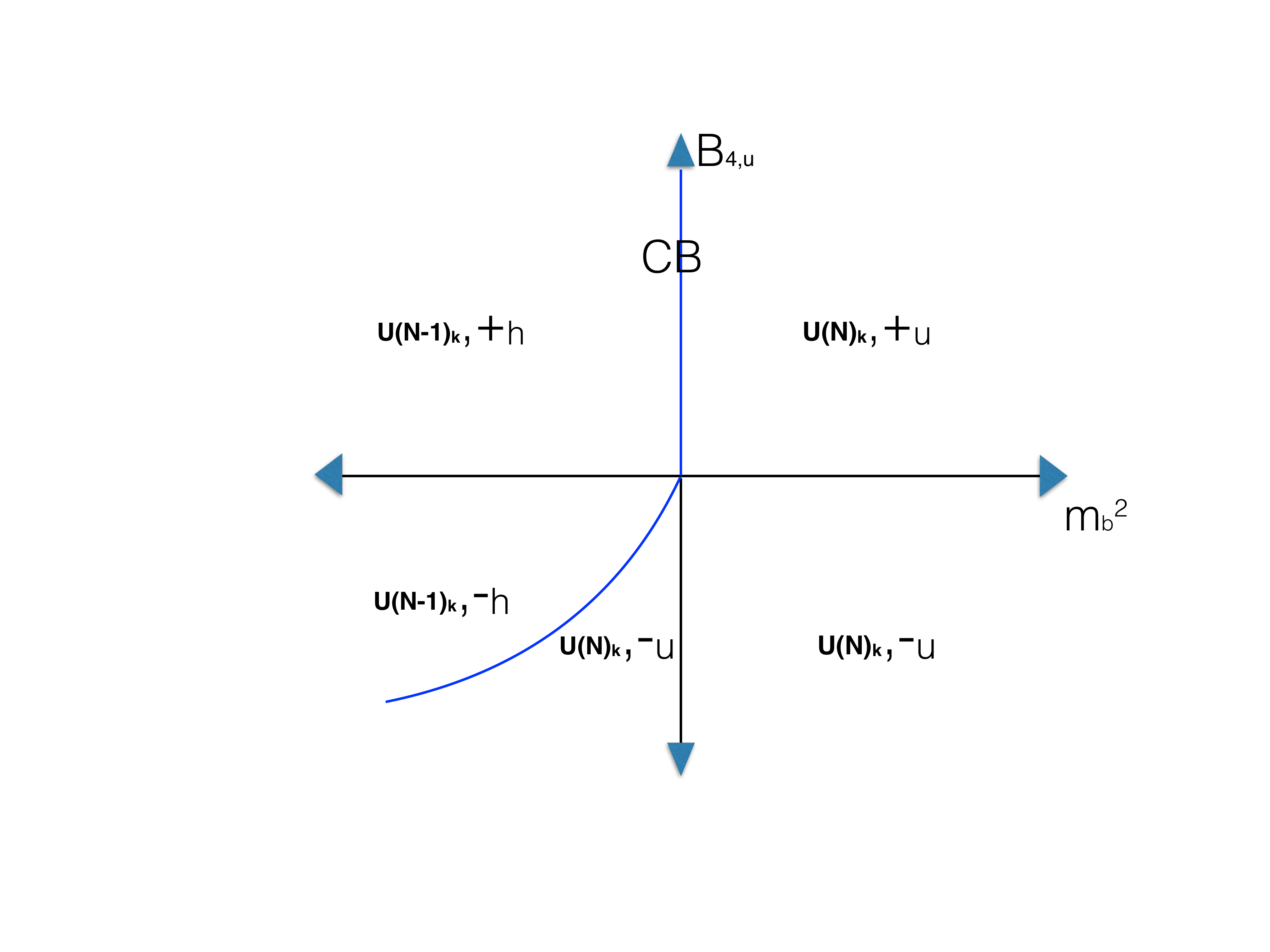}
  \caption{\label{figctb} The phase diagram for $\phi_1 < x_6 < \phi_2$. For $B_{4,u} > 0$ we have a second order phase transition at $m_b^2=0$ between $+_u$ and $+_h$ phases. For $B_{4,u} < 0$, the blue line is a first order phase transition between the region in which the solution $-_u$ 
 is the dominant phase and the region in which the $-_h$ solution dominates. Note that the first order phase transition is at $m_b^2 > 0$ for $x_6 > (\phi_1 + \phi_2) / 2$,
 but it is at $m_b^2 < 0$ otherwise. 
 Our schematic phase diagram has been sketched for the latter case. }
 \end{center}
\end{figure}

For $x_6 > \phi_2$, $A_h$ remains  negative as it was for $\phi_1 < x_6<\phi_2$. It follows that  the space of Higgsed solutions 
in this parameter regime is the same as for the case $ \phi_1 <x_6 < \phi_2$ that we have just analyzed above. 
However $A_u$, which was positive in the earlier parametric regime, turns negative when $x_6>\phi_2$. It follows that the space of unHiggsed solutions changes in the manner we now describe. 

For $B_{4,u} > 0$, the $-_u$ solution exists for  $m_b^2 < -B_{4,u}^2 / A_u$ while the $+_u$ solution exists in 
the range $0 < m_b^2 < -B_{4,u}^2 / A_u$. These two solutions merge smoothly at $m_b^2 = -B_{4,u}^2 / A_u$.  For $m_b^2 > -B_{4,u}^2 / A_u$ there is no solution, unHiggsed or Higgsed, to the gap equations.  Below this value there are always two solutions ($-_u$ and either $+_u$ or $+_h$ depending on whether $m_b^2$ is positive or negative). It turns out that the dominant solutions are the $+_u$ and $+_h$ solutions, so that the theory has a second order phase transition between these two solutions at $m_b^2=0$. The long distance behaviour of the phase transition point is again governed by the 
CB theory. 

For $B_{4,u} < 0$, the only unHiggsed solution is $+_u$, and this solution exists only when  $m_b^2 < 0$. The Higgsed solutions are the same as above, so now we have no solutions when $m_b^2 > -B_{4,h}^2 / A_h$. Below this value we have the $-_h$ solution and either the $+_u$ or the $+_h$ solution, depending on whether $m_b^2$ is negative or positive (with a smooth transition between the latter when $m_b^2=0$). In this case it turns out that the $-_h$ solution dominates, so our theory is always in the Higgsed phase for negative $B_{4,u}$ and there are no phase transitions as $m_b^2 \to 0$.

\begin{figure}
   \includegraphics[width=12.5cm,height=9cm]{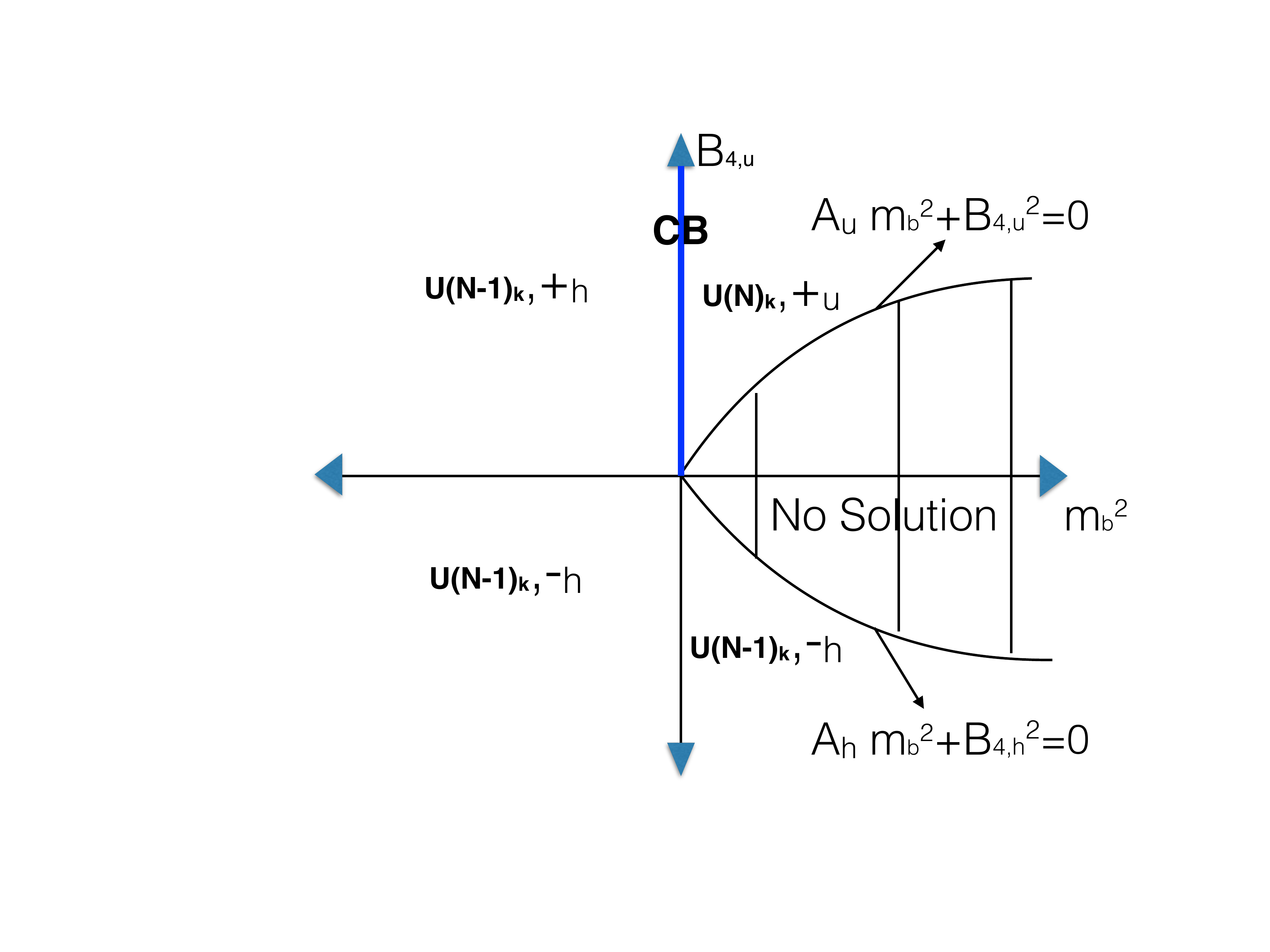}
  \caption{The phase diagram for $x_6 > \phi_2$. For $B_{4,u} > 0$ we have a second order phase transition at $m_b^2=0$ between $+_u$ and $+_h$ phases, while for $B_{4,u} < 0$ the $-_h$ solution dominates whenever it exists. There are no solutions for large enough values of $m_b^2$.}
\label{figA1B1ph}
\end{figure}

Again we find that nothing special happens as $b_4 \to 0$ (with no solutions for $m_b^2 > 0$, and with the $-_h$ solution smoothly going into the $+_h$ solution for $m_b^2 < 0$). The full phase structure for this range of values of $x_6$ is drawn in figure \ref{figA1B1ph}.

Next, let us consider what happens when we increase $x_6$ to $\phi_2$, such that $A_u$ goes to zero and then becomes negative, while nothing special happens to the Higgsed solutions. Figures \ref{figctb} and \ref{figA1B1ph} suggest that in this limit nothing dramatic happens except in the lower right quadrant, and this is indeed the case. As $x_6 \to \phi_2$, the $-_u$ solution has $|c_B| = -2B_{4,u} / A_u$, so it exists in the full region $B_{4,u} < 0$ and its $|c_B|$ goes to infinity, and then for larger $x_6$ its $|c_B|$ becomes negative so that this solution no longer exists, leaving no solutions for $B_{4,u} < 0$ and $m_b^2 > -B_{4,h}^2 / A_h$. As we pass $x_6 = \phi_2$, the $-_u$ solution starts existing (as a sub-dominant solution) on the other side $B_{4,u} > 0$, gradually coming down from $|c_B|=\infty$. On the other hand, the $+_u$ solution in this limit has $|c_B| = m_b^2 / 4 B_{4,u}$, so it exists whenever this ratio is positive, and smoothly goes through $x_6 = \phi_2$.

\begin{figure}
    \includegraphics[width=12.5cm,height=9cm]{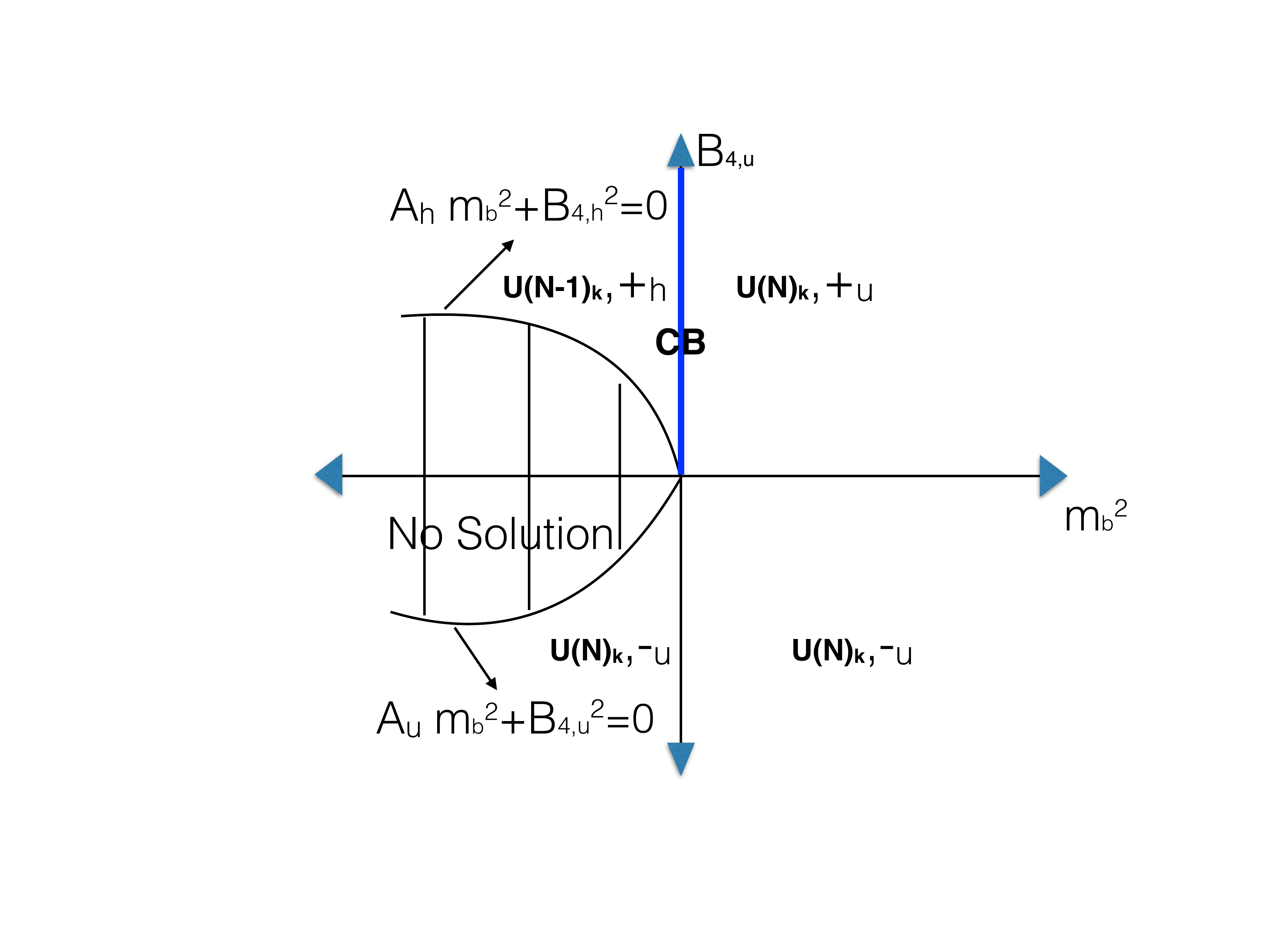}
  \caption{The phase diagram for $x_6 < \phi_1$. For $B_{4,u} > 0$ we have a second order phase transition at $m_b^2=0$ between $+_u$ and $+_h$ phases, while for $B_{4,u} < 0$ the $-_u$ solution dominates whenever it exists. There are no solutions for small enough values of $m_b^2$.}
\label{figA3B3}
\end{figure}

For $x_6 < \phi_1$ the picture is essentially the same, just exchanging the unHiggsed and Higgsed phases, and the sign of $m_b^2$ -- see figure \ref{figA3B3} and the Appendix for more details. 

\subsection{A curious limit}
\label{curious}

Our discussion above implies (as we checked also in section \ref{rd} for finite temperatures) that for $B_{4,u} > 0$ and $m_b^2=0$ the RB theory flows to the CB theory (for any value of $x_6$). Somewhat surprisingly, there is also another limit where this seems to be the case. Consider the limit\footnote{A similar limit exists also for $x_6 \to -\infty$.}
\begin{equation}\label{scal123}
b_4=0 ~,~x_6\rightarrow\infty ,~m_b^2\rightarrow -\infty,
\end{equation}
keeping the pole mass $c_B$ and $m_B^{cri}$ fixed, where 
\begin{equation} \label{mbcrinew}
(m_B^{cri})^2\equiv -\frac{4 m_b^2}{3 x_6 \lambda_B^2}.
\end{equation}
 In this limit we have two solutions to the RB theory's finite temperature (and zero temperature) gap equations (above we called these solutions at zero temperature by different names when $B_{4,u} > 0$ and when $B_{4,u} < 0$, but they go smoothly through $B_{4,u}=0$). The first solution is to the gap equation \eqref{scalargapequation} in the unHiggsed phase, and it is given by
${\tilde S} =\frac{| m_{B}^{cri} |}{2}$ ; note that this is the same as the gap equation of the CB theory in its unHiggsed phase (equation \eqref{offshellfe} with ${\rm sgn}(X_B)=-{\rm sgn}(\lambda_B)$).  At the same values of the RB theory's microscopic parameters, the equations for the Higgsed phase \eqref{offb} reduce to 
\begin{equation}
\label{curioushiggsed} 
|c_B|=|\lambda_B| \cS + \frac{| m_B^{cri}  \lambda_B|}{2},
\end{equation}
which agrees with the gap equation of the CB theory in its Higgsed phase (equation \eqref{offshellfe} with ${\rm sgn}(X_B)={\rm sgn}(\lambda_B)$). 

The observations of the last paragraph suggest that there is a close connection between the scaling limit 
\eqref{scal123} and the critical boson theory; naively it seems that we flow to the latter theory at low energies. However, the relationship between the deformed RB theory and the CB  theory is a curious one. 
To see this let us focus on the zero temperature limit. 

In a deformed critical boson theory we are either in the Higgsed or 
unHiggsed phase depending on the value of the microscopic parameter $m_B^{cri}$; positive values of this parameter 
put us in the unHiggsed phase while negative values of this parameter put us in the Higgsed phase. This is clear, for instance, from figure \ref{figA1B1ph}. 

In the scaling 
limit \eqref{scal123} of the RB theory, on the other hand, we found above that at the {\it same} value of the microscopic parameters there are two different phases, which both lie on the negative $m_b^2$ axis in figure \ref{figA1B1ph}. Recall that there are two solutions of the gap equation in all allowed regions of this figure; in figure \ref{figA1B1ph} we drew only the thermodynamically dominant phase, while in fact for every value of $m_b^2$ and $b_4$ there is also another, thermodynamically sub-dominant phase; the two phases are analyzed in appendix \ref{phasecomp} and depicted in figures \ref{figA1} and \ref{figB1}. These two phases do not have the same free energy (the free energy 
of the Higgs phase is negative and dominates) and so at any finite $x_6$ the unHiggsed phase is unstable and 
decays. However, in the limit \eqref{scal123} the decay amplitude between these phases vanishes, so the deformed RB theory has two 
different effective superselection sectors (the Higgsed and unHiggsed saddle points of the RB theory, which are superselection sectors because tunneling switches off at large $x_6$). We can then identify these by the relations described above with the Higgsed and unHiggsed deformations of the 
CB theory, depending on our choice of the sign of $m_B^{cri}$ in \eqref{mbcrinew}. It seems that in the limit \eqref{scal123}, the CB theory controls not just the region near the positive $B_{4,u}$ axis in figure \ref{figA1B1ph}, but also the region near the negative $m_b^2$ axis (note that in the $x_6 \to \infty$ limit there are no solutions to the gap equation with finite $B_{4,u}$ and $m_b^2 > 0$).

It is clear that the relationship we found between the scaling limit \eqref{scal123} of the RB theory and the CB theory, for finite values of $m_B^{cri}$, cannot hold when  $m_B^{cri}=0$. For one thing the conformal  RB theory, which sits at the origin of figure \ref{figA1B1ph}, has 
a deformation (along the $m_b^2$ axis to the right) for which it has no solution to the gap equations. 
There is no analogue of this phenomenon in the CB theory. Also, the RB theory has a dimension 1 scalar operator at all finite values of $x_6$, while that is not the case for the CB theory. So the precise statement is that as $x_6$ becomes large at infinite $N$, the CB theory controls also flows along the negative $m_b^2$ axis, but not the point at the origin. 

For finite $N$, as we have seen above,
$x_6$ is not really a marginal parameter; RG flows dynamically drive large values of $x_6$ to infinity. So we can speculate that perhaps when we go to finite values of $N$, the relation between these theories can be extended also to the origin of the parameter space, when we start in the UV with $x_6 \gg 0$.
The naive expectations we described above for the behaviour of the RB theory may be significantly modified at $x_6 \sim N_B^{\frac{1}{3}}$ (when the $\beta$ 
function of $x_6$ ceases to be small), and then the true end point of the RG flow along positive $x_6$ may 
genuinely be the CB theory. It is also possible that this is not the case. We leave further discussion of this 
issue to future work.

\subsection{Interplay between the phase diagram and the RG fixed points} \label{interplay}

In section \ref{phasediag} we have worked out the phase diagram of the RB theory at infinite $N_B$; we discovered 
that the phase diagram changes qualitatively depending on where $x_6$ lies in relation to the distinguished 
values $\phi_1$ and $\phi_2$ defined in \eqref{phdef}. 

In section \ref{betafunc} we saw that $1/N_B$ corrections cause $x_6$ to run. The running is governed by 
a $\beta$ function with three fixed points. In other words we can treat $x_6$ as a free parameter only in the 
strict large $N_B$ limit. At any finite $N_B$, no matter how large,  we have true fixed points of the renormalization 
group only at these 3 particular values of $x_6$. The phase diagrams that we have evaluated in section \ref{phasediag} 
(which are approximately correct even at large but finite $N_B$, at any given value of $x_6$) are of most interest 
at the fixed point values of $x_6$. Thus it is of interest to understand where the three fixed point 
values of $x_6$ lie in relation to $\phi_1$ and $\phi_2$. 

\begin{figure}
    \includegraphics[width=14.5cm,height=8cm]{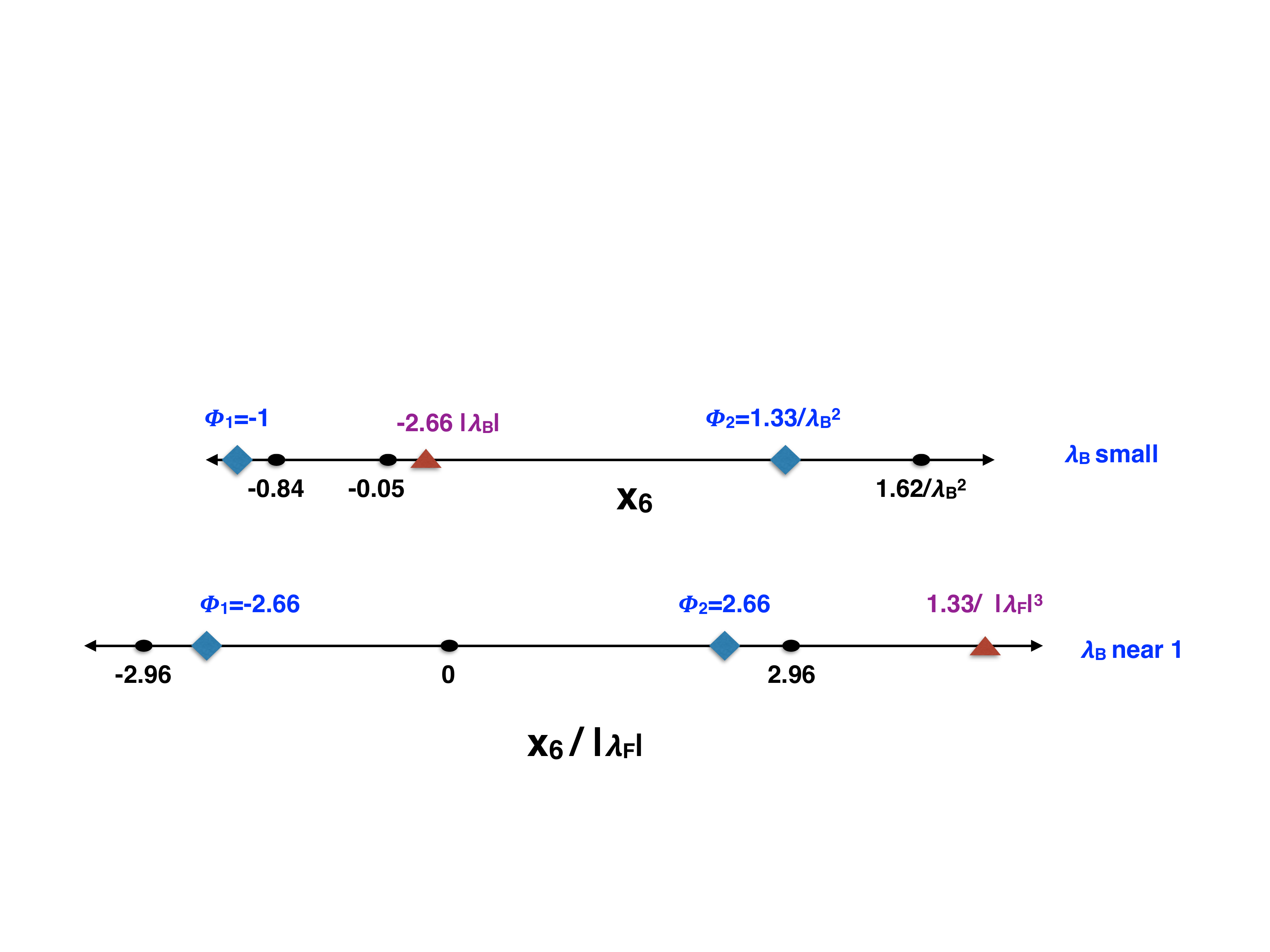}
  \caption{A sketch of the $x_6$ axis at small $|\lambda_B|$ and at small $1-|\lambda_B|$. In each case we have plotted 
  the three fixed points of the renormalization group as black dots, $\phi_1$ and $\phi_2$ as blue diamonds,  and 
  the end point of SUSY RG flows (discussed in the next section) as a red triangle. In the case of small $|\lambda_B|$ the fixed points of the RG flow 
  are listed in \eqref{betzero}, $\phi_1=-1$, $\phi_2 \approx {4}/{(3|\lambda_B|^2)}$, and the end point of the SUSY RG flow occurs at  $-{8|\lambda_B|}/{3}$. At small $1-|\lambda_B|$ the fixed points of the RG flow 
  are listed in \eqref{sls}, $\phi_1$ and $\phi_2$ are $\approx \pm {8|\lambda_F|}/{3}$, and the end point 
  of the SUSY RG flow occurs at  ${4}/{(3|\lambda_F|^2)}$.} 
\label{figphse12a}
\end{figure}

While $\phi_1$ and $\phi_2$ are known analytically at all $\lambda_B$ (see \eqref{phdef}),  we do not have closed form expressions for the fixed point values of $x_6$ at arbitrary $\lambda_B$. However the corresponding expressions 
are available at small $|\lambda_B|$ (see \eqref{betzero}) and at small $1-|\lambda_B|$ (see \eqref{sls}). For these 
special cases we have sketched the relative positions of $\phi_1$, $\phi_2$ (blue stars) and of the three fixed points of the 
renormalization group (black dots) in Figure \ref{figphse12a}.\footnote{In the same figures we have 
also depicted the end point of RG flows starting at the supersymmetric theory (red triangles), which we will describe in the next 
section.} We see from Figure  \ref{figphse12a} that in both of these cases the middle fixed point (the attractive fixed point about which the $x_6$ deformation is irrelevant, which has a total of two relevant deformations) lies in between $\phi_1$ and $\phi_2$. It is natural to conjecture that this continues to be the case at all values of 
$\lambda_B$, and thus that the phase structure of this attractive fixed point for large finite $N_B$ is given by figure \ref{figctb}. 

Similarly, the rightmost fixed point is to the right of $\phi_2$ in both limits; there seems no 
reason to expect this to change at any value of $\lambda_B$. However the leftmost fixed point lies 
between $\phi_1$ and $\phi_2$ when $|\lambda_B|$ is small, but lies to the left of $\phi_1$ when $|\lambda_B|$
is near unity. It is likely that this fixed point will lie between $\phi_1$ and $\phi_2$ for all $|\lambda_B|$ less than 
some critical value that lies between zero and unity, but that it will lie to the left of $\phi_1$ for all $|\lambda_B|$ 
larger than this critical value. 

\subsection{Flows from Yang-Mills-Chern-Simons theories to Chern-Simons-matter theories}
\label{ymcs}

In this subsection we make some general remarks about 
the phase diagram one may expect for the RB theory
from a particular point of view. The discussion in this 
subsection is very general, makes no reference to particular 
computations, and should apply even at finite $N_B$. 
At the end of this subsection we discuss the relationship between the general remarks of this subsection and the explicit phase 
diagrams presented earlier in this section. 

As we mentioned in section \ref{prp}, one way to analyze the existence of the fixed points \eqref{cst1}-\eqref{cft1} is to start at high energies with Yang-Mills-Chern-Simons (YM-CS) theories, with some gauge coupling $g_{YM}$ and level $\kappa$. At energies much larger than $g_{YM}^2 N$ and $g_{YM}^2 \kappa$ these theories are weakly coupled, so they provide good starting points for RG flows. In this section we analyze this flow in detail, and relate the resulting phase structure (which is reliable at energies above $g_{YM}^2$) to our analysis above of the phase structure of the fixed points (these arise here at energies well below $g_{YM}^2$). To be explicit we discuss here the $U(N)_{\kappa}$ theory, but the same general statements apply for $SU(N)_{\kappa}$ theories and for $U(N)$ theories with different levels for the $SU(N)$ and $U(1)$ factors; for finite $N$ and $\kappa$ the precise behaviour may be different for these other cases.

In order to analyze the theories we are interested in, it is useful to start from scalar fundamental fields coupled to the YM-CS theory. In this case we can add at high energies a scalar potential
\begin{equation} \label{scalarpot}
V(\phi) = g_2 |\phi|^2 + g_4 (|\phi|^2)^2 + g_6 (|\phi^2|)^3.
\end{equation}
For $g_6 \neq 0$ the theory remains interacting also at high energies and it is not clear if it is well-defined, but if we choose $g_6=0$ then these theories are well-defined, for any value of $g_2$ and $g_4$ (generally $g_6$ is generated along the RG flow). Moreover, if we choose the values of $g_2$ and $g_4$ such that the scalar field becomes massive at a high scale (above $g_{YM}^2 N$ and $g_{YM}^2 \kappa$), then at that scale we can trust the analysis of the scalar using the potential \eqref{scalarpot}, and below this scale we have a pure YM-CS theory which is believed to flow to a pure CS theory (for these values the Yang-Mills interactions are always small; the scalar self-interactions may be large, but they are not expected to change the qualitative form of \eqref{scalarpot}). Note that all these statements do not depend on taking any large $N$ or large $\kappa$ limits.

Let us begin with the case $g_4 > 0$ (and large in units of $g_{YM}^2 N$ or $g_{YM}^2 \kappa$; this is what we will mean by large throughout this subsection). If $g_2 > 0$ is also large, then $\phi$ is massive and has no expectation value, so that the low energy theory is a pure $U(N)_{\kappa}$ CS theory. On the other hand, if $g_2 < 0$ is large (in its absolute value), $\phi$ gets a large expectation value, such that at low energies we have a pure $U(N-1)_{\kappa}$ CS theory. As we change $g_2$ from large and negative to large and positive there has to be a phase transition. This may be a second (or higher) order phase transition as suggested by the classical potential \eqref{scalarpot}, or a first order phase transition; since quantum corrections are important when $\phi$ is light and near the origin, this is difficult to analyze. Whenever we have a second order phase transition it is described by the CB theory, while for values of $N$ and $\kappa$ for which there is a first order transition it is not clear if the CB theory exists. 

Next, consider the case of $g_4 \ll 0$. The behaviour at large $\phi$ now depends on $g_6$, but since it vanishes (at the cutoff scale) it seems that there would be a runaway behaviour in this regime (the classical potential should be reliable at very large $\phi$ since quantum corrections are small there)\footnote{Perhaps this implies that this quantum field theory does not really make sense, since it has no stable vacuum.}. If $g_2 \gg 0$ we now have a massive meta-stable vacuum at the origin, in which the low-energy theory is given by the $U(N)_{\kappa}$ CS theory. For very large $g_2$ the life-time of this vacuum goes to infinity, so its existence is reliable. If $g_2 \ll 0$ there is no meta-stable vacuum, and $\phi$ runs away to some large finite or infinite value. So again there is some phase transition as a function of $g_2$, and in this case it is clearly of first order (whenever stable vacua exist at large $\phi$).

\begin{figure}[h]
\begin{center}
\includegraphics[width=14.5cm,height=7.5cm]{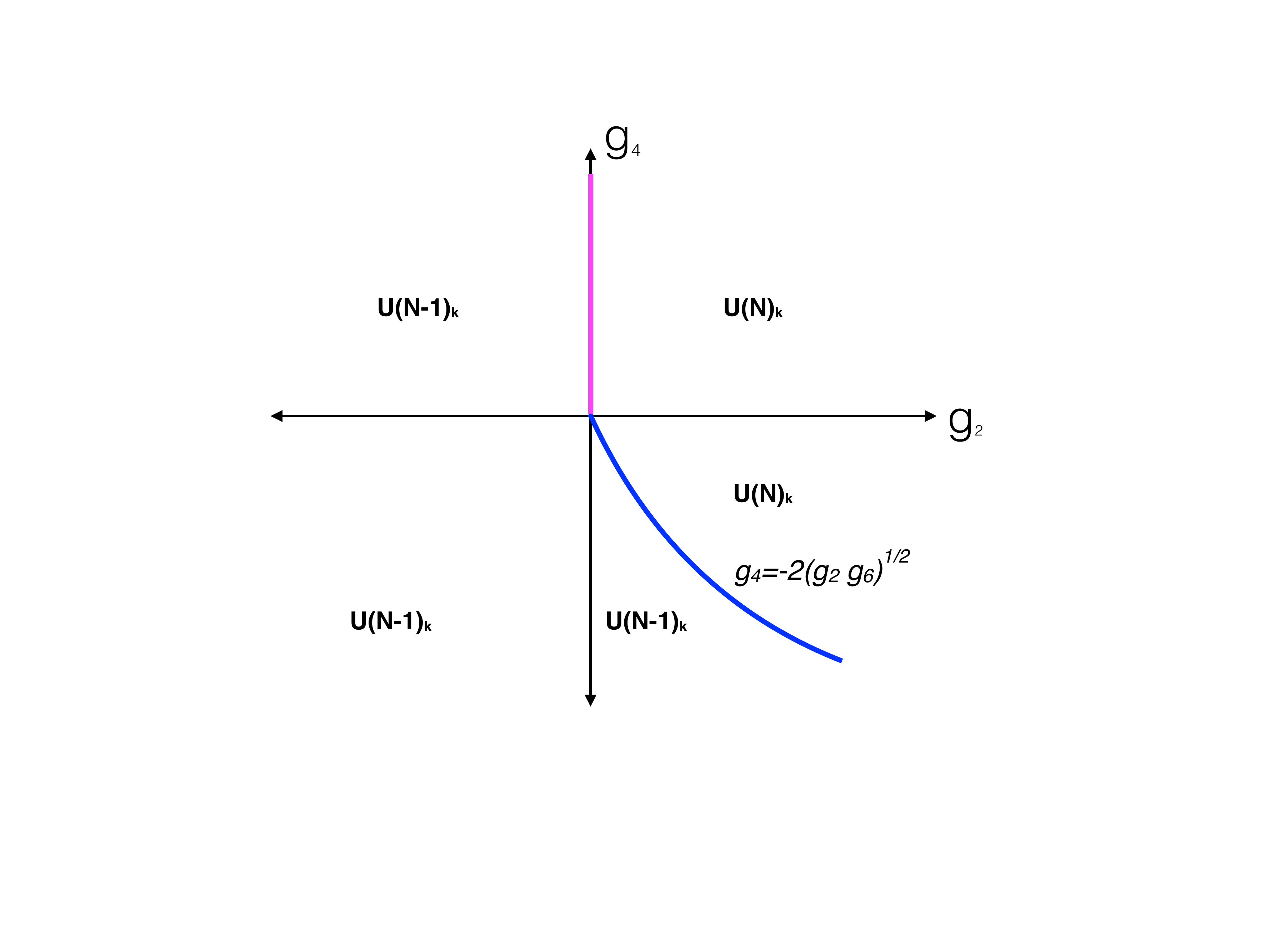}
  \caption{\label{rgflgv} The phase structure of YM-CS theories as a function of the parameters of the scalar potential \eqref{scalarpot}. This structure is reliable far away from the two axes, when both $|g_2|$ and $|g_4|$ are large; the regions marked by $U(N-1)_{\kappa}$ may instead involve runaway behaviour with no stable vacuum.}
 \end{center}
\end{figure} 

See figure \ref{rgflgv} for the phase structure, assuming that there is a range of parameters where we flow to the CB theory; this is reliable when both $|g_2|$ and $|g_4|$ are large.

For all values of $g_4$ we thus have a transition as a function of $g_2$, between a vacuum at the origin and a situation where $\phi$ has an expectation value (or runs away to large $\phi$). Suppose that for $g_4 \gg 0$ we have a second order phase transition and thus a CB theory. As we decrease $g_4$, at some point this line of second order phase transitions should end and become a line of first order phase transitions. Classically the end is at $g_4=0$ where we get an RB theory with $g_6=0$; in the quantum theory, with the gauge interactions and the running $g_6$, it is difficult to say where the line ends, but it ends somewhere. One option is that this end is a CB fixed point, but in this case it is difficult to see how a small deformation of this theory could lead to a theory with two stable vacua or with a runaway (as we have near the line of first order phase transitions); recall that the CB theory has a single stable vacuum for either sign of its deformation parameter. Thus, a more likely option seems to be that the end is an RB theory; since this theory has parameters corresponding to $g_2$ and $g_4$, it naturally leads to the correct phase diagram around it. This argument naively implies that whenever the CB theory exists as an IR limit of the corresponding YM-CS theory, the RB theory should exist as well. Note that in particular this is the case in the ungauged scalar theory, where the role of the RB theory is simply played by free massless scalars.

In the large $N$ limit we can try to relate the analysis here to our discussion of the running coupling. The coupling $g_6$ runs (at least below the scale $g_{YM}^2 \kappa$), but for large $N$ it runs slowly (when we take $g_6 = \lambda_6 / N^2$ for fixed $\lambda_6$) since its beta function goes as $1/N$, so one can approximate the RG flow to the end-point of the CB line, described in the previous paragraph, as leading to an RB-like theory with some value of $g_6$, and then $g_6$ continuing to slowly flow. Note that in this context we cannot choose the value of $g_6$, since there is a specific value that follows from the high-energy theory with $g_6=0$ that we specified. If $g_6$ flows to a fixed point then we get precisely the picture of the previous paragraph; in particular this picture requires that there is an IR-stable fixed point for $g_6$. However, from the point of view of the flow as a function of $g_6$ it is possible also to have $g_6$ flow to $(+\infty)$ or $(-\infty)$, and then we cannot use the Yang-Mills-Chern-Simons theories to define the quasi-bosonic fixed points. Note that at leading order in $1/N$ our discussion of the beta function implies that large enough positive values of $\lambda_6$ will always flow to $(+\infty)$, and negative to $(-\infty)$, but for finite large $N$ this analysis is only valid when $\lambda_6$ does not scale with $N$, so it is possible that even large values of $|\lambda_6|$ may end up at a fixed point.

In section \ref{phasediag} we analyzed the large $N$ phase structure of the quasi-bosonic theories at large $N$ and zero temperature. This analysis is valid in an opposite regime from the one described above -- namely when the deformation parameters $g_2$ and $g_4$ are much smaller than $g_{YM}^2 N$ and $g_{YM}^2 \kappa$ (which are taken to infinity in this paper). At infinite $N$, $g_6$ (or equivalently the parameter $x_6$ of the low-energy theory \eqref{rbt1n}) is exactly marginal and we can compute the phase structure of \eqref{scalarpot} as a function of this parameter. When it is small and positive we find the same phase structure as in our computation of section \ref{phasediag} (see figure \ref{figctb}), consistent with having the Yang-Mills-Chern-Simons theory flow to the RB fixed point at large $N$. In particular the RB theory flows after a mass deformation of the appropriate sign to the CB theory. For large $|x_6|$ we found in section \ref{phasediag} that the phase structure changes. There is still a CB line as described above, but for the other sign of the mass deformation there seems to be no stable vacuum (this agrees with the semi-classical expectation from \eqref{scalarpot} when $x_6 < 0$, but we find this behaviour also for positive $x_6$). Of course, when the coupling is not small there is no reason for the semi-classical analysis above to be valid. In the upcoming paper \cite{new} an exact Landau 
Ginzburg description of the phase structure is presented  which is consistent with our large $N$ results.

Note that the issues we discuss in this section are generally similar to the ones arising in the flow from the SUSY theory that we described in section \ref{prp}; in both cases we have a well-defined UV starting point, but it is not clear what happens near the putative RB point. In the SUSY flow for large $N$ we could compute at which value of $x_6$ we end up, so that in some cases we know that we end up at a stable fixed point, while in this section it is not clear how to analyze this. It seems likely that the phase structure near the putative RB point is the same in both cases, but we do not know how to prove this. We can also discuss a similar UV completion for the CF theories \eqref{cft1}, that should include a kinetic term for the scalar field $\zeta$ in addition to the Yang-Mills term; we expect this to be analogous to our discussion here.

\section{The ${\cal N}=2$ superconformal Chern-Simons-matter theory} \label{susysec}

In this section we study RG flows emanating from the ${\cal N} =2$ superconformal Chern-Simons-matter theory with a 
single fundamental chiral multiplet. In components, the matter content of this theory  is 
a single fundamental complex boson $\phi$ plus a single fundamental complex fermion $\psi$, 
each minimally coupled to a Chern-Simons gauge field. As in the previous sections the gauge group can be either $SU(N)_{\kappas}$ or $U(N)_{\kappas, \kappas_2}$ (see Appendix \ref{levels} for 
notation). This will affect the precise form of the duality in the SUSY theory (and in the resulting non-SUSY theories), but it will not matter for our leading-order (in $1/N$) computations.

We will use the results of \cite{Jain:2013gza} to study the RG flows 
seeded by relevant deformations from this fixed point. In particular we begin with a pair of superconformal field theories that are dual to each other by the generalizations of the Giveon-Kutasov duality \cite{Benini:2011mf}, and construct fine-tuned large $N$ RG flows 
that begin at these superconformal theories and end on the RB/CF
theories at a particular value of $x_6$. The flows from the 
superconformal fixed point to the RB/CF
theories are seeded by operators that are relevant even in the strict 
large $N$ limit and so are `fast flows' in the sense that they 
take place over ratios of scales of order unity. Once these flows have 
settled down into the RB/CF manifolds, 
the much slower flow of $x_6$ -- governed by the $\beta$ functions 
computed in detail in section \ref{betafunc} -- ensues over 
RG flow `times' of order $N$, and our analysis in this section does not add any information about the end-point of these flows. However, the flows in this section are well-defined even for finite $N$, at least if $N$ is large enough, so one can conjecture that they may end (for appropriate fine-tunings) at RB/CF fixed points even for small values of $N$.

\subsection{Lagrangian}

After integrating out all auxiliary fields, and working 
with the dimensional reduction scheme (which is convenient for actual computations), the Lagrangian 
of a general  $U(N)_{\kappas, \kappas_2}$  theory with the same field content as the SUSY theory takes the form   
\beal{
S  &= \int d^3 x  \biggl[i \varepsilon^{\mu\nu\rho} {\kappa \over 4 \pi}
\Tr( A_\mu\partial_\nu A_\rho -{2 i\over3}  A_\mu A_\nu A_\rho) + i \varepsilon^{\mu\nu\rho} {\kappa_2 \over 4 \pi}
B_\mu\partial_\nu B_\rho \nn
&+ D_\mu \bar \phi D^\mu\phi + \bar\psi \gamma^\mu D_\mu \psi 
+m_B^2 \bar\phi \phi +m_F \bar\psi \psi+ {4\pi b_4 \over \kappa} (\bar\phi \phi)^2 
+ \frac{4 \pi^2 x_6}{\kappa^2} (\bar\phi\phi)^3 \nn
&
+ \frac{4 \pi x_4}{\kappa} (\bar\psi \psi) (\bar\phi\phi)
+ \frac{2 \pi y_4'}{\kappa} (\bar\psi\phi)( \bar\phi \psi)
+ \frac{2 \pi y_4''}{\kappa} \left((\bar\psi \phi)( \bar \psi \phi ) 
+(\bar \phi \psi)( \bar \phi \psi )\right) \biggl],
\label{generalaction}
}
with 
\begin{equation}\label{kis}
\kappa=\kappas, ~~~\kappa_2=\kappas_2.
\end{equation}
In the ${\cal N}=2$ superconformal theory the coupling constants are given by
\begin{equation}\label{susyvalues}
m_F=m_B=b_4=y_4''=0, ~~~x_4=x_6=y_4'=1.
\end{equation}
Here $A_\mu$ is an $SU(N)$ gauge field, $B_\mu$ is a $U(1)$ gauge field and 
the covariant derivatives are defined as in \eqref{covder}.
The Lagrangian for the $SU(N)_{\kappas}$ theory is the same as \eqref{generalaction}, but with $B_{\mu}$ set to zero. We review the SUSY dualities between these theories in appendix \ref{mapping}.

\subsection{Marginal and relevant deformations} \label{mrd}

The ${\cal N}=2$ supersymmetric theory defined in the previous subsection enjoys 
invariance under ${\cal N}=2$ superconformal symmetry. In particular it is a fixed point under 
the renormalization group. In this section we will enumerate the relevant and marginal
deformations about this fixed point. 

In the large $N$ limit this fixed point has 3 relevant deformations 
($m_B^2$, $m_F$ and $b_4$ in \eqref{generalaction}) together with four exactly marginal 
deformations ($x_6$, $x_4$, $y_4'$ and $y_4''$). At large but finite $N$
these operators get anomalous dimensions. Actually these anomalous dimensions are not all independent;
 some are related to others by supersymmetry as 
we now explain.

Consider the superconformal operator $ O=\left( \Tr ({\bar \phi} \phi) \right)^2$. 
The supersymmetric descendents $Q_\alpha {\bar Q}^\alpha O$ and 
$ \left( Q_\alpha Q^\alpha + {\bar Q}_\alpha {\bar Q}^\alpha\right) O$ 
are linear combinations of the four operators parameterized by $x_6$, $x_4$, $y_4'$ and $y_4''$.
It follows that the anomalous dimensions of the four large $N$ marginal operators are determined 
by those of $O$ plus two additional independent anomalous dimensions. Explicit calculation
at weak coupling (to order $\lambda^2$) reveals that all these three independent 
anomalous dimensions are positive (see Table \ref{tab1} for a listing of the leading weak coupling 
results quoted from \cite{Avdeev:1992jt}).

\begin{table}
 \begin{tabular}{ l | c  }
    \hline
    Eigen Vector & Eigen Value  \\ \hline
      ${\delta m_B}$& $0$  \\ \hline
   $-\frac{{\delta m_F} }{4\pi\lambda}\frac{N^2}{ N-1}+\frac{4 \pi \lambda   }{N}{\delta b_4}$& $0$  \\ \hline
    $\frac{4 \pi   }{k}{\delta b_4}$& $\frac{16 \pi ^2 \lambda ^2 (N-1) \left(8 N^2+7 N-16\right)}{N^4}$  \\ \hline
    $-\frac{4 \pi ^2 {\delta x_6} \lambda ^2}{N^2}$& $\frac{48 \pi^2 \lambda^2 (N-1) (N (8 N+17)-26)}{N^4}$  \\ \hline
    $\frac{8\pi\lambda}{N}{\delta y}_4''$& $\frac{16 \pi ^2 \lambda ^2 (N-1) (N (8 N+7)-16)}{N^4}$  \\ \hline
    $\frac{N (2 {\delta x}_4+{\delta y}_4')}{2 (N-1)}-\frac{4
   \pi ^2 {\delta x}_6 \lambda ^2}{N^2}$& $\frac{16 \pi ^2 \lambda ^2 (N-1) (N (8 N+7)-16)}{N^4}$  \\ \hline
    $ \frac{N (4 N (8 N (N+1)-47)+127) (2 {\delta x}_4-\delta
   y_4')}{6 (N-1) (2 N-1) (4 N-3)}-\frac{4 \pi ^2 {\delta x}_6
   \lambda ^2}{N^2}$& $\frac{16 \pi ^2 \lambda ^2 (N+1) (N (8 N+9)-20)}{3 N^4}$  \\ \hline
  \end{tabular}
  \caption{The leading weak coupling anomalous dimensions about the 
${\cal N}=2$ $SU(N)$ superconformal point. The first column lists the flow eigenvectors
in the notation of \eqref{kis}, while the second column lists the eigenvalues 
(corrections to the classical dimensions of the corresponding operators). 
The first two lines assert that the mass operators, which preserve SUSY, receive no anomalous 
dimensions. The third line of the table lists the positive anomalous dimension 
of the superconformal primary $(\Tr ({\bar \phi }\phi))^2$. The fifth and 
sixth lines assert that the descendants of this operator share the same 
anomalous dimensions. The fourth line lists the positive anomalous dimension of 
the primary operator $(\Tr ({\bar \phi }\phi))^3$. The seventh line of the 
table lists the positive anomalous dimension of a superconformal 
primary which is built out of linear combinations of products of descendants of  
$\Tr ({\bar \phi} \phi)$.}
\label{tab1}    
\end{table}

At weak gauge coupling it follows that all the large $N$ marginal operators are in fact irrelevant,
and the ${\cal N}=2$ fixed point has three relevant deformations. Although it will not be 
very important for the rest of this paper, there is no reason to believe that this 
qualitative pattern does not persist at all values of the level and the rank.

It follows that at any finite $N$ the Lagrangian \eqref{generalaction} 
really defines a quantum field theory only upon setting 
\begin{equation}\label{spcval}
y_4''=0, ~~~x_4=x_6=y_4'=1.
\end{equation}
As the anomalous dimensions that reveal this fact are all of order 
$\frac{1}{N}$, however, calculations performed in the strict large 
$N$ limit are well-defined for all values of these four parameters. 
Below we will present results for the free energy of these theories 
in the strict large $N$ limit for arbitrary values of these parameters 
simply because we can, and then insert \eqref{spcval} into 
our final results.

\subsection{The partition function along general large $N$ RG flows}

At leading order in the large $N$ limit the operators parameterized by $x_6$, $x_4$, $y_4'$ and $y_4''$
are all exactly marginal, and there exists a 6 parameter set of RG flows (corresponding to a seven 
parameter set of quantum field theories) emanating out of the ${\cal N}=2$ supersymmetric fixed point. 
This seven parameter set of theories was listed in \eqref{generalaction}.
Note that the generic theory in this 7 parameter space is not supersymmetric. 
Several quantities in these theories can be computed in the large $N$ limit as we now review. 

The simplest physical quantities one can compute from this Lagrangian are the pole masses 
of the boson and the fermion. One can demonstrate  (see \cite{Jain:2013gza} for details)  that the zero temperature 
bosonic/fermionic propagators have poles at $p^2+ c_{B,0}^2=0$ and $p^2+c_{F,0}^2=0$, respectively, where 
$c_{B,0}$ and $c_{F,0}$ obey the following equations:
\begin{equation} \label{bbm}
c^2_{B,0} = \lambda^2 (1+3x_6) \frac{ c^2_{B,0}}{  4} -2\lambda  b_4 |c_{B,0}|  
 +  x_4 \frac{ \lambda (-\lambda + 2 ~\sgn(\rm{X_0}))}{   (\lambda -\sgn(\rm{X_0}) )^2} ( - m_F + x_4 \lambda |c_{B,0}| )^2 + m_B^2 ,
\end{equation}
\begin{equation} \label{fpm}
c^2_{F,0}=\left( {m_F - x_4 |c_{B,0}| \lambda  \over -\lambda + \sgn(\rm{X_0})} \right)^2,
\end{equation}
where
\begin{equation}
X_0= \lambda \left(c_{F,0}-x_4 c_{B,0} \right)+m_F.
\end{equation}

Following the discussion of subsection \ref{du} one can also compute the 
finite temperature sphere partition function of these theories. The 
relevant gap equations for the theory of this subsection turn out to be 
\begin{equation}\label{fTgp}
c_F^2 = (2\lambda (\cC-x_4 \cS) +{\hat m}_F)^2,
\end{equation}
\begin{equation}\label{bTgp}
c_{B}^2= \lambda^2 (1+3 x_6)\cS^2-4\lambda {\hat b}_4 \cS+ 4 x_4 (\lambda^2 \cC^2 -2 x_4 \lambda^2 \cS \cC+ {\hat m}_F \lambda \cC)+{\hat m_B}^2
\end{equation}
(see \eqref{ss} for definitions of $\cC$ and $\cS$). 
These gap equations were demonstrated in \cite{Jain:2013gza} to enjoy 
a self duality under the transformations 
\begin{equation}\label{dmap}
\begin{split}
& N'=|\kappa|-N, ~~~\kappa'=-\kappa,~~~x_4' = {1\over x_4}, \quad  m_F'= -\frac{  m_F }{ x_4},\quad \lambda' = \lambda - \sgn(\lambda),\\
&x_6'=1+\frac{1-x_6} { x_4^3}, \quad  b_4'=-\frac{1}{ x_4^2}( b_4 + \frac{3}{ 4}\frac{1-x_6}{ x_4}  m_F),\\
& m_B'^2=- \frac{1}{ x_4}  m_B^2 + \frac{3}{ 4} \frac{1-x_6}{  x_4^3}  m_F^2 
+\frac {2}{x_4^2}  b_4  m_F. 
\end{split}
\end{equation}
More concretely the gap equations transform back to themselves, 
provided we also make the identifications
\begin{equation}\begin{split} \label{dualofq}
&c_B'= c_F, ~~~c_F'=c_B,\\
&\lambda' \cS'=-\frac{\sgn(\lambda)}{2}|c_F|+\lambda \cC,\\
&\lambda' \cC'=-\frac{\sgn(\lambda)}{2}|c_B|+\lambda \cS
\end{split}\end{equation}
(see subsection \ref{du} for more discussion). 

The first line in \eqref{dualofq} reflects the fact that bosons and 
fermions are interchanged under the conjectured self duality of these 
deformed ${\cal N}=2$ theories. Note that the ${\cal N}=2$ theories
\eqref{susyvalues} map to themselves by \eqref{dmap}, which agrees with the known ${\cal N}=2$ dualities \cite{Benini:2011mf}. 
In particular, the restriction to the 
renormalizable manifold \eqref{spcval} is preserved by the duality 
map \eqref{dmap}.

\subsection{Flows to purely bosonic or fermionic theories}

The authors of \cite{Jain:2013gza} 
demonstrated that under a one parameter fine-tuning, the gap equations 
\eqref{fTgp} and \eqref{bTgp} of the typical RG flow away from the ${\cal N}=2$ fixed point reduce in the low-energy limit to the gap equation for the critical boson and regular 
fermion theories, respectively, providing strong evidence for such RG flows at large $N$. RG flows from the theories \eqref{generalaction} are parameterized by the ratios of $m_B$, $m_F$ and $b_4$ (which all have the same dimension in the large $N$ limit), so we have two parameters to play around with, which we will denote by $a_0 = m_B^2 / m_F^2$ and $g_0 = b_4 / m_F$. The one parameter fine-tuning was needed because 
each of these IR theories has a single relevant deformation. 

In the rest of this section we will demonstrate that the gap equations \eqref{fTgp} and \eqref{bTgp} 
of the generic RG flow away from the ${\cal N}=2$ fixed 
point also reduce at low energies to the gap equation for the regular boson and critical fermion
theories, respectively, after a two parameter fine-tuning (namely, choosing specific values for $a_0$ and $g_0$). 
We need a two parameter fine-tuning because these IR theories each have two 
relevant deformations. As discussed in great detail above, the IR theories in the strict large $N$ 
limit each also possess a marginal operator parameterized by $x_6$. 
The two parameter fine tuned large $N$ flows end up in the IR theories at values of $x_6$ that depend on the 4 
(large $N$) marginal parameters of the starting UV theory. If, on the other hand, we start our flow in the UV at the 
SUSY theory \eqref{spcval} (as we appear to be forced to do, if we want flows that extend all the way to the UV and 
make sense at any finite $N$ no matter how large) all these 4 `marginal' parameters are fixed. We have no extra
parameters to tune, so that we end up in the IR at a specific value $x_6$ that depends on $\lambda$ (see below).

Let us repeat the  bose-fermi gap equations \eqref{fTgp} and \eqref{bTgp} for convenience:
\begin{equation} \label{bfge}
\begin{split}
c_F^2 &= (2\lambda (\cC-x_4 \cS) +{\hat m}_F)^2,\\
c_{B}^2&= \lambda^2 (1+3 x_6)\cS^2-4\lambda {\hat b}_4 \cS+ 4 x_4 (\lambda^2 \cC^2 -2 x_4 \lambda^2 \cS \cC+ {\hat m}_F \lambda \cC)+{\hat m}_B^2.\\
\end{split}\end{equation}
We will be interested in flows in which one of $c_B$ or $c_F$ is held
fixed, with the other one taken to infinity. In taking the large $c_B$ or $c_F$ limit it is useful to note 
that
\begin{equation}\begin{split} \label{scaa}
&\cC \rightarrow \frac{|c_F|}{2}~~~~~~~~~ {\rm {in ~the ~limit }}~~ c_F\rightarrow \infty, \\
&\cS \rightarrow \frac{|c_B|}{2}~~~~~~~~~ {\rm {in ~the ~limit }}~~ c_B\rightarrow \infty .
\end{split}\end{equation}

\subsection{Fermionic scaling limits}

In order to analyze the end-points of the RG flows from the theories \eqref{generalaction} we want to take some or all of $m_F$, $m_B$ and $b_4$ to infinity with fixed ratios. We begin by analyzing fermionic scaling limits, in which we end up in the IR in a fermionic theory, namely with a finite value of $c_F$ and an infinite value of $c_B$.

Let us begin by considering a flow ending up at a fine-tuned fermionic fixed point with $c_{F,0}=0$. Using \eqref{fpm}, obtaining $c_{F,0}=0$ is the same as obtaining $|c_{B,0}| = m_F / x_4 \lambda$, and plugging this into \eqref{bbm} gives us a relation between the parameters $m_B^2$, $b_4$ and $m_F$. Equivalently, we can use \eqref{fpm} to determine $|c_{B,0}|$ in terms of $|c_{F,0}|$ 
(and other parameters). Plugging this expression into \eqref{bbm}, we obtain a quadratic equation for $|c_{F,0}|$ 
which can be cast into the schematic form  
\begin{equation}\label{schematicform}
\left(|c_{F,0}|-f_1(a_0, g_0, x_4, x_6, \lambda) \right) \left(|c_{F,0}|-f_2(a_0, g_0, x_4, x_6, \lambda) \right)=0,
\end{equation}
where we have defined 
\begin{equation}\label{aobf}
a_0 \equiv \frac{m_B^2}{m_F^2}, ~~~g_0 \equiv \frac{b_4}{m_F}.
\end{equation}
It is easy to check that $|c_{F,0}|=0$ is a solution to \eqref{schematicform} (i.e. that either $f_1$ or $f_2$ vanish)
if and only if 
\begin{equation} \label{fermflow}
a_0=\frac{\lambda^2 \left( 8 g_0 x_4-3 x_6-1\right)+4}{4\lambda^2 x_4^2}.
\end{equation}
This is the same fine-tuning that was performed in \cite{Jain:2013gza}.
Specializing to the case \eqref{spcval} in which the theory is supersymmetric in the UV, \eqref{fermflow} simplifies to
\begin{equation}\label{fff}
a_0 = 2g_0 - 1 + {1\over \lambda^2}.
\end{equation}

It is easy to check that for generic values of $g_0$, $c_{B,0}$ diverges as $m_F \to \infty$. So for such generic values we expect 
to land in the RF theory, and we expect (and will soon check) that a small deviation in $a_0$ away from \eqref{fermflow} 
(of order $1/m_F$) will generate a finite mass deformation of the RF theory. However, there may be a special fine-tuned value of 
$g_0$ for which we end up in the CF theory. If this turns out to be the case, then near this value small deviations of
 $a_0$ and $g_0$ (of order $1/m_F$) will 
 correspond to turning on the two relevant deformations of the CF theory ($\sigma$ and $\sigma^2$ in the language of the
previous sections). In particular, one combination of these deformations - the one that turns on $\sigma^2$ but not $\sigma$ - 
should leave $c_{F,0}=0$. This can only happen if \eqref{schematicform} has a double zero at the special tuned values of 
$a_0$ and $g_0$, i.e. if $f_1$ and $f_2$ in \eqref{schematicform} both vanish. In that case the deformation that preserves 
\eqref{fermflow} could leave, say, $f_1=0$ while changing $f_2$ (this is the $\sigma^2$ deformation). 
It is straightforward to check that \eqref{schematicform}  has a double zero when
\begin{equation} \label{flowtocf}
g_0 = -{1\over {x_4 \lambda^2}} + {{1 + 3x_6}\over {4 x_4}}, \qquad\qquad a_0 = {g_0 \over x_4}.
\end{equation}
In the case that the theory is SUSY in the UV \eqref{spcval} the values are
\begin{equation}
a_0 = g_0 = 1 - {1\over \lambda^2}.
\end{equation}
Thus, for these specially tuned values it is plausible that the RG flow ends up in the IR at the CF theory 
rather than the RF theory. 

We will now test this guess by demonstrating that in the limits above (and small deviations thereof), the thermodynamic expressions reduce to those of the (deformed) RF and CF theories. In order to show this, let us parameterize the small deviations from \eqref{fermflow} by taking
%
\begin{equation}\label{scal1}
m_F\rightarrow \infty,\qquad
m_B^2 = a_1 m_F^2+a_2 m_F+a_3,\qquad
b_4=g_1 m_F +g_2,
\end{equation}
where
\begin{equation}\label{alot}
\begin{split}
&a_1=\frac{\lambda^2 \left( 8 g_1 x_4-3 x_6-1\right)+4}{4\lambda^2 x_4^2},~~a_2=\frac{2 g_2}{x_4} -\frac{m_F^{\rm{reg}} \left( \lambda^2 \left(4 g_1 x_4-3 x_6-1\right)  +4\right) }{2\lambda^2 x_4^2},\\
&a_3=(m_F^{\rm{reg}} )^2 \left( x_4 \frac{\lambda \left(\lambda -2 \sgn(m_F^{\rm{reg}}) \right)}{\left(\sgn(m_F^{\rm{reg}}) -\lambda\right)^2}  +\frac{\frac{4}{\lambda^2} -3 x_6-1}{4 x_4^2} \right)-\frac{2 g_2 m_F^{\rm{reg}} }{x_4}.
\end{split}
\end{equation}
This reproduces \eqref{fermflow} for infinite $m_F$ (where $a_0 \to a_1$, $g_0 \to g_1$), but now we include also small deviations of $a_0$ and $g_0$. As we take $m_F\to \infty$ we keep fixed (in addition to $x_4$, $x_6$ and $g_1$)
also the two parameters of dimension mass, 
$g_2$ and $m_F^{\rm{reg}}$.

Let us analyze the thermodynamics and the gap equations in the limit \eqref{scal1}. We will soon see that, in the limit under consideration, $c_B \to \infty$. Consequently, we can use the second 
of \eqref{scaa} to simplify the gap equations \eqref{bfge}. 
We find that the second of \eqref{bfge} is obeyed at 
 order ${\cal O}(m_F^2)$ provided we choose 
\begin{equation}\label{cbho}
c_B=\frac{{\hat m}_F}{x_4 \lambda} + \delta c_B,
\end{equation} 
where $\delta c_B$ is of order $m_F^0$ (recall that ${\hat m}_F$ is $m_F$ measured in units of the temperature).
Plugging \eqref{cbho} into the second of \eqref{bfge}, we find that the term in this equation of order ${\cal O}(m_F)$ is proportional to 
\begin{equation}\label{tomf}
\left(g_1 + \frac{3}{4} x_4^2 x_6^F   \right) \left( \delta c_B+ \frac{ {\hat m}_F^{\rm{reg}}}{x_4 \lambda} \right) ,
\end{equation}
where 
\begin{equation}\label{ysixeff}
x_6^F\equiv  \frac{4}{3 x_4^3 \lambda_F^2}- \frac{1+3 x_6}{3 x_4^3}.
\end{equation} 

Now we see that we have two options; either the expression in the first parenthesis in \eqref{tomf} or the second one should vanish. Above we identified the first case as the CF flow \eqref{flowtocf}, so let us start with the second case,
\begin{equation}\label{gone}
g_1 \neq  -\frac{3}{4} x_4^2 x_6^F,
\end{equation}
which we expect to lead to the RF theory; we must then have
\begin{equation}
\delta c_B=- \frac{ {\hat m}_F^{\rm{reg}}}{x_4 \lambda},
\end{equation}
or
 \begin{equation} \label{wicb}
c_B=\frac{{\hat m}_F-{\hat m}_F^{\rm{reg}}}{x_4 \lambda}+{\cal O}\left(\frac{1}{{\hat m}_F}\right).
\end{equation}
Plugging \eqref{wicb} into the second of \eqref{bfge} we find that the terms involving ${\hat m_F}$ in that equation 
all cancel (indeed the scalings \eqref{scal1}, \eqref{alot} were chosen to ensure precisely this cancellation) 
and the first of \eqref{bfge}
turns into 
\begin{equation} \label{fio}
\cC=\frac{\sgn(\lambda) |c_F|- {\hat m}_F^{\rm{reg}}}{2\lambda}+{\cal O}\left(\frac{1}{{\hat m}_F}\right).
\end{equation}
\eqref{fio} is the same as \eqref{mdrf}, establishing that in the scaling limit \eqref{scal1} with \eqref{gone}, the thermodynamics of the deformed 
theory \eqref{generalaction} reduces to (a finite deformation of) the regular fermion theory. Recall that the pole mass of the regular fermion theory, $c_{F,0}$, is related to 
$m_F^{\rm{reg}}$ by the equation
\begin{equation}
|c_{F,0}| = \frac{m_F^{\rm{reg}}}{{\sgn(m_F^{\rm{reg}})}-\lambda}.
\end{equation}

\subsubsection{Critical fermionic limit}%

Next, let us check what happens when
$g_1$ is tuned to take the special value 
\begin{equation}\label{spcchg}
g_1 =-\frac{3}{4} x_4^2 x_6^F,
\end{equation}
which we expect to lead to (a finite deformation of) the CF theory.

In studying this particular case we find it convenient to reexpress the parameter $g_2$ of the previous subsubsection 
in terms of another parameter $y_4$ defined by
\begin{equation}\label{gtyf}
g_2 = -x_4^2 y_4
\end{equation}
(note that $g_2$ and $y_4$ both have the dimension of mass). We also find it useful to reexpress the parameter 
$m_F^{\rm{reg}}$ of the previous subsubsection in terms of yet another parameter $y_2$ (also with the dimension of a mass), defined by
\begin{equation}\label{mfregyt}
a_3 = -x_4 y_2^2.\quad
\end{equation}
With these choices and definitions we have
\begin{equation} \label{gval}
a_1=-\frac{3}{4} x_4 x_6^F,\quad
a_2 = -2 x_4 y_4.\quad
\end{equation}

As above we define the order unity quantity $\delta c_B$ by \eqref{cbho}, and determine it 
by requiring that the second of \eqref{bfge} be obeyed at subleading order. As above, the 
leading term in \eqref{cbho} ensures that the second of \eqref{bfge} is obeyed to order $m_F^2$. It follows from 
\eqref{tomf} that, at our special value of $g_1$, this equation is also automatically obeyed at order $m_F$ for 
any choice of $\delta c_B$. In order to determine $\delta c_B$ we then require that the equation be obeyed at order unity. 
Let 
\begin{equation}\label{cbsol}
c_{B}= \frac{1}{x_4 \lambda}{\hat m}_F- \frac{{\hat \sigma}_F}{x_4 \lambda},
\end{equation}
where the order unity quantity ${\hat \sigma}_F$ is a rescaled version of $\delta c_B$. It may be shown that the 
second of \eqref{bfge} is obeyed at order unity provided   
\begin{equation}\label{sigeq}
0= \frac{3}{4} \hat\sigma_F^2 x_6^F-4 \hat\sigma_F \lambda\cC+2 \hat\sigma_F \hat y_4
-4 \lambda^2\cC^2+\hat y_2^2.
\end{equation}

Using \eqref{cbsol} and \eqref{sigeq}, the fermionic gap equation then reduces to 
\begin{equation}\label{fge}
c_F^2= \left(2 \lambda\cC +{\hat \sigma}_F \right)^2.
\end{equation}
\eqref{fge} and \eqref{sigeq} match perfectly with
\eqref{cf} and \eqref{sigsol}, respectively, if we identify ${\hat y}_4$ and ${\hat y}_2$ with the two relevant deformations of the CF theory, and we identify $x_6^F$ of that theory with $x_6^F$ defined by \eqref{ysixeff}. It follows that the flow defined in this section yields the critical fermion theory with these parameters. 

In particular, deforming away from the CF point along the line \eqref{fermflow} corresponds to turning on the $y_4$ deformation, that for one sign induces a flow towards the RF theory. However, as discussed in section \ref{phasestruct}, for the other sign we expect that this flow will lead to a situation where for every $y_2$ we have a gapped theory (perhaps with a first order phase transition) or no vacuum at all, rather than a second order phase transition, so we do not expect any fixed points for these values of $y_4$ (see the phase diagrams sketched in Figs \ref{figctb},  Figure \ref{figA1B1ph} 
and Figure \ref{figA3B3} ). The coefficient of the $\zeta^2$ term that we add to the CF Euclidean action in \eqref{csfnonlinear} is proportional to $(-y_4 \lambda)$, and we expect that when this is large and positive we will reach the RF point. This means that $g_2 \sgn(\lambda)$ should be above some upper bound, which means that our change in $g_0 \simeq g_1 + g_2 / m_F$ should have the same sign as that of $\lambda m_F$. Translating back to our original variables, we conclude that the RG flow should only reach the RF theory when
\begin{equation} \label{flowtorf}
\sgn(\lambda m_F) g_0 > \sgn(\lambda m_F) \left[ -{1\over {x_4 \lambda^2}} + {{1 + 3x_6}\over {4 x_4}} \right]
\end{equation}
and $a_0$ is given by \eqref{fermflow},
namely along a half-infinite-line in the $(a_0, g_0)$ plane. 
In the SUSY case this becomes 
\begin{equation}\label{mpo}
\sgn(\lambda m_F) g_0 > \sgn(\lambda m_F) \left[ 1 - {1\over \lambda^2} \right], \qquad\qquad a_0 = 2g_0 - 1 + {1\over \lambda^2}.
\end{equation}


Let us now focus on the special case $y_2=y_4=0$. In this limit 
\eqref{gval} defines an RG flow that ends up precisely on the line 
of critical fermion theories. Restricting attention to RG flows that 
originate on the renormalizable (SUSY at high energies) manifold \eqref{spcval}, it follows from 
\eqref{ysixeff} that we land on this line of theories at 
\begin{equation}\label{ysixeffp}
x^F_6= \frac{4}{3} \left( \frac{1}{\lambda^2} -1 \right).
\end{equation}
At small 
values of $\lambda$, the value of $x_6^F$ listed in \eqref{ysixeffp} lies 
far outside the domain of attraction \eqref{daar} of the stable fixed 
point of the RG flow, so in this case, at large $N$, the RG flow of this section does not reach this stable fixed point. For other values of $x_4$ and $x_6$ we end up at other values of $x_6^F$ in the IR, including values within this domain of attraction, but this observation appears to be of limited significance for reasons discussed above. 

\subsection{Bosonic scaling Limits}

Let us now repeat the analysis of the previous section for bosonic scaling limits, in which we want to end up at a finite value  of $c_B$ but an infinite value of $c_F$. We begin by considering a flow that should land precisely on a bosonic critical point with $c_{B,0}=0$. Using \eqref{bbm} this gives a simple relation between $m_B$ and $m_F$, such that
\begin{equation}\label{azerobosonic}
a_0=x_4 -\frac{x_4}{\left(\sgn(m_F) -\lambda\right)^2},
\end{equation}
while $g_0$ can take any value. This is the same as the bosonic flow discussed in \cite{Jain:2013gza}.

As in the discussion above, we expect generic values of $g_0$ to end up in the CB theory, but there may be a special value where we end up in the RB theory, and for this value we expect $c_{B,0}=0$ to be a double root of \eqref{bbm}. Requiring that the linear term in this equation vanish gives us the extra condition
\begin{equation} \label{flowtorb}
g_0= x_4^2 \left(1-\frac{1}{(\sgn(m_F) - \lambda)^2}\right),
\end{equation}
with $a_0$ still given by \eqref{azerobosonic}; note that again we find $g_0 = x_4 a_0$, just like in the CF flow.

Let us again check these expectations by matching the thermodynamics of the limit theories to that of the CB and RB theories. We now choose to parameterize the small deviations from the flows mentioned above by (using the same notations as in \eqref{scal1})
\begin{equation}\label{scalab}
\begin{split}
&a_1=x_4 -\frac{x_4}{\left(\sgn(m_F) -\lambda\right)^2},~~a_2=2 m_B^{\rm{cri}} \lambda \left( x_4^2 \left( \frac{1}{\left(\sgn(m_F) -\lambda\right)^2} -1\right) +g_1  \right),\\
&a_3=\frac{1}{4} \left(m_B^{\rm{cri}}\right)^2  \left(\lambda^2 x_4^3\left( 4-\frac{4}{\left(\sgn(m_F) -\lambda\right)^2 }\right) -(3 x_6+1)\lambda^2 +4\right)  +2 m_B^{\rm{cri}} g_2 \lambda.
\end{split}
\end{equation}

We then take the limit $m_F \to \infty$ with the dimensionless parameters 
$x_4$ and  $x_6$ together with $g_1$, $g_2$ and $m_B^{\rm{cri}}$ held fixed.
In this limit, as we will see below, $c_F \to \infty$. Consequently we can use the first of \eqref{scaa} 
to simplify the gap equations \eqref{bfge}.  
To leading order in $m_F$ we find 
\begin{equation} \label{cffo}
c_F=\frac{{\hat m}_F}{\sgn(m_F) - \lambda} +{\cal O}(1).
\end{equation}
When we plug \eqref{cffo} into the first of \eqref{bfge} we find that all terms at leading order (order $m_F^2$) in this 
equation cancel, while the equation is not satisfied at first subleading order (order $m_F$). We find it convenient to 
parameterize the ${\cal O}(1)$ correction to the solution \eqref{cffo} by the order unity quantity $\delta c_F$ defined by  
\begin{equation} \label{cff}
c_F=\frac{{\hat m}_F- \delta c_F x_4 \lambda}{\sgn(m_F) - \lambda}.
\end{equation}
When we plug \eqref{cff} into the first of \eqref{bfge}, collect all terms of order $m_F$, and equate their sum to zero, we find 
the equation
\begin{equation}\label{Scb1}
\cS=\frac{ \delta c_F}{2}.
\end{equation}
Now using \eqref{scalab} and \eqref{Scb1}, the second equation of \eqref{bfge} is automatically solved at leading order 
(order $m_F^2$) but is nontrivial at first subleading order (order $m_F$). At this order the equation takes the form
\begin{equation}\label{fobe}
\left( g_1- x_4^2 \left(1-\frac{1}{(\sgn(m_F) - \lambda)^2}\right) \right) 
\left (2 \cS-{\hat m}_B^{\rm{cri}}  \right) =0.
\end{equation}

Let us begin by looking at the solutions to \eqref{fobe} with
\begin{equation} \label{gonne}
g_1\neq x_4^2 \left(1-\frac{1}{(\sgn(m_F) - \lambda)^2}\right),
\end{equation}
which we expect to flow to the CB theory. In this case we find
\begin{equation} \label{bgaa}
\cS=\frac{{\hat m}_B^{\rm{cri}}}{2}+{\cal O}\left(\frac{1}{{\hat m}_F}\right),
\end{equation}
which agrees with \eqref{mdcb}, establishing that the general theory reduces, in the scaling limit 
of the present subsubsection, to the critical boson theory with mass parameter ${\hat m}_B^{\rm{cri}}$. 

\subsubsection{Regular bosonic scaling limit} \label{rbl}

Let us now study the special case 
\begin{equation}\label{goc}
g_1= x_4^2 \left(1-\frac{1}{(\sgn(m_F) - \lambda)^2}\right),
\end{equation}
which as discussed above we expect to flow to the RB theory.
We find it convenient to rename the parameter $g_2$ (with the dimensions of mass) as
\begin{equation}\label{gtr}
g_2=b_4.
\end{equation}
We will also trade the fixed mass $m_B^{\rm{cri}}$ used in above for another fixed mass ${\tilde m}_B$ defined by
\begin{equation}\label{at}
 a_3 = {\tilde  m}_B^2.
 \end{equation}
 With these choices we have 
\begin{equation}\label{aoat}
\begin{split}
a_1 &= x_4 \left(1-\frac{1}{(\sgn(m_F) -\lambda)^2}\right),
\qquad a_2 = 0, ,\\
\end{split}
\end{equation}
Once again we insert the expansion \eqref{cff} into the fermionic 
gap equation and find \eqref{Scb1}, so that
\begin{equation} \label{cfsoll}
c_F= \frac{{\hat m}_F}{\sgn(m_F)-\lambda} - 2 {\mathcal S} \frac{x_4 \lambda_B}{\sgn(m_F)-\lambda}.
\end{equation}
As explained above, once we plug \eqref{cfsoll} into the bosonic gap equation
 ( the second of \eqref{bfge}) at the special value of $g_1$ \eqref{goc}, we find that this equation is automatically 
 satisfied both at leading order (order $m_F^2$) and at first subleading order (order $m_F$). The equation is first 
 nontrivial at order unity, and yields 
\begin{equation} \label{ige}
{c_B^2} 
= (4+3x_6^B)\lambda^2 \cS^2 -4\lambda \hat b_4 \cS  +\hat m_B^2,
\end{equation}
with
\begin{equation}\label{bTgp1}
x_6^B =(x_6-1) -\frac{4}{3}x_4^3  \left(1-\frac{1}{(\sgn(m_F) - \lambda)^2}\right).
\end{equation} 
Note that \eqref{bTgp1} agrees perfectly with \eqref{scalargapequation}, if we identify the parameters of this subsection with those describing the RB theory, establishing that the scaling limit of the theory \eqref{generalaction} considered in this subsubsection is simply the regular boson theory described in previous sections. 

As in our discussion of the previous subsection, deformations away from the RB point by the parameter $b_4$ should change $g_0$ while preserving \eqref{azerobosonic}, and by looking at the RB action we expect a non-trivial CB fixed point to arise when $(b_4 \lambda)$ is large and positive (this is justified by the analysis of section \ref{phasestruct}. Mapping this to the change in the sign of $g_0$ using \eqref{gtr}, we expect a flow to the CB theory when
\begin{equation}\label{flowtocb}
\sgn(m_F \lambda) g_0 >  \sgn(m_F \lambda) x_4^2 \left(1-\frac{1}{(\sgn(m_F) - \lambda)^2}\right).
\end{equation}
For other values of $g_0$ we do not expect to land at any critical point. As expected (naively) given \eqref{generalaction}, we find using $g_0 = b_4 / m_F$ that in the regime where $(b_4 \lambda)$ of our deformation is large and positive we end up at a non-critical fixed point, while when it is large and negative we do not.

When our 
RG flows start on the renormalizable (SUSY) manifold \eqref{spcval}, \eqref{bTgp1} simplifies to
\begin{equation}\label{endpf}
x_6^B =-\frac{4}{3}\left(1-\frac{1}{(1- |\lambda|)^2}\right) =-\frac{8 |\lambda|}{3} + {\cal O}(\lambda^2).
\end{equation}
In particular, setting $b_4={\tilde m}_B^2=0$, we have  a flow 
from the renormalizable sub-manifold of \eqref{generalaction} to 
the point on the regular boson manifold of theories with $x_6^B$ 
listed in \eqref{endpf}. Notice that this flow at small $\lambda$
lands well within the domain of attraction of the stable fixed point 
\eqref{doa}. It follows that the subsequent RG flow of the regular boson theory, which occurs over 
an RG flow time of order $N$, lands up in the stable fixed point of the finite $N$ regular boson theory 
at small $\lambda_B$.
Once again, for infinite $N$ we can make other choices of $x_4$ and $x_6$ giving different values of $x_6^B$, but such flows are not UV-complete at finite values of $N$.

\subsection{Duality of the  scaling limits} \label{dotsl}

We can use \eqref{dmap} to show that the duality of the boson+fermion theories \eqref{generalaction} maps a flow labeled by $(a_0, g_0)$ at some coupling $\lambda$, to a flow labeled by
\begin{equation} \label{flowduality}
a_0' = -x_4 a_0 + \frac{3(1-x_6)}{x_4} + 2 g_0, \qquad\qquad g_0' = \frac{g_0}{x_4} + \frac{3(1-x_6)}{4x_4^2}
\end{equation}
in the theory with coupling $\lambda' = \lambda - \sgn(\lambda)$, with $x_4$ and $x_6$ of the two sides related as in \eqref{dmap}. In particular, the deformation away from the SUSY theory labeled by $(a_0, g_0)$ at some coupling $\lambda$ maps in the dual theory to a deformation by $(2g_0-a_0, g_0)$. 

As expected, this mapping takes the general fermionic scaling limit \eqref{fermflow} to the general bosonic scaling limit \eqref{azerobosonic}. Moreover, it also maps the flow \eqref{flowtocf} to the CF point to the flow \eqref{flowtorb} to the RB point, and, moreover, it maps the second-order-phase-transition range \eqref{flowtorf} of the fermionic flow to the analogous range \eqref{flowtocb} in the bosonic flow. Thus, starting from the general duality of \eqref{generalaction}, and in particular the SUSY duality, and following our flows, leads to a duality between the RF and CB points as shown in \cite{Jain:2013gza}, and also to a duality between the CF and RB points. At infinite $N$ our computation derives this, and we will discuss the finite $N$ situation in the next subsection.

Comparing in more detail the flows that end up slightly away from the RF/CB points, we find that in addition to relating $g_1'$ and $g_1$ by \eqref{flowduality}, we should also take $g_2' = -g_2 / x_4^2$, and then we obtain $-\lambda_B m_B^{{\rm cri}} = m_F^{\rm reg}$ as expected from the RF/CB duality \eqref{relm}.

Similarly, if we follow the flows that go close to the CF/RB points, we find that the duality mapping of their parameters leading to a mapping taking
\begin{equation}\label{dmapcf}
y_4=b_4,~~y_2^{2}={\tilde m}_B^2,
\end{equation}
in perfect agreement with \eqref{dualitytransform}.

\subsection{The end-points of RG flows}

Starting from the SUSY theory, the  RG flows \eqref{fff} and \eqref{azerobosonic} lead to `phases' 
which are governed by the RF and CB theories, respectively. This point was noted in \cite{Jain:2013gza}, where 
it was also noticed that the lines \eqref{fff} and  \eqref{azerobosonic} cross each other. At this crossing point 
the authors of \cite{Jain:2013gza} proposed (see section 3.3 of that paper) that the dynamics is that of a critical boson and a regular fermion interacting with each other. We believe that this conclusion is incorrect; the analysis of section 3.3  of \cite{Jain:2013gza} ignored a subtlety that we now explain. 

\begin{figure}
    \includegraphics[width=14.5cm,height=8cm]{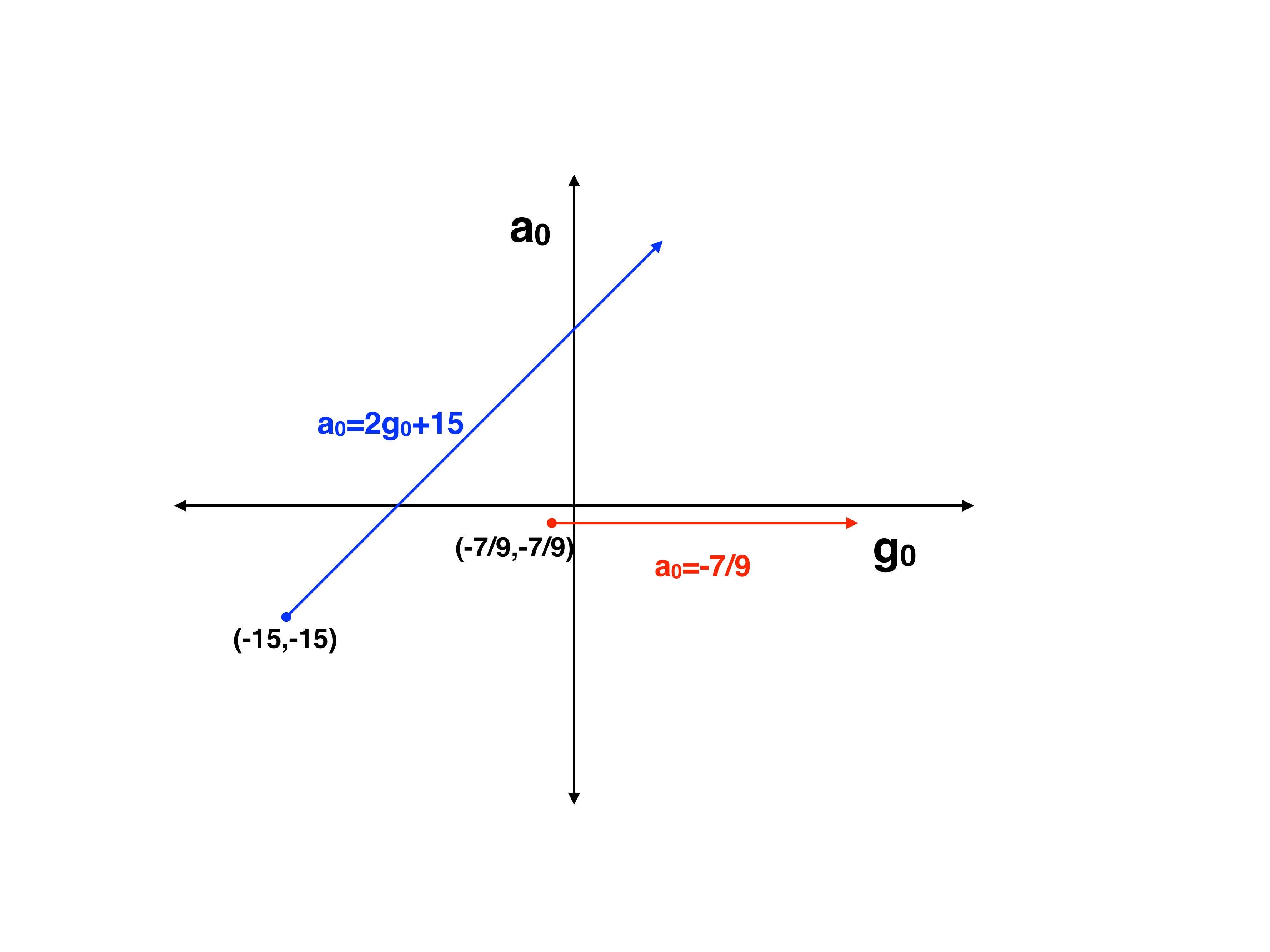}
  \caption{A sketch of the RG flows originating in the SUSY theory at $\lambda=1/4$ and with $m_F>0$. 
 On a generic point along the blue ray \eqref{fff} we flow to the RF theory; at the beginning of this ray, the point 
 $(-15, -15)$, we flow to the CF theory. On a generic point on the horizontal red ray \eqref{azerobosonic} we flow to the CB theory; at the 
 beginning of this ray we flow to the RB theory.  Note that the two rays nowhere intersect, so the SUSY theory never flows to the CB+RF theory. } 
\label{figag}
\end{figure}

 As we have analyzed in great detail above, in certain ranges of parameters the CB theory (and so the RF theory and the supersymmetric theory) has multiple solutions to the gap equation. The RF fixed point is the dominant fixed 
point along the line \eqref{fff} when the inequality \eqref{mpo} is obeyed (when the inequality is saturated we 
find the CF theory with $x_6^F$ given by \eqref{ysixeff}). Similarly the CB theory is the 
dominant solution to the gap equation along the curve \eqref{azerobosonic} only when the inequality 
\eqref{flowtocb} is obeyed (when the inequality is saturated we find the RB theory  with $x_6^B$ given by \eqref{bTgp1}). In other words we actually have RF and CB dynamics along rays (semi-infinite lines) rather than infinite lines in the $(g_0, a_0)$ plane. In Figure \ref{figag} we have sketched the half lines \eqref{fff} subject to the inequality 
\eqref{mpo}, and \eqref{azerobosonic} subject to the  inequality \eqref{flowtocb}, in the special case ${\rm sgn} (m_F) \lambda=\frac{1}{4}$. As is apparent from the figure, the two half lines do not, in fact, meet. 

This conclusion holds at generic values of $\lambda$. It turns out that the intersection of the lines \eqref{fermflow} and \eqref{azerobosonic} happens at a value of $g_0$ that is exactly half of the sum of the critical $g_0$ \eqref{flowtocf} for the CF flow and the critical $g_0$ \eqref{flowtorb} for the RB flow. It follows that unless these two critical values are equal to each other, we are always above one of them and below the other, such that exactly one of the inequalities \eqref{flowtorf} and \eqref{flowtocb} is not satisfied. So we never end up at a RF+CB point. This does not mean that such a point, that was recently analyzed in \cite{Benini:2017aed,Jensen:2017bjo}, does not exist, but just that it does not arise as the end-point of an RG flow from our boson+fermion theories \eqref{generalaction}; it can perhaps arise from more general flows in which we have more independent fine-tunings of the deformations away from such a fixed point. There is a special choice of the parameters for which the two critical values of $g_0$ are equal to each other, so that we do have a flow that ends at a CF+RB point; this happens when
\begin{equation}
{1\over 4}(3x_6 + 1) = {1\over \lambda^2} + x_4^3 \left[ 1 - {1\over {(\sgn(\lambda) - \lambda)^2}} \right].
\end{equation}
In the supersymmetric case $x_4=x_6=1$ this happens for $\lambda = \pm {1\over 2}$.

At finite $N$ several things change. For large but finite values of $N$, one change is that the only UV-complete flow starts from the SUSY theory with $x_4=x_6=1$; but in practice we can also consider other values of $x_4$ and $x_6$, treating them as effective field theories that have some UV completion at higher energies (recall that the beta functions of $x_4$ and $x_6$ vanish at large $N$, so these effective actions can be valid over a large range of energies). Near the IR we should have a small shift in the position of the RF and CB lines, but they should still exist, and end at some point. The big difference is that now this point looks approximately like a CF/RB theory with some value of $x_6$, and this value now starts slowly running. This flow may approach a non-trivial CF/RB fixed point, or it make take $x_6$ to infinity\footnote{As we discussed in section \ref{curious}, this may perhaps  be interpreted as an RF/CB point.}. We expect to have some range of values of $\lambda$ for which we flow to a CF/RB point, assuming that it exists, while for other values we do not flow to this point (but it may still exist). For $x_4=x_6=1$ we drew in figure \ref{figphse12a} the position of the end-point of the `fast flow' at large $N$, which sometimes flows to a stable CF/RB point and sometimes does not.

At smaller values of $N$, the flow could change more dramatically. It is possible that one of the three relevant operators that we deform by could become irrelevant, and then we would only have a one-parameter set of flows, that can perhaps be fine-tuned to land at an RF/CB fixed point, but not at a CF/RB point. Conversely, it is also possible that more operators become relevant, but this would not change the number of required fine-tunings. We do not know for which range of values of $N$ and $k$ the RF/CB point exists; it is possible that for some values there is no second order phase transition at all. When this fixed point exists we do expect to have a flow from the SUSY theory to it as in large $N$, and then the flow line has to end as in the figure above. But as we discussed, it is not obvious that this end-point is a new fixed point, and it could also be a RF/CB point.

\section{Discussion and future directions}

The three main results of this paper are the computation of the large $N_B$ beta function for the parameter $x_6$ of quasi-bosonic theories, the elaboration of the zero temperature phase structure of these theories, and the study of flows from
SUSY theories to quasi-bosonic theories. We begin this discussion section by considering each of these topics in 
turn. 

Our final result for the beta function of RB theories, at leading nontrivial order in $\frac{1}{N_B}$, is presented in \eqref{rbt} and \eqref{cft}. Unfortunately this result is not completely explicit. We have demonstrated that the beta function is 
a cubic polynomial in $x_6$ and have computed two of the four coefficients of this polynomial at all orders in 
$\lambda_B$. However the remaining coefficients of this polynomial are known only in terms of three functions of 
$\lambda_B$ that characterize aspects of correlation functions of ${\tilde J}_0$ in the critical boson theory; the 
corresponding functions are only known explicitly at small $\lambda_B$ or small $\lambda_F$, but are not explicitly 
known at arbitrary $\lambda_B$. It would be very satisfying to complete the computation initiated in this paper by 
explicitly determining these three unknown functions.

One of the functions that we need in order to complete the evaluation of our $\beta$ function is the quantity 
$g_{(5,0)}$ defined in \eqref{formg43}. This coefficient characterizes a particular kinematical limit of the 
5-point function of scalar operators in the CB theory at leading order in the large $N_B$ limit. We suspect that
the evaluation of this coefficient at all values of $\lambda_B$ may not prove too difficult a task (see 
\cite{Yacoby:2018yvy} for recent progress). The remaining two coefficients that we need deal with $1/N_B$ corrections 
to correlators. The first of these quantities is the anomalous dimension of ${\tilde J}_0$ at first order in the 
$1/N_B$ expansion (see \eqref{sdopb}). The second of these is the quantity $r_B$ that governs the  splitting of the three point function of three ${\tilde J}_0$ operators into a contact piece and a power law piece (see \eqref{thptdet}). As both of these quantities 
refer to subleading orders in the $1/N_B$ expansion they may not be easy to compute by direct diagrammatic evaluation. 
Other approaches, such as the conformal bootstrap, may prove to be better for the evaluation 
of these quantities
(see \cite{Turiaci:2018nua, Aharony:2018npf} for some recent progress on the bootstrap approach).

The deformed supersymmetric theory that we have studied in section \ref{susysec} has a rich phase structure as a function of two 
dimensionless parameters (dimensionless ratios of the coefficients of the three relevant operators of the theory). 
While we have already analyzed several aspects of the phase structure of this theory in Section \ref{susysec}, 
it would be useful to complete this analysis in a completely systematic manner. The end point of such an analysis 
would be a detailed phase diagram of the theory along the lines of our analysis of the phase diagram of 
the RB theory section \ref{phasestruct}. In order to accomplish this we would need to generalize the free energy 
computation of \cite{Jain:2013gza} to include Higgsed phases, along the lines of the recent computation 
\cite{Choudhury:2018iwf}.

The CB and RF theories are believed to be dual to Vasiliev theories of gravity with a specific boundary condition for the scalar field on $AdS_4$, such that its classical dimension is $\Delta=2$. At finite values of $N$ the quantum gravitational theories can be defined using the three dimensional CFTs. Our discussion suggests that we can use the RB and CF theories to define quantum gravitational theories with other boundary conditions for the bulk scalar field, in which its classical dimension is $\Delta=1$. In the bulk the coupling $x_6^B$ (or $x_6^F$) in \eqref{rbt1} (or \eqref{cft1}) is a ``multi-trace'' coupling that defines the precise boundary condition for the bulk scalar field, and our results imply that only specific values for this coupling really lead to consistent quantum theories on $AdS_4$. It would be nice to verify this directly by bulk computations; the bulk computation of beta functions for multi-trace couplings was discussed at the classical level in \cite{Witten:2001ua, Grozdanov:2011aa, Das:2013qea, Aharony:2015afa}, but in our case we expect the leading contribution to arise at one-loop order in the bulk.

There are many possible generalizations of our results :

\begin{itemize}
\item In this paper we only discuss theories with a single matter field in the fundamental representation. However, a similar picture is believed to hold also for larger numbers $N_f$ of matter fields, at least when these numbers are small enough compared to $N$ and $\kappa$ \cite{Seiberg:2016gmd,Hsin:2016blu}. For $N_f > 1$ the flow diagrams are more complicated, since the RB and CF theories contain one additional relevant deformation (with a different contraction of the flavor indices), and two additional classically marginal deformations. We conjecture that there still exist fixed points leading to a duality between RB and CF theories, with one or both of the extra relevant deformations tuned to zero, but additional evidence is required to bring this conjecture to the same level as the $N_f=1$ conjecture that we discuss here. In particular one has to compute the beta functions for all three classically marginal operators.

\item In this paper we discussed the phase structure of the regular boson theories at zero temperature, and found an intricate structure, where for some parameters there is no solution to the gap equations, suggesting a runaway behaviour. It would be interesting to generalize this analysis to finite temperature, and to see if this stabilizes the cases with no stable vacuum, and how the phase structure is deformed.

\item It would be interesting to generalize the flow from theory S to the bosonic and fermionic theories to $SO(N)$ gauge groups, in order to provide more evidence for the $SO(N)$ dualities \cite{Aharony:2016jvv} at finite $N$. This is more complicated than the $U(N)$ case because the supersymmetric theory has complex matter fields, while the simplest $SO(N)$ duality involves real matter fields.

\item It would be interesting to study the duality between the RB and CF theories for small values of $N$, as in \cite{Seiberg:2016gmd,Hsin:2016blu}, to provide evidence for or against their validity for these values. Note that for the $SU(1)$ theories the RB theory is free (at low energies), as was the case for the RF theory in \cite{Seiberg:2016gmd,Hsin:2016blu}. It would, of course, 
be fascinating if the theories discussed in this paper continue to be well-defined (and to enjoy invariance under duality) at  small finite values of $N$. It would be even more satisfying 
if these small $N$ theories made an appearance in experimentally relevant condensed matter systems (which allow for two fine-tunings of parameters). 
\end{itemize}
We leave all these issues for future study.

\section*{Acknowledgments}
We would like to thank A. Bissi,  K. Damle,  R. Gopakumar, T. Hartman, I. Klebanov, Z. Komargodski, S. Kundu, A. Nizami, 
S. Prakash and S. Wadia for useful discussions, and especially F. Benini, S. Giombi, G. Gur-Ari, N. Seiberg and R. Yacoby for useful discussions and comments on a draft of this paper.  O.A. and S.M. were supported
 by an Indo-Israeli grant (no.\ 1200/14) within the ISF-UGC joint research program framework. 
The work of O.A. was also supported in part  by the I-CORE program of the Planning and Budgeting Committee and the Israel Science Foundation (grant number 1937/12), by an Israel Science Foundation center for excellence grant, and by the Minerva foundation with funding from the Federal German Ministry for Education and Research. O.A. is the Samuel Sebba Professorial Chair of Pure and Applied Physics. The work of S.M. was supported by the Infosys Endowment for the study of the Quantum Structure of Spacetime. Some part of the paper was done when  S.J. was a postdoc at Cornell and his research was supported by grant No:488643 from the Simons Foundation. S.J. is also supported by Ramanujan Fellowship.
S.M. and S.J. would like to acknowledge their debt to the people of India 
for their steady and generous support to research in the basic sciences.

\appendix

\section{Constraints on the ${\tilde J}_0$ 
three point function from conformal invariance} \label{confform}

As we have explained in the main text, the three point function of ${\tilde J}_0$ takes the 
form \eqref{thpln}. In this subsection we will study the expression for $F(q_1, q_2, q_3, \epsilon )$, 
and in particular find the leading dependence of this quantity on $\ln (\Lambda)$ at small 
\begin{equation}
\epsilon\equiv \frac{\delta_B(\lambda_B)}{\kappa_B},
\end{equation}
which is the anomalous dimension of ${\tilde J}_0$ to first subleading order in $\frac{1}{N_B}$.

The explicit expression for 
$F(q_1, q_2, q_3, \frac{\delta_B}{\kappa_B} )$ is given by
\cite{Bzowski:2013sza} :
\begin{equation}\label{expF}
F(q_1, q_2, q_3, \epsilon)= {\cal N}\Lambda^{-3 \epsilon} (q_1q_2q_3)^{ \frac{1}{2} + \epsilon}
\int_0^\infty dx  \sqrt{x} K_{\frac{1}{2} + \epsilon}(x q_1)
K_{\frac{1}{2} + \epsilon}(x q_2) K_{\frac{1}{2} + \epsilon}(x q_3),
\end{equation}
where $\Lambda$ is a UV cutoff, and ${\cal N}$ is a normalization 
constant of order unity that we will tune for later convenience. As mentioned above, $F(q_1, q_2, q_3, \epsilon)$ is the 
Fourier transform of the expression 
\begin{equation}
\frac{ \Lambda^{-3 \epsilon} }{(x_1-x_2)^{2+ \epsilon} (x_2-x_3)^{2+ \epsilon} (x_3-x_1)^{2+ \epsilon}}
\end{equation}
up to a normalization  that tends to a finite non-zero number in the limit $\epsilon \to 0$.  

The integral in \eqref{expF} is convergent for $\epsilon <0$, and is defined for arbitrary values of 
$\epsilon$ by analytic continuation from these values. As the integral diverges logarithmically at 
$\epsilon =0$ (the divergence comes the neighborhood of $x=0$), the expression for $F$ has a pole at $\epsilon =0$.
Using the fact that $K_\frac{1}{2}(x) \propto \frac{e^{-x}}{\sqrt{x}}$ 
together with the small argument expansion of the Bessel function, it is easy to convince oneself that 
\begin{equation}\label{expFD}
F(q_1, q_2, q_3, \epsilon) = \frac{1}{\epsilon} \left( \frac{\Lambda}{q_1+q_2 +q_3} \right)^{-3\epsilon}
\left( 1+ {\cal O}(\epsilon) \right) .
\end{equation}
Recall that $\epsilon={\cal O}(\frac{1}{N}).$ It follows that the function $\epsilon F$ admits the expansion 
\begin{equation} \label{appf}
\epsilon~ F = 1- 3 \epsilon~ \ln\left( \frac{\Lambda}{q_1+q_2+q_3}\right)+{\cal O}(\epsilon^2).
\end{equation}

\section{The critical boson theory at $\lambda_B=0$} \label{fbt}

In this Appendix we compute two, three and four point functions of 
the operator ${\tilde J}_0= \sigma$ in the critical boson theory 
in the limit $\lambda_B \to 0$, at leading order in the large $N_B$ expansion. 
 As a check on our results we 
compute the anomalous dimension of $\sigma$ to first order in the 
expansion in $\frac{1}{N_B}$, and recover the known result for this 
quantity. 

In order to compute $\sigma$ correlators, we start 
with \eqref{cst}, and integrate out the (free) boson fields to obtain an effective 
action for $\sigma$. Tree graphs of this effectively classical action then generate 
correlators of $\sigma$ at leading order in $N_B$. Loops of the same effective action 
determine the anomalous dimension of $\sigma$.

\subsection{$S_{eff}(\sigma)$ and $\sigma$ Green's functions at leading 
order in $\frac{1}{N}$}

In this subsection we study $S_{eff}(\sigma)$ and the Green's functions of 
$\sigma$ at quadratic, cubic, quartic and pentic order. All through this 
appendix we work with the theory at $\lambda_B=0$. In this subsection we 
focus on the leading order in an expansion in $\frac{1}{N_B}$. In the next 
subsection we turn to a discussion of the first corrections in $\frac{1}{N_B}$. 

 At $\lambda_B=0$ the scalar field $\phi$ is easily integrated 
out of \eqref{cst} so that 
\begin{equation}\label{sigea} \begin{split} 
&\int D\phi e^{-  \int \left( \partial {\bar \phi} \partial \phi + 
\sigma {\bar \phi} \phi \right) }
= e^{-S_{eff}(\sigma)}\\
&S_{eff}(\sigma)= \sum_{n=2}^{\infty} \frac{1}{n!} \int dp_n {\tilde S}_n(-p_i) \sigma(p_1) \ldots
\sigma(p_n),
\end{split}
\end{equation}
where $dp_n$ was defined in \eqref{rbtzeta} (and includes a momentum delta function). 

\subsubsection{Quadratic and cubic order}

\begin{figure}[h]
\begin{center}
\includegraphics[width=8.5cm,height=4.5cm]{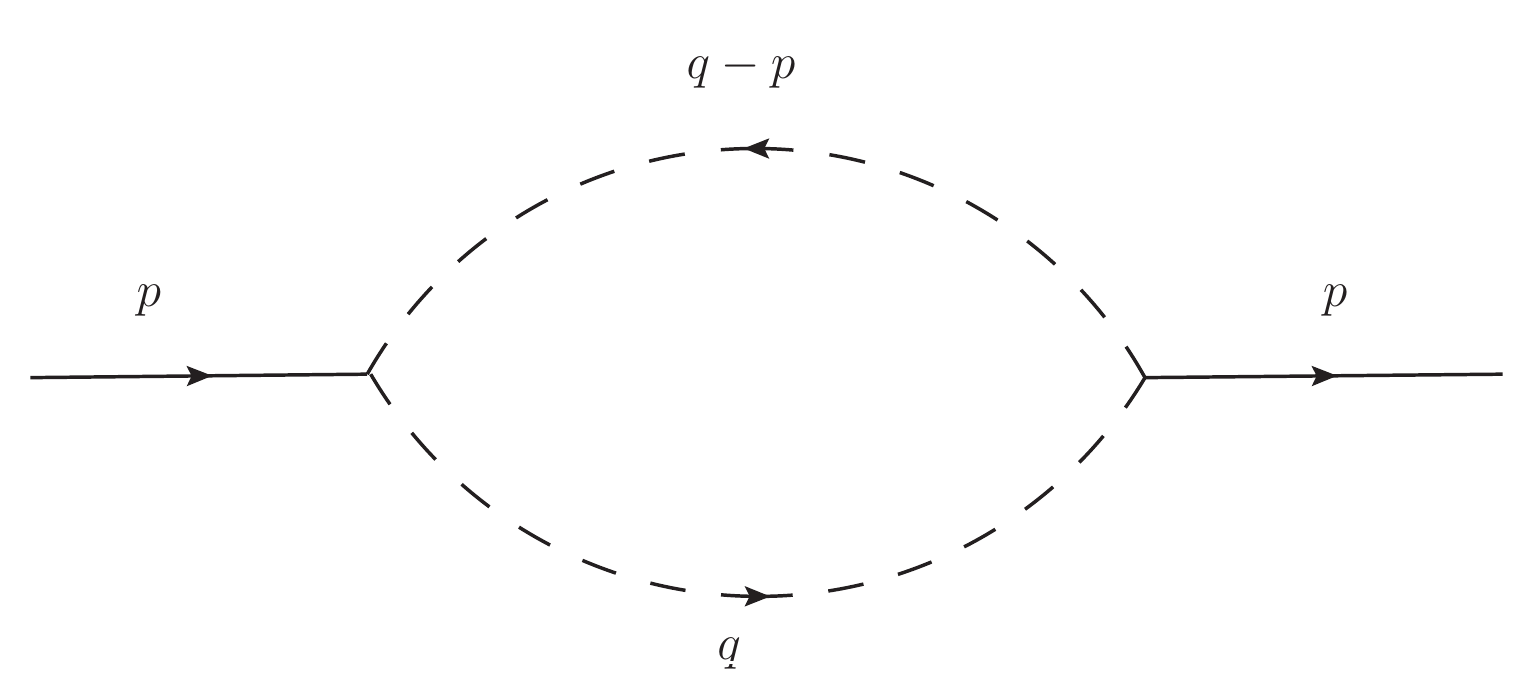}
  \caption{\label{twop} This diagram defines ${\tilde S}_2(p)$ in \eqref{vars}. }
 \end{center}
\end{figure}

\begin{figure}[h]
\begin{center}
\includegraphics[width=8.5cm,height=4.5cm]{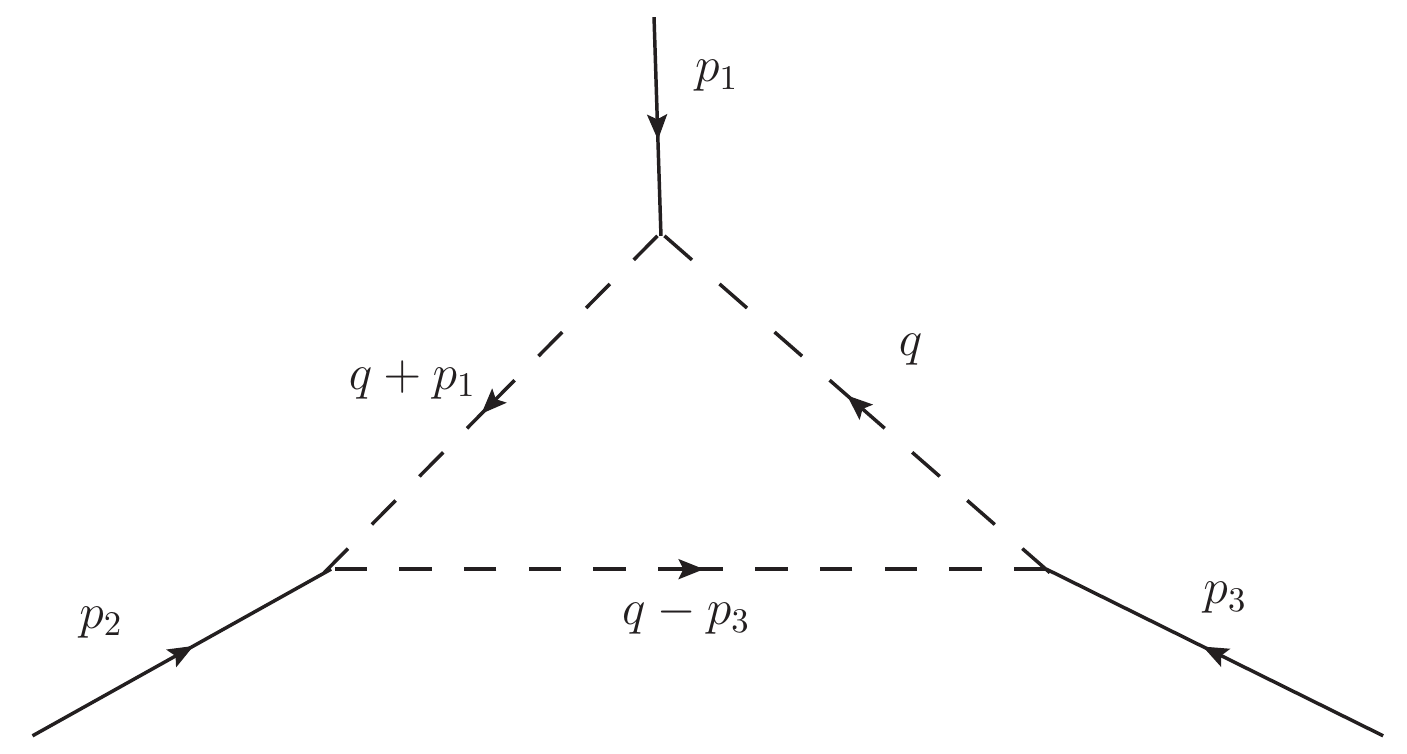}
  \caption{\label{threep} This diagram defines ${\tilde S}_3(p_1,p_2,p_3)$ in \eqref{vars}. }
 \end{center}
\end{figure}

The quadratic and cubic terms in this effective action are very easily 
computed. The relevant graph at quadratic order in presented in 
Figure \ref{twop}; the relevant graph at cubic order is presented in 
Figure \ref{threep}. Both graphs are easily evaluated and we find  
\begin{equation} \label{vars} \begin{split}
&{\tilde S}_2(p) = - N_B \int \frac{d^3 q}{(2 \pi )^3} \frac{1}{q^2 (p-q)^2} 
= - N_B\frac{1}{8|p|}, \\
&{\tilde S}_3(p_1, p_2, p_3) =2 N_B\int \frac{d^3 q}{(2 \pi )^3} \frac{1}{q^2 (p_1+q)^2 (q- p_3)^2} = \frac{N_B}{4}\frac{1}{|p_1||p_2||p_3|}.
\end{split}
\end{equation}

The two and three point Green's functions for $\sigma$ follow immediately 
from these results. The two point function is simply the inverse of 
${\tilde S}_2(p)$ and agrees with \eqref{tplnlw} at $\lambda=0$. 
The three point function is given by 
\begin{equation}
\frac{{\tilde S}_3(p_1, p_2, p_3)}{{\tilde S}_2(p_1) 
{\tilde S}_2(p_2){\tilde S}_2(p_3)}
\end{equation}
and reduces to \eqref{thplnlw} upon setting
$\lambda_B=0$.

\subsubsection{The quartic term}

\begin{figure}[h]
\begin{center}
\includegraphics[width=9.5cm,height=5.5cm]{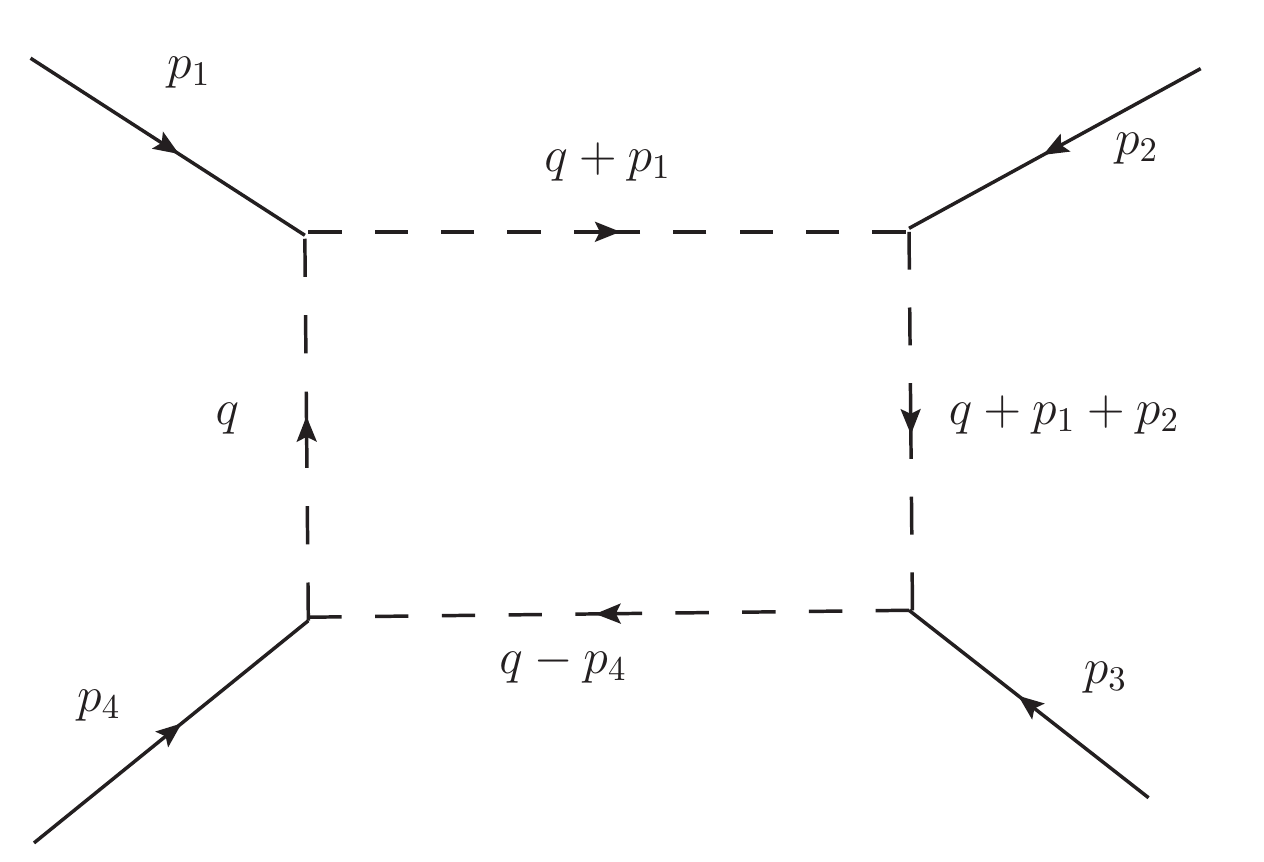}
  \caption{\label{4pta} This diagram defines $W(p_1,p_2,p_3,p_4)$ in \eqref{Fdef}. }
 \end{center}
\end{figure}

The quartic term in the effective action is given by 
\begin{equation}\begin{split}\label{quartic}
&{\tilde S}_4(p_1,p_2,p_3,p_4)=-N_B\Bigg( W(p_1,p_2,p_3,p_4)+ W(p_1,p_2,p_4,p_3) +W(p_1,p_4,p_3,p_2)\\
&\qquad\qquad\qquad\qquad\qquad+W(p_1,p_4,p_2,p_3)+W(p_1,p_3,p_4,p_2)+ W(p_1,p_3,p_2,p_4)\Bigg),\\
\end{split}
\end{equation}
where (see Figure \ref{4pta})
\begin{equation}\label{Fdef}
W(p_1,p_2,p_3,p_4)\equiv \int \frac{d^3q}{(2\pi)^3} \frac{1}{q^2 (q+p_1)^2 (q+p_1+p_2)^2 (q-p_4)^2}.
\end{equation}

The expression for ${\tilde S}_4(p_1,p_2,p_3,p_4)$ can be
somewhat simplified by noting that\footnote{ This identity follows by making the variable change $q= -q'-p$ 
in the integral that defines $W$. It simply reflects the fact that there is no preferred orientation to the graph, and the graph cares about the cyclic order of 
momentum insertions only up to reflections: a particular cyclic order 
in the clockwise direction gives the same graph as the same cyclic 
order in the anticlockwise direction.}
\begin{equation}\label{reflection}
W(p_1, a, b, c)= W(p_1, c, b, a),
\end{equation}
so that
\begin{equation}\label{s4simp}
{\tilde S}_4(p_1,p_2,p_3,p_4)=-2N_B\left( W(p_1,p_2,p_3,p_4)+ W(p_1,p_2,p_4,p_3) +W(p_1,p_3,p_2,p_4) \right).
\end{equation}

\begin{figure}[h]
\begin{center}
\includegraphics[width=9.5cm,height=5.5cm]{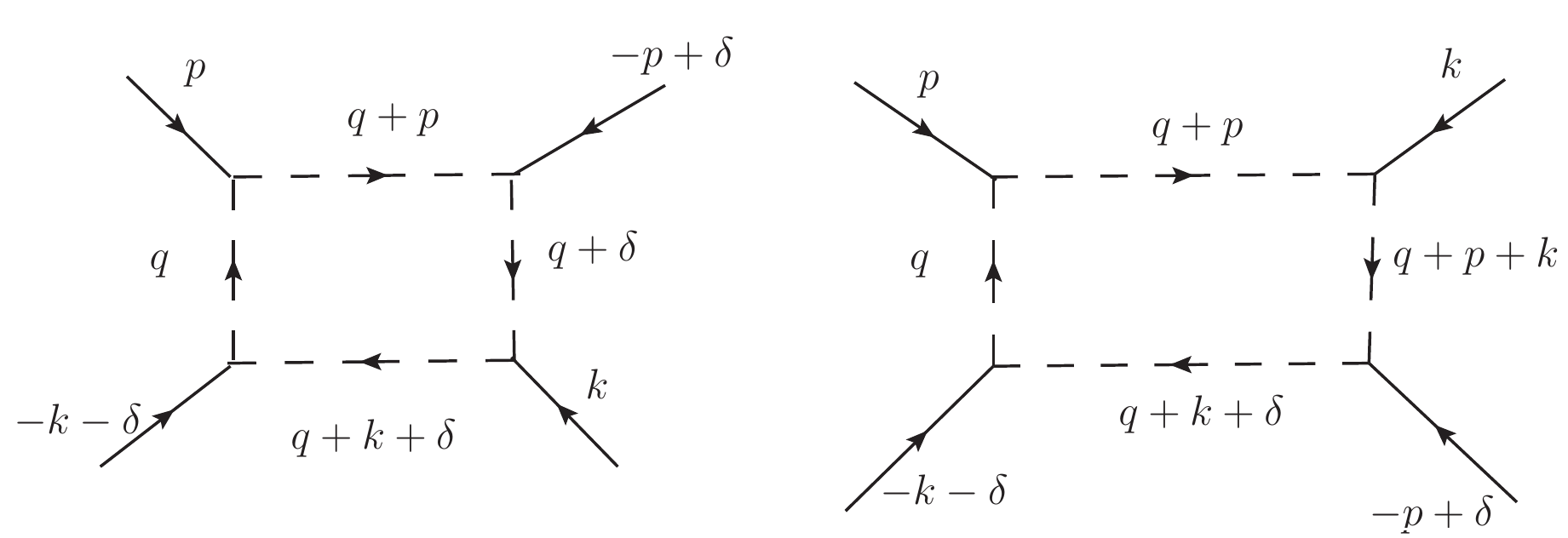}
  \caption{\label{4ptvari} This diagram shows contributions to ${\tilde S}(p,-p+\delta,k,-k-\delta)$ }
 \end{center}
\end{figure}

In the rest of this subsubsection we  will  proceed to further simplify \eqref{s4simp} in the kinematical  regime of most interest to us, namely
${\tilde S}_4(p, -p+\delta, k, -k-\delta)$ in the limit $\delta \to 0$ (see Figure \ref{4ptvari}). 
It follows from \eqref{s4simp} that  
\begin{equation} \label{fsi} \begin{split}
-\frac{1}{N_B} {\tilde S}_4(p, -p+\delta, k, -k-\delta)= &
2 W(p, -p+\delta, k, -k-\delta) + 2 W(p, -p+\delta, -k-\delta, k) \\
& + 2 W(p, k, -p+ \delta , -k-\delta).
\end{split}
\end{equation}
The expressions on the first line of \eqref{fsi} blow up in the limit 
$\delta \to 0$, while the expression on the second line is finite in the same limit. We will first proceed to evaluate 
the expressions on the two lines separately, and then physically interpret 
the divergence as $\delta \to 0$ and explain how it is dealt with. 

Let us first deal with the second line of \eqref{fsi}. In this expression 
we can simply set $\delta=0$, obtaining
\begin{equation}
2 \int \frac{d^3 q}{(2 \pi)^3} 
\frac{1}{q^2 (q+p)^2 (q+p+k)^2 (q+k)^2}.
\end{equation}
This integral is still rather complicated; however it simplifies in the 
limit that $\frac{|k|}{|p|}$ is small. In this limit we can separate the integral over $q$ into two balls and the rest. The first ball $B_1$ is given by  $|q| < M |k|$. The second ball $B_2$ is the region $|p+q| < M |k|$. Here $M$ is a number chosen so that 
\begin{equation}\label{limonm}
 1 \ll M \ll \frac{|p|}{|k|} .
 \end{equation}

Let the contribution to the integral from points $q \in B_1$ be denoted by 
$C_B(p, k)$. The contribution to the integral 
from points $q$ in $B_2$ is then given by $C_B(-p, k)= C_B(p, -k)$. To leading order 
in the small $|k|$ limit we have 
\begin{equation}
C_B(p, k)= \frac{2}{|k| |p|^4} \left( \frac{1}{8} - \frac{1}{2 \pi^2 M} 
\right),
\end{equation}
where we have used
\begin{equation} \label{bi}
\int \frac{d^3q}{(2\pi)^3} \frac{1}{q^2 (q-p)^2} =\frac{1}{8|p|}.
\end{equation} 
On the other hand the contribution of the rest of space, $C_R(p, k)$ to this integral is 
\begin{equation}
C_R(p,k)= \frac{4}{|k| |p|^4} \left( \frac{1}{2 \pi^2 M} + 
{\cal O} (\frac{|k|}{|p|})
\right).
\end{equation}
Adding together the contributions of the two balls and the rest, we 
find that the second line of \eqref{fsi} is given by 
\begin{equation} \label{intere}
 \frac{1}{2 |k| |p|^4} \left(1 + O\left(\frac{|k|}{|p|}\right) \right).
\end{equation}

Let us now turn to the first line of  \eqref{fsi}. We have 
\begin{equation} \label{process} \begin{split}
&2 W(p, -p+\delta, k, -k-\delta) + 2 W(p, -p+\delta, -k-\delta, k)\\
&=2\int \frac{d^3 q}{(2 \pi )^3}\frac{1}{q^2  (\delta +q)^2 (p+q)^2} \left( \frac{1}{(\delta +k+q)^2}+\frac{1}{(q-k)^2}  \right)\\
&=2\int \frac{d^3 q}{(2 \pi )^3}\frac{1}{q^2  (\delta +q)^2 } 
\left( \frac{1}{(p+q)^2(\delta +k+q)^2}+\frac{1}{(p+q)^2 (q-k)^2}  
- \frac{2}{|p| |p-\delta| |k| |k+\delta|} \right) \\
&+ \frac{1}{2 |\delta|~ |p| |p-\delta| |k| |k+\delta|},
\end{split}
\end{equation}
where we have used \eqref{bi} to obtain the last line of \eqref{process}.
The final integral in \eqref{process} is finite in the limit $\delta \to 0$; 
it follows that 
\begin{equation}\label{mproc} \begin{split}
&\lim_{\delta \to 0} 
\left( 2 W(p, -p+\delta, k, -k-\delta) + 2 W(p, -p+\delta, -k-\delta, k)
- \frac{1}{2 |\delta| |p| |p-\delta| |k| |k+\delta|} \right) \\
&=2 \lim_{\delta \to 0}   \int \frac{d^3q}{(2\pi)^3} \frac{1}{q^2 (q+\delta)^2}\left( \frac{1}{(p+q)^2 (k+q+\delta)^2} + \frac{1}{(p+q)^2(k-q)^2}- \frac{2}{|p| |p-\delta| |k| |k+\delta|}\right).\\
\end{split}
\end{equation}

As above we now explore the limit $|k| \ll |p|$. In this limit it is useful,  
as above, to view the integral in \eqref{mproc} as the sum of two integrals. In the first integral $q$ is contained in a ball of radius $M |k|$ centered 
around $q=0$, where $M$ obeys the inequality \eqref{limonm}. 
In this region $|q| \ll |p|$ and we can simplify the integrand by 
Taylor expanding the integrand in a Taylor series expansion in $|q|/|p|$ 
as well as $|k|/|p|$. In the complement of the ball, on the other hand, 
$|k|/ |q|$ is small so the integrand can be simplified by Taylor expanding
in $|k|/|q|$ as well as $|k|/|p|$. As above it turns out that the contribution 
of the complement of the ball scales like $\frac{1}{M}$ and so can be ignored
in the large $M$ limit of actual interest to us. We focus entirely on the 
contribution of the interior of the ball in what follows.

The first potential concern about the integral \eqref{mproc} is whether the 
limit $\delta \to 0$ is nonsingular. In order to address this concern we expand the integral in the second line of 
\eqref{mproc} in a Taylor series expansion in $\frac{1}{|p|}$. As $|\delta|$, 
$|q|$ and $|k|$ are all much less than $|p|$ in the interior of the ball, 
this Taylor expansion can be legitimately carried out inside the integral 
and we find that the integral \eqref{mproc} reduces to  
\begin{equation}\begin{split} \label{szi}
&2\int \frac{d^3q}{(2\pi)^3} \frac{1}{p^2 q^2 (q+\delta)^2}\left( \frac{1}{ (k+q+\delta)^2} + \frac{1}{(k-q)^2}- \frac{2}{|k| |k+\delta|}\right)\\
&-\frac{4}{p^4} \int \frac{d^3q}{(2\pi)^3} \frac{q.p}{ q^2 (q+\delta)^2}\left( \frac{1}{ (k+q+\delta)^2} + \frac{1}{(k-q)^2}\right) - \frac{p.\delta}{2 p^4 |\delta| |k| |k+\delta|} \\
& + 4\int \frac{d^3q}{(2\pi)^3} \left( \frac{ \frac{q^2}{p^2} -4 \frac{(p\cdot q)^2}{p^4} }
{(q^2)^2 (k+q)^2 p^2} \right)
+{\cal O}(1/|p|^5).
\end{split}
\end{equation}
In the first line of \eqref{szi} we have collected all terms in the integrand 
of \eqref{mproc} of 
order $\frac{1}{|p|^2}$. Terms of order $\frac{1}{|p|^3}$ and order 
$\frac{1}{|p|^4}$, respectively, are  collected on the second and third lines. 
All terms of higher order in $\frac{1}{|p|}$ are ignored.

Using the integrals
\begin{equation}
\label{usflint}
\begin{split}
&\int \frac{d^3q}{(2\pi)^3} \frac{1}{q^2 (q+p_1)^2 (q+p_2)^2} =\frac{1}{|p_1| |p_2| |p_1-p_2|},\\
&\int \frac{d^3q}{(2\pi)^3} \frac{q_\mu}{q^2 (q+\alpha)^2 (q+\beta)^2} =\frac{1}{2}\left(a~ \alpha_{\mu} +b~  \beta_{\mu}\right),\\
&a =-\frac{1}{8 |\alpha| |\alpha-\beta|} \left( \frac{|\alpha|+|\beta|-|\alpha-\beta|}{|\alpha||\beta|+\alpha\cdot \beta}  \right),\\
&b =-\frac{1}{8 |\beta| |\alpha-\beta|} \left( \frac{|\alpha|+|\beta|-|\alpha-\beta|}{|\alpha||\beta|+\alpha \cdot \beta}  \right),
\end{split}
\end{equation}
it is easily verified that the first two lines of \eqref{szi} simply vanish. 
This is a very important point; individual terms on the first line of 
\eqref{szi} integrate to expressions proportional to $\frac{1}{|\delta|}$, 
and individual terms on the second line of \eqref{szi} evaluate to 
expressions of the schematic form $\frac{\alpha \cdot \delta}{|\delta|}$; these 
terms all have singular $|\delta| \to 0$ limits. The fact that the first 
and second lines of \eqref{szi} simply vanish implies that the $\delta \to 0$ 
limit in the integral \eqref{mproc} is well-defined. 

The fourth line in \eqref{szi} does not vanish. This integral is the sum of 
two terms. The first of these is an integral proportional to \eqref{bi} and 
is easily evaluated. The second is proportional to the integral 
\begin{equation}
 \int \frac{d^3q}{(2\pi)^3} \frac{ \frac{(p\cdot q)^2}{p^4} }
{(q^2)^2 (k+q)^2 p^2} 
\end{equation}
In order to evaluate this integral we first note that by symmetry
\begin{equation}\label{B01}
 \int \frac{d^3q}{(2\pi)^3}  \frac{q_{\mu} q_{\nu}}
{(q^2)^2 (k+q)^2 } = B_0(|k|) \delta_{\mu\nu} + B_1(|k|)\frac{ k_{\mu}k_{\nu}}{k^2}.
\end{equation}
A short calculation yields 
\footnote{One way to obtain this result is by orienting the $z$ axis along 
$k$ and computing the integrals in polar coordinates, 
evaluating all angular integrals first. Another way is to separate this 
convergent integral into a sum of simpler divergent integrals and then 
compute each individual integral using dimensional regularization. Both 
methods yield the same answer.}
\begin{equation}
B_0(|k|)= \frac{1}{32 |k|},~~~ B_1(|k|)= \frac{1}{32 |k|}.
\end{equation}

Putting everything together we find that the third line in \eqref{szi} -- 
and so our final result for the integral \eqref{mproc} up to terms of order 
$\frac{1}{|p|^5}$ -- is 
\begin{equation} \label{fcont}
 \frac{(p\cdot k)^2}{2 k^2 p^6}.
\end{equation}
Adding this to \eqref{intere} we obtain our final result which may be 
expressed as follows. Let us define 
\begin{equation}\label{effer}
{\tilde S}_4^E(p_1-\delta_1-\delta_2, -p, \delta_1, \delta_2)= 
{\tilde S}_4(p-\delta_1-\delta_2,-p,\delta_1,\delta_2)-\frac{S_3(p-\delta_1-\delta_2,-p,\delta_1+\delta_2) S_3(\delta_1,\delta_2,-\delta_1-\delta_2)}{S_2(\delta_1+\delta_2)}.
\end{equation}
It follows that 
\begin{equation}\label{sfe}
{\tilde S}_4^E(p, -p+\delta, k, -k-\delta)
= 
 \left( {\tilde S}_4(p, -p+\delta, k, -k-\delta)+ 
 \frac{1}{2 |\delta| |p| |p-\delta| |k| |k+\delta|}  \right).
\end{equation} 
Note that ${\tilde S}_4^E$ is simply the effective vertex that determines 
the leading large $N$ four point Green's function of four $\sigma$ operators. 
We get the Green's function by multiplying  ${\tilde S}_4^E$ by 
\begin{equation}
P=\prod_{i=1}^4 \frac{1}{{\tilde S}_2(p_i)}.
\end{equation}
The final result of our computations is best worded in terms of 
${\tilde S}_4^E$ and is 
\begin{equation}\label{fpvert}
\begin{split}
&\lim_{\delta \to 0} \frac{1}{N_B}
\left( {\tilde S}_4^E  \right) = 
-\frac{1}{2}\frac{1}{|p|^4 |k|}-\frac{1}{2}\frac{(p\cdot k)^2}{|p|^6 k^3} 
+{\cal O}(1/|p|^5),\\
\end{split}
\end{equation}
where the right-hand side is presented in a Taylor expansion in $\frac{|k|}{|p|}$, with 
all terms of the same homogeneity in $k$ being thought of as of the same 
order. 

It follows that the Green's function of four $\sigma$ operators is given by 
\begin{equation}\label{fptgffree}
\lim_{\delta \to 0} G_4^0(p,-p+\delta,k,-k-\delta)= 
-\frac{8^4}{N_B^3} \left( \frac{1}{2 |k+p|}+\frac{1}{2 |k-p|} -\frac{1}{2}\frac{|k|}{ |p|^2} -\frac{1}{2} \frac{(p\cdot k)^2}{|p|^4 |k|}  + {\cal O}(k^2/p^3) \right).
\end{equation}

\subsubsection{The quintic term}
\label{quintterm}

\begin{figure}[h]
\begin{center}
\includegraphics[width=9.5cm,height=5.5cm]{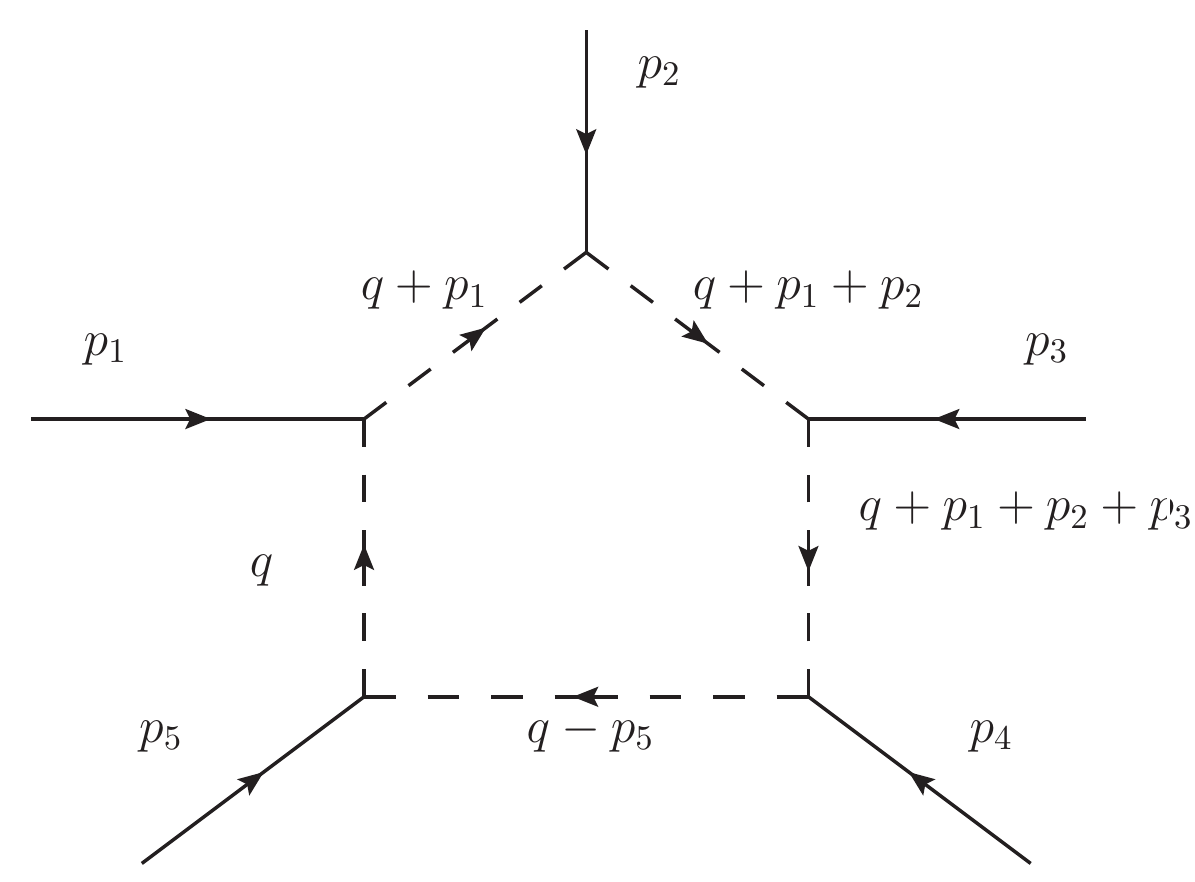}
  \caption{\label{5pt} This diagram defines $H(p_1,p_2,p_3,p_4,p_5)$ in \eqref{Hdef}. Note that by momentum conservation $p_1+p_2+p_3=-p_4-p_5.$}
 \end{center}
\end{figure}

The fifth order term in the effective action is given by 
\begin{equation}\begin{split}\label{fi}
&{\tilde S}_5(p_1,p_2,p_3,p_4)=2N_B\Bigg( H(p_1,p_2,p_3,p_4, p_5)+ H(p_1,p_2,p_3, p_5, p_4) +H(p_1,p_2,p_4,p_3, p_5)\\
&\qquad\qquad\qquad\qquad\qquad+H(p_1,p_2,p_4,p_5, p_3)+H(p_1,p_2,p_5,p_3, p_4)+ H(p_1,p_2,p_5,p_4, p_3)\\
&\qquad\qquad\qquad\qquad\qquad+H(p_1,p_3,p_2,p_4, p_5)+H(p_1,p_3,p_2,p_5, p_4)+ H(p_1,p_4,p_2,p_3, p_5)\\
&\qquad\qquad\qquad\qquad\qquad+H(p_1,p_4,p_2,p_5, p_3)+H(p_1,p_5,p_2,p_3, p_4)+ H(p_1,p_5,p_2,p_4, p_3)\Bigg)
\end{split}
\end{equation}
(one first finds 24 terms, but then uses `reflection symmetry' of the graphs to simplify this to a sum of 12 terms with an 
additional factor of 2, as in the case of the quartic expression).
Here (see figure \ref{5pt})
\begin{equation}\label{Hdef}
H(p_1,p_2,p_3,p_4, p_5)\equiv \int \frac{d^3q}{(2\pi)^3} \frac{1}{q^2 (q+p_1)^2 (q+p_1+p_2)^2 (q-p_4-p_5)^2 (q-p_5)^2}.
\end{equation}

In the rest of this subsubsection we  will  proceed to simplify \eqref{fi} in the kinematical  regime of most interest to us, namely 
${\tilde S}_5(p -\delta_1-\delta_2-\delta_3, -p, \delta_1, \delta_2, \delta_3)$ in the limit $\delta_i \to 0$ for all $i= 1,2, 3$. 
${\tilde S}_5$ is highly singular in this limit. Note that on dimensional grounds ${\tilde S}_5$ scales like $(momentum)^{-7}$. 
For the physical purposes of interest to this paper, we are not interested in ${\tilde S}_5$ itself, but in the tree level 5 point function 
computed from the action formed from ${\tilde S}_5$, ${\tilde S}_4$, ${\tilde S}_3$ and the propagator derived from 
${\tilde S}_2$. We expect this five point function to have a smooth limit as $\delta_i \to 0$. 

\begin{figure}[h]
\begin{center}
\includegraphics[width=9.5cm,height=5.5cm]{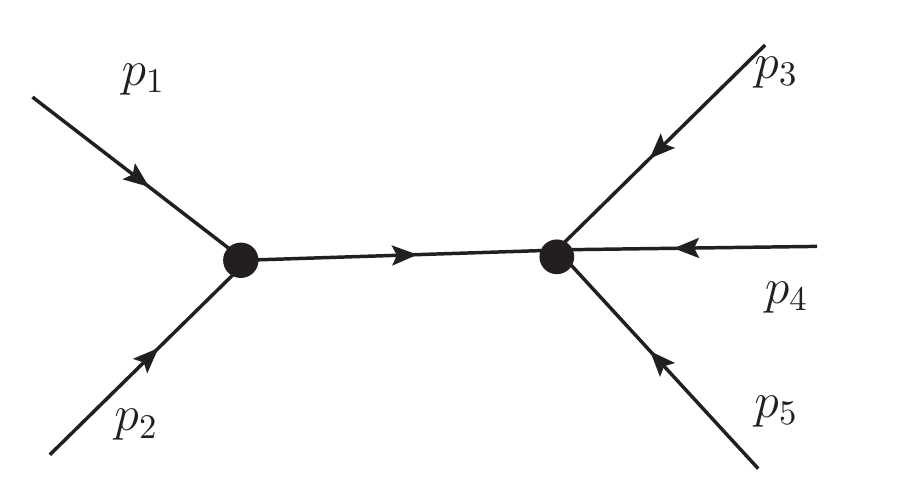}
  \caption{\label{5ptn} This diagram defines $J(p_1,p_2,p_3,p_4,p_5)$ in \eqref{Jdef}. }
 \end{center}
\end{figure}

The contribution of tree diagrams involving ${\tilde S}_4$, ${\tilde S}_3$ effectively renormalizes ${\tilde S}_5$ in an additive 
manner. Graphs with one ${\tilde S}_4$ vertex and one ${\tilde S}_3$ vertex contribute as follows. Define (see Figure \ref{5ptn})
\begin{equation}\label{Jdef}
J(p_1, p_2, p_3, p_4, p_5)= -
\frac{ {\tilde S}_3(p_1, p_2, -(p_1+p_2)) {\tilde S}_4(p_1+p_2, p_3, p_4, p_5) } {{\tilde S}_2(p_1+p_2)}.
\end{equation}

\begin{figure}[h]
\begin{center}
\includegraphics[width=9.5cm,height=5.5cm]{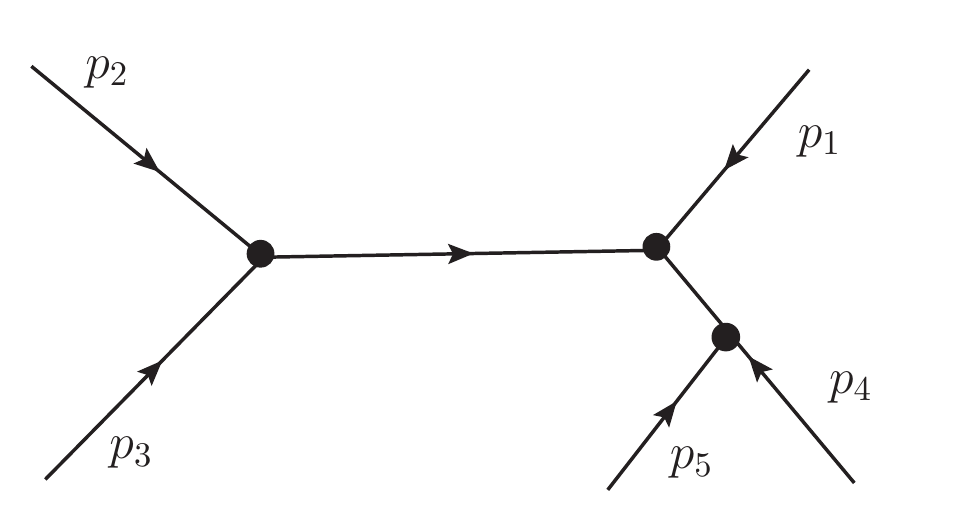}
  \caption{\label{5pt11} This diagram defines $K(p_1,p_2,p_3,p_4,p_5)$ in \eqref{Kdef}. }
 \end{center}
\end{figure}

Then the contribution of such mixed graphs to the effective value of ${\tilde S}_5$ is 
\begin{equation}\label{contemg} \begin{split}
& J(p_1, p_2, p_3, p_4, p_5) + J(p_1, p_3, p_2, p_4, p_5) + J(p_1, p_4, p_3, p_2, p_5) + J(p_1, p_5, p_3, p_4, p_2) \\
&+ J(p_3, p_2, p_1, p_4, p_5) + J(p_4, p_2, p_3, p_1, p_5) + J(p_5, p_2, p_3, p_4, p_1) + J(p_3, p_4, p_1, p_2, p_5)\\
 &+ J(p_3, p_5, p_1, p_4, p_2) +J(p_4, p_5, p_3, p_1, p_2). \\
\end{split}
\end{equation}
In a similar manner, the contribution from tree graphs involving three cubic vertices is given as follows. Let us define (see Figure \ref{5pt11})
\begin{equation}\label{Kdef}
K(p_1, p_2, p_3, p_4, p_5)= \frac{{\tilde S}_3(p_1, p_2+p_3,p_4+p_5) {\tilde S}_3(p_2, p_3, -(p_2+p_3) )
{\tilde S}_3(p_4, p_5, -(p_4+p_5) ) } {{\tilde S}_2(p_2+p_3) {\tilde S}_2(p_4+p_5) }.
\end{equation}

The effective additive shift to ${\tilde S_5}$ from such graphs is given by 
\begin{equation}\label{conthpg} \begin{split}
& K(p_1, p_2, p_3, p_4, p_5) + K(p_1, p_2, p_4, p_3, p_5) + K(p_1, p_2, p_5, p_3, p_4)+ \\
& K(p_2, p_1, p_3, p_4, p_5) + K(p_2, p_1, p_4, p_3, p_5) + K(p_2, p_1, p_5, p_3, p_4)+ \\
& K(p_3, p_2, p_1, p_4, p_5) + K(p_3, p_2, p_4, p_1, p_5) + K(p_3, p_2, p_5, p_1, p_4)+ \\
& K(p_4, p_2, p_3, p_1, p_5) + K(p_4, p_2, p_1, p_3, p_5) + K(p_4, p_2, p_5, p_3, p_1)+ \\
& K(p_5, p_2, p_3, p_4, p_1) + K(p_5, p_2, p_4, p_3, p_1) + K(p_5, p_2, p_1, p_3, p_4). \\
\end{split}
\end{equation}
The final effective five point function, ${\tilde S}^E_5$ is obtained by 
adding \eqref{contemg} and \eqref{conthpg} to ${\tilde S}_5$. In order to 
obtain the 5 point Green's function for $\sigma$ we must
then multiply this effective five point function by  
\footnote{We expect this final result to be finite in the limit 
$\delta_i \to 0$, and therefore to be proportional to $\frac{1}{p^2}$. 
In order for this to be the case it must be that the most singular piece in 
${\tilde S}^E_5$ is $\propto \frac{1}{p^2 |\delta_1| |\delta_2| |\delta_3|}$. }
\begin{equation}
P=\prod_{i=1}^5 \frac{1}{{\tilde S}_2(p_i)}.
\end{equation}

\subsubsection{$SO(N)$}

Throughout this subsection we have so far assumed that the scalar fields in 
the initial action \eqref{cst} are all complex. All our results for 
the effective action as a function of $\sigma$ are easily modified 
to the case that the fields $\phi$ are real (so that we have an $SO(N_B)$ 
symmetry rather than a unitary symmetry).  
If we replace the second line in \eqref{cst} by the action 
\begin{equation} \label{actson}
S= \int d^3x \left( \frac{1}{2} \partial_{\mu}\phi \partial^\mu \phi + 
\sigma \phi^2  \right)
\end{equation}
for real fields $\phi_m$, we find that 
\begin{equation} \label{corss}
{\tilde S}_n^{SO(N)} =2^{n-1}{\tilde S}_n^{SU(N)}.
\end{equation}

\subsection{First subleading corrections to the quadratic and cubic vertices}

In this subsection we continue to work with the critical boson theory at 
$\lambda_B=0$. Using the explicit results of the previous subsection, we 
proceed to evaluate the first corrections, in an expansion in $\frac{1}{N_B}$, 
to the effective action for $\sigma$ at quadratic order, and study the 
same correction at cubic order. Our explicit evaluation of the quadratic 
correction allows us to compute the anomalous dimension of $\sigma$ at 
order $\frac{1}{N_B}$; we find agreement with results previously computed 
in the literature. This serves as a check of the formalism developed in this 
paper. At cubic order we find a formal expression for the divergent part of 
the correction to the cubic effective action for $\sigma$. The expressions 
we derive will allow us to demonstrate that the constant part of the 
$\beta$ function  for $x_6$ -- computed earlier in this paper -- 
indeed vanishes for the regular boson theory at $\lambda_B=0$ as it must on 
physical grounds, providing an additional check of the results of this paper.

\subsubsection{Anomalous dimension of $\sigma$ at first subleading order 
in $\frac{1}{N}$}

It is useful to think of $S_{eff}(\sigma)$ as a classical action which 
then receives loop corrections; the loop expansion parameter is 
$\frac{1}{N_B}$. At quadratic order it is given by 
\begin{equation} \label{quadclact}
S_{eff}(\sigma)= \frac{1}{2} 
\int \frac{d^3p}{(2\pi)^3} {\tilde S}_2(p) \sigma(p) \sigma(-p)
= \frac{1}{2} 
\int \frac{d^3p}{(2\pi)^3} ( \frac{-N_B}{8 |p|}) \sigma(p) \sigma(-p).
\end{equation}

\begin{figure}[h]
\begin{center}
\includegraphics[width=8.5cm,height=4.5cm]{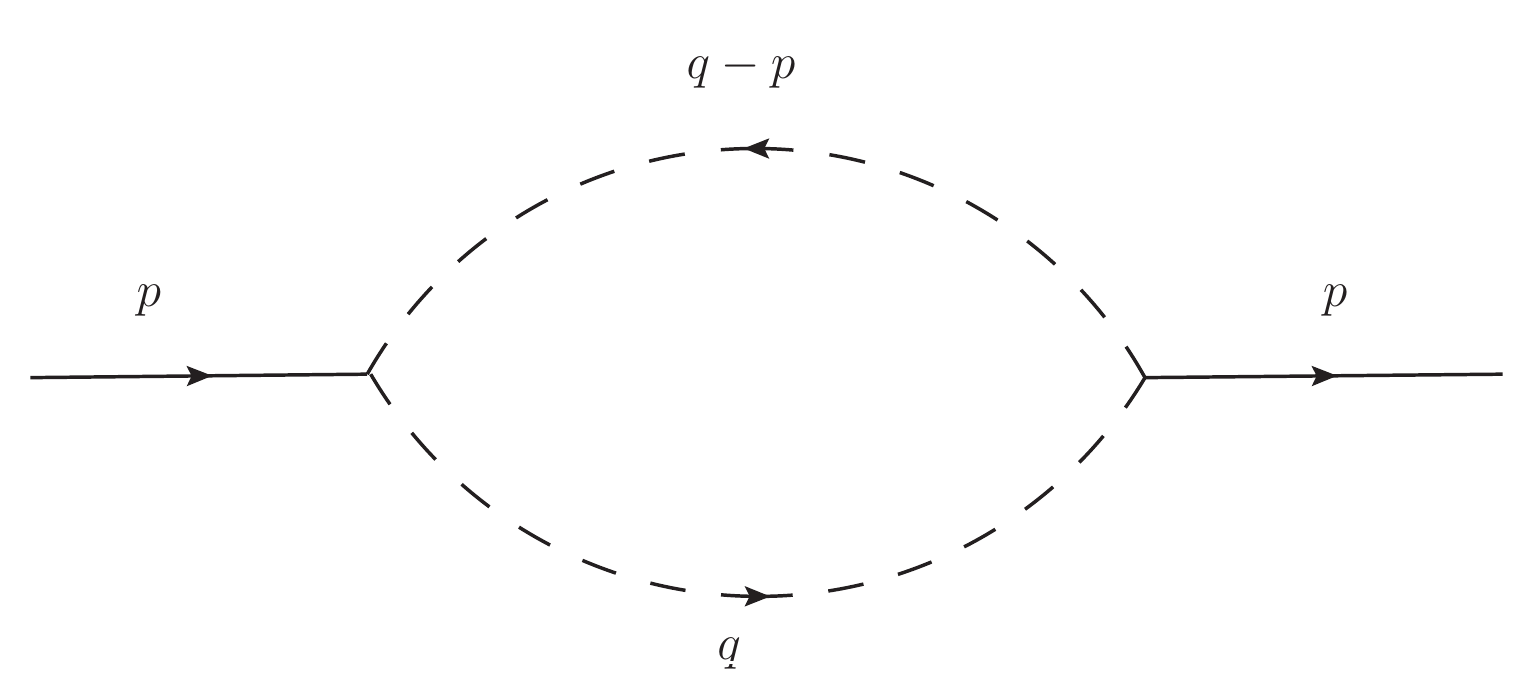}
  \caption{\label{onecolor} One of the three one loop corrections to 
the quadratic effective action for $\sigma$.}
 \end{center}
\end{figure}
The quantum effective action for $\sigma$ includes the classical term 
\eqref{quadclact} together with quantum corrections. At first 
subleading order in $\frac{1}{N_B}$ it is given by
\begin{equation} \label{quadclactn}
S_{eff}(\sigma)= \frac{1}{2} 
\int \frac{d^3p}{(2\pi)^3} \left( {\tilde S}_2(p) + \delta S_2 
\right) \sigma(p) \sigma(-p)
= \frac{1}{2} 
\int \frac{d^3p}{(2\pi)^3} ( \frac{-N_B}{8 |p|} + \delta S_2(p)) 
\sigma(p) \sigma(-p),
\end{equation}
where the self energy correction $\delta S_2$ receives contributions from three distinct one loop graphs.
The first graph describes a $\sigma$ `particle' splitting into two 
via a cubic vertex (see figure \ref{onecolor}); the resultant `particles' 
then rejoin into a single 
$\sigma$ particle through another cubic vertex. We call this contribution 
$A_3(p)$ below. The remaining two graphs are both tadpole type graphs. The
first of these is the contribution of a four point vertex with 
two its legs sewn together. 
The second graph has the two external $\sigma$ lines `colliding' 
to form a single internal sigma propagator, whose other end meets a cubic 
vertex. The remaining two lines in the cubic vertex contract. We denote the combined contribution of these two graphs 
as $A_4(p)$ below (the graphs are similar to those of
figure \ref{2pt1loop} above). We find 
\begin{equation} \begin{split} \label{atafexp}
\delta S_2(p) &= \left(A_3(p)+ A_4(p) \right), \\
A_3(p) & = -\frac{1}{2} \int \frac{d^3q}{(2\pi)^3} \frac{S_3(p,q,-(p+q)) S_3(-p,-q,(p+q))}{S_2(q)S_2(p+q)}= \frac{1}{2\pi^2 |p|},\\
A_4(p) &=\frac{1}{2} \int \frac{d^3q}{(2\pi)^3} 
\frac{ \lim_{\delta \to 0} \left( S_4(p,-p+ \delta,q,-q - \delta) 
- \frac{S_3(p, -p -\delta, \delta) S_3(q, -q - \delta , \delta )}{S_2(\delta)} \right)  }{S_2(q)}
\\
&=\frac{1}{2} \int \frac{d^3q}{(2\pi)^3} 
\lim_{\delta \to 0} \left({\tilde S}^E_4(p,-p+ \delta,q,-q - \delta) 
\right)\\
&=\frac{4}{3\pi^2 |p|} \left ( \alpha \ln (\frac{\Lambda}{|p|})  + \beta
\right) .
\end{split}
\end{equation}
The explicit result for $A_3(p)$ above is obtained by substituting 
the explicit expression for ${\tilde S}_3(p_1, p_2, p_3)$  and using  
\eqref{bi}. As we do not have an explicit expression for 
${\tilde S}_4(p_1, p_2, p_3, p_4)$, we evaluate the integral $A_4(p)$ in 
a more indirect manner. Dimensional analysis together with power counting imply that $A_4(p)$ takes the form reported in the last line of \eqref{atafexp},
where $\Lambda$ is a UV cutoff and $\alpha$ and $\beta$ are pure numbers. 
The number $\beta$ depends on the precise definition of the UV cutoff. 
On the other hand the number $\alpha$ is universal (in the sense 
that it is insensitive to the details of the UV cutoff). In order to 
compute $\alpha$ we need only compute the coefficient of $\ln (\Lambda)$ 
in $A_4$. By power counting, only those terms that scale like 
$\frac{1}{q^4}$ in the numerator of the  integrand in the second last line of 
\eqref{atafexp} give rise to a logarithmic divergence in $A_4(p)$. 
However all such terms were listed in \eqref{fpvert}. Substituting 
\eqref{fpvert} into the last line of \eqref{atafexp} we find
\begin{equation} \label{workout}
\begin{split}
A_4(p)&=\frac{1}{2} 
\left( \int \frac{d^3k}{(2\pi)^3} \frac{-\frac{1}{2}\frac{1}{|k|^4 |p|}-\frac{1}{2}\frac{(p\cdot k)^2}{|k|^6 p^3}  }{ - \frac{1}{8|k|}} + ...\right) \\
&= \frac{4}{3 \pi^2 |p|}  \left( \int \frac{dk}{k} + ...\right) = \frac{4}{3\pi^2 |p|}\ln (\Lambda) + \cdots
\\
\end{split}
\end{equation}
The terms $\cdots$ denote those parts of the integrand whose contribution 
to the integral is finite in the UV. In going from the first to the second line of \eqref{workout} we have 
performed the angular integral and used the fact that $\cos^2 (\theta)$ 
averages to $\frac{1}{3}$.

It follows that $\alpha=1$, so that 
\begin{equation} \label{dpqa}
{\tilde S}_2(p) + \delta S_2 (p)= 
- \frac{N_B}{8 |p|} \left( 1 + \frac{32}{3\pi^2 N_B}\ln\left(\frac{|p|}{|\Lambda|}\right)
+ \frac{c}{N_B} \right) = - 
( 1 + \frac{c}{N_B}) \frac{N_B}{8 |p|} \left( \frac{\Lambda}{|p|} \right)^{ -\frac{32}{3\pi^2 N_B}},
\end{equation}
where $c$ is a number of order unity that we will 
not need. 

The Green's function for $\sigma$ is given by inverting the quantum 
effective action for $\sigma$, which gives
\begin{equation} \label{zfgf}
G_2(p) \propto -8 |p|  \left( \frac{\Lambda}{|p|} \right)^\frac{32}{3\pi^2 N_B}.
\end{equation}
It follows that the scaling dimension\footnote{  Note that
\begin{equation}
|p|-|p| \frac{32}{ 3\pi^2 N_B}\log\left(|p|\right)\simeq p^{1- \frac{32}{ 3\pi^2 N_B}}.
\end{equation}
The Fourier transform of 
\begin{equation}
p^{1+\delta}\sim \frac{1}{x^{4+\delta}}
\end{equation}  
implies an anomalous dimension $\frac{\delta}{2}$, leading to \eqref{sdfsl}.}
of $\sigma$ is given at first subleading order in $\frac{1}{N_B}$ by 
\begin{equation}\label{sdfsl}
\Delta_{\sigma}= 2  -  \frac{16}{3\pi^2 N_B}.
\end{equation} 

We have so far focused on the case of the $U(N_B)$ theory, in which 
the basic scalar fields are all complex. The analogous results for the 
$SO(N_B)$ theory are, however, easily obtained. The first line of 
\eqref{atafexp} still applies to the $SO(N_B)$ theory. The second and 
fourth line of that equation also apply provided we use the values of 
$S_4$, $S_3$, and $S_2$ appropriate to the $SO(N_B)$ theory. Using \eqref{corss}
it follows that 
\begin{equation}\label{souad}
A_3(p)^{SO(N_B)}= 4 A_3(p)^{U(N_B)}, ~~~~A_4(p)^{SO(N_B)}= 4 A_4(p)^{U(N_B)}.
\end{equation}
It follows that 
\begin{equation}\label{soud}
\delta S_2^{SO(N_B)}(p)= 4 \delta S_2^{U(N_B)}(p) .
\end{equation}
On the other hand \eqref{corss} asserts that 
\begin{equation}\label{soudtwo}
S_2^{SO(N_B)}(p)= 2 S_2^{U(N_B)}(p) .
\end{equation}
As the anomalous dimension of $\sigma$ is determined by the relative 
factor between $\delta S_2$ and $S_2$, it follows that the anomalous
dimension of $\sigma$ in the $SO(N_B)$ theory is twice the anomalous 
dimension in the $U(N_B)$ theory, so that the dimension of $\sigma$ 
in the $SO(N_B)$ theory is given by 
\begin{equation}\label{sodfsl}
\Delta_{\sigma}= 2  -  \frac{32}{3\pi^2 N_B}.
\end{equation} 

\subsubsection{Correction to the divergent part of the 
Green's function $G_3$ at first order in $\frac{1}{N}$}

As in the previous section we now proceed to compute the first correction to ${\tilde S_3}$ 
in the $\frac{1}{N_B}$ expansion. Focusing on terms at cubic order, the $\sigma$ effective action 
takes the schematic form 
\begin{equation}
S_{eff}(\sigma)= \left( {\tilde S}_3 + \delta S_3 \right) \sigma^3,
\end{equation}
where ${\tilde S}_3$ was listed in \eqref{vars} and 
\begin{equation}\label{effthres}
\begin{split}
\delta S_3(q_1,q_2,q_3)
&= \int \frac{d^3 q}{(2\pi)^3} \frac{{\tilde S}_3(q_1,q,-q-q_1) {\tilde S}_3(q_2,-q+q_3,q+q_1) {\tilde S}_3(q_3,-q,q-q_3)}{ {\tilde S}_2(q+q_1) {\tilde S}_2(q-q_3) {\tilde S}_2(q)}\\
&\qquad-\frac{3}{2}\int \frac{d^3 q}{(2\pi)^3} \frac{{\tilde S}_4(q_1,q_2,-q+q_3,q) {\tilde S}_3(q-q_3,-q,q_3) }{{\tilde S}_2(q) {\tilde S}_2(q-q_3)} \\
&\qquad+\frac{1}{2}\int \frac{d^3 q}{(2\pi)^3}   \frac{{\tilde S}_5(q_1,q_2,q_3,q,-q) }{{\tilde S}_2(q)}
\end{split}
\end{equation}
(where the second line should be symmetrized over the three external momenta).

Accurate to first subleading order, the three-point function of three $\sigma$ operators is given by the formula
\begin{equation}\label{tpf}
\langle \sigma(q_1) \sigma(q_2) \sigma(q_3) \rangle
= - \prod_{i=1}^3 \left( {\tilde S_2}(q_i) + \delta S_2(q_i) \right)^{-1}
\left({\tilde S_3}(q_1, q_2, q_3) + \delta S_3(q_1, q_2, q_3) \right) .
\end{equation}
It follows that $\delta G_3$, the first correction to ${\tilde G}_3$ (the leading 
order three point function for $\sigma$) in an expansion in $\frac{1}{N_B}$,  is  given by
\begin{equation}\label{dgt}
\delta G_3 = \frac{1}{{\tilde S}_2(q_1) {\tilde S}_2(q_2) {\tilde S}_2(q_3) }
\left( - \delta S_3 +  {\tilde S}_3 \sum_{i=1}^3 \frac{\delta { S}_2(q_i)}{{\tilde S}_2(q_i)} \right).
\end{equation}
Joining everything together we have 
\begin{equation}\label{effthrees}
\begin{split}
&\delta G_3(q_1,q_2,q_3) \\
&= -\frac{1}{{\tilde S}_2(q_1) {\tilde S}_2(q_2) {\tilde S}_2(q_3)} \bigg(\int \frac{d^3 q}{(2\pi)^3} \frac{{\tilde S}_3(q_1,q,-q-q_1) {\tilde S}_3(q_2,-q+q_3,q+q_1) {\tilde S}_3(q_3,-q,q-q_3)}{ {\tilde S}_2(q+q_1) {\tilde S}_2(q-q_3) {\tilde S}_2(q)}\\
&\qquad\qquad-\frac{3}{2}\int \frac{d^3 q}{(2\pi)^3} \frac{{\tilde S}_4(q_1,q_2,-q+q_3,q) {\tilde S}_3(q-q_3,-q,q_3) }{{\tilde S}_2(q) {\tilde S}_2(q-q_3)} \\
&\qquad\qquad-\frac{3}{2}\int \frac{d^3 q}{(2\pi)^3} \frac{{\tilde S}_4(q_3,q,-q,q_1+q_2) {\tilde S}_3(-(q_1+q_2),q_1,q_2) }{{\tilde S}_2(q) {\tilde S}_2(q_1+q_2)} \\
&\qquad\qquad+\frac{3}{2} \int \frac{d^3 q}{(2\pi)^3} \frac{  {\tilde S}_3(q_1,q_2,q_3) {\tilde S}_3(q_3,q,q-q_3) {\tilde S}_3(q,q_3,q-q_3)}{ {\tilde S}_2(q_3) {\tilde S}_2(q) {\tilde S}_2(q-q_3)}\\
&\qquad\qquad+\frac{1}{2}\int \frac{d^3 q}{(2\pi)^3}   \frac{{\tilde S}_5(q_1,q_2,q_3,q,-q) }{{\tilde S}_2(q)} \bigg)
\end{split}
\end{equation}
(where the middle three lines should be symmetrized over the three external momenta).
The third and fourth lines in \eqref{effthrees} account for the second term (proportional to $\delta S_2$) in \eqref{dgt}. 
\begin{figure}[h]
\begin{center}
\includegraphics[width=14.5cm,height=7.5cm]{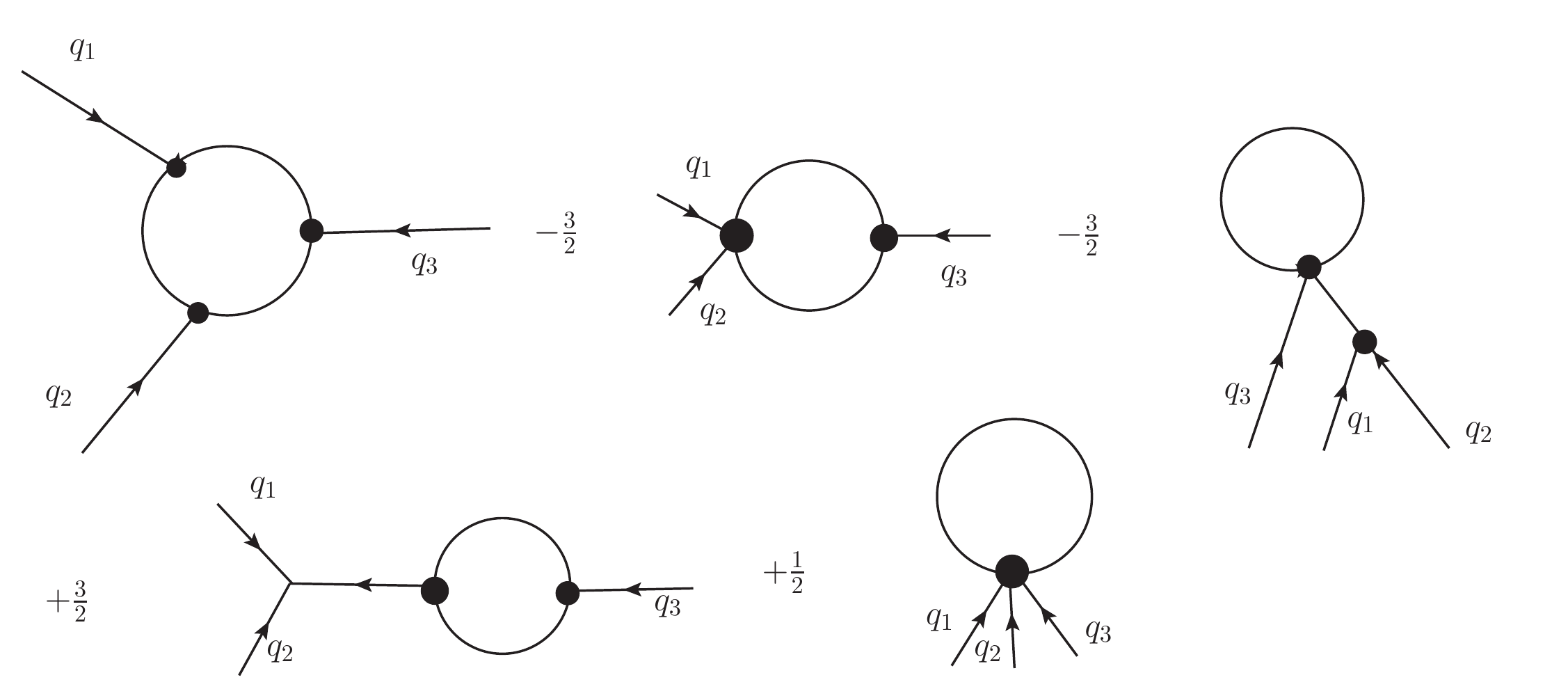}
  \caption{\label{effective} Contributions to the effective action \eqref{effthrees}.}
 \end{center}
\end{figure}

The different contributions to \eqref{effthrees}  together with their symmetry factors  are expressed in terms of graphs in  figure \ref{effective}.

The value of the parameter $r_B$ (see \eqref{thptdet})  at $\lambda_B=0$  may be obtained by evaluating
the terms in \eqref{effthrees} that are proportional to $\ln (\Lambda)$. More directly the part of \eqref{effthrees} that is 
proportional to $\ln (\Lambda)$ should be identified with the negative of the third line in \eqref{termlisy}.

The equation \eqref{dgt} is slightly formal because the expressions presented in \eqref{effthrees} for $\delta S_3$ 
are expressed in terms of quite formidable integrals that have not been explicitly evaluated. Fortunately we will not need
explicit expressions for the relevant expressions; the formal expression \eqref{effthrees} will prove sufficient for 
our purposes below. 

\subsubsection{Vanishing of the $x_6$ independent piece of the $\beta$ function at $x_6=0$} \label{vipb}

The beta function \eqref{finalbetafn} has three distinct kinds of contributions. One contribution, proportional to $\delta'_B$, has 
its origins in the anomalous dimension of the $\zeta$ field. This anomalous dimension vanishes in the limit $\lambda_B=0$ -- the 
limit of interest in this section. The second contribution -- proportional to ${\tilde g}_3$ -- has its origin in the $1/N_B$ correction 
to the correlator of 3 $\sigma$ fields. This piece was computed in the previous subsection;  as explained above 
it is determined by  the coefficient of 
$\ln (\Lambda) $ in \eqref{dgt} above. The last contribution comes from the coefficient of the $\ln (\Lambda)$ terms in the 
graphs reported in \eqref{sopith}, i.e. from the computation of $S^{IPI}(\zeta) - S^{eff}(\zeta)$.

In this subsection we will demonstrate that the second and third contributions described in the paragraph above cancel each other 
when $x_6=0$ (recall that we assume $\lambda_B=0$ throughout this section). In order to demonstrate cancellation we will not have 
to evaluate any integrals - it turns out that the cancellation happens at the formal level of unevaluated integrals. 

As we have mentioned above, the second contribution to the $\beta$ function has already been listed in \eqref{effthrees} in terms 
of integrals involving the vertex factors $S_2$, $S_3$, $S_4$ and $S_5$. In order to demonstrate cancellation we simply rewrite the 
integrals in \eqref{sopith} at $x_6=0$ 
 and check that the resulting expressions cancel those listed in \eqref{effthrees} term by term. Terms involving only the 
vertices  ${\tilde G}_3$ and ${\tilde G}_4$ can be rewritten in terms of ${\tilde S_3}$, ${\tilde S}_4$, and ${\tilde S_2}$
in a quite straightforward manner; we find  
\begin{equation}\label{th}
\begin{split}
& \int \frac{d^3 q}{(2\pi)^3} \frac{G_3^0(q_1,q,-q-q_1) G_3^0(q_2,-q+q_3,q+q_1) G_3^0(q_3,-q,q-q_3)}{ G_2^0(q_1) G_2^0(q_2) G_2^0(q_3) G_2^0(q+q_1) G_2^0(q-q_3) G_2^0(q)}\\
&\qquad= -\int \frac{d^3 q}{(2\pi)^3} \frac{{\tilde S}_3(q_1,q,-q-q_1) {\tilde S}_3(q_2,-q+q_3,q+q_1) {\tilde S}_3(q_3,-q,q-q_3)}{ {\tilde S}_2(q+q_1) {\tilde S}_2(q-q_3) {\tilde S}_2(q)},
\end{split}
\end{equation}

\begin{equation} \label{fou}
\begin{split}
&\int \frac{d^3 q}{(2\pi)^3} \frac{G_4^0(q_1,q_2,-q+q_3,q) G_3^0(q-q_3,-q,q_3) }{G_2^0(q_1) G_2^0(q_2) G_2^0(q_3) G_2^0(q) G_2^0(q-q_3)}\\
&=\int \frac{d^3 q}{(2\pi)^3} \frac{{\tilde S}_4(q_1,q_2,-q+q_3,q) {\tilde S}_3(q-q_3,-q,q_3) }{{\tilde S}_2(q) {\tilde S}_2(q-q_3)} \\
&\qquad-\int \frac{d^3 q}{(2\pi)^3} \frac{{\tilde S}_3(q_1,q_2,-(q_1+q_2)) {\tilde S}_3(q-q_3,-q,q_3) {\tilde S}_3(q-q_3,-q,q_3) }{{\tilde S}_2(q_1+q_2){\tilde S}_2(q) {\tilde S}_2(q-q_3)}\\
&\qquad-2\int \frac{d^3 q}{(2\pi)^3} \frac{{\tilde S}_3(q_1,q,-q-q_1) {\tilde S}_3(q_2,-q+q_3,q+q_1) {\tilde S}_3(q_3,-q,q-q_3)}{ {\tilde S}_2(q+q_1) {\tilde S}_2(q-q_3) {\tilde S}_2(q)}.
\end{split}
\end{equation} 

The terms involving ${\tilde G}_5$ are more delicate, and require considerable care with signs to accurately keep track of. 
We find 
\begin{equation}\label{fiv}
\begin{split}
&\int \frac{d^3 q}{(2\pi)^3}   \frac{G_5^0(q_1,q_2,q_3,q,-q) }{G_2^0(q_1) G_2^0(q_2) G_2^0(q_3) G_2^0(q)}\\
&=-\int \frac{d^3 q}{(2\pi)^3}   \frac{{\tilde S}_5(q_1,q_2,q_3,q,-q) }{{\tilde S}_2(q)}+6\int \frac{d^3 q}{(2\pi)^3} \frac{{\tilde S}_4(q_1,q_2,-q+q_3,q) {\tilde S}_3(q-q_3,-q,q_3) }{{\tilde S}_2(q) {\tilde S}_2(q-q_3)}\\
&\qquad+3\int \frac{d^3 q}{(2\pi)^3} \frac{{\tilde S}_4(q_3,q,-q,q_1+q_2) {\tilde S}_3(-(q_1+q_2),q_1,q_2) }{{\tilde S}_2(q) {\tilde S}_2(q_1+q_2)}\\
&\qquad-6\int \frac{d^3 q}{(2\pi)^3} \frac{{\tilde S}_3(q_1,q_2,-(q_1+q_2)) {\tilde S}_3(q-q_3,-q,q_3) {\tilde S}_3(q-q_3,-q,q_3) }{{\tilde S}_2(q_1+q_2){\tilde S}_2(q) {\tilde S}_2(q-q_3)}\\
&\qquad-6\int \frac{d^3 q}{(2\pi)^3} \frac{{\tilde S}_3(q_1,q,-q-q_1) {\tilde S}_3(q_2,-q+q_3,q+q_1) {\tilde S}_3(q_3,-q,q-q_3)}{ {\tilde S}_2(q+q_1) {\tilde S}_2(q-q_3) {\tilde S}_2(q)}.
\end{split}\end{equation}
In the expressions  \eqref{fiv}, various factors of $3$ and $6$ should be thought of as different terms leading to same kind of integral.
Adding the expressions \eqref{th}, \eqref{fou} and \eqref{fiv} together we find that the net expression for the third contribution 
(terms in $S^{1PI}- S^{eff}$) is 
\begin{equation}\label{tot}
\begin{split}
&\int \frac{d^3 q}{(2\pi)^3} \frac{G_3^0(q_1,q,-q-q_1) G_3^0(q_2,-q+q_3,q+q_1) G_3^0(q_3,-q,q-q_3)}{ G_2^0(q+q_1) G_2^0(q-q_3) G_2^0(q)}\\
&-\frac{3}{2}\int \frac{d^3 q}{(2\pi)^3} \frac{G_4^0(q_1,q_2,-q+q_3,q) G_3^0(q-q_3,-q,q_3) }{G_2^0(q) G_2^0(q-q_3)} 
+\frac{1}{2}\int \frac{d^3 q}{(2\pi)^3}   \frac{G_5^0(q_1,q_2,q_3,q,-q) }{G_2^0(q)}\\
&= -\frac{1}{{\tilde S}_2(q_1) {\tilde S}_2(q_2) {\tilde S}_2(q_3)} 
\bigg(-\int \frac{d^3 q}{(2\pi)^3} \frac{{\tilde S}_3(q_1,q,-q-q_1) {\tilde S}_3(q_2,-q+q_3,q+q_1) {\tilde S}_3(q_3,-q,q-q_3)}{ {\tilde S}_2(q+q_1) {\tilde S}_2(q-q_3) {\tilde S}_2(q)}\\
&\qquad+\frac{3}{2}\int \frac{d^3 q}{(2\pi)^3} \frac{{\tilde S}_4(q_1,q_2,-q+q_3,q) {\tilde S}_3(q-q_3,-q,q_3) }{{\tilde S}_2(q) {\tilde S}_2(q-q_3)}\\
&\qquad+\frac{3}{2}\int \frac{d^3 q}{(2\pi)^3} \frac{{\tilde S}_4(q_3,q,-q,q_1+q_2) {\tilde S}_3(-(q_1+q_2),q_1,q_2) }{{\tilde S}_2(q) {\tilde S}_2(q_1+q_2)}\\
&\qquad-\frac{3}{2} \int \frac{d^3 q}{(2\pi)^3} \frac{{\tilde S}_3(q_1,q_2,-(q_1+q_2)) {\tilde S}_3(q-q_3,-q,q_3) {\tilde S}_3(q-q_3,-q,q_3) }{{\tilde S}_2(q_1+q_2){\tilde S}_2(q) {\tilde S}_2(q-q_3)}\\
&\qquad-\frac{1}{2}\int \frac{d^3 q}{(2\pi)^3}   \frac{{\tilde S}_5(q_1,q_2,q_3,q,-q) }{{\tilde S}_2(q)} \bigg)\\
&=\delta G_3.
\end{split}
\end{equation}

The fact that $S^{1PI}-S^{eff}$ equals $\delta G_3$ establishes that the contribution to $S^{1PI}$ from the correction to 
$G_3$ exactly cancels the loop contributions to $S^{1PI}$ at $x_6=0$. It follows that the $\beta$ function vanishes at $x_6=0$.

Of course the vanishing of this $\beta$ function is obvious on physical grounds (a free theory does not have a $\beta$ function). However 
the fact that this works out algebraically within our formalism is a consistency check of the formulae presented in this paper.

\section{Correlators for free fermions} \label{fft}

In this Appendix we present computations of the two, three and 
(the connected part of) four point functions of the operator $J_0$ 
in the free fermion theory. 

Recall that 
\begin{equation}\label{normj}
J_0= \frac{4 \pi {\bar \psi }\psi }{{\tilde \kappa}_F}.
\end{equation}
It follows that in the limit ${\tilde \kappa}_F \to \infty$ of 
interest to this Appendix, the $n$ point functions of $J_0$ are given 
by $(\frac{4 \pi}{{\tilde \kappa}_F})^{n}$ times the Wick contractions of 
$n$ copies of ${\bar \psi}\psi$. The Wick contractions are very easy 
to compute at arbitrary values of $N_F$. We now proceed to study two, 
three and four point functions in turn. 
We follow the same conventions as in \cite{Jain:2014nza}.
The two point function is given by 
\begin{equation} \label{tpff}
\langle J^F_0 (q)  J^F_0 (-q') \rangle = (2 \pi)^3 \delta^3(q-q') 
N_F \left( \frac{ 4\pi}{{\tilde \kappa_F}} \right)^2
\Tr \left(  \int \frac{d^{d}q}{(2\pi)^d} \left(  \frac{\slashed{q} \left(\slashed{q}+\slashed{p}\right)}{q^2 (q+p)^2} \right) \right),
\end{equation}
where $d=3$. The integral on the right-hand side of \eqref{tpff} is linearly divergent; we give it meaning by using dimensional regularization. Using
$ \Tr \left( \gamma^{\mu}\gamma^{\nu} \right)= 2 \eta^{\mu\nu} $, we find that
within the dimensional regularization scheme
\begin{equation}\begin{split}
& \int \frac{d^{d}q}{(2\pi)^d} \left(  \frac{{\rm Tr}(\slashed{q} \left(\slashed{q}+\slashed{p})\right)}{q^2 (q+p)^2} \right)
=- \int \frac{d^{d}q}{(2\pi)^d}\left( \frac{p^2}{(p+q)^2 q^2}-\frac{1}{q^2}  -\frac{1}{(p+q)^2} \right)\\
&\qquad=- \int \frac{d^{d}q}{(2\pi)^d}\left( \frac{p^2}{(p+q)^2 q^2}-\frac{2}{q^2} \right)=-\frac{|p|}{8}.
\end{split}\end{equation}
We have used the fact that in dimensional regularization
\begin{equation}\begin{split}
 \int \frac{d^{d}q}{(2\pi)^d} \frac{1}{q^2} &= \int \frac{d^{d}q}{(2\pi)^d} \frac{1}{(q+p)^2} =0,
\end{split}
\end{equation} 
and the convergent integral \eqref{bi}.
It follows that 
\begin{equation} \label{tpfff}
\langle J^F_0 (q)  J^F_0 (-q') \rangle = (2 \pi)^3 \delta^3(q-q') 
\frac{ (-2 N_F) \pi^2 |q|}{{\tilde \kappa}_F^2} .
\end{equation}
Equation \eqref{tpfff} -- which is accurate in the limit ${\tilde \kappa_F} 
\to \infty$ but at all values of $N_F$ -- matches perfectly with 
\eqref{tplnfw}.

The three point function of three $J_0^F$ operators simply vanishes. This 
fact can be seen algebraically from the fact that the diagram 
that computes this correlator is proportional to 
\begin{equation}\label{tpfnint}
\int \frac{d^3 q}{(2 \pi)^3} 
\frac{\epsilon_{\mu \nu \rho} q^\mu (p_1 +q)^\nu (-p_2 +q)^\rho}
{q^2 (p_1+q)^2 (q-p_3)^2} .
\end{equation}
But clearly
\begin{equation}\label{woin}
\int \frac{d^3 q}{(2 \pi)^3} 
\frac{q^\mu}
{q^2 (p_1+q)^2 (q-p_3)^2} = a p_1^\mu + b p_3^\mu,
\end{equation}
where $a$ and $b$ are scalar functions of $p_1$ and $p_2$. Substituting
\eqref{woin} into \eqref{tpfnint} we see that the three point 
function of three $J_0^F$ operators vanishes. 

The vanishing of the three point function could have been anticipated as 
follows. Recall that parity is a symmetry of the free fermion theory, 
and that $J_0^F$ is odd under a parity transformation. It follows that 
the three point function must also be odd under parity. The position 
dependence of the three point function of three $J_0^F$ operators, however, 
is completely fixed by conformal invariance to a functional form 
that is even under parity transformations of the three coordinates.
It follows that the three point function must vanish\footnote{This is actually true in the 't Hooft large $N_F$ limit for all values of the 't Hooft coupling.}. 

Finally let us turn to the four point function \eqref{fplnfer}. The 
function $G^0_4(p_1, p_2, p_3, p_4)$ defined in that equation is given by 
\begin{equation} \label{fggfp}
\begin{split}
G^0_4(p_1,p_2,p_3,p_4)&=N_F \left( \frac{4 \pi}{ {\tilde \kappa_F} }
\right)^4 \Bigg( Y(p_1,p_2,p_3,p_4)+ Y(p_1,p_2,p_4,p_3) +Y(p_1,p_4,p_3,p_2)\\
&+Y(p_1,p_4,p_2,p_3)+Y(p_1,p_3,p_4,p_2)+ Y(p_1,p_3,p_2,p_4)\Bigg),
\end{split}\end{equation}
where 
\begin{equation} \label{fpmin}
\begin{split}
Y(p_1,p_2,p_3,p_4)&= - \int  \frac{d^{d}q}{(2\pi)^d} {\rm Tr} \left( \frac{\slashed{q}}{q^2}  \frac{\slashed{q}+\slashed{p_1}}{(q+p_1)^2}  \frac{ \slashed{q} + \slashed{p_1} + \slashed{p_2} } {(q+p_1+p_2)^2}  \frac{ \slashed{q}-\slashed{p_4}  }{(q-p_4)^2}\right) \\
&=-2\int  \frac{d^{d}q}{(2\pi)^d}\frac{X(q,p_1,p_2,p_3,p_4)}{q^2 (q+p_1)^2 (q+p_1+p_2)^2 (q-p_4)^2},\\
X(q,p_1,p_2,p_3,p_4)&=q\cdot \left( q+p_1 \right) \left(q+p_1+p_2\right)\cdot \left( q-p_4 \right) - q\cdot \left(q+p_1+p_2\right) \left(q+p_1\right)\cdot \left( q-p_4 \right)\\
& + q\cdot \left( q-p_4 \right) \left(q+p_1+p_2\right)\cdot \left( q+p_1 \right).
\end{split}
\end{equation}
In going from the first to the second line of \eqref{fpmin} we have used
\begin{equation}
\begin{split}
{\rm tr}\left( \gamma^{\mu}  \gamma^{\nu}\gamma^{\rho}\gamma^{\sigma}\right)&=2 \left(\eta^{\mu\nu}\eta^{\rho\sigma}-\eta^{\mu\rho}\eta^{\nu\sigma}+ \eta^{\mu\sigma}\eta^{\nu\rho}\right).
\end{split}
\end{equation}
Note that
\begin{equation}
Y(p_1,a,b,c) = Y(p_1,c,b,a),
\end{equation}
which gives
\begin{equation} \label{fggfp1}
\begin{split}
G^0_4(p_1,p_2,p_3,p_4)&=2N_F \left( \frac{4 \pi}{ {\tilde \kappa_F} }
\right)^4 \Bigg( Y(p_1,p_2,p_3,p_4)+ Y(p_1,p_2,p_4,p_3) +Y(p_1,p_3,p_2,p_4)\Bigg).
\end{split}\end{equation}

The integral on the second line of \eqref{fpmin} is our final result for 
the connected four point function of four $J_0^F$ operators. 
This integral 
is complicated in general, but simplifies in the kinematical regimes 
of interest to this paper.
For our purpose, we need to compute
\begin{equation}
\begin{split}
&G^0_4(p,-p,k,-k)=2N_F \left( \frac{4 \pi}{ {\tilde \kappa_F} }
\right)^4 \Bigg( Y(p,-p,k,-k)+ Y(p,-p,-k,k) +Y(p,k,-p,-k)\Bigg).
\end{split}
\end{equation}
The above expressions can be simplified as
\begin{equation}
\begin{split}
&-\frac{1}{N_F} \left( \frac{ {\tilde \kappa_F} }{4 \pi}
\right)^4G^0_4(p,-p,k,-k) \\
&=\int \frac{d^{3}q}{(2\pi)^3} \Bigg( \frac{2}{q^4}- \frac{2 p\cdot q}{q^4 (p+q)^2}+\frac{2}{(q+p)^4}-\frac{2 p^2}{q^2 (q+p)^4}-\frac{2 k^2}{q^4 (q+k)^2}+\frac{2 k^2 p^2}{q^4 (q+k)^2 (q+p)^2} \\
&~-\frac{2 k^2}{(q+p)^4 (q+p+k)^2}+\frac{2 k^2 p^2}{q^2 (q+p)^4 (q+p+k)^2}\\
&+\frac{4}{(p+q)^2 (k+q)^2}+\frac{4}{q^2 (q+p+k)^2}-\frac{6 p\cdot k}{q^2 (k+q)^2 (p+q)^2}-\frac{2 p\cdot k}{(p+q)^2 (k+q)^2 (k+p+q)^2}\\
&+\frac{2 p\cdot k}{q^2 (k+q)^2 (k+p+q)^2}+\frac{6 p\cdot k}{q^2 (p+q)^2 (k+p+q)^2}\\
&+\frac{4 (p\cdot k)^2}{q^2 (k+q)^2 (p+q)^2 (p+k+q)^2}-\frac{2 k^2 p^2}{q^2 (q+p)^2 (q+k)^2 (q+p+k)^2}\Bigg)\\
&= \frac{1}{2 |p+k|}+ \frac{1}{2 |p-k|} -\frac{|k|}{2 p^2}+\frac{3}{2}\frac{(p\cdot k)^2}{p^4 |k|}+\frac{p\cdot k}{|p||k|}\left(\frac{1}{|p+k|}- \frac{1}{|p-k|}\right)\\
&= \frac{1}{2 |p+k|}+ \frac{1}{2 |p-k|} -\frac{|k|}{2 p^2}-\frac{1}{2}\frac{(p\cdot k)^2}{p^4 |k|}\\
&=\frac{1}{|p|} -\frac{|k|}{2 p^2}-\frac{1}{2}\frac{(p\cdot k)^2}{p^4 |k|},
\end{split}
\end{equation}
where we have used
\begin{equation}
\int \frac{d^{3}q}{(2\pi)^3} \frac{1}{q^2 (q-p_1)^2 (q+p_2)^2} = \frac{1}{8 |p_1||p_2||p_1+p_2|}.
\end{equation}

In the limit $|p|>>|k|$ we have
\begin{equation}
\begin{split}
&\int \frac{d^{3}q}{(2\pi)^3}  \frac{1}{ q^2 (q+k)^2 (q+p)^2 (q+p+k)^2} = \frac{1}{4 p^4 |k|},\\
&\int \frac{d^{3}q}{(2\pi)^3} \left( \frac{1}{ q^4 (q+k)^2 (q+p)^2 } + \frac{1}{ q^4 (q+k)^2 (q-p)^2 }\right)=\frac{1}{2}\frac{(p\cdot k)^2}{|k|^3 |p|^6}.
\end{split}
\end{equation}
This gives
\begin{equation} \label{fermfn}
{\tilde g}_{(4,2)} =N_F\left( \frac{4 \pi}{ {\tilde \kappa_F} }
\right)^4\frac{1}{2},~~~{\tilde g}_{(4,3)} =N_F\left( \frac{4 \pi}{ {\tilde \kappa_F} }
\right)^4\frac{1}{2},
\end{equation}
which gives
\begin{equation}
\delta_F=-\frac{16}{3\pi^2 \lambda_F}.
\end{equation}
For $SO(N_F)$ we obtain
\begin{equation}
\delta_F=-\frac{32}{3\pi^2 \lambda_F}
\end{equation}
This matches with equation 3.12 of \cite{Giombi:2017rhm} up to a sign, using
$4 \frac{\Gamma(d)\sin(\frac{\pi d}{2})}{\pi d \Gamma(\frac{d}{2})^2 } \Bigg{|}_{d=3} =-\frac{32}{3\pi^2}.$

\section{Conventions for Chern-Simons levels}\label{levels}

In this Appendix we explain our notations for Chern-Simons levels. 

\subsection{${\cal N}=2$ Chern-Simons levels}

We use the symbol $\kappas$ to denote the level of an ${\cal N}=2$ supersymmetric Chern-Simons theory. 
The level is defined as the coefficient of the Chern-Simons term {\it before} 
the gluino is integrated out (we use a scheme in which divergences are regulated by 
adding an infinitesimal ${\cal N}=2$ Yang-Mills term to the action). 

We use the notation $SU(N)_{\kappas}$ to denote an $SU(N)$ ${\cal N}=2$ supersymmetric theory 
with supersymmetric level $\kappas$. The notation $U(N)_{\kappas_1, \kappas_2}$ is used 
to denote an ${\cal N}=2$ supersymmetric theory with supersymmetric $SU(N)$ and $U(1)$ 
levels given by $\kappa_1^{{\cal N}=2}$ and $\kappa_2^{{\cal N}=2}$, respectively. 

\subsection{(Non-supersymmetric) Chern-Simons levels}

We use the notation $k$ to denote the level of a matter Chern-Simons theory once all auxiliary 
fields -- if any -- have been integrated out. We work in a regulation scheme in which divergences are 
regulated by adding an infinitesimal Yang-Mills term to the action. We denote by $SU(N)_k$ an $SU(N)$ theory at 
level $k$. The notation $U(N)_{k_1, k_2}$ is used 
to denote a matter theory with $SU(N)$ and $U(1)$ 
levels given by $k_1$ and $k_2$, respectively.

We pause to note the following subtlety. A Chern-Simons-matter $SU(N)$ theory at level $\kappa$ 
in the dimensional regulation scheme is the same as an identical theory with a Yang-Mills regulator
at level $k$, with 
\begin{equation}\label{kkap}
\kappa={\sgn}(k) \left( |k|+N \right).
\end{equation}
In a similar manner the $U(N)_{k_1, k_2}$ theory is equivalent to a $U(N)$ theory with levels 
$\kappa_1$ and $\kappa_2$ in the dimensional regulation scheme with 
\begin{equation}\label{kkaptwo}
\kappa_1={\sgn}(k_1) \left( |k_1|+N \right), ~~~\kappa_2=k_2.
\end{equation}

In the special case of an ${\cal N}=2$ theory 
the gluino is a non propagating field and can be integrated out. 
This process shifts the level of the Chern-Simons term according to the following rule.
The $SU(N)_{\kappas}$ pure Chern-Simons theory is in fact an
$SU(N)_{k}$ Chern-Simons theory with
\begin{equation}\label{ss1}
k={\sgn} (\kappas) \left( |\kappas|-  N \right), ~~~\kappa=\kappas.
\end{equation}
Similarly the $U(N)_{\kappas_1, \kappas_2}$ theory may be rewritten as a $U(N)_{k_1, k_2}$ theory with 
\begin{equation}\label{suns}
k_1= {\sgn}( \kappas_1) \left( |\kappas_1|-  N \right) ,~~~ k_2= \kappas_2, ~~~\kappa_1=\kappas_1, ~~
\kappa_2=\kappas_2.
\end{equation}

\subsection{Integrating out massive fundamental fermions}

Integrating out a massive fundamental fermion of mass $m$ increases both the $SU(N)$ and 
the $U(1)$ levels of a Chern-Simons-matter theory by $\frac{1}{2}{\sgn} (m) $. 

Consider an $SU(N)_\kappas$ theory with $N_f$ fermions. If we integrate out the gluino, and give masses to the matter bosons and fermions and integrate them out as well, at long distances 
we are left with a pure Chern-Simons theory with level  
\begin{equation}\label{pcsl}
k= {\sgn} (\kappas) \left( |\kappas|-  N \right) +{\sgn} (m) \frac{N_f}{2},
\end{equation}
where we assume that the masses of all fermions have the same sign ${\sgn}(m)$.
The level $k$ in \eqref{pcsl} is the coefficient of the Chern-Simons term in 
the pure Chern-Simons action with Yang-Mills regulator. $k$ may also be identified 
with the level of the WZW theory `dual' to the pure Chern-Simons theory.

In a similar manner the level of the pure Chern-Simons theory obtained by integrating out 
the gluino and the matter fields from the $U(N)_{\kappas_1, \kappas_2}$ theory is the pure 
$U(N)_{k_1, k_2}$ theory with 
\begin{equation}\label{purecs}
k_1={\sgn}( \kappas_1) \left( |\kappas_1|-  N \right) + {\sgn} (m) \frac{N_f}{2}, ~~~
k_2= \kappas_2 + {\sgn} (m) \frac{N_f}{2},
\end{equation}
where we have once again assumed that the masses of all fermions have the same sign ${\sgn}(m)$.
Once again $k_1$ and $k_2$ in \eqref{purecs} represent the levels of the dual WZW theory.

\section{The mapping of ranks and levels under duality}
\label{mapping}

In section \ref{dotsl} we `derived' the duality between regular bosons and critical fermions from a supersymmetric duality, making 
certain assumptions about the structure of RG flows around these two theories. In this section 
we will assume these assumptions are correct and use the construction of our paper to 
derive the precise form of the map of levels and ranks between conjecturally dual theories. 
Our analysis in this section is essentially identical to that of \cite{Aharony:2015mjs} 
and yields the same results. Nonetheless we present it here for completeness.

The starting points of our analysis are the well known dualities between the ${\cal N}=2$ 
supersymmetric theories that are the parents of the RG flows studied in this paper. 
The supersymmetric dualities that are of interest to us in this paper relate $SU(N)$ or 
$U(N)$ supersymmetric theories with different levels and ranks. On both sides of 
the duality the gauge fields are coupled to $N_f$ fundamental chiral multiplets.

The first supersymmetric duality identifies the theories with gauge groups and levels 
\cite{Kapustin:2013hpk,Aharony:2014uya}
\begin{equation}\label{susyd1}
\begin{split}
&SU(N)_{k^{{\cal N}=2}}  \equiv  U(|k^{{\cal N}=2}|-N+\frac{N_f}{2})_{    {-k^{{\cal N}=2}}, k_2^{{\cal N}=2}} ~  ({\rm for}   ~N_f \le 2 |k^{{\cal N}=2}| ),  \\
&k_2^{{\cal N}=2}=-\sgn( k^{{\cal N}=2}) \left(N-\frac{N_f}{2}\right).
\end{split}\end{equation}
The second such duality identifies theories with gauge groups and levels as \cite{Giveon:2008zn,Benini:2011mf}
\begin{equation}\label{susyd2}
\begin{split}
&U(N)_{{-k^{{\cal N}=2}},-k^{{\cal N}=2}} \equiv U(|k^{{\cal N}=2}| -N+\frac{N_f}{2})_{k^{{\cal N}=2},k^{{\cal N}=2}} ~  ({\rm for}   
~N_f \le 2 |k^{{\cal N}=2}| ) . \\
\end{split}\end{equation}
In \eqref{susyd1} and \eqref{susyd2}, and in Table \ref{susyd}, we have quoted all Chern-Simons levels before integrating out 
the respective gauginos, and with Yang-Mills regularization, according to the standard convention for ${\cal N}=2$ theories
(see Appendix \ref{levels} for notation). Note that these are the same levels that one would get in the dimensional regulation convention after integrating out the gauginos, so they will be connected in a simple way to the non-supersymmetric levels $\kappa$ after the RG flow in this convention.

\begin{table}
 \begin{tabular}{ l | c  }
    \hline
    Chern-Simons gauge group& Chern-Simons gauge group  \\ \hline
      $SU(N)_{k^{{\cal N}=2}} $& $U(|k^{{\cal N}=2}|-N+\frac{N_f}{2})_{-k^{{\cal N}=2},-{\sgn(k^{{\cal N}=2})}\left(N-\frac{N_f}{2}\right) }$  \\ \hline
   $U(N)_{k^{{\cal N}=2},k^{{\cal N}=2}}$& $U(|k^{{\cal N}=2}|-N+\frac{N_f}{2})_{-k^{{\cal N}=2},-k^{{\cal N}=2}}$  \\ \hline
  \end{tabular}
  \caption{We list the supersymmetric dualities of \eqref{susyd1},\eqref{susyd2}. In the table, we only list the gauge group on the two sides of the duality. On both sides we have a gauge field coupled to $N_f$ fundamental chiral multiplets. The dualities for opposite signs of the CS levels may be obtained by a parity transformation. These dualities hold for $N_f \leq 2|k^{{\cal N}=2}|$ and for positive ranks on both sides.}
\label{susyd}    
\end{table}

Starting with the supersymmetric dualities listed in Table \ref{susyd}, we can now follow through the 
RG flows constructed in this paper. As emphasized in \cite{Jain:2013gza}, our construction of the 
RG flows (and assertion of their dual equivalence) is reliable only when the condition
\begin{equation}\label{sgnkmf}
\sgn{(\kappa ~m_F)}>0
\end{equation}
 is satisfied.
Here $\kappa$ is the level of the gauge group; note that the $SU(N)$ and $U(1)$ levels in 
Table \ref{susyd} always have the same sign, so the condition \eqref{sgnkmf} applies to either 
of these levels. When this condition is not obeyed, the dual scalars appear to condense in the 
vacuum \cite{Jain:2013gza}. As the partition function for this condensed phase has not been determined, 
all statements about duality based on the analysis of partition functions become unreliable. 
For that reason we will always assume that the condition \eqref{sgnkmf} is obeyed. In particular
when we give a mass to and integrate out fermions, we will always do so with the understanding that 
the fermion masses are chosen to obey \eqref{sgnkmf}. Recall (see Appendix \ref{levels}) that integrating out a massive fundamental fermion of mass $m$ shifts the Chern-Simons level by 
$\frac{{\sgn }(m)}{2}$. Keeping in mind the condition \eqref{sgnkmf}, this shift may equally
well be written as $\frac{{\sgn }(k)}{2}$ where $k$ is the level of the gauge group before 
integrating out the fermions. 

Using these rules, the RG flows constructed in this paper allow us to derive the dualities 
listed in Table \ref{dr}.
\begin{table}
 \begin{tabular}{ l | c  }
    \hline
    Chern-Simons+  $N_f$ fundamental boson&  Chern-Simons +  $N_f$ fundamental fermion \\ \hline
      $ U(|\kappa|-N+\frac{N_f}{2})_{-{\sgn (\kappa)}\left(|\kappa|+\frac{N_f}{2}\right),-{\sgn (\kappa)} N} $& $SU(N)_{\sgn(\kappa)|\kappa|}$  \\ \hline
   $SU(N)_{\sgn (\kappa)\left(|\kappa|+\frac{N_f}{2}\right)}$& $U(|\kappa|-N+\frac{N_f}{2})_{-\sgn (\kappa)|\kappa|,-\sgn (\kappa)\left(N-\frac{N_f}{2}\right)}$  \\ \hline
  $U(|\kappa|-N+\frac{N_f}{2})_{-{\sgn (\kappa)}\left(|\kappa|+\frac{N_f}{2}\right),-{\sgn (\kappa)}\left(|\kappa|+\frac{N_f}{2}\right) }$& $U(N)_{{\sgn (\kappa)}|\kappa|,{\sgn (\kappa)}|\kappa|}$  \\ \hline
  \end{tabular}
  \caption{We list non-supersymmetric dualities obtained by flowing from \eqref{susyd1}, \eqref{susyd2}. 
}
\label{dr}    
\end{table}

We obtain the first line of Table \ref{dr} by starting with the dual pair of theories listed in the first line of Table \ref{susyd} and following an RG flow that integrates out the fermion on the right-hand side 
of this duality. In a similar manner we get the second equation of Table \ref{dr} starting with 
the first line of Table \ref{susyd} but this time following a flow that integrates out the 
fermions on the left-hand side of this duality. Finally we get the last line of Table \ref{dr} starting with 
the second line of Table \ref{susyd} and following an RG flow that integrates out the fermions 
from the right-hand side of this duality. Integrating out the fermions from the left-hand side of the second line of 
Table \ref{susyd} yields the same duality.

Let us emphasize that the analysis of this paper has argued for the dualities listed in 
Table \ref{dr} only in the special case $N_f=1$. We have listed the formulae 
at general $N_f$ with a view to possible generalizations in the future. 

As a final consistency check of the dualities proposed in Table \ref{dr} let us 
also integrate out the remaining matter in these theories. In order to avoid 
having to deal with bosons with negative squared masses that 
condense, we once again restrict attention to fermions whose mass obeys 
the condition \eqref{sgnkmf}. Imposing this condition and integrating out all matter 
fields, the dualities listed in Table \ref{dr} reduce to the
dualities between pure Chern-Simons theories listed in Table \ref{pcsd}.

\begin{table}
 \begin{tabular}{ l | c  }
    \hline
    Pure Chern-Simons gauge group&  Pure Chern-Simons gauge group \\ \hline
      $ U(|\kappa|-N+\frac{N_f}{2})_{-{\sgn (\kappa)}\left(|\kappa|+\frac{N_f}{2}\right),-{\sgn (\kappa)} N} $& $SU(N)_{\sgn(\kappa)\left(|\kappa| +\frac{N_f}{2}\right)}$  \\ \hline
   $U(N)_{-{\sgn(\kappa)}\left(|\kappa|+\frac{N_f}{2}\right),-{\sgn(\kappa)} \left(|\kappa|+\frac{N_f}{2}\right)}$& $U(|\kappa|-N+\frac{N_f}{2})_{\sgn(\kappa)\left(|\kappa|+\frac{N_f}{2}\right),\sgn(\kappa)\left(|\kappa|+\frac{N_f}{2}\right)}$  \\ \hline
  \end{tabular}
  \caption{By integrating out the remaining matter content from Table \ref{dr}, we obtain  dual pairs of Pure Chern-Simons gauge theories.}
\label{pcsd}    
\end{table}

It is easy to see that the dualities of Table \ref{pcsd} are indeed true by 
virtue of the usual well known level rank dualities between pure Chern-Simons 
theories \cite{Naculich:1990pa,Camperi:1990dk,Nakanishi:1990hj,Naculich:2007nc} (in the dimensional regulation convention)
\begin{equation}\label{purecsdual1}
\begin{split}
&SU(N)_{{\sgn(\kappa)}|\kappa|} \equiv  U(|\kappa|-N)_{-{\sgn(\kappa)}|\kappa|,-{\sgn(\kappa)}N}\\
&U(N)_{-{\sgn(\kappa)}|\kappa|,-\sgn(\kappa)|\kappa|}\equiv  U(|\kappa|-N)_{{\sgn(\kappa)} |\kappa|, {\sgn(\kappa)}|\kappa|}.
\end{split}\end{equation}

\section{Detailed free energy computations for the zero temperature phase structure}
\label{phasecomp}

In this appendix we compute the free energies of the different phases described in section \ref{phasestruct}, to determine the zero temperature phase structure of the regular boson/critical fermion theory. We analyze each of the three parameter ranges of $x_6$ listed in \eqref{axc} in turn. 

\subsection{ $x_6> \phi_2$}

\begin{figure}
  \begin{subfigure}[b]{0.55\textwidth}
    \includegraphics[width=\textwidth]{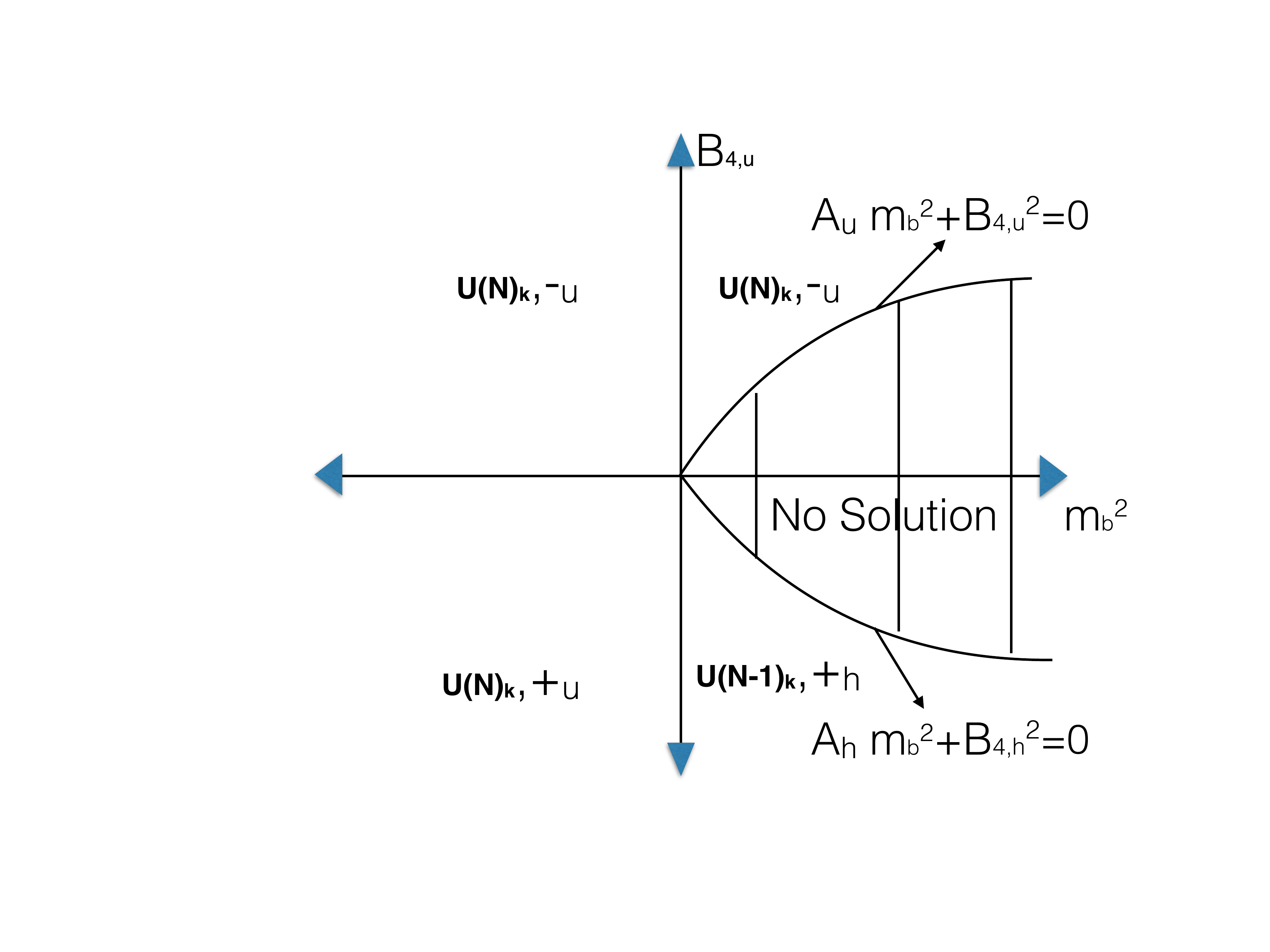}
    \caption{Subdominant solutions}
    \label{figA1}
  \end{subfigure}
  \begin{subfigure}[b]{0.55\textwidth}
    \includegraphics[width=\textwidth]{Case-IAa.pdf}
    \caption{Dominant solutions}
    \label{figB1}
  \end{subfigure}
  \caption{Graphs of the domain of existence of the dominant and subdominant solutions to the gap equation for $x_6>\phi_2$. The dominant and 
  subdominant solutions merge along the two parabolas in the graphs above. The solution space may thus be 
  thought of as a double cover of the compliment of the unshaded portion of the  $m_b^2$-$B_{4,u}$ 
  plane. }\label{figA1B1ph1}
\end{figure}

The set of solutions to the gap equation in this case was described in section \ref{phasestruct} and is schematically plotted in Figure \ref{figA1B1ph1}. In this subsection we will demonstrate that the solutions depicted in Figure \ref{figB1} are free-energetically dominant over those in Figure \ref{figA1}, and so represent  the true phase diagram of the theory. 

In order to analyze the free energies it is useful define the dimensionless parameter $\nu$ by
\begin{equation} \label{defnu}
 \nu \equiv \frac{m_b^2}{B_{4,u}^2}.
\end{equation}
This definition is useful as it follows from dimensional analysis that the free energy in each phase -- and hence the difference of free energies between any two phases -- is given by $|B_{4,u}|^3$ times a function of the dimensionless parameter $\nu$. 
In other words, the phase structure of our theory depends on $B_{4,u}$ and $m_b^2$ only through the dimensionless 
combination $\nu$. Given any two phases $X$ and $Y$ with free energies $F_X$ and $F_Y$ let us define 
\begin{equation}\label{fediff}
{\cal F}_{XY}\equiv \frac{F_{X}- F_{Y}}{|B_{4,u}|^3};
\end{equation}
this depends only on $\nu$
(of course it also depends on $\lambda_B$ and $x_6$, and on the sign of $B_{4,u}$, but this dependence is suppressed in our notation).



For $B_{4,u} > 0$ and $m_b^2 > 0$ a short computation reveals that 
\begin{equation} \label{ffqu}
{\cal F}_{-_u +_u}(\nu)=-4\frac{\lambda_B}{\lambda_F} \frac{\left(4+\nu (4-(4+3 x_6)\lambda_B^2)\right)^{\frac{3}{2}} }{\left(4-(4+3 x_6)\lambda_B^2\right)^2}.
\end{equation}
As $\frac{\lambda_B}{\lambda_F}<0$ it follows immediately that this quantity is positive, and so the phase $+_u$ is
dominant. 

For $B_{4,u} < 0$ and $m_b^2 > 0$ a similar computation yields 
\begin{equation} \label{fffee}
{\cal F}_{+_h -_h}(\nu)=-4\frac{\lambda_B}{\lambda_F} \frac{\left(4+\nu (4-(4+3 x_6)\lambda_B^2)\right)^{\frac{3}{2}} }{\left(4-(4+3 x_6)\lambda_B^2\right)^2}.
\end{equation}
Once again using the fact that $\frac{\lambda_B}{\lambda_F}<0$ it follows that ${\cal F}_{+_h -_h}(\nu)$ is positive, so that 
the phase $-_h$ is dominant. 

\begin{figure}
  \begin{subfigure}[b]{0.4\textwidth}
    \includegraphics[width=\textwidth]{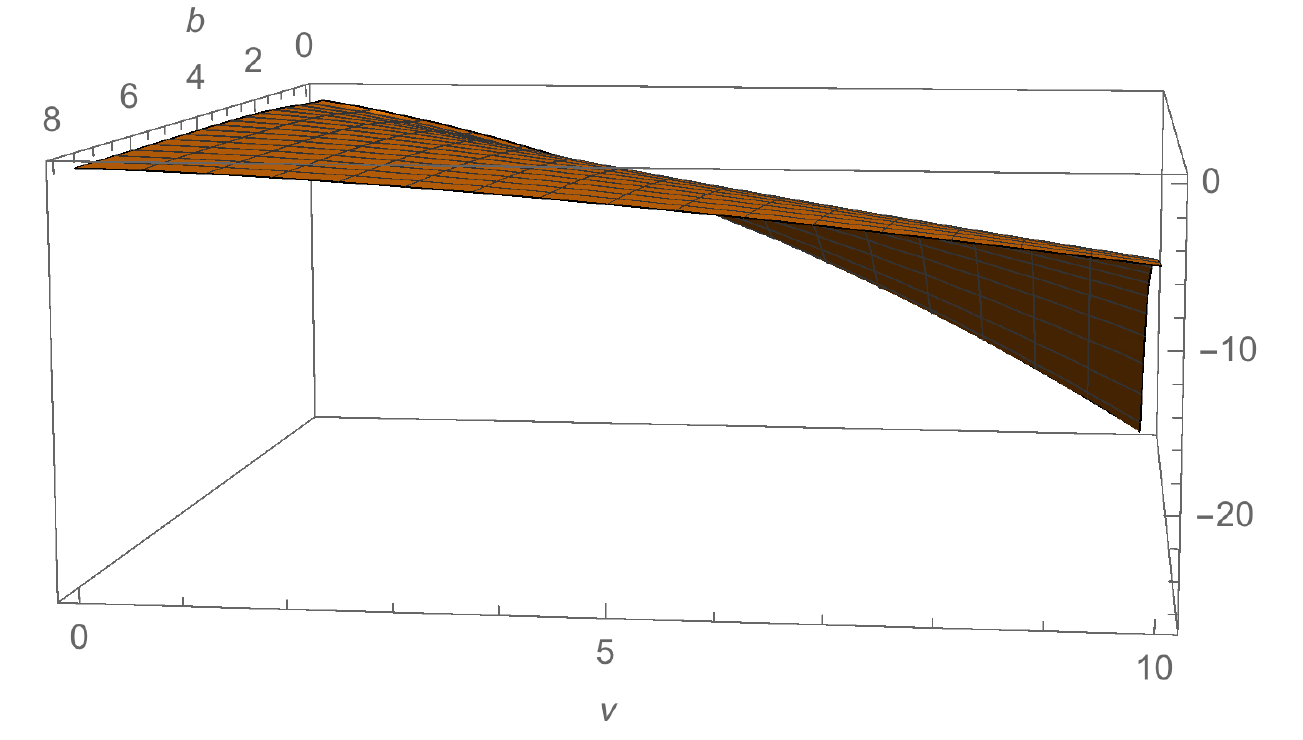}
    \caption{{\color {blue} A plot of  $-{\cal F}_{-_u +_h}$ at $\lambda_B=\frac{1}{2}$, when $B_{4,u} > 0$ and $m_b^2 < 0$. The axes of our plot 
    are $\nu$ and $b$, where $b^2=x_6-\phi_2$.  We see that everywhere $-{\cal F}_{-_u +_h}< 0$, demonstrating that $+_h$ is the 
    dominant phase.}}
    \label{fig1nup}
  \end{subfigure}\hspace{2cm}
  \begin{subfigure}[b]{0.4\textwidth}
    \includegraphics[width=\textwidth]{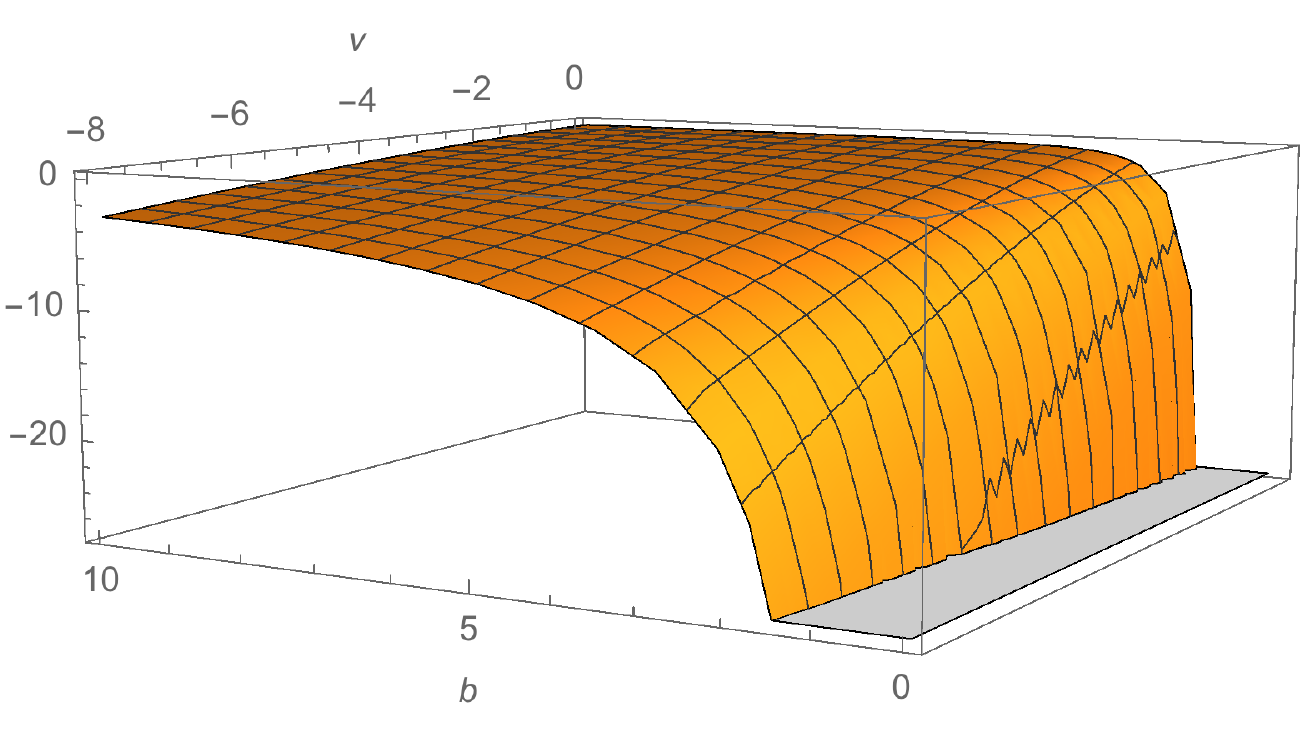}
    \caption{{\color{red} A plot of $-{\cal F}_{+_u -_h}$ at $\lambda_B=\frac{1}{2}$, when $B_{4,u} < 0$ and $m_b^2 < 0$. Once again the axes are 
    $\nu$ and $b$, where $b$ was defined in the caption to the left.  We see that $-{\cal F}_{+_u -_h}<0$,
    so that $-_h$ is the dominant phase.}}
    \label{fig1num}
  \end{subfigure}
\caption{}
\label{figA1B1}
\end{figure}

For $m_b^2 < 0$, the expressions for ${\cal F}_{-_u +_h}$ (when $B_{4,u} > 0$) and for ${\cal F}_{+_u -_h}$ (when $B_{4,u} < 0$) are messier.
Rather than analytically 
massaging them into a manifestly positive definite form, we found it easier to simply plot them 
using Mathematica. Our plots are presented in Figure \ref{figA1B1}. The plots clearly reveal that both of these are positive for all $\nu$ and at all allowed values of $x_6$, at $\lambda_B=\frac{1}{2}$, consistent with our statements above (we have generated similar plots at various values of $\lambda_B$ and always find similar results).

\subsection{ $\phi_1<x_6< \phi_2$}

\begin{figure}
    \includegraphics[width=\textwidth]{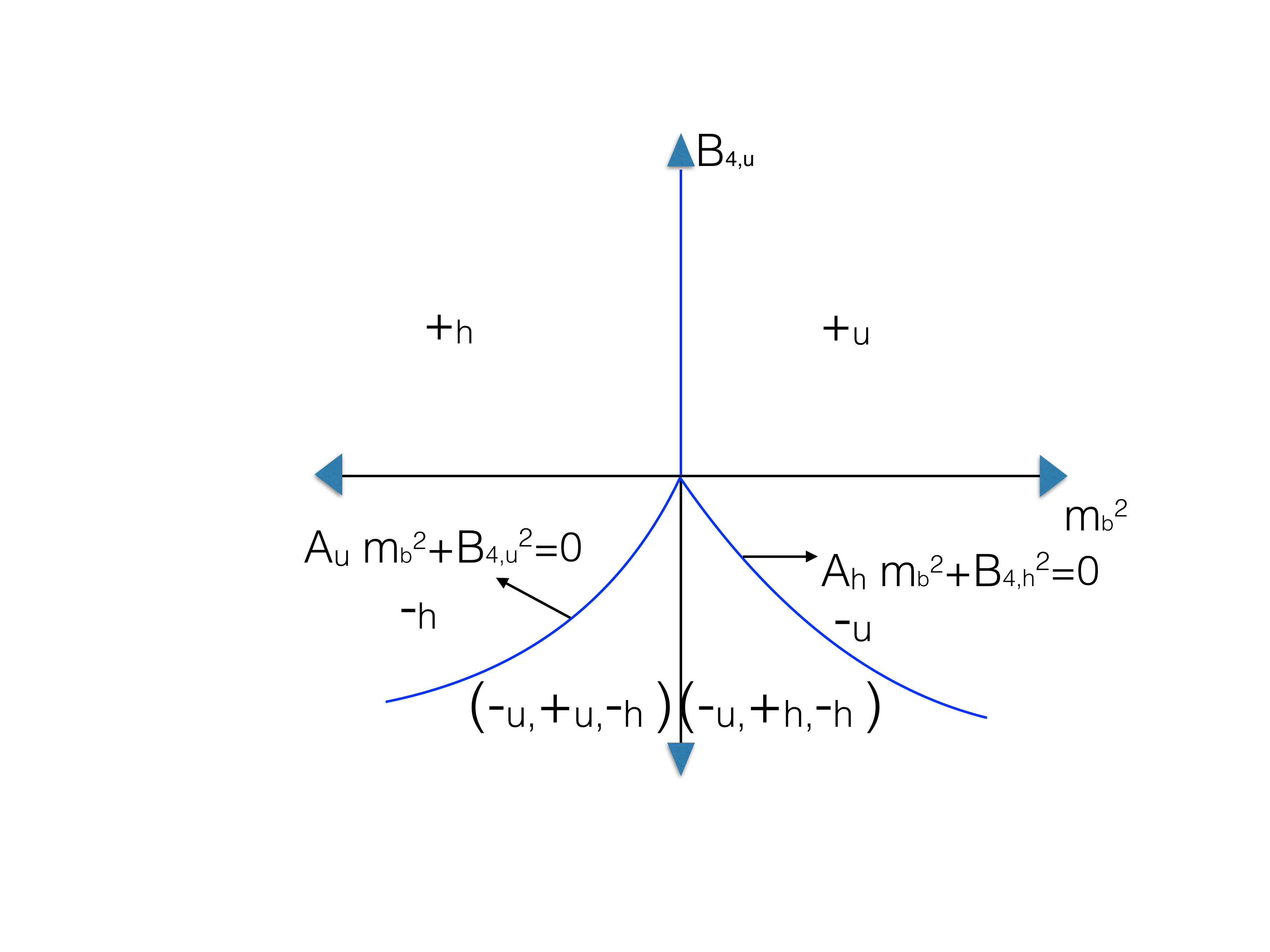}
  \caption{A graph of the domain of existence of solutions to the gap equation for $\phi_2>x_6>\phi_1$. 
  We have a single solution outside the two parabolas, but three different solutions between these two parabolas. 
  The $+_u$ and $-_u$ solutions reduce to each other on the left parabola while the $+_h$ and $-_h$ solutions 
  reduce to each other on the right parabola. Moreover the $+_u$ and $+_h$ solutions continue into each 
  other through a second order phase transition along the negative $B_{4,u}$ axis. It follows that the space of solutions 
  represents a single cover of parameter space outside the two parabolas, but a triple cover of parameter space 
  inside the two parabolas. Note that the $(+_u,-_u)$  and $(+_h,-_h)$ phases correspond to $U(N)_\kappa$  and $U(N-1)_\kappa$ theories, respectively.}\label{figallphs}
\end{figure}

Next we study the zero temperature phase diagram of our theory in the intermediate range $\phi_1<x_6<\phi_2$, as a function of
$B_{4,u}$ and $m_b^2$.  As noted in \eqref{axc}, in this range of parameters $A_u$ is positive while $A_h$ is negative. For most parameters here there is just a single solution to the gap equation, but for $B_{4,u} < 0$ and in the region where $-B_{4,u}^2 / A_u < m_b^2 < -B_{4,h}^2 / A_h$ there are three solutions, so we need to check which one dominates. The solutions are the $-_h$ and $-_u$ solutions, and in addition the $+_u$ solution for $m_b^2 < 0$, and the $+_h$ solution for $m_b^2 > 0$.  All of this is summarized in Figure \ref{figallphs}. We will now demonstrate that the $+$ solutions never dominate in this region, and that there is a first order phase transition somewhere in the middle of this region, between the $-_u$ and $-_h$ solutions.

In the region where three solutions exist and $m_b^2 < 0$ we find that
\begin{equation} \label{loss}
{\cal F}_{-_u +_u}=4\frac{\lambda_B}{\lambda_F} \frac{\left(4+\nu (4-(4+3 x_6)\lambda_B^2)\right)^{\frac{3}{2}} }{\left(4-(4+3 x_6)\lambda_B^2\right)^2}.
\end{equation} 
As the right-hand side of \eqref{loss} is negative, it follows that $-_u$ always has a lower free energy than (and so is thermodynamically 
dominant compared to) $+_u$. 

Similarly, for $m_b^2 > 0$ we find that 
\begin{equation} \label{moss}
\begin{split}
&{\cal F}_{-_h +_h}\\
&=-\frac{2\lambda_B^2}{(1-\alpha)^2 (\lambda_B^2-{\hat \lambda}_B^2)^2}
\sqrt{\frac{{\hat\lambda}_B^2 +(1-\alpha)\nu (\lambda_B^2-{\hat\lambda}_B^2)}{\lambda_B^2}}
\frac{{\hat\lambda}_B^3+(1-\alpha)\nu{\hat \lambda}_B \left(\lambda_B^2-{\hat\lambda}_B^2\right)}{\lambda_F}.
\end{split}\end{equation}
The right-hand side of \eqref{moss} is everywhere negative, establishing that the solutions $-_h$ are always thermodynamically 
preferred compared to $+_h$. 

\begin{figure}[h]
\begin{center}
\includegraphics[width=8.5cm,height=5cm]{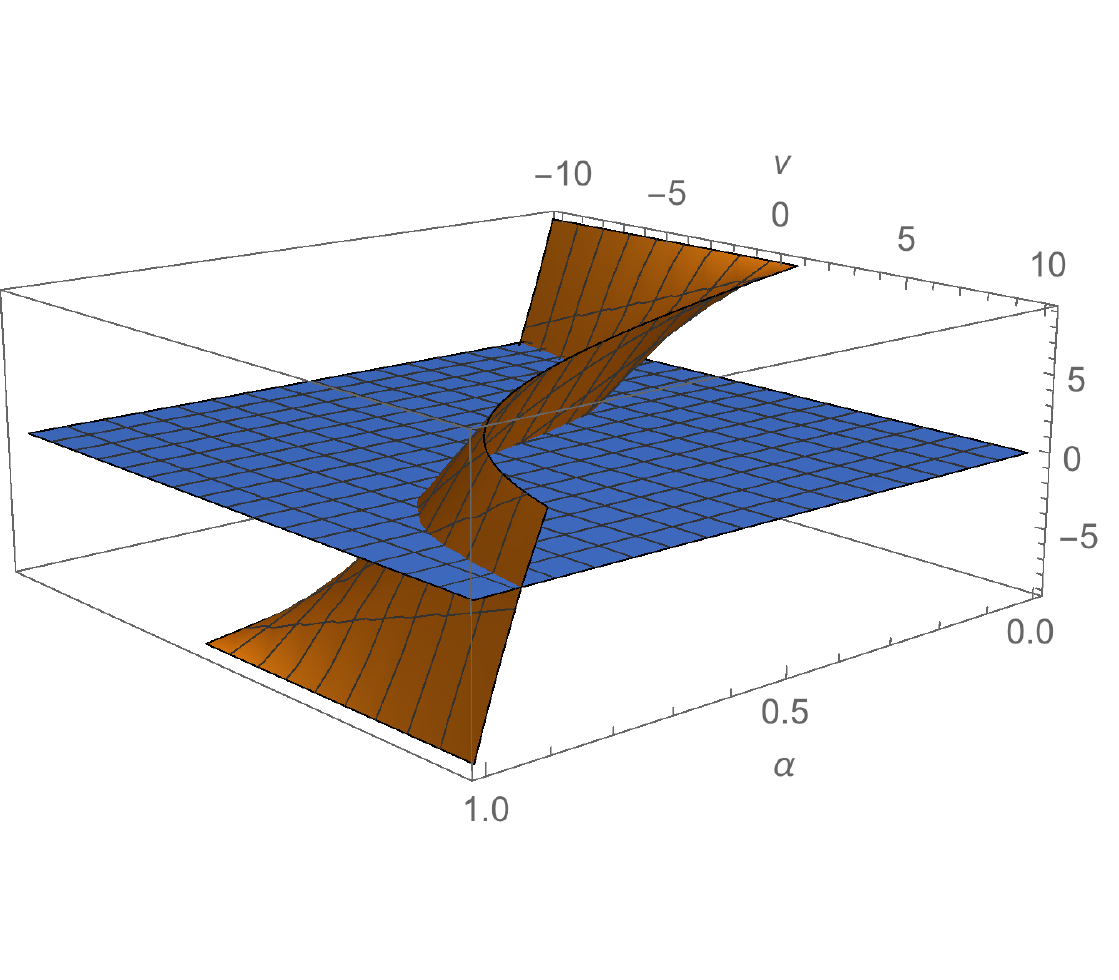}
  \caption{\label{figctb1} ${\cal F}_{-_h -_u}$ in the $\alpha$ and $\nu$ plane at $\lambda_B=\frac{1}{2}$. The plane with blue color is to indicate when ${\cal F}_{-_h -_u}$ is positive or negative.}
 \end{center}
\end{figure}

It follows that in the intersection region, the dominant phase is everywhere 
either $-_u$ (which clearly dominates also for $m_b^2 > -B_{4,h}^2 / A_h$ where it is the only solution) or $-_h$ (which clearly dominates also for $m_b^2 < -B_{4,u}^2 / A_u$). In order to see which phase dominates, we have plotted in Figure \ref{figctb1} ${\cal F}_{-_h -_u}$ as a function of 
$\nu$ and $\alpha$ at $\lambda_B=\frac{1}{2}$ (the analogous plots at different values of $\lambda_B$ are all 
similar), where $\alpha$ is defined by the relation 
\begin{equation}\label{defalph}
 x_6= \alpha \phi_1 + (1-\alpha) \phi_2,
\end{equation}
so that it goes between zero and one as we increase $x_6$.

\begin{figure}[h]
\begin{center}
\includegraphics[width=8.5cm,height=5cm]{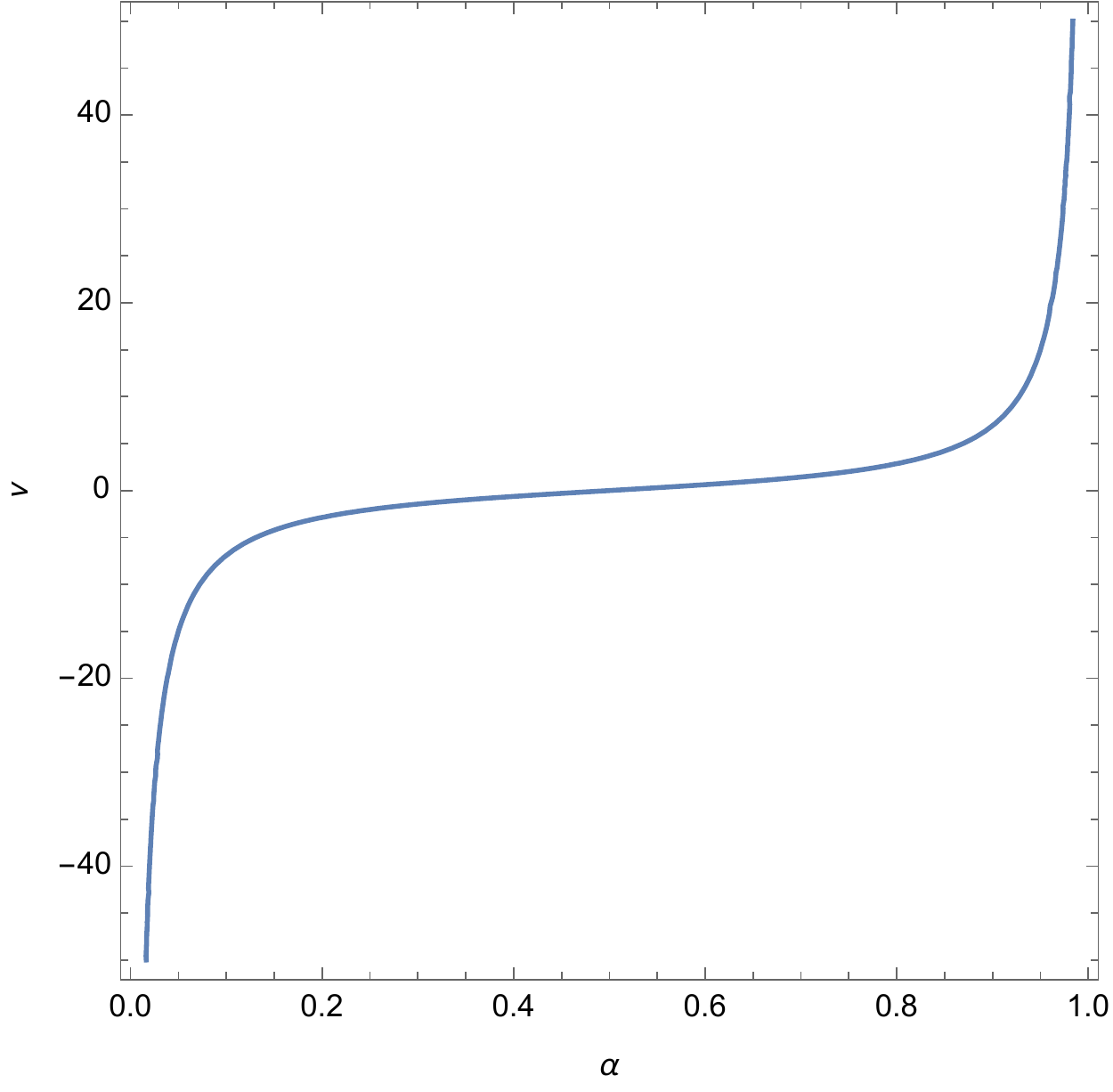}
  \caption{\label{figcta} $\nu_{c}$ as a function of $\alpha$ at $\lambda_B=\frac{1}{2}.$}
 \end{center}
\end{figure}

We see that for every value of $\alpha$ there is a phase transition as we change $\nu$ (or $m_b^2$). We see that the phase $-_u$ is thermodynamically dominant for $\nu \leq \nu_c$, while $-_h$ is dominant for 
$\nu \geq \nu_c$, where the quantity $\nu_c(\alpha)$ is plotted for $\lambda_B=\frac{1}{2}$  in Figure \ref{figcta}. 

Note that $\nu_{c} \to -\infty$ as $\alpha \to 0$ (i.e. as $x_6 \to \phi_2$). On the other hand $\nu_{c} \to +\infty$ as $\alpha \to 1$. The resulting phase diagram is drawn in Figure \ref{figctb} in Section \ref{phasestruct}.

\subsection{ $x_6< \phi_1$}

\begin{figure}
  \begin{subfigure}[b]{0.55\textwidth}
    \includegraphics[width=\textwidth]{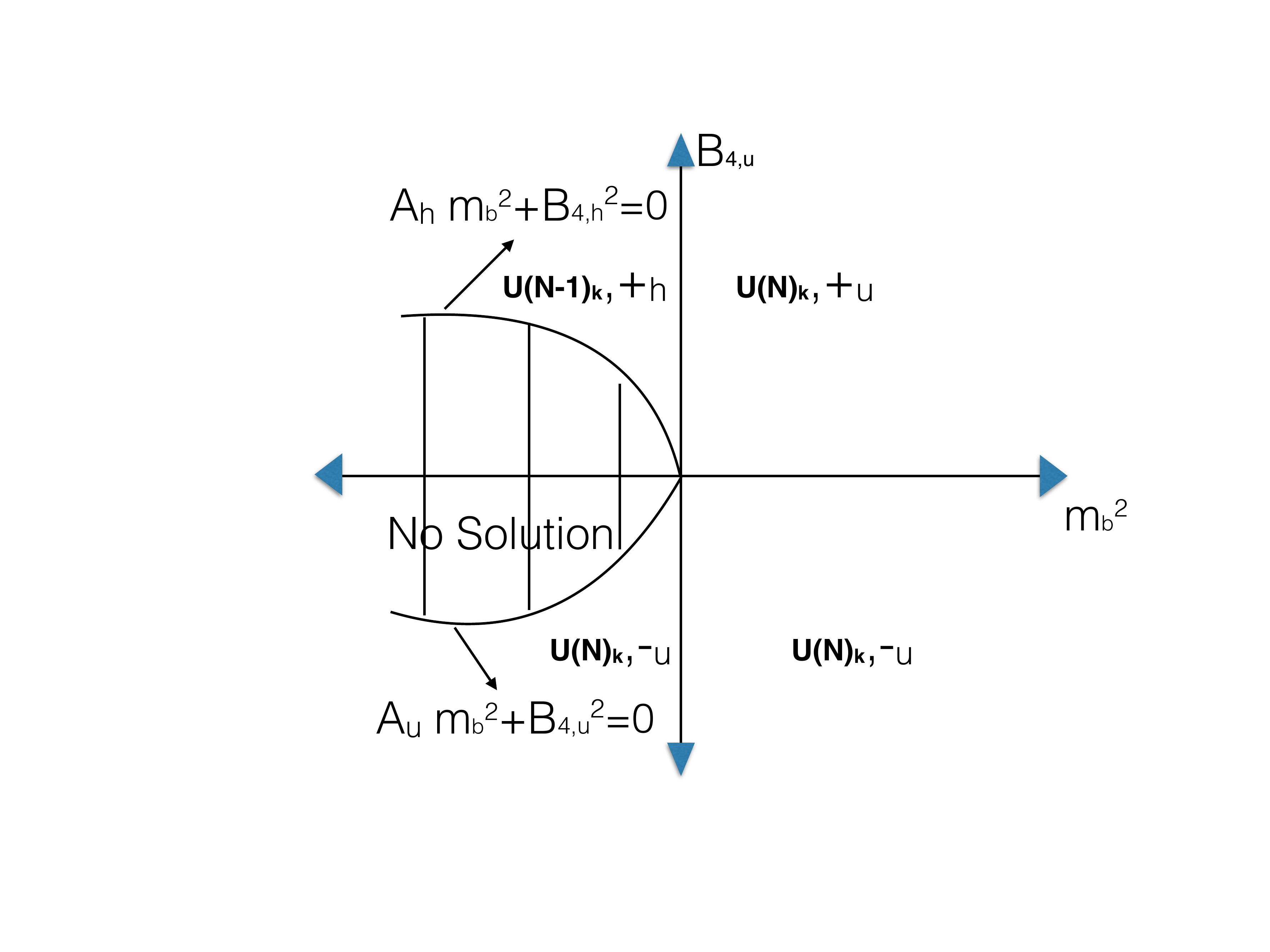}
    \caption{Dominant Solutions}
    \label{figc31}
  \end{subfigure}
  \begin{subfigure}[b]{0.55\textwidth}
    \includegraphics[width=\textwidth]{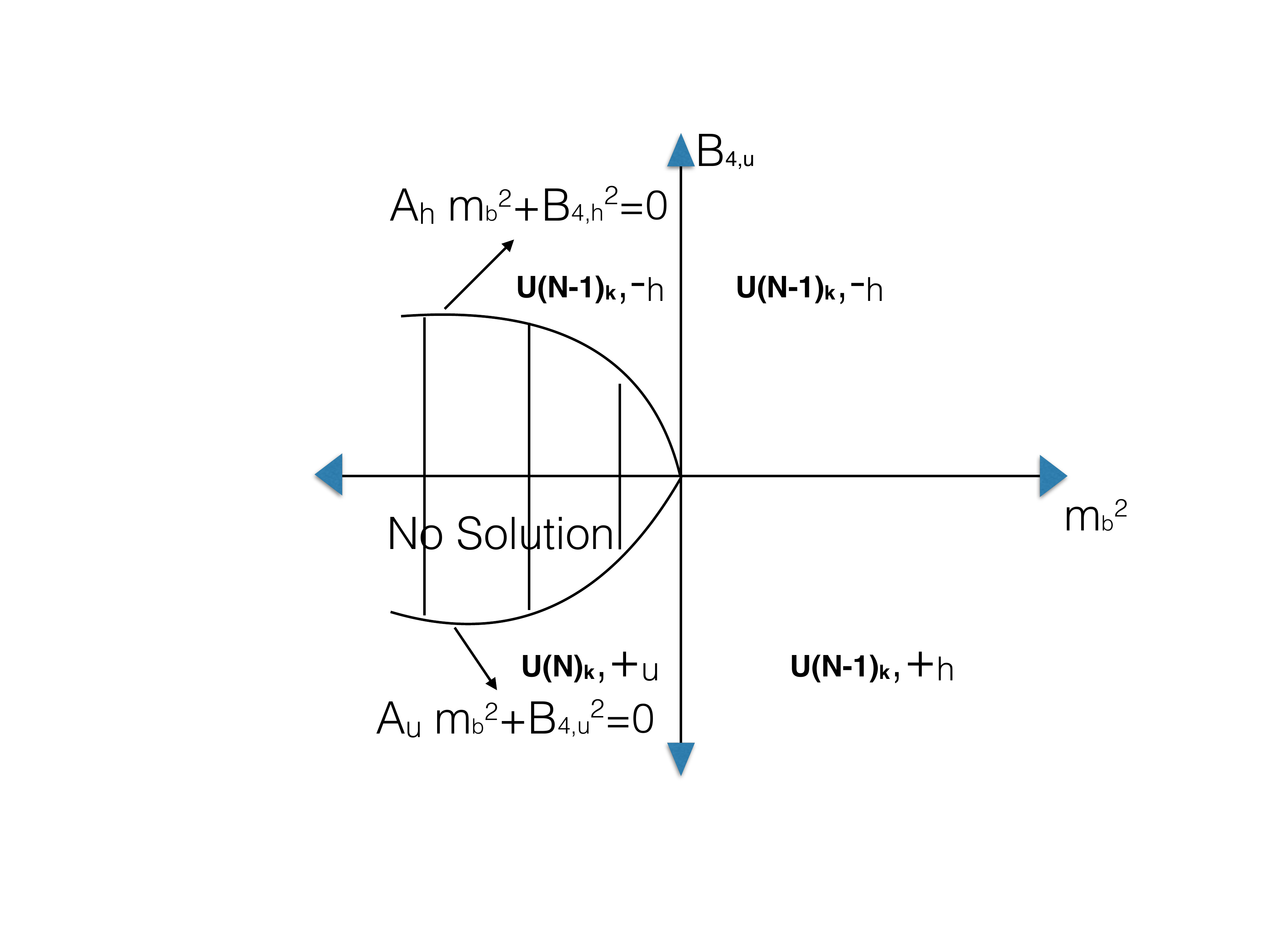}
    \caption{Subdominant Solutions}
    \label{figc32}
  \end{subfigure}
  \caption{ Graphs of the domain of existence of the dominant and subdominant solutions to the gap equation for $x_6<\phi_1$. The dominant and 
  subdominant solutions merge along the two parabolas in the graphs above. The solution space may thus be 
  thought of as a double cover of the compliment of the unshaded portion of the  $m_b^2$-$B_{4,u}$ 
  plane.}\label{figA3B312}
\end{figure}

The situation in this range of parameters is very similar to that of $x_6 > \phi_2$, with two solutions or none in all regions of the parameter space (see Figure \ref{figA3B312}). For $B_{4,u} > 0$ the $+_u$ or $+_h$ phase dominates, with a second order phase transition between them at $m_b^2=0$, while for $B_{4,u} < 0$ the $-_u$ phase dominates.

When $B_{4,u} < 0$ and $m_b^2 < 0$ we have
\begin{equation} \label{atbt}
{\cal F}_{-_u +_u}=4\frac{\lambda_B}{\lambda_F} \frac{\left(4+\nu (4-(4+3 x_6)\lambda_B^2)\right)^{\frac{3}{2}} }{\left(4-(4+3 x_6)\lambda_B^2\right)^2}.
\end{equation}
The right-hand side of \eqref{atbt} is always negative, establishing that $-_u$ is the dominant phase. 

In a similar manner, for $B_{4,u} > 0$ and $m_b^2 < 0$ we find
\begin{equation}
{\cal F}_{+_h -_h}=-4\frac{{\hat\lambda}_B}{\lambda_F} \frac{\left(4\frac{{\hat\lambda}_B^2}{\lambda_B^2}+\nu (-4+(4+3 x_6){\hat\lambda}_B^2)\right)^{\frac{3}{2}} }{\left(4-(4+3 x_6){\hat\lambda}_B^2\right)^2}
\end{equation}
which is again negative, establishing the thermodynamical dominance of $+_h$ over $-_h$.

\begin{figure}
  \begin{subfigure}[b]{0.4\textwidth}
    \includegraphics[width=\textwidth]{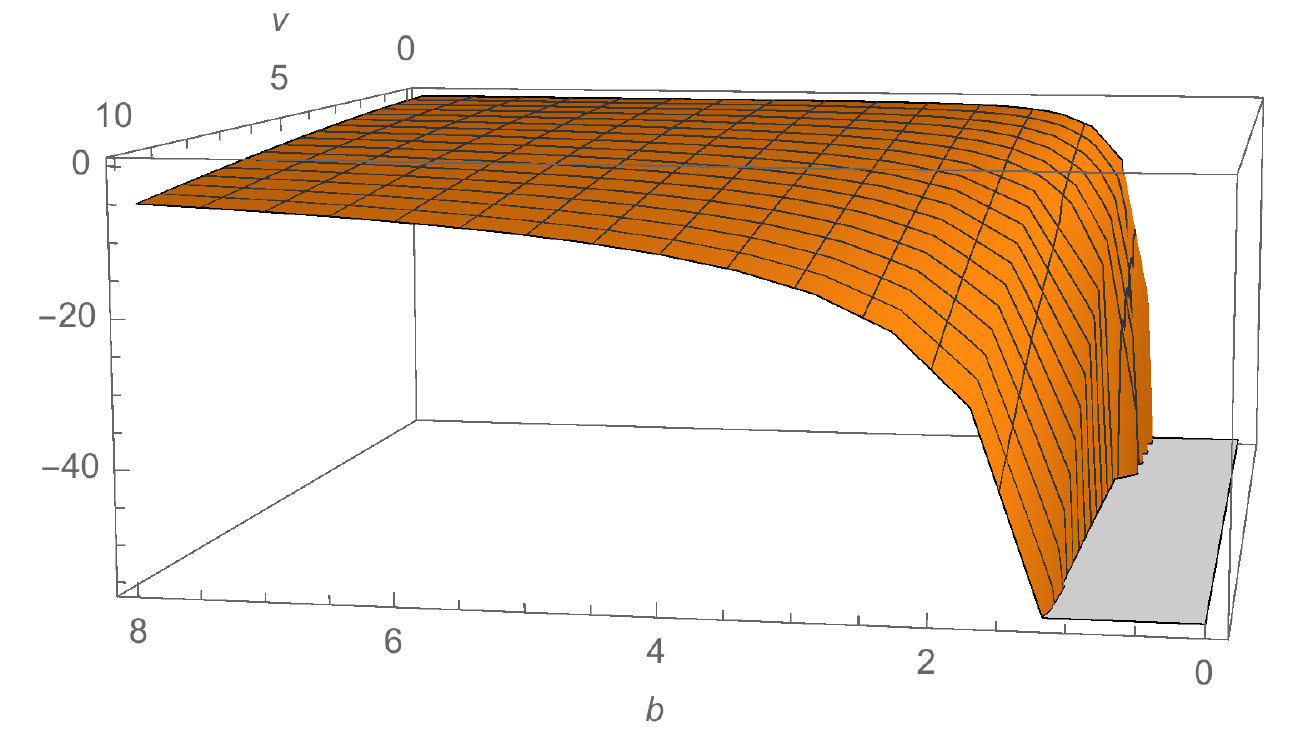}
    \caption{ {\color{blue} A plot of ${\cal F}_{+_u -_h}$ for $B_{4,u} > 0$, $m_b^2 > 0$. The axes are $\nu$ and $b$ where 
    $x_6= \phi_1 -b^2$. Note that ${\cal F}_{+_u -_h}$ is everywhere negative.}}
    \label{figp31}
  \end{subfigure}\hspace{2cm}
  \begin{subfigure}[b]{0.4\textwidth}
    \includegraphics[width=\textwidth]{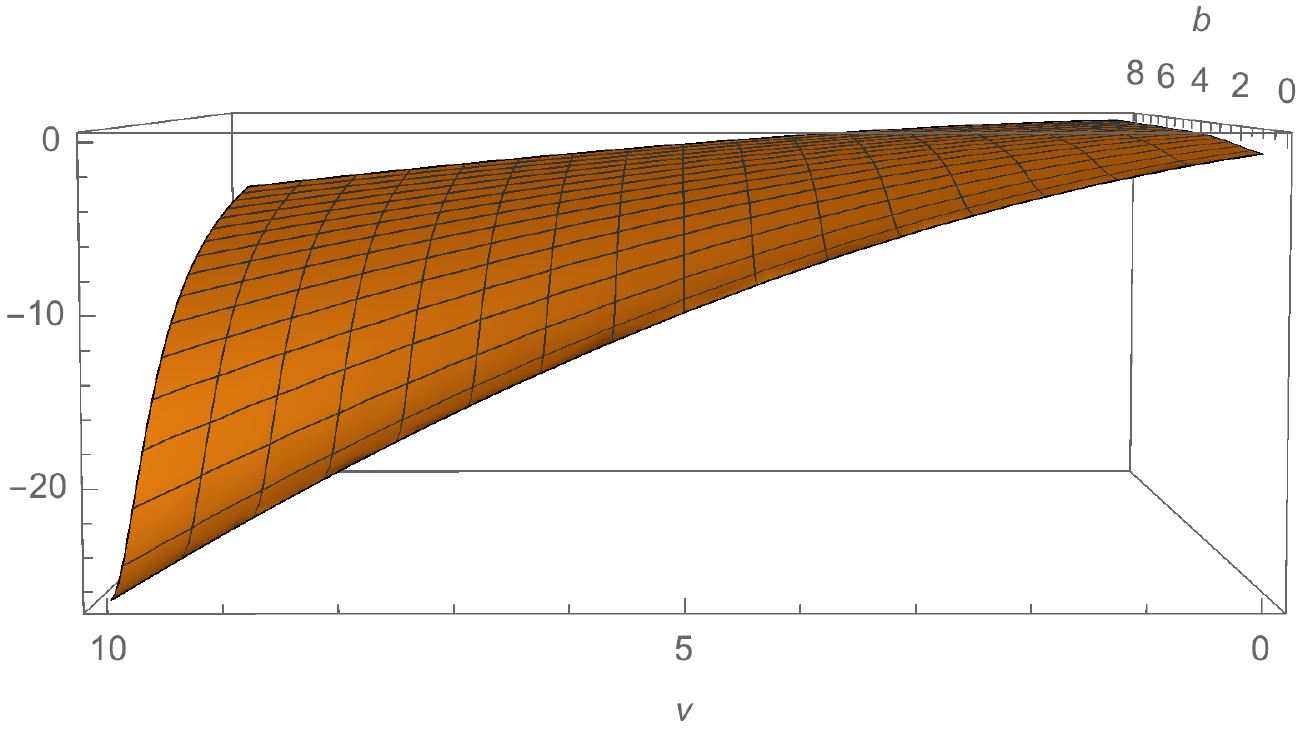}
    \caption{ {\color{red} A plot of ${\cal F}_{-_u +_h}$ for $B_{4,u} < 0$, $m_b^2 > 0$, with the same axes.  Note that ${\cal F}_{-_u +_h}$ is everywhere negative.}}
    \label{figp32}
  \end{subfigure}
\caption{}
\label{figAA}
\end{figure}

For $m_b^2 > 0$ we do not have analytic expressions. We present a numerical evaluation of the free energy differences in Figure \ref{figAA}, 
for the particular 
value $\lambda_B=\frac{1}{2}$ (these graphs are  qualitatively similar at all values of $\lambda_B$). It justifies the phase structure described above. 

\bigskip
\bibliography{cf}
\end{document}